\documentclass[aps,nofootinbib,floatfix]{revtex4}
\usepackage{bm}
\usepackage{graphicx,tabularx}
\usepackage{latexsym,color}
\usepackage{physics}
\usepackage{centernot}
\usepackage{relsize}
\usepackage{mathtools}
\usepackage{stmaryrd}
\usepackage{rotating}
\usepackage{amssymb}
\usepackage[T1]{fontenc}
\usepackage{natbib}
\usepackage{amsmath,amssymb,amsthm,mathrsfs,amsfonts,dsfont}
\usepackage{color}
\usepackage[hyperfootnotes=true]{hyperref}

\usepackage{enumitem}

\begin{document}
\title{Disclosing connections between black holes and naked singularities: Horizon remnants,  Killing throats and bottlenecks}
\author{Daniela  Pugliese and Hernando Quevedo}
\email{d.pugliese.physics@gmail.com}
\affiliation{\vspace{3mm}
Institute of Physics and Research Centre of Theoretical Physics and Astrophysics, Faculty of Philosophy \& Science,
  Silesian University in Opava,
 Bezru\v{c}ovo n\'{a}m\v{e}st\'{i} 13, CZ-74601 Opava, Czech Republic \\
          Dipartimento di Fisica, Universit\`a di Roma ``La Sapienza", I-00185 Roma, Italy \\
					Instituto de Ciencias Nucleares, Universidad Nacional Aut\'onoma de M\'exico,  AP 70543, M\'exico, DF 04510, Mexico \\
					Department of Theoretical and Nuclear Physics,
Kazakh National University,
Almaty 050040, Kazakhstan
}
\date{\today}
\begin{abstract}
We study the  properties of black holes and naked singularities by  considering stationary observers and light surfaces in Kerr spacetimes.
We  reconsider the notion of  Killing horizons from a special perspective by  exploring   the entire family of Kerr metrics. To this end, we introduce the concepts of extended plane,  Killing  throats  and  bottlenecks for weak (slowly spinning) naked  singularities.
Killing bottlenecks  (or horizon remnants in analogy  with the corresponding definition of throats in black holes) are restrictions of  the Killing throats appearing in special classes of slowly spinning naked singularities. Killing bottlenecks appear in association with the concept of  pre-horizon regime introduced in \cite{de-Felice1-frirdtforstati,de-Felice-first-Kerr}.
In the extended plane of the Kerr spacetime, we introduce particular sets, \emph{metric bundles}, of  metric tensors which allow us to reinterpret the concept of horizon and to find connections between black holes and naked singularities throughout the horizons.
 To evaluate the effects of frame-dragging on the formation and structure of Killing bottlenecks and horizons in the extended plane,
we consider  also the  Kerr-Newman and the  Reissner--Norstr\"om spacetimes.
We argue that these results might be significant for the comprehension of processes that lead to the  formation and eventually destruction of Killing horizons.
\end{abstract}

\maketitle

\date{\today}
\def\be{\begin{equation}}
\def\ee{\end{equation}}
\def\bea{\begin{eqnarray}}
\def\eea{\end{eqnarray}}
\newcommand{\bt}[1]{\mathbf{\mathtt{#1}}}
\newcommand{\tb}[1]{\textbf{{#1}}}
\newcommand{\rtb}[1]{\textcolor[rgb]{1.00,0.00,0.00}{\tb{#1}}}
\newcommand{\btb}[1]{\textcolor[rgb]{0.00,0.00,1.00}{\tb{#1}}}
\newcommand{\otb}[1]{\textcolor[rgb]{1.00,0.50,0.00}{\tb{#1}}}
\newcommand{\gtb}[1]{\textcolor[rgb]{0.00,.50,0.00}{\tb{#1}}}
\newcommand{\ptb}[1]{\textcolor[rgb]{0.70,0.00,0.70}{\tb{#1}}}
\newcommand{\il}{~}
\newcommand{\Qa}{\mathcal{Q}}

\section{Introduction}
\label{sec:int}
One of the most important exact solutions of Einstein vacuum field equations is the Kerr metric, which in Boyer-Lindquist (BL) coordinates can be expressed as
\bea
\nonumber
&& ds^2=-\frac{\Delta-a^2 \sin ^2\theta}{\rho^2}dt^2+\frac{\rho^2}{\Delta}dr^2+\rho^2
d\theta^2+\frac{\sin^2\theta\left(\left(a^2+r^2\right)^2-a^2 \Delta \sin^2\theta\right)}{\rho^2}d\phi^2\\\label{alai}&&-2\frac{a M \sin^2(\theta ) \left(a^2-\Delta+r^2\right)}{\rho^2}d\phi dt\ ,
\\
&&
\Delta\equiv r^2-2Mr+a^2,\quad\mbox{and}\quad\rho^2\equiv r^2+a^2\cos^2\theta \ .
\eea
It describes an axisymmetric,   stationary, asymptotically flat spacetime.
The parameter $M\geq0$  is      interpreted as  the mass  of the gravitational source, while  the rotation parameter  $a\equiv J/M $
(\emph{spin})
is  the  \emph{specific} angular momentum, and   $J$ is the
\emph{total} angular momentum of the source.
The spherically symmetric (static)  Schwarzschild solution  corresponds to the  limiting case with $a=0$.
If the spin-mass ratio is within the range $a/M\in ]0,1[ $, the spacetime corresponds to a Kerr black hole (\textbf{BH}). The extreme black hole case is defined by the relation $a=M$, whereas a super-spinner Kerr compact object or  a naked singularity (\textbf{NS}) geometry occurs  when $a/M>1$.

The Kerr metric tensor (\ref{alai})  has several remarkably properties.

\textbf{(i)} The metric  (\ref{alai}) is invariant under the application of any two different transformations of the form
\(\mathbf{\mathcal{P}}_{\mathbf{Q}}:\mathbf{Q}\rightarrow-\mathbf{Q},
\)
where $\mathbf{Q}$  is one of the coordinates $(t,\phi)$ or the metric parameter $a$: a single transformation
leads to a spacetime with an opposite  rotation with respect to the unchanged metric.

\textbf{(ii)} The Kerr solution is stationary and axisymmetric due to the presence of the   {Killing} fields  $\xi_{t}=\partial_{t} $ and
$\xi_{\phi}=\partial_{\phi} $, respectively.

An observer moving with uniform angular velocity along the curves $r=$constant and $\theta=$constant will see a spacetime which does not change at all (therefore, the covariant
components $p_{\phi}$ and $p_{t}$ of the particle four--momentum are
conserved along the circular   geodesics).\footnote{We use geometrical  units with $c=1=G$ and  the  signature $(-,+,+,+)$,
Greek indices run in $\{0,1,2,3\}$.  The   four-velocity  satisfies the condition $u^\alpha u_\alpha=-1$.
The radius $r$ has units of
mass $[M]$, and the angular momentum  units of $[M]^2$, the velocities  $[u^t]=[u^r]=1$
and $[u^{\phi}]=[u^{\theta}]=[M]^{-1}$ with $[u^{\phi}/u^{t}]=[M]^{-1}$ and
$[u_{\phi }/u_{t}]=[M]$. For the sake of convenience, we  consider a
dimensionless  energy and an angular momentum per
unit of mass $[L]/[M]=[M]$.}

\textbf{ (iii)}
As the metric is invariant under reflections with respect to  the  equatorial hyperplane $\theta=\pi/2$,  equatorial trajectories   are confined in the equatorial geodesic plane.

 For black hole and extreme black hole spacetimes, the radii
%
\bea
r_{\pm}\equiv M \pm\sqrt{M^2-a^2},
\eea
solutions of $g^{rr}=0$, are the radii of the   outer and inner Killing  horizons,
 whereas
\bea
r_{\epsilon}^{\pm}\equiv M\pm
\sqrt{M^2-a^2cos^2\theta},
\eea
solutions of $g_{tt}=0$,
are the outer and inner \emph{ergosurfaces},  respectively,
with $r_{\epsilon}^{-}\leq r_-\leq r_+\leq r_{\epsilon}^{+}$.
In an  extreme \textbf{BH} geometry, the horizons coincide, $r_-=r_+ =M$, and  the relation
$
r_{\epsilon}^{\pm}=r_{\pm}$ is valid on the rotational axis (i.e., when  $\cos^2\theta=1$).
In the  Kerr \textbf{BH} spacetime,  the Killing vector representing  time translations at infinity,  $\partial_t$,  becomes null  on the outer  ergosurface, $r_{\epsilon}^{+}$, which is, however, a \emph{timelike}  surface.
On the contrary, a Killing horizon is a  lightlike hypersurface (generated by the flow of a Killing vector) on
which the norm of a Killing vector vanishes. That is, the Kerr  horizons are  \emph{null} surfaces, $\mathcal{S}_0$,
whose \emph{null} generators coincide with the orbits of an
one-parameter group of isometries, i.e., in general    there exists a Killing field $\mathcal{L}$, which is normal to $\mathcal{S}_0$.

Some additional properties of the Kerr spacetime include:

\textbf{ (iv)}
In the limiting case of the Schwarzschild spacetime ($a=0$), $r=2M$ is the  Killing horizon with \emph{respect} to the  Killing vector
$\partial_t$. In general, in   the special case of static (and spherically symmetric) \textbf{BH} spacetimes, the
event, apparent, and Killing horizons  with respect to the  Killing field   $\xi_t$ coincide.

\textbf{ (v)}
The event horizons  of a spinning \textbf{BH}  are   Killing horizons   with respect to  the Killing field
$\mathcal{L}_H=\partial_t +\omega_H \partial_{\phi}$, where  $\omega_H$ is the angular velocity of the horizon.

\medskip

In this work,  we  extensively discuss the properties of the Killing vector $\mathcal{L}=\partial_t +\omega \partial_{\phi}$
in the case of \textbf{NS} geometries. In  \textbf{BH} spacetimes, this vector plays a crucial role in defining  thermodynamic variables.
As we will see below, the velocity $\omega$ (and its limit $\omega_H$) and the vector $\mathcal{L}$ (and its limit $\mathcal{L}_H$) are important for defining horizons and establishing relations between black holes and extreme black holes. In fact, it can be shown that:
\textbf{(a)} In the context of the rigidity theorem, $\omega_H$  represents  the \textbf{BH} rigid rotation.
Stated differently, the (strong) rigidity theorem  connects  the event horizon with a Killing  horizon. In fact, under certain
conditions,   the event horizon of a stationary asymptotically flat solution (with matter satisfying suitable hyperbolic equations)
 is a Killing horizon.
 \textbf{(b)} The \textbf{BH} event horizon of this
stationary  solution
is moreover a  Killing horizon with constant surface gravity (zeroth \textbf{BH}  law-area theorem--
the surface gravity is constant on the horizon of stationary black holes)
\cite{Chrusciel:2012jk,Wald:1999xu}.
  \textbf{(c)} Finally, the surface area of the \textbf{BH} event horizon
is non-decreasing in time, which is the content of the second \textbf{BH}  law (the laws state also the impossibility  to achieve
by a physical
process a \textbf{BH} state with surface gravity $\kappa=0$.)

We note here that the surface gravity of a \textbf{BH} may be defined as the  rate at which the norm of the Killing vector vanishes from the outside.
(The surface gravity is related to
the acceleration of a particle corotating with the \textbf{BH} at the horizon and it can be written as
 ($\mathcal{SG}_{Kerr}= (r_+-r_-)/2(r_+^2+a^2)$). It is, therefore, a conformal invariant of the metric).

Possibly, we could  isolate the contribution of the rotation in the expression of the surface  gravity  by comparing it  with the static (and spherically symmetric) metric of Schwarzschild.
In fact, the   Kerr  \textbf{BH} surface gravity  can be written as the combination  $\kappa =\kappa_s-\gamma_a$, where $\kappa_s\equiv { {1}/{4M}}$ is the Schwarzschild surface gravity, while  $\gamma_a=M\omega_{H}^{2}$ is the contribution due to the additional component of the
\textbf{BH} intrinsic spin; $\omega_{H}$ is,  therefore, the  angular velocity (in units of $1/M$) on the event horizon.

These laws, which depend also on the horizon angular velocity, impose important constraints on any physical
 process in the  \textbf{BH} spacetime, but they also allow  to distinguish the  static solution,  $a=0$, from the Kerr \textbf{BH} solution.
The first law of \textbf{BH} thermodynamics, {applied to a Kerr \textbf{BH} spacetime,} actually relates the variation of the \textbf{BH}
mass, horizon area and angular momentum, including the surface gravity and angular velocity on the horizon, i.e.,
$\delta M = (1/8\pi)\kappa \delta A + \omega_H \delta J$.
In here, the term dependent on the \textbf{BH}  angular velocity represents the ``work term'' of the first law, while  the fact that the surface gravity  is  constant on the \textbf{BH} horizon, together with  other considerations, allows us to associate it with
the concept of temperature.
This aspect  tends to emphasize  the difference (also topological) between  Kerr's \textbf{BH} and its extreme solution:
in the extreme case, where  ($r_{\pm}=M$), it is easy to see that the surface gravity  is zero and, considering the association with the temperature, there is  $T_H = 0$, with consequences also with respect to the stability
 against Hawking radiation. Nevertheless,   the  entropy (or \textbf{BH} area) of an
extremal \textbf{BH} is not null\cite{Chrusciel:2012jk,Wald:1999xu,WW,Li:2013sea,Jacobson:2010iu,Sha-teu91,Goswami:2005fu,ze}.
(An analogue  implication of the  third law it is said that a  non-extremal
\textbf{BH} cannot reach an   extremal case in a finite number of steps.)

We investigate the properties of Kerr \textbf{BHs} and \textbf{NSs} from the point of view of stationary observers.
In particular, we explore the characteristics of light surfaces, which correspond to the limiting frequencies of stationary observers.
From the analysis of these orbital  frequencies (and associated orbits),  we introduce the concept of
 Killing throats,  arising  in \textbf{NS}  spacetimes   as the ``opening'' and disappearance of Killing horizons.
  More precisely,  the Killing throat is a region bounded by a particular set of curves
	that we identify with the  frequency of a stationary observer, which depends on the radial distance and the spin parameter
$a$ of the source.
{We define a Killing bottleneck as a particular case of a of Killing throat that appears in weak naked singularities (\textbf{WNS}).
Thus, bottlenecks can be interpreted as throats ``restrictions" that characterize \textbf{WNS}
}.
	{The concept of strong and weak \textbf{NSs} depends on the value of the spin parameter and  has been explored in several works \cite{Pugliese:2010he,Pu:Kerr,Pu:Neutral,Pu:Charged,
Pu:KN,ergon,Pu:class,observers}. However, in general,  they are also differently defined as strong curvature singularities, for example,
	in \cite{strong}. Regarding various \textbf{NSs} properties  and characteristics of the gravitational collapse, possible formation and stability of naked singularities as well as other analysis concerning   observational phenomena related to possible  \textbf{NSs} existence, we refer also to \cite{Bini,deFelice,FDEFELICE1,Chakraborty:2016mhx,Blaschke:2016uyo,Bejger:2012yb,Stuchlik:2004wk,Stuchlik:2012zza,
Nakao:2017rgv,Gao:2012ca,Evo,Pradhan:2012yx,Kolos:2013bca,Stuchlik:2010zz,
Esitenza,Joshi-Book,Giacomazzo:2011cv,miller,Wald:1991zz}.
In this work, \textbf{WNSs} are characterized by  spin-mass ratios  close to the value of the extreme \textbf{BH}.
To explore these \textbf{NS} effects and  to compare \textbf{BHs} with \textbf{NSs}, it is convenient to  introduce
the concept of  ``metric bundles'' and ``extended planes''.  A metric bundle is a curve on the extended plane,   i.e.,
a family of spacetimes  defined by one characteristic photon orbital frequency $\omega$  and  characterized by a particular relation
between the metrics  parameters. This turns  out to establish a relation between \textbf{BHs} and \textbf{NSs} in the extended plane.
All the metric bundles are tangent to the horizon curve in the extended plane. Then, the horizon curve emerges as the
envelope surface of the set of metric bundles.
As a consequence,  \textbf{WNSs} turn out to be related to a part of the inner horizon, whereas strong naked
singularities \textbf(\textbf{SNSs}) with $a>2M$ are related  to the outer  horizon.

\medskip

This work is organized as follows.
In Sec.\il(\ref{Sec:1-st}), we study the main definitions and properties of  stationary observers and light surfaces
 in \textbf{BH} and \textbf{NS} Kerr geometries.
Killing  throats and bottlenecks  are the focus of
Sec.\il(\ref{Sec:bottlem}). In Sec.\il(\ref{Sef:aomega}), we introduce the concept of metric bundles and discuss the resulting connections between \textbf{BHs} and \textbf{NSs}.
These results are generalized to include the cases of the  Kerr-Newman  and Reissner-Nordstr\"om spacetimes in Sec.(\ref{Sec:RN-KN-cases}).
Concluding remarks and future perspectives follow in Sec.\il(\ref{Sec:con-RAEM-FP}).
This article closes with  two  Appendices.  The off-equatorial case in the Kerr  and Kerr-Newman geometries is considered in Appendix \ref{Eq:off-e}.
In Appendix \ref{Sec:appendix-area}, we study the areas of the horizons and regions of the extended plane delimited by
different  metric bundles.}
{Throughout this work, we introduce a considerable number of symbols and notations which are necessary to
explain all the details of the results we will obtain. For clarity, we list in Table\il\ref{Table:pol-cy-multi} the main symbols and
 their definitions.}
\begin{table*}
\caption{{Lookup table with the main symbols and relevant notations  used throughout the article.}}
\label{Table:pol-cy-multi}
\centering
\begin{tabular}{ll}
 \hline \hline
$ \omega_{\pm}$&  limiting frequencies for stationary observers:  Eqs.(\ref{Eq:ex-ce})-(\ref{Eq:b-y-proc})\\
$\omega_0=M/a$&  limiting frequency $\omega_{\pm}$ at the singularity and frequency of the metric bundle: Eq.\il(\ref{Eq:omega000})-Sec.\il(\ref{Sef:aomega})
\\
$\mathcal{L}_{\pm}$& null  Killing vector
 (generators of Killing event  horizons): Eq.\il(\ref{Eq:comb}) \\
 $r_{s}^{\pm}$&  light surfaces radii: Eq.\il(\ref{Eq:rspm}) \\
$\omega_H^{\pm}$& frequencies at the horizons $r_{\pm}$: Eq.\il(\ref{Eq:freq-BH}) \\
$r^{\mp}_{\mp}$& photon orbits  with   frequencies  $\omega_H^{\pm}$ at the horizons:
Eq.\il(\ref{Eq:mart-re})--Fig.\il\ref{Fig:MaSuorVIRMa}\\
$g_{\omega}^{\pm}$& metric bundles in the extended plane $\pi_a$: Sec.\il(\ref{Sef:aomega}) \\
$a_{\omega}^{\pm}(r,\omega;M)$& $g_{\omega}^{\pm}$  in terms of the bundle frequency $\omega$: Eq.\il(\ref{Eq.lcospis}) \\
$a_{\pm}$& horizon curve in the extended plane: Fig.\il(\ref{Fig:SoOv}) \\
$r_{\partial}^{\pm}(\omega)$& closing radii  of the metric bundle  in $\pi_a^+$ (=$\pi_a$ for $a>0$): Eq.\il(\ref{Eq:rela-a-partialv})\\
$a_g$& spin of metric bundle tangent to the horizons in $\pi_a^+$: Eq.\il(\ref{Eq:agar})-- Fig.\il(\ref{Fig:RoyalS}) \\
$a_p$&  bundle origin, i.e., $a_g(a_0)=a_g(a_0^{\prime})$ with $a^{\prime}_0=a_p\equiv 4M^2/a_0$: Figs.\il\ref{Fig:pesoMa},
\ref{Fig:PostMerge},  \ref{Fig:manet}; Tables \ref{Table:ABCDmodels} and \ref{Table:number-cicle} \\
horizons relations \textbf{I}& $\omega_0^{-1}\equiv a_0^{\pm}/M=\frac{2 r_{\pm}(a_g)}{a_g}\equiv \omega_H^{-1}(a_g)$,  $\omega_H^+(r_g,a_g)=\omega_0=Ma_0^{-1}$, $\omega_H^-(r^{\prime}_g,a_g)=\omega^{\prime}_0=M/a_0^{\prime}$ \\
&
 $r^{\prime}_g\in r_-$ ($r_+=r_g$,  $r_-=r_g^{\prime}$): Fig.\il(\ref{Fig:RoyalS})
 \\
 horizons relations \textbf{II}& $\omega^{\prime}_0=\frac{1}{4 \omega_0}$, $ \omega_H^+\omega_H^-=\frac{1}{4}$, ($a_0^+(a_g)a_0^-(a_g)=4M^2$), $a_0^{\pm}/M=\frac{2 r_{\pm}(a_g)}{a_g}$  where $a=a_0$ and  $a=a_p$: Fig.\il\ref{Fig:pesoMa}
\\
\textbf{SNS}&(=$\textbf{SNS}^+\cup\textbf{SNS}^-$) strong naked singularities  $a_0>2M$,
$\textbf{SNS}^+$ for $a_0>4M$\\
& $\textbf{SNS}^-$ for $a_0\in[2M,4M[$: Fig.\il(\ref{Fig:RoyalS})\\
\\
\textbf{WNS}&weak naked singularities  $a_0\in]M,2M[$: Fig.\il(\ref{Fig:RoyalS})\\
$\textbf{BH}=\textbf{BH}^+\cup\textbf{BH}^-$&  $\textbf{BH}^+$ for $a\in[a_{g}^1,M]$, $a_{g}^1=3/4 M$ and
$\textbf{BH}^-$ for $a\in[0,a_{g_1}]$: Fig.\il(\ref{Fig:RoyalS})
\\
\textbf{\emph{Left region}} Fig.\il(\ref{Fig:RoyalS}) & $a_0\in[0,2M]$
\\
\textbf{\emph{Right  region}}Fig.\il(\ref{Fig:RoyalS})& $a_0>2M$
\\
\textbf{\emph{up-sector}} Fig.\il(\ref{Fig:RoyalS})& $a_g>a_{g}^1$
\\
\textbf{\emph{down-sector}}  Fig.\il(\ref{Fig:RoyalS})& $a_g<a_{g}^1$
\\
 $\varpi_\pm$& second frequency of a  metric bundle: Eq.\il(\ref{Eq:gen2-dat})
\\
$a_{\omega}^{\mathbf{(\natural)}}(\omega_H^{\mathbf{\flat}})$& metric bundles  parameterized  for the tangent point $a_g$:
Eq.\il(\ref{Eq:UE-dif})
\\
$a_{tangent}(r)$&  tangent curve to the horizon in terms of  $r_g$: Eq.\il(\ref{Eq:a-tangent})
\\
$(r^{real}_g, r^{\checkmark}_g, r_g^\mp)$& solution of the tangency condition
$a_g=a_{\pm}$: Eq.\il(\ref{Eq:rgpmrg}), Fig.\il(\ref{Fig:Bollet})
\\
$
(Q_{\omega}^{\pm})^2$& metric bundles in terms of the charge $Q$: Eq.\il(\ref{Eq:Q-para-metric})
\\
\hline
\end{tabular}
\end{table*}
\section{Stationary observers and light surfaces}
\label{Sec:1-st}
%
\emph{Stationary observers} have
 a tangent vector which is  a spacetime  Killing vector; their  four-velocity  is, therefore, a
linear combination of the two Killing vectors $\xi_{\phi}$ and $\xi_{t}$ as:
\bea&&\label{Eq:spectrum}
{d\phi}/{dt}={u^{\phi}}/{u^t}\equiv\omega,\quad\mbox{or}\quad
u^\alpha=\gamma(\xi_t^\alpha+\omega \xi_\phi^\alpha),
\\&&\label{Eq:sic-six}
\mbox{with}\quad\gamma^{-2}\equiv-\varepsilon(\omega^2 g_{\phi\phi}+2\omega g_{t\phi}+g_{tt}) ,
\eea
where  $\omega$  (a dimensionless quantity) is the (uniform) \emph{angular velocity}, while $\gamma$ is a normalization factor
 ($g_{\alpha\beta} u^\alpha u^\beta=-\varepsilon$).

Because of the symmetries, the coordinates  $r$ and $\theta$  of a stationary observer are constants along  its worldline, consequently  a stationary observer does not see the spacetime changing along its trajectory.
Timelike stationary  observers have  angular velocity  bounded in the range
{
\bea\label{Eq:ex-ce}
\omega\in]\omega_-,\omega_+[ \quad\mbox{where}\quad \omega_{\pm}\equiv \omega_{Z}\pm\sqrt{\omega_{Z}^2-\wp^2},
\\\nonumber
\wp^2\equiv \frac{g_{tt}}{g_{\phi \phi}}=\frac{g^{\phi \phi}}{g^{tt}},\quad
\wp^{-2}\equiv \frac{g^{tt}}{g^{\phi\phi}}=\frac{g_{\phi\phi}}{g_{tt}},\quad \omega_{Z}\equiv-\frac{g_{\phi t}}{g_{\phi\phi}},
\eea}
($\varepsilon=+1$)\footnote{{This equation corrects a typo in Eq. (8) of Ref. \cite{observers}}}.
Zero Angular Momentum Observers (ZAMOs) are defined by the condition
$\mathcal{L_{ZAMO}}=0$ and  have  angular velocities $\omega_Z$, which depend on the spin. A ZAMO, well defined in the ergoregion, corotates with the \textbf{BH}. ZAMOs have interesting properties in the case of slowly spinning naked singularities and certainly offer   a particularly appropriate and convenient  description of the spacetime in the ergoregion --see for example   \cite{observers,ergon} and \cite{Gariel:2014ara,Pelavas:2000za,Herdeiro:2014jaa,Cardoso:2007az,CoSch,Frolov:2014dta,Stu80,CCQ-suer,Schee:2013bya,
Torok:2005ct,BiStuBa89,Patil,erg1}.
 Static observers are defined by the limiting condition  $\omega=0$  and  cannot exist in the ergoregion.
The particular frequencies $\omega_{\pm}$ provide an alternative definition of the  horizons.
Since the horizons are null surfaces, it should hold that $\omega_+=\omega_-$, which is the limiting angular velocity for physical observers corresponding in fact to orbital photon frequencies.
  The quantity in parenthesis in the r.h.s. of Eq.\il(\ref{Eq:sic-six}) becomes null  for photon-like particles and the rotational frequencies $\omega_{\pm}$, as in Eq.\il(\ref{Eq:ex-ce}). On
the equatorial plane,  the limiting orbital frequencies are
\be\label{Eq:b-y-proc}
\omega_{\pm}\equiv\frac{2 aM^2\pm M\sqrt{r^2 \Delta}}{r^3+a^2 (2M+r)}\ , \ \
  \omega_{\pm}(r_+)= \omega_{Z}(r_+)=\omega_{H}\equiv\frac{a}{2 r_+}\equiv \frac{M}{2\omega_0 r_+}.
 \ee
The following limits are valid
 \be\label{Eq:omega000}
 \lim_{r\rightarrow\infty}\omega_{\pm}=0,\quad \lim_{a\rightarrow\infty}\omega_{\pm}=0,\quad\lim_{r\rightarrow0}\omega_{\pm}=
\omega_0\equiv\frac{M}{a}\ .
\ee
The limit $a\to\infty$ is used to formally explore the behavior in the strong \textbf{NS} singularity regime, for a given constant value  of $M$, and more generally in the limit  $a\gg M$.
 As already mentioned in Sec.\il(\ref{sec:int}), the  Killing vector
\be\label{Eq:comb}
\mathcal{L}_{\pm}\equiv \xi_{t}+\omega_{\pm}\xi_{\phi}
\ee
{can be read as generator of  null curves ($g_{\alpha\beta}\mathcal{L}^\alpha_{\pm}\mathcal{L}^\beta_{\pm}=0$)
as the Killing vectors $\mathcal{L}_{\pm} $, null at r = $r_+$, are also generators of Killing event  horizons.

The expression (\ref{Eq:ex-ce}) for the frequency of a stationary observer can be considered as an equation for the radii of the
light surfaces $r_{s}^{\pm}$. The solutions are then given as functions of the  frequency $\omega$ and can be written as \cite{observers}
\bea\label{Eq:rspm}
&&\frac{r_s^-}{M} \equiv \frac{2 \beta_1 \sin \left(\frac{1}{3} \arcsin\beta_0\right)}{\sqrt{3}}
,\quad
\frac{r_s^+}{M} \equiv \frac{2 \beta_1 \cos \left(\frac{1}{3}\arccos(-\beta_0)\right)}{\sqrt{3}}\\\nonumber
&&
\mbox{where}\quad
\beta_1\equiv\sqrt{\frac{1}{\omega ^2}-\frac{1}{\omega_0^2}},\quad \beta_0\equiv\frac{3 \sqrt{3} \beta_1 \omega ^2}{\left(\frac{\omega }{\omega_0}+1\right)^2}.
\eea
\section{Killing  throats and bottlenecks}
\label{Sec:bottlem}
The concept of Killing  throats emerges   through the analysis of the radii $r_{s}^{\pm}(\omega,a)$  (the frequencies $\omega_{\pm}(r,a)$)
with respect to the orbital  frequency  $\omega$ (the radius $r$)  of  light-like particles -- see
Figs.\il\ref{Fig:cOALESCENzE} and \ref{Fig:QPlot}.
 A  Killing  throat  in \textbf{NS} geometries   is a connected region in the $r-\omega$ plane, which is bounded by
   $r_{s}^{\pm}(\omega,a)$  or equivalently $\omega_{\pm}(r,a)$,  and contains all the stationary observers allowed within the
	limiting  frequencies  $]\omega_-, \omega_+[$. Conversely, a \textbf{BH} Killing  throat is either a disconnected  region in the Kerr spacetime {(in a sense similar to the concept of a path-connected space)}
	or a  region bounded by  non-regular surfaces in the  extreme Kerr \textbf{BH} spacetime.
	A   \emph{Killing bottleneck in a \textbf{NS} spacetime  is a narrowing of the  Killing  throat which appears only for specific naked singularities and  involves a narrow range of frequencies and orbits--Figs\il(\ref{Fig:MaSuorVIRMa}).}
   The limiting case of a     Killing bottleneck occurs in  the extreme Kerr spacetime, as seen in the BL frame, where the narrowing actually closes\footnote{Killing throats and bottlenecks, represented Figs\il(\ref{Fig:MaSuorVIRMa}) and (\ref{Fig:cOALESCENzE}), were grouped in \cite{Tanatarov:2016mcs}  in   structures named ``whale  diagrams'', considering the  escape cones, particles motion and collisionals problems  in the Kerr and
Kerr-Newman spacetimes--see also
\cite{Mukherjee:2018cbu,Zaslavskii:2018kix}.} at the \textbf{BH} horizon $r=M$.

More generally,   the Kerr   horizons  determine the  following frequencies:
\bea\label{Eq:freq-BH}
\omega_H^{\pm}\equiv\frac{a}{2 r_{\pm}}\equiv \omega_{\pm}(r_{\pm})\quad \mbox{with}\quad\omega_H^+<\omega_H^-\quad \mbox{for}\quad a>0\quad\mbox{and}\quad \omega_H^{\pm}=\frac{1}{2}\quad\mbox{ for}\quad
a=M.
\eea
It should be noted then that while the horizons radii are functions of the metric parameters  only,    meaning that  there is only one frequency, $\omega_H^{\pm}$, on the horizons $r_{\pm}$, respectively,  the   Killing bottlenecks depend on  both  frequency and radius, corresponding to the fact that the throat never closes, but in the limit of the extreme Kerr spacetime.

{
Considering again the horizons  frequencies  $\omega_H^{\pm}$, we introduce the
radii $r^{\mp}_{\mp}$  representing   the set  of photon orbits  with   frequencies  $\omega_H^{\pm}$ at
the $\textbf{BH}$ horizons
\bea\label{Eq:mart-re}
r^{-}_{-}:\quad  \omega_-(r_-^-)=\omega_-(r_-)=\omega_H^- \quad\mbox{where}\quad r^{-}_{-}=\frac{1}{2} \left(\sqrt{\frac{32 M^3 r_-}{a^2}-a^2+6 M\sqrt{M^2-a^2}-22M^2}-r_-\right)
 \\
r_+^+:\quad\omega_+(r_+^+)=\omega_+(r_+)=\omega_H^+,\quad\mbox{where}\quad  r_+^+=\frac{1}{2} \left(\sqrt{\frac{32 M^3 r_+}{a^2}-a^2-6 M \sqrt{M^2-a^2}-22M^2}-r_+\right),
\eea
with $ r_-^-<r_-<r_+<r_+^+$ -- see Fig.\il\ref{Fig:MaSuorVIRMa}.}
    In Sec.\il\ref{Sef:aomega},  this property is  displayed in a different context, showing a close connection between \textbf{BHs} and \textbf{NSs}. Moreover, the Killing  horizons,  $r_{\pm}$, are defined  as the tangent  (envelope surfaces)  of the curves defined  by the conditions $\omega_{\pm}(r,a)=$constant.
  \begin{figure}
  \includegraphics[width=5.6cm]{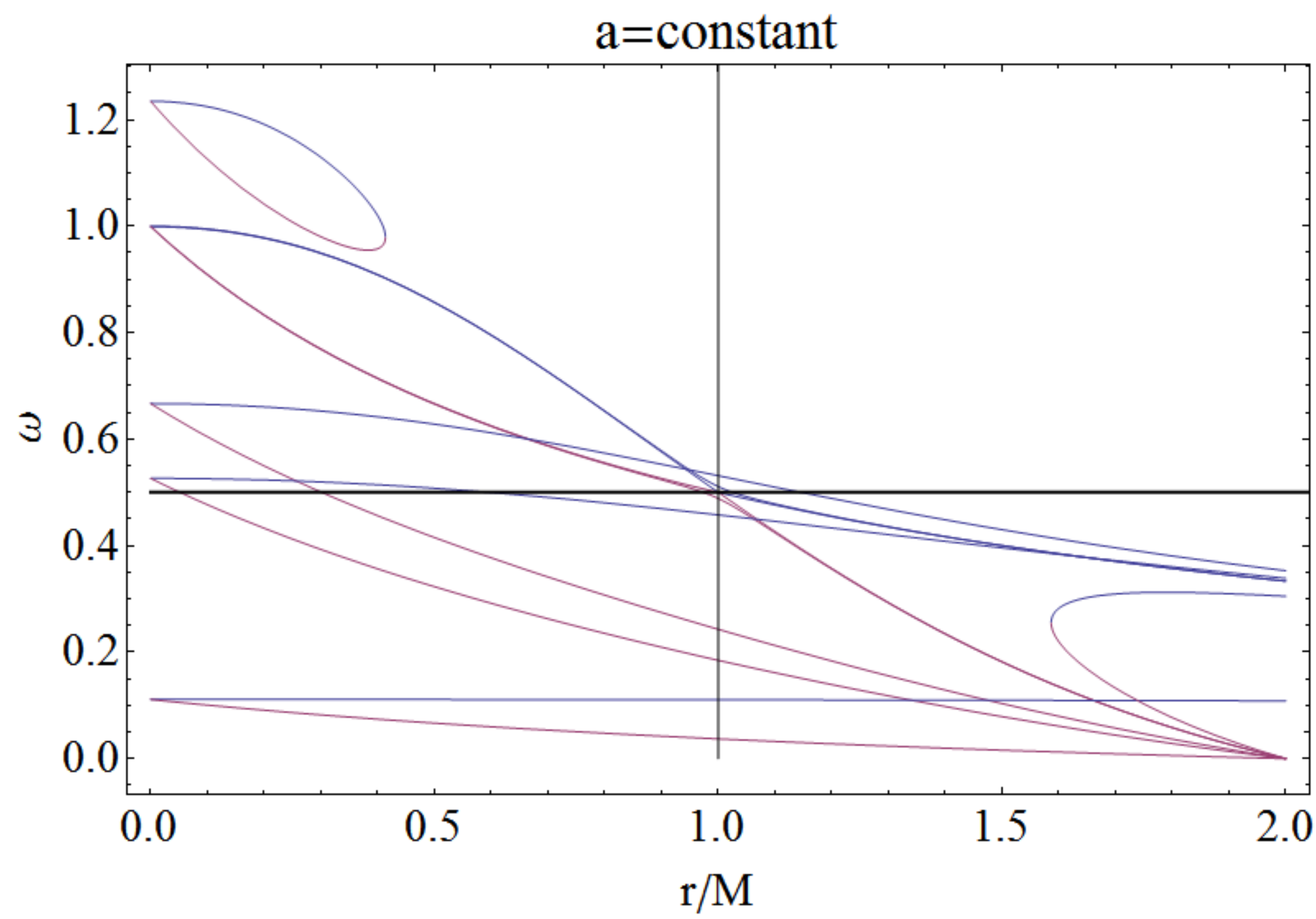}
  \includegraphics[width=5.6cm]{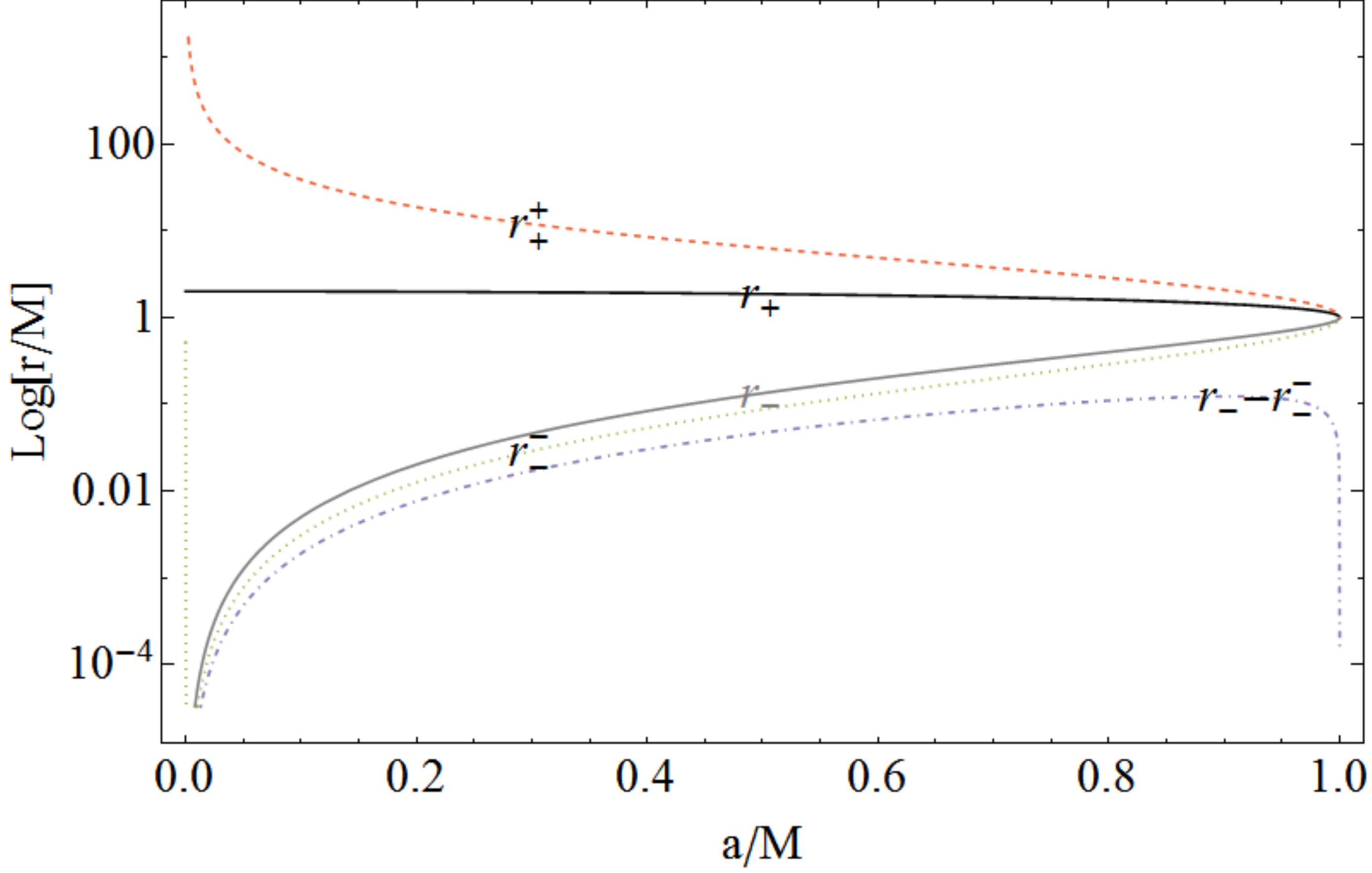}
  \includegraphics[width=5.6cm]{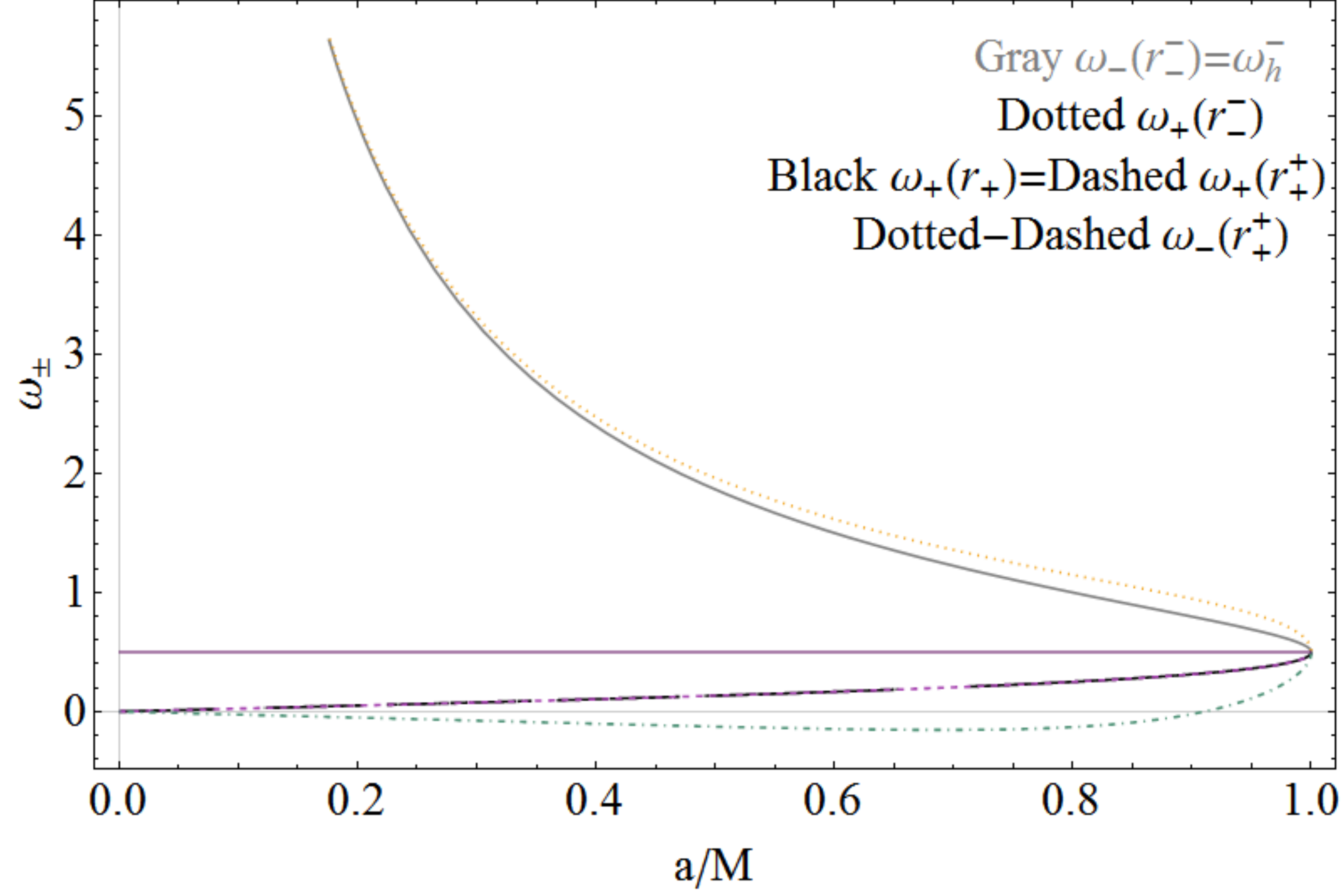}
  \caption{Left panel: Frequencies $\omega_{\pm}$ for fixed values of the spin  $a/M$ for \textbf{BHs} and \textbf{NSs}.
	The coalescence of the Killing horizons $r_{+}$  and $r_-$ in the extreme black hole geometries and the emergence of a  Killing throat  and a Killing bottleneck in the \textbf{NS} geometries are clear --see also Figs\il(\ref{Fig:cOALESCENzE},\ref{Fig:CORESUNO},\ref{Fig:possov131},\ref{Fig:possov132}), and  Figs\il(\ref{Fig:Gpossov4}). Center panel: Killing horizons, $r_{\pm}$, and radii $(r_{-}^-,r_+^+)$
	as given in Eq.\il(\ref{Eq:mart-re}). Right panel: Frequencies $\omega_{\pm}$ in $r_{\pm}$ and  $(r_{-}^-,r_+^+)$ as functions of $a/M$.}\label{Fig:MaSuorVIRMa}
\end{figure}
The existence of $r_+^+>r_+$ implies that an  observer could  eventually measure  the  frequency  of the outer horizon $\omega_H^{+}$ on the equatorial plane,  while \emph{no information}  can be obtained from  $r_{-}^-$ for  the inner horizon frequency $\omega_H^-$.     In this sense, we may call this property as \emph{inner horizon confinement}.
This situation can be used to distinguish between slow rotating \textbf{BHs} and
fast spinning \textbf{BHs} since the distance  $(r_+^+-r_+)$ decreases with the spin.
The existence of these radii may be related to the bottleneck presence.

Figure\il\ref{Fig:cOALESCENzE} shows  the formation of the Killing throat as the spin of the naked singularity varies. The emergence of the Killing bottleneck in terms of the  frequency ($\omega-r$ plane) and of the  radius  ($r-\omega$ plane)  is  evident in the case of weak naked singularities, i.e., slow spinning singularities. We specify below the limiting spin values which define weak naked singularities.
\begin{figure}[h!]
\centering
\begin{tabular}{lcr}
\includegraphics[scale=.5]{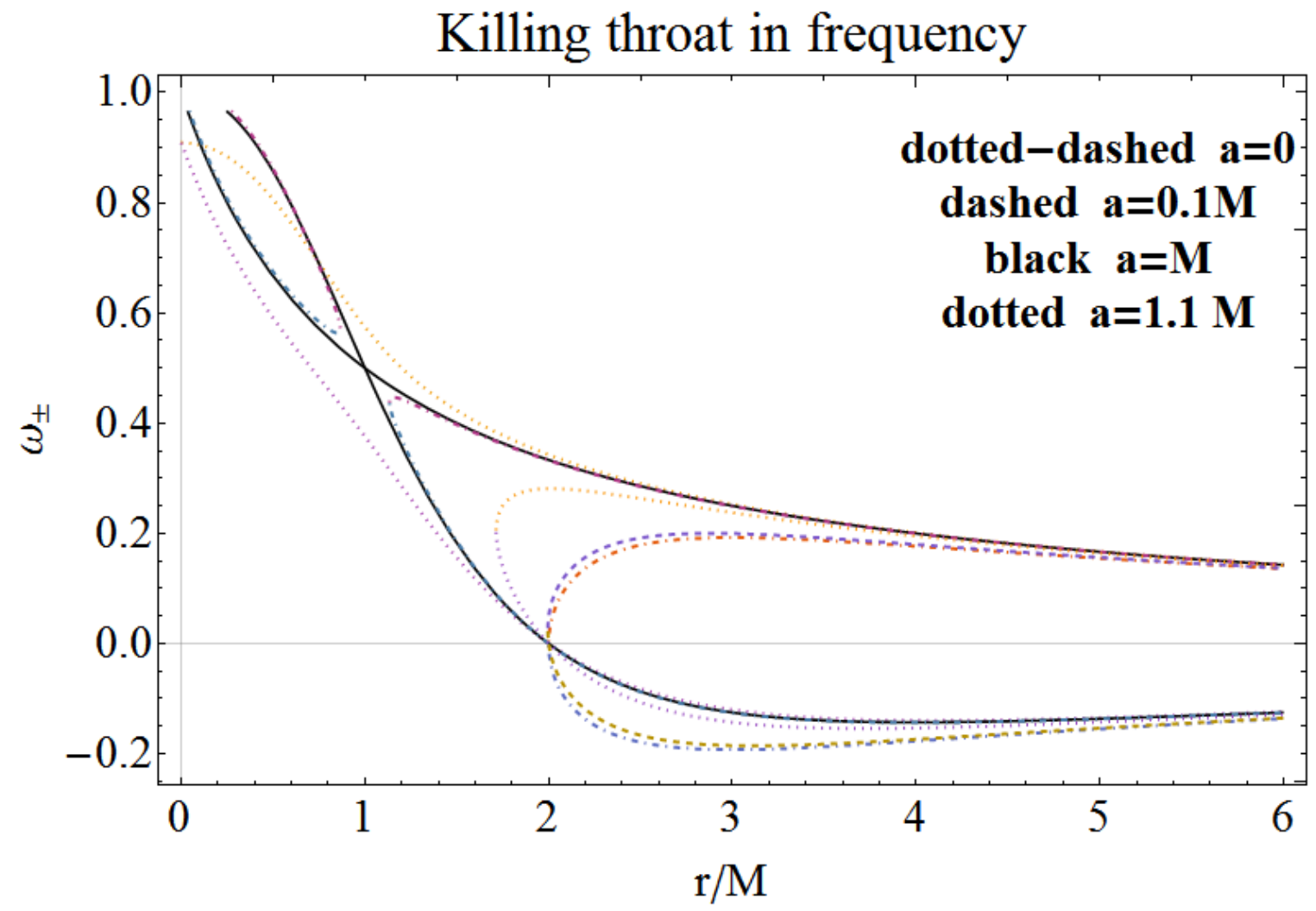}
\includegraphics[scale=.5]{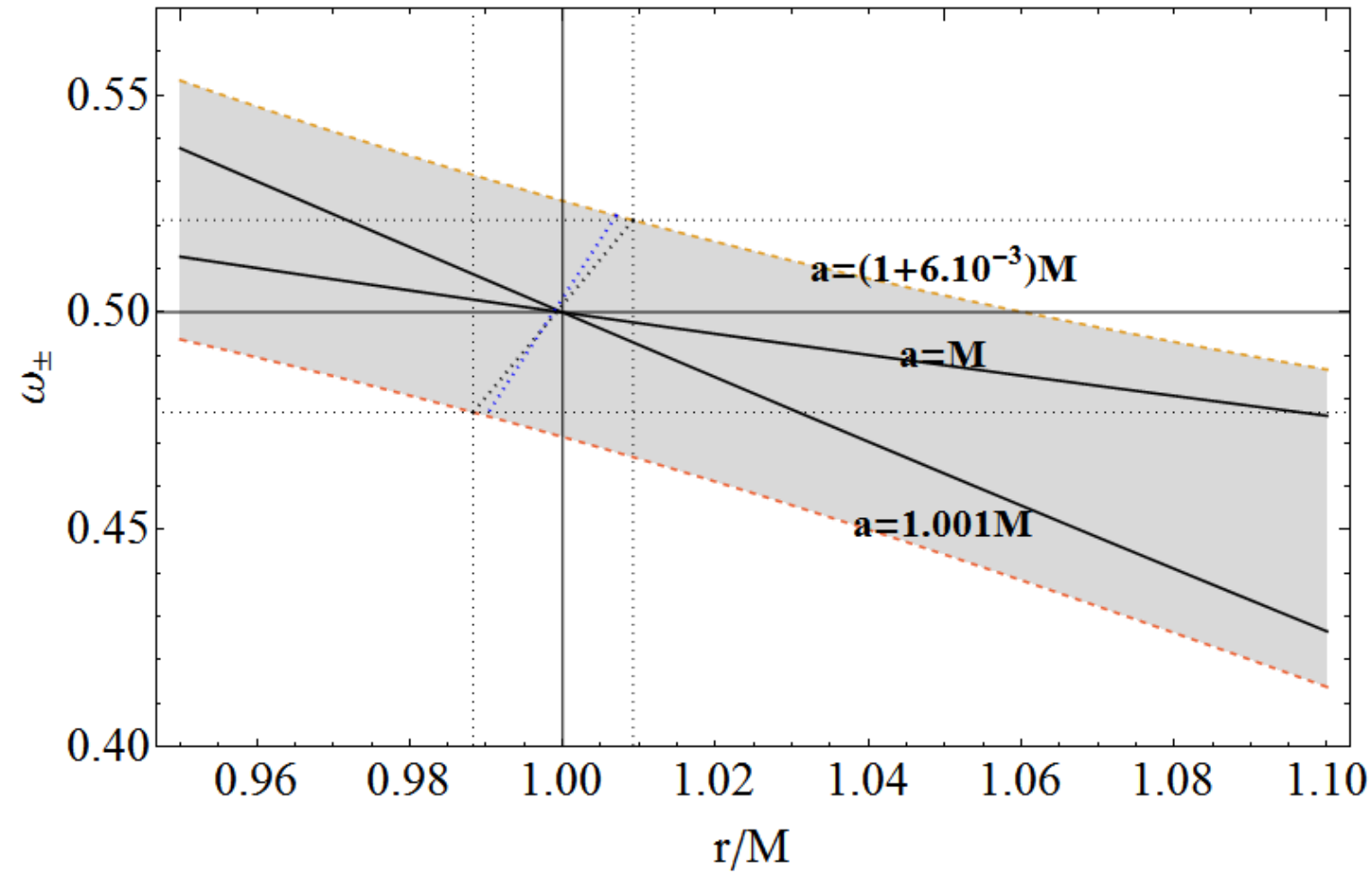}
\\
\includegraphics[scale=.3]{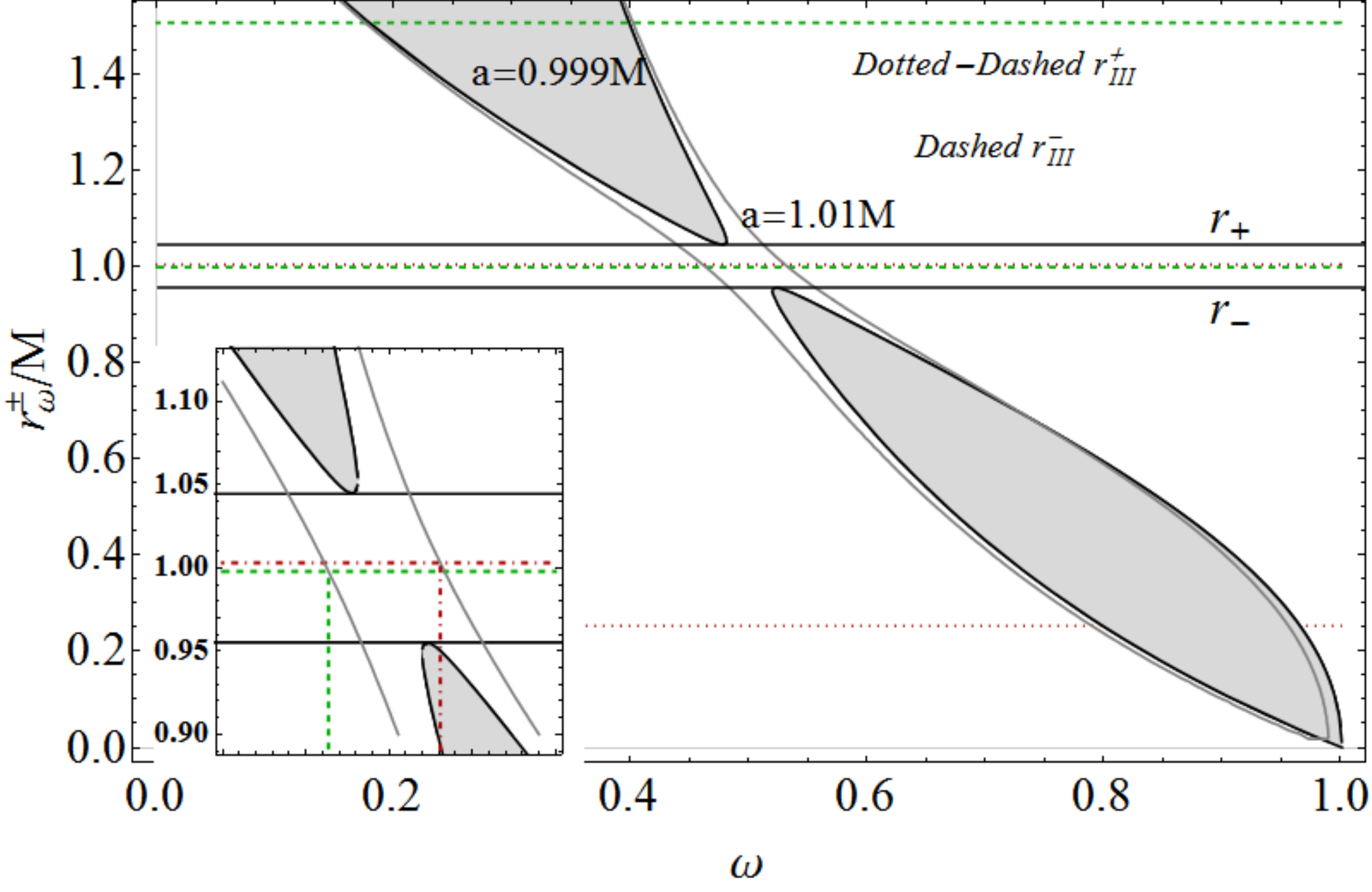}
\includegraphics[scale=.5]{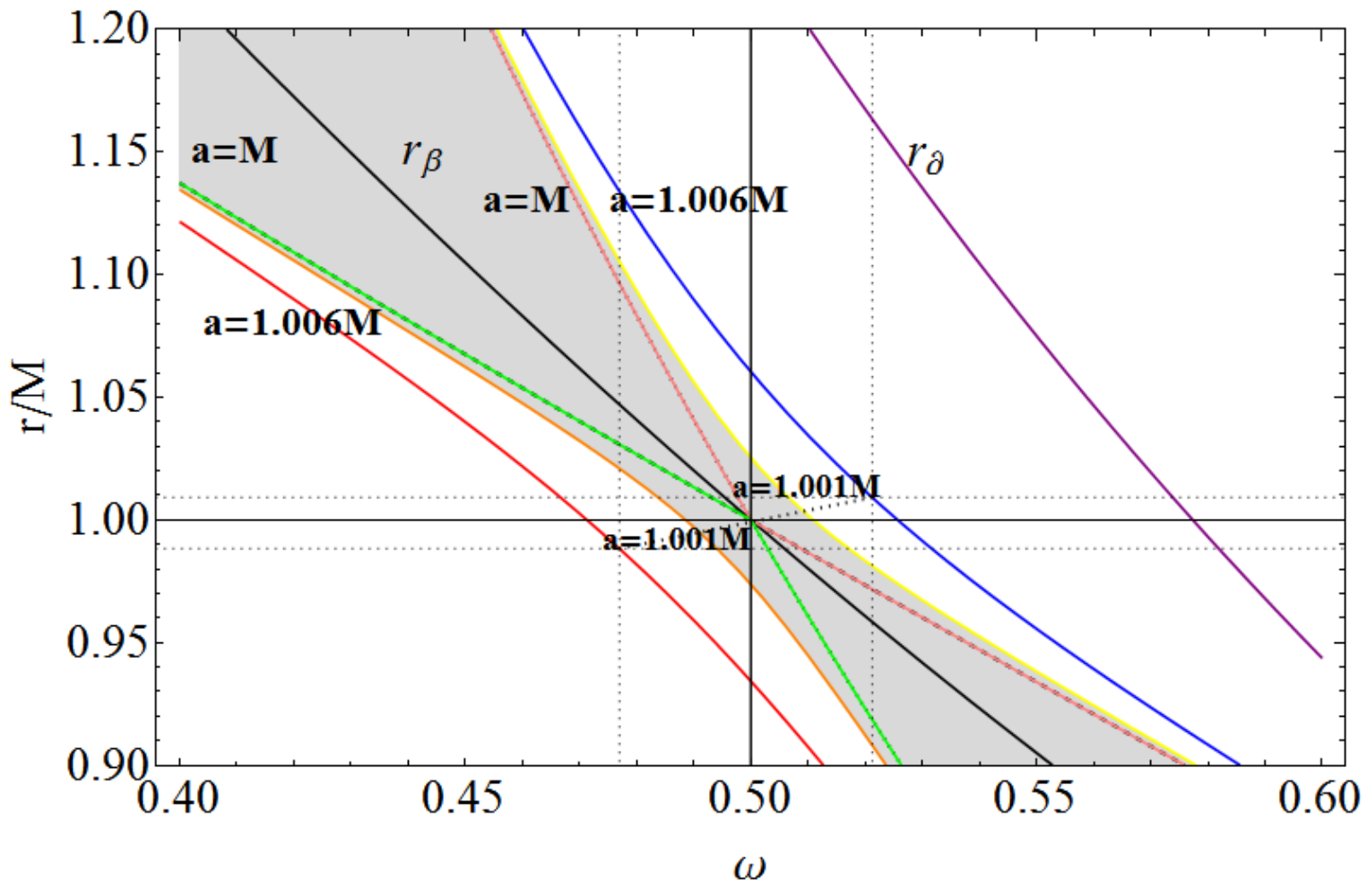}\\
\includegraphics[scale=.5]{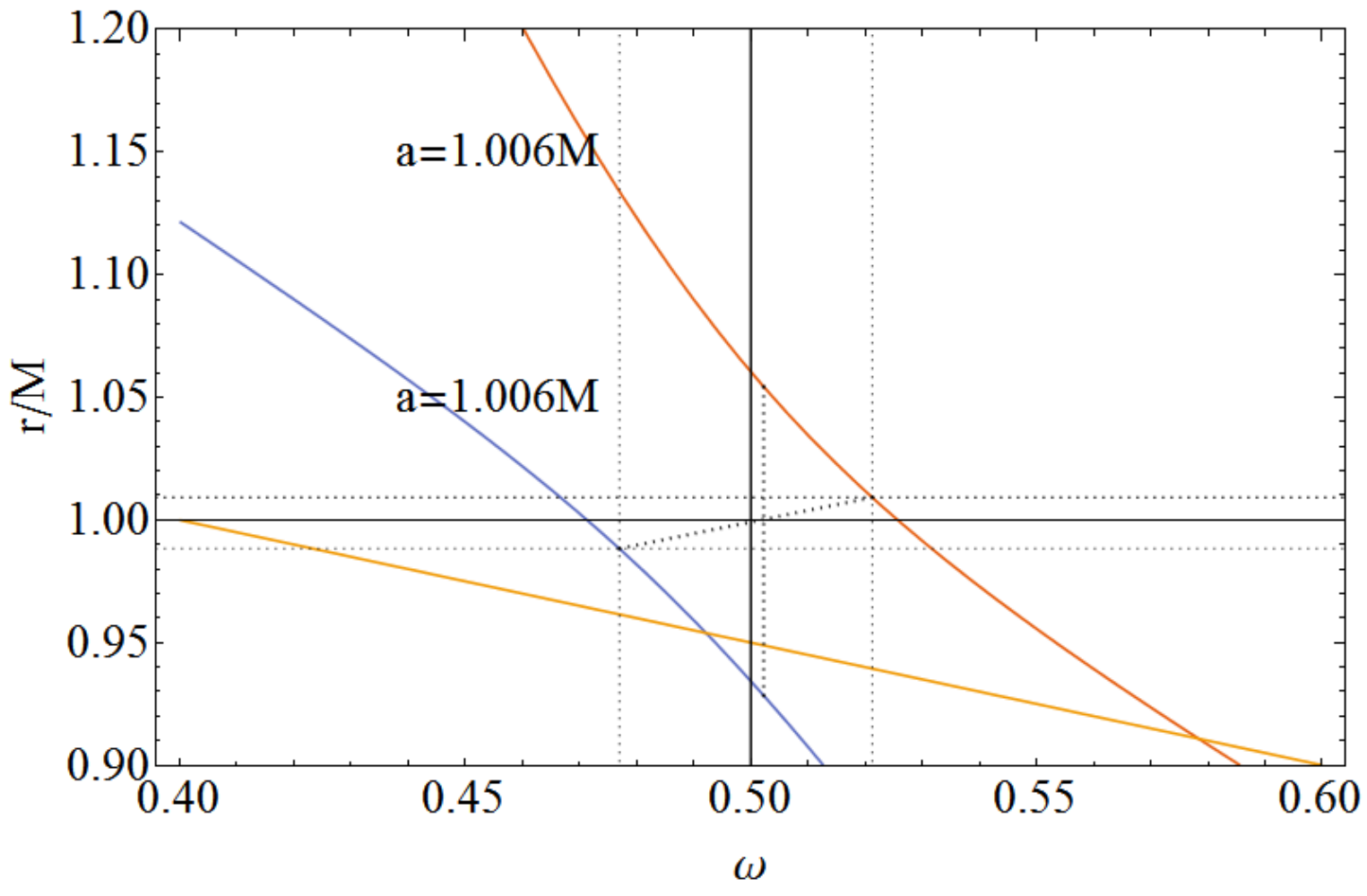}
\end{tabular}
\caption[font={footnotesize,it}]{Equatorial plane of the Kerr spacetime. Killing throat: zoom of the Killing bottleneck.
Upper panels: The plane $\omega$-$r/M$ with plots of $\omega_{\pm}$ for different spins $a/M$.
Left upper panel: Light surfaces for different \textbf{BHs} to \textbf{NSs}. As $a/M$ changes for weak naked singularities,  the narrowing of the light surfaces is shown explicitly.
Right upper panel: Light surfaces in the bottleneck and the corresponding relevant radii.
Bottom panels: The plane $r/M$-$\omega$ with plots of the radii of stationary observers $r_{s}^{\pm}$ for different spins $a/M$.
Center and right bottom panels: Zoom on the Killing bottleneck and relevant radii. The radii $r_{III}^{\pm}$ (solutions of
   $\partial_r^3\omega_{\pm}=0$) are shown as functions of the spin $a/M$. The radii
	$r_{\partial}$ and $r_{\beta}$ are defined in Eq.\il(\ref{Eq:rela-a-partialv}) and Eq.\il(\ref{Eq:rom-grad-mira}),  respectively. }
\label{Fig:cOALESCENzE}
\end{figure}
To evaluate the effects of the spacetime dragging  on the formation of a Killing bottleneck, we investigate in
Sec.\il(\ref{Sec:RN-KN-cases}) the
Kerr-Newman geometry, the limiting  static case of the  Reissner Nordstr\"om geometry,  and the off-equatorial case.

Three distinct phases are significant in the process of formation of  bottlenecks:

\textbf{(1)} Coalescence of the Killing  horizons, which occurs in the extreme Kerr \textbf{BH} solution;
	
\textbf{(2)} Formation of  the Killing  throat and emergence of the  bottleneck  in  weak \textbf{NSs};

\textbf{(3)} Disappearance of the Killing bottleneck in strong \textbf{NSs}.

The analysis carried out in Figs.\il\ref{Fig:cOALESCENzE} and  \ref{Fig:MaSuorVIRMscri} suggests that the Killing bottlenecks can be defined through the  conditions   $r_{III}^{\pm}:\; \partial_{\omega}^3r_s^{\pm}=0$
and   $\omega_{III}^{\pm}:\; \partial_{\omega}^3r_s^{\pm}=0$.
On the other hand,  the radii $r_{II}^{\pm}:\; \partial_{\omega}^2r_s^{\pm}=0$  and, analogously, $\omega_{II}^{\pm}:\;
\partial_{r}^2\omega_{\pm}=0$ characterize the curvatures of the curves $\omega_{\pm}$  and $r_s^{\pm}$.

{Killing bottlenecks, identified  in \cite{observers} as ripples  in the
$r-\omega$ plane (see  Figs.\il\ref{Fig:QPlot} and \ref{Fig:QPlot1}),
  were interpreted in the \textbf{BL} frame as ``\emph{remnants}''  of the disconnection between the Killing throat  present in \textbf{BH}-geometries  and  the singular bottleneck of the extreme \textbf{BH}}.
\begin{figure*}[ht!]
\begin{tabular}{llcr}
\includegraphics[scale=.22]{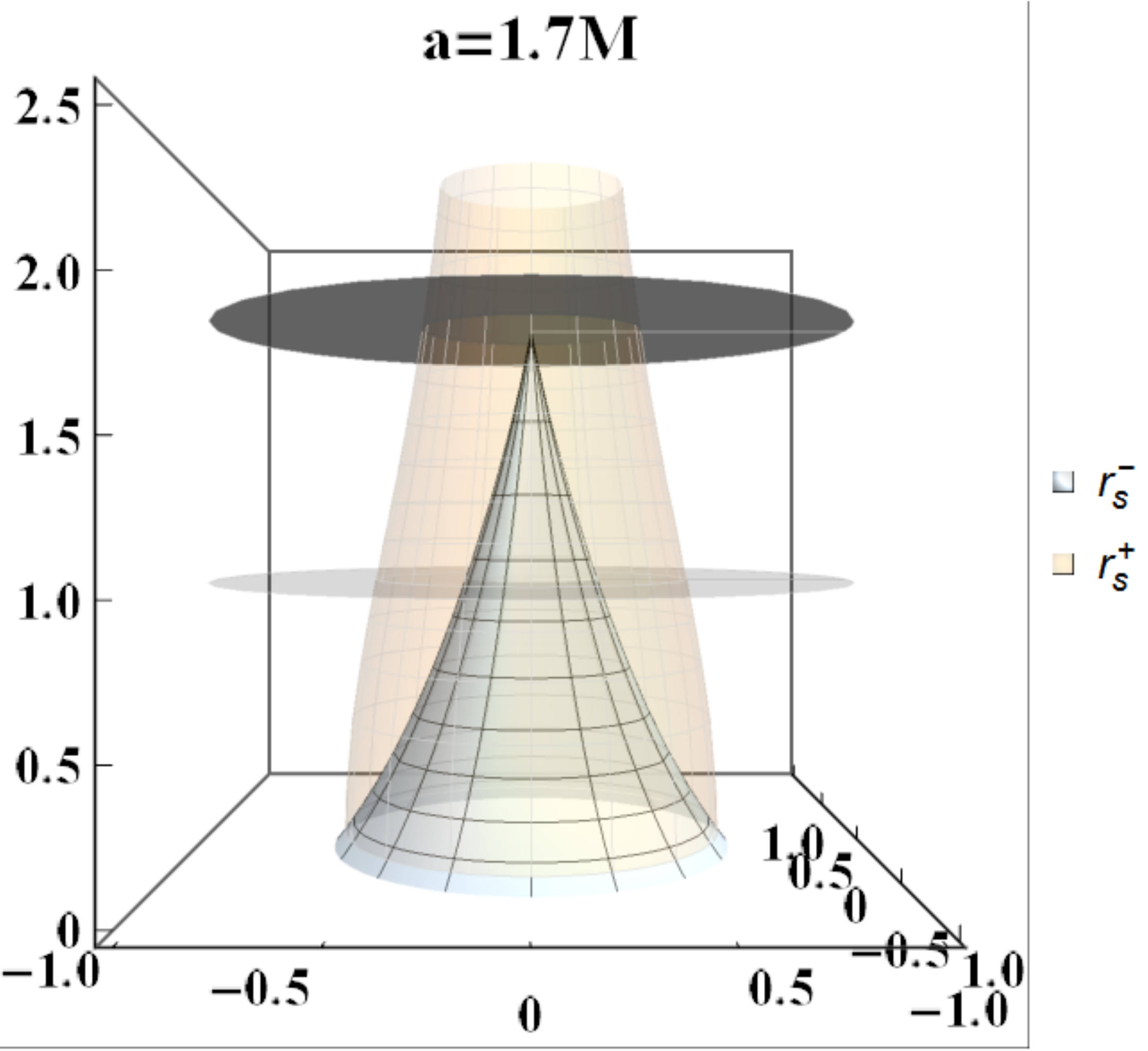}%
\includegraphics[scale=.22]{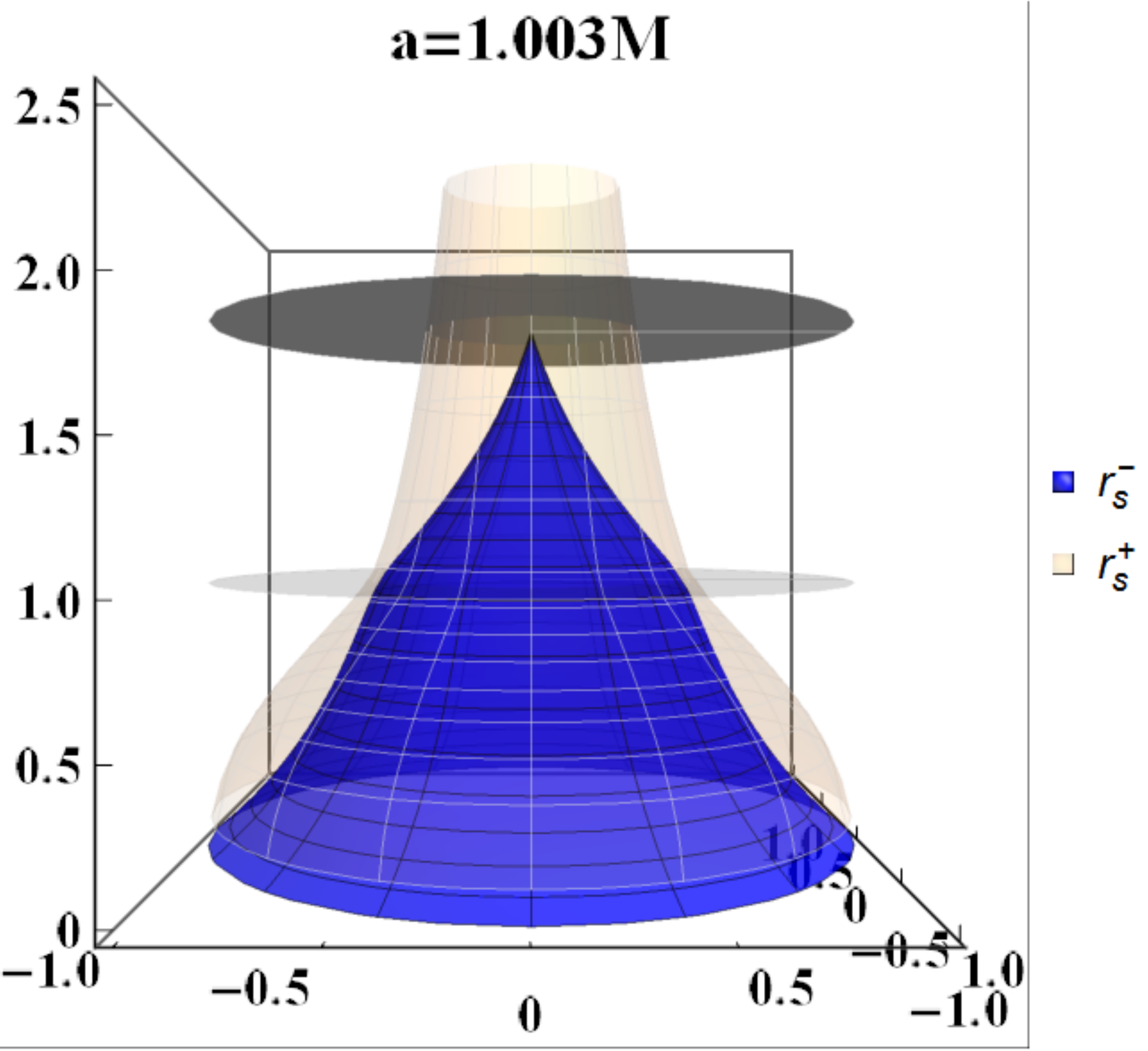}
\includegraphics[scale=.22]{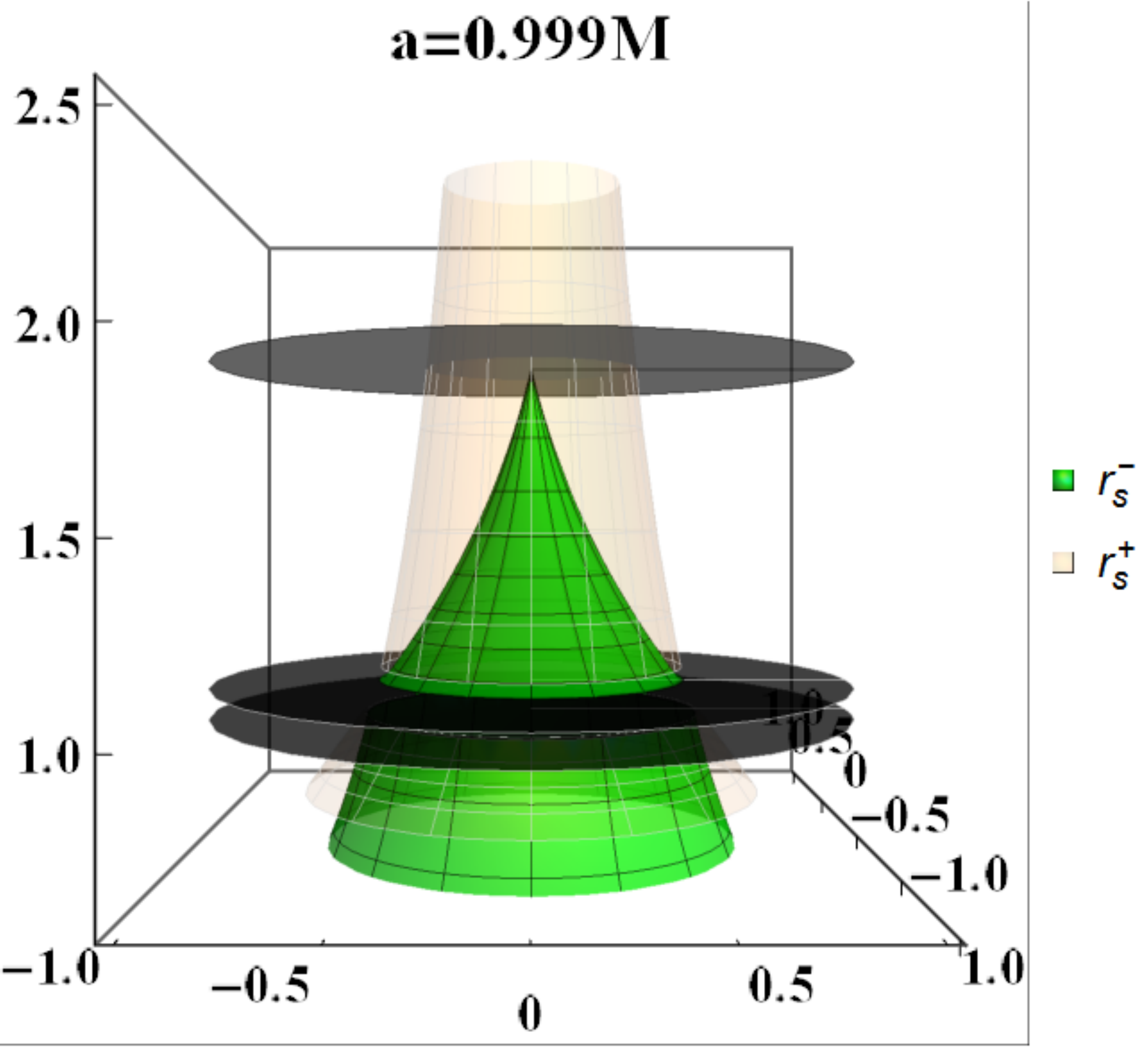}
\includegraphics[scale=.22]{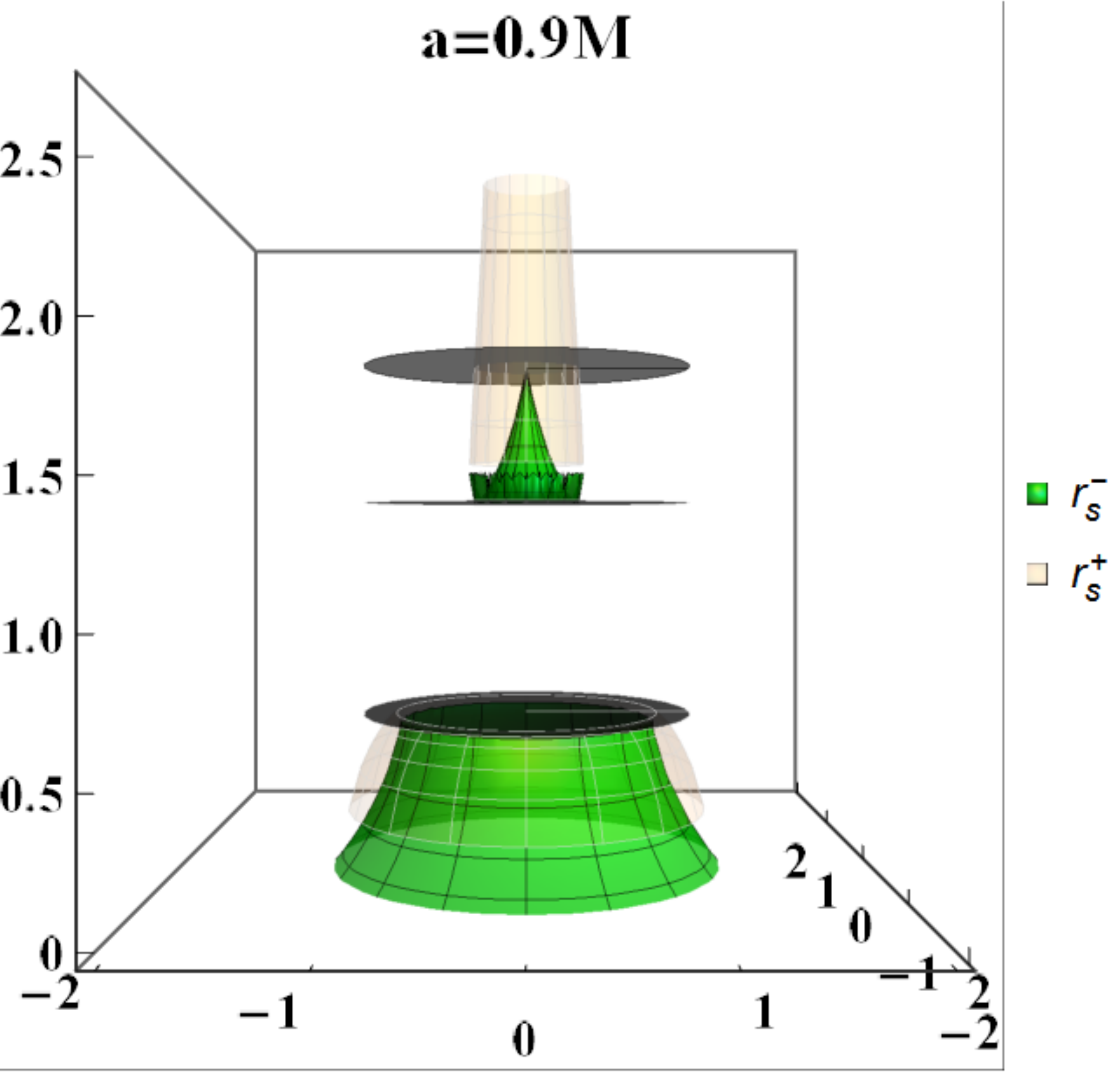}
\end{tabular}
    \caption[font={footnotesize,it}]{{Plots of the surfaces $r_{s}^{\pm}$ (in units of mass) versus the frequency
		$\omega$ for different spin values $a/M$, including \textbf{BH} and \textbf{NS} geometries.  The surfaces $r_{s}^{\pm}$ are represented    as  revolution surfaces   with height
		$r_{s}^{\pm}$ (\emph{vertical axes}) and radius $\omega$ (horizontal plane).
		Surfaces are generated by rotating  the two-dimensional curves $r_{s}^{\pm}$ around an axis
		(revolution of the  function curves $r_{s}^{\pm}$ around the ``z'' axis). Thus,
		$r=$constant with respect to the frequency $\omega$ is represented by a circle.
		The disks in the plots are either $r=M$, $r=r_\pm$(black) or $r=r_{\epsilon}^+=2M$.}}
\label{Fig:QPlot}
\end{figure*}
\begin{figure*}[ht!]
\begin{tabular}{lcr}
\includegraphics[scale=.23]{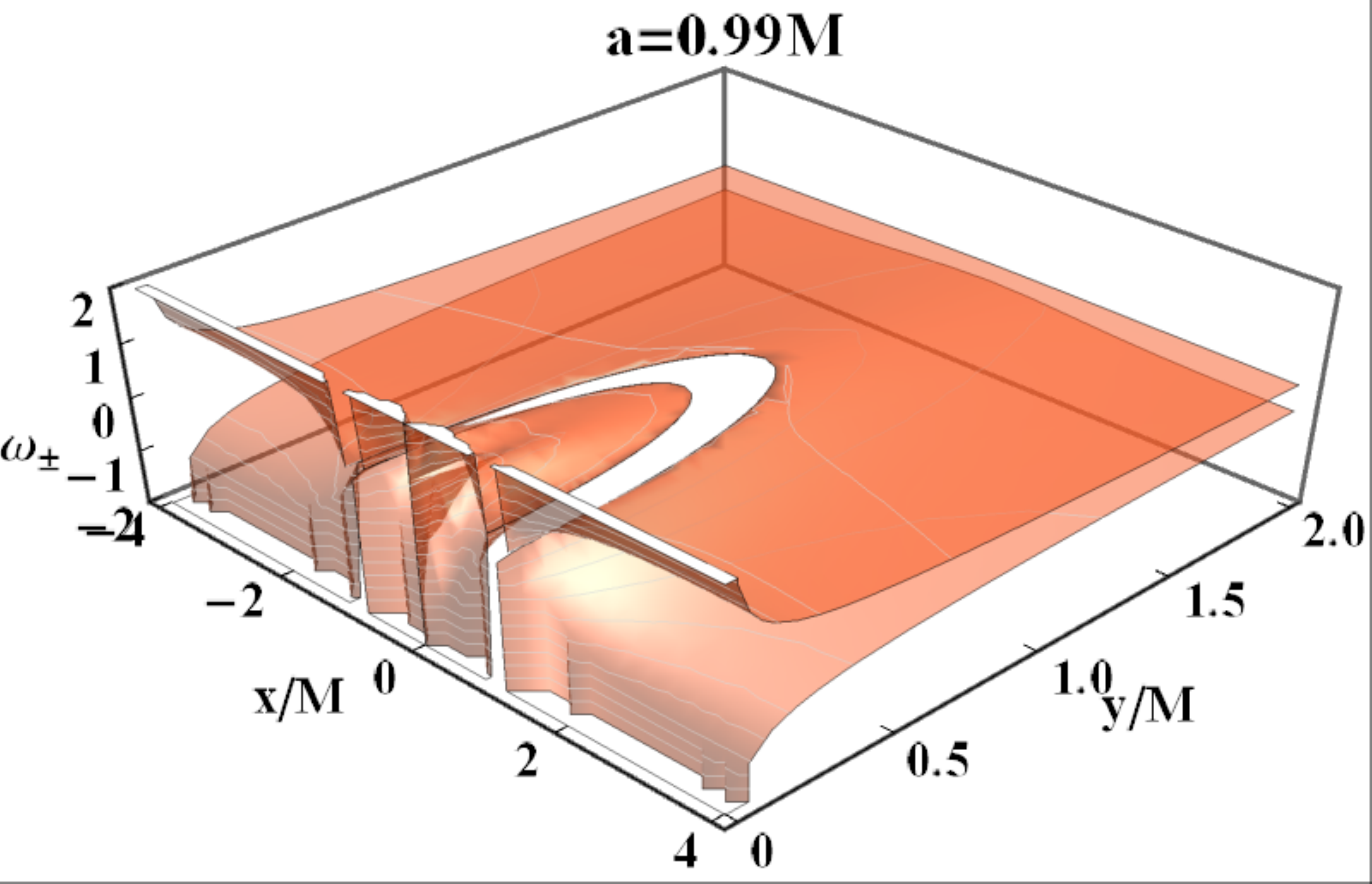}
\includegraphics[scale=.23]{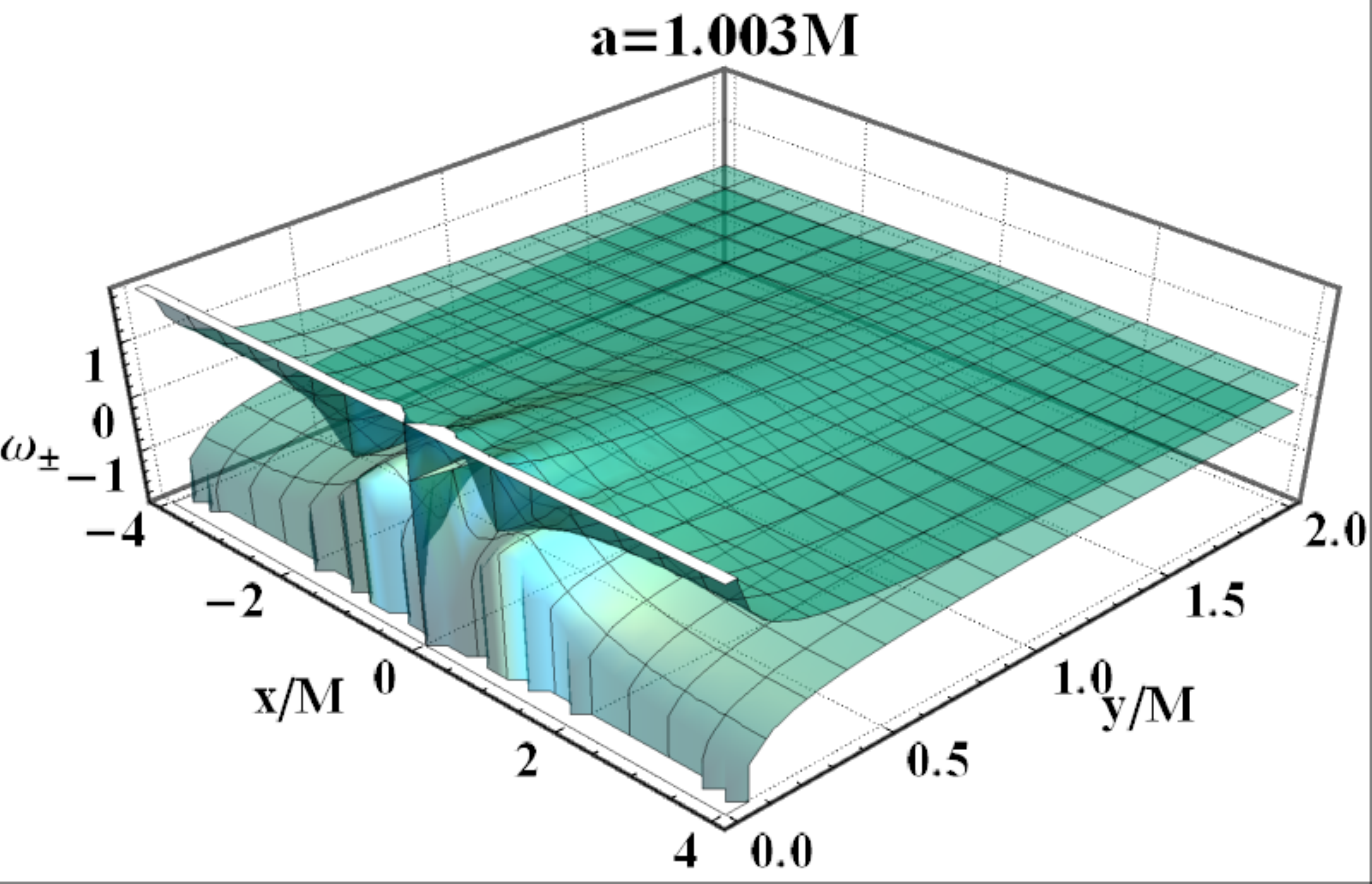}%
\includegraphics[scale=.23]{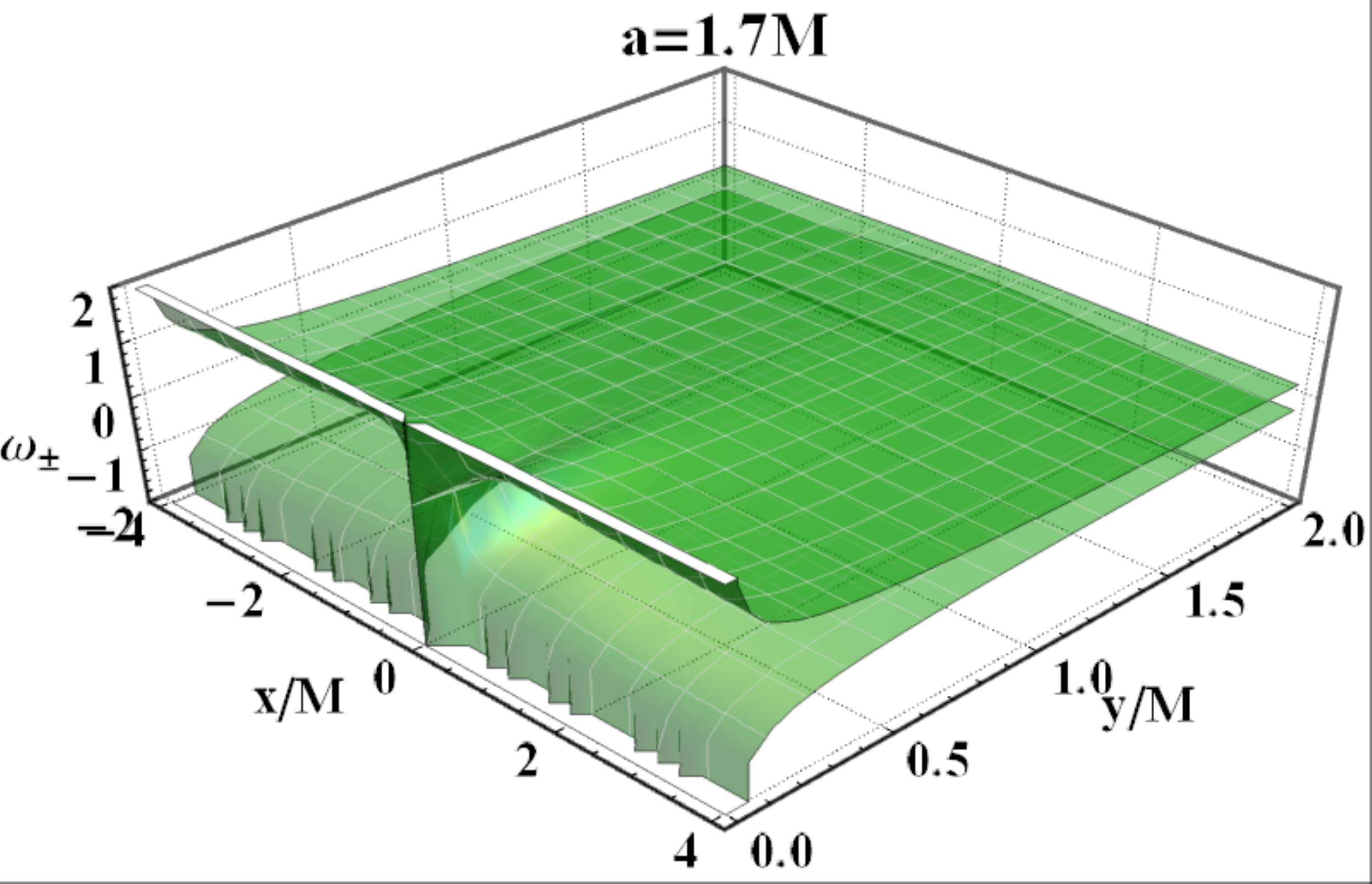}%
\end{tabular}
    \caption[font={footnotesize,it}]{Plots of the frequency surfaces $\omega_{\pm}(r,\theta)$ as functions of the radial distance $r/M$
		in Cartesian coordinates $(x,y)$ for different spin values $a$, including \textbf{BH}s and \textbf{NSs}.}
\label{Fig:QPlot1}
\end{figure*}
\begin{figure}
  \includegraphics[width=5.5cm]{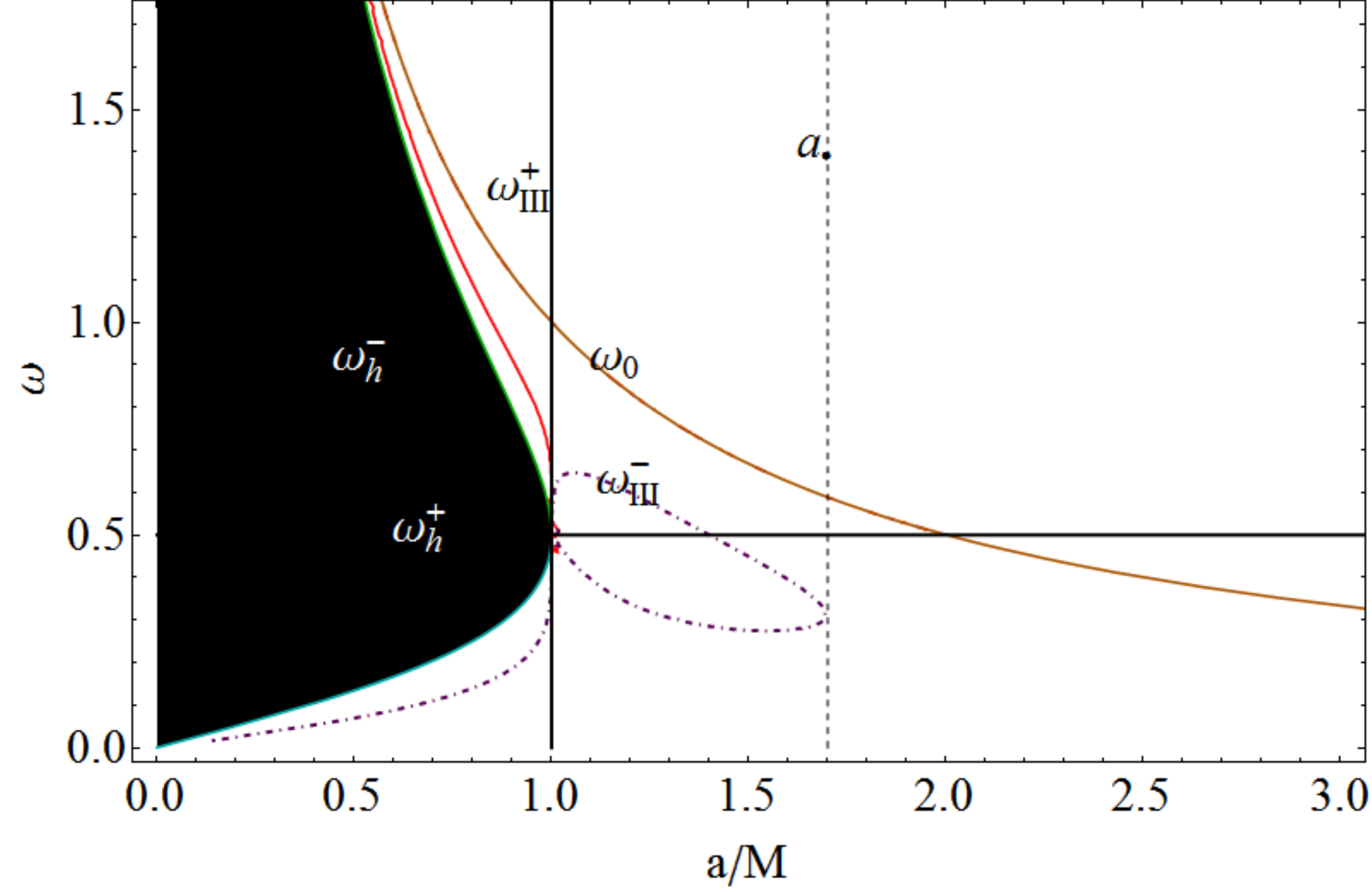}
  \includegraphics[width=5.5cm]{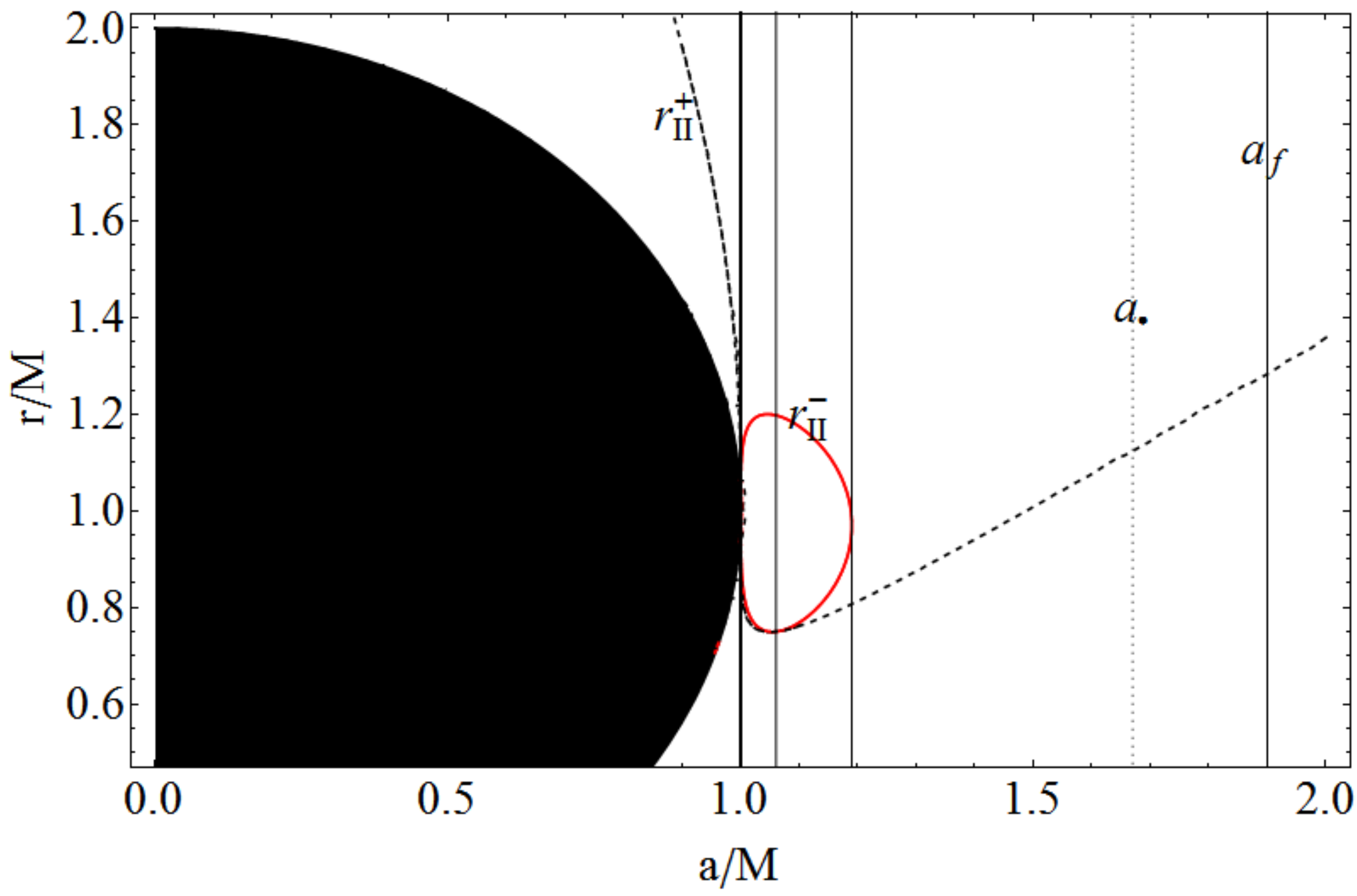}
  \includegraphics[width=5.5cm]{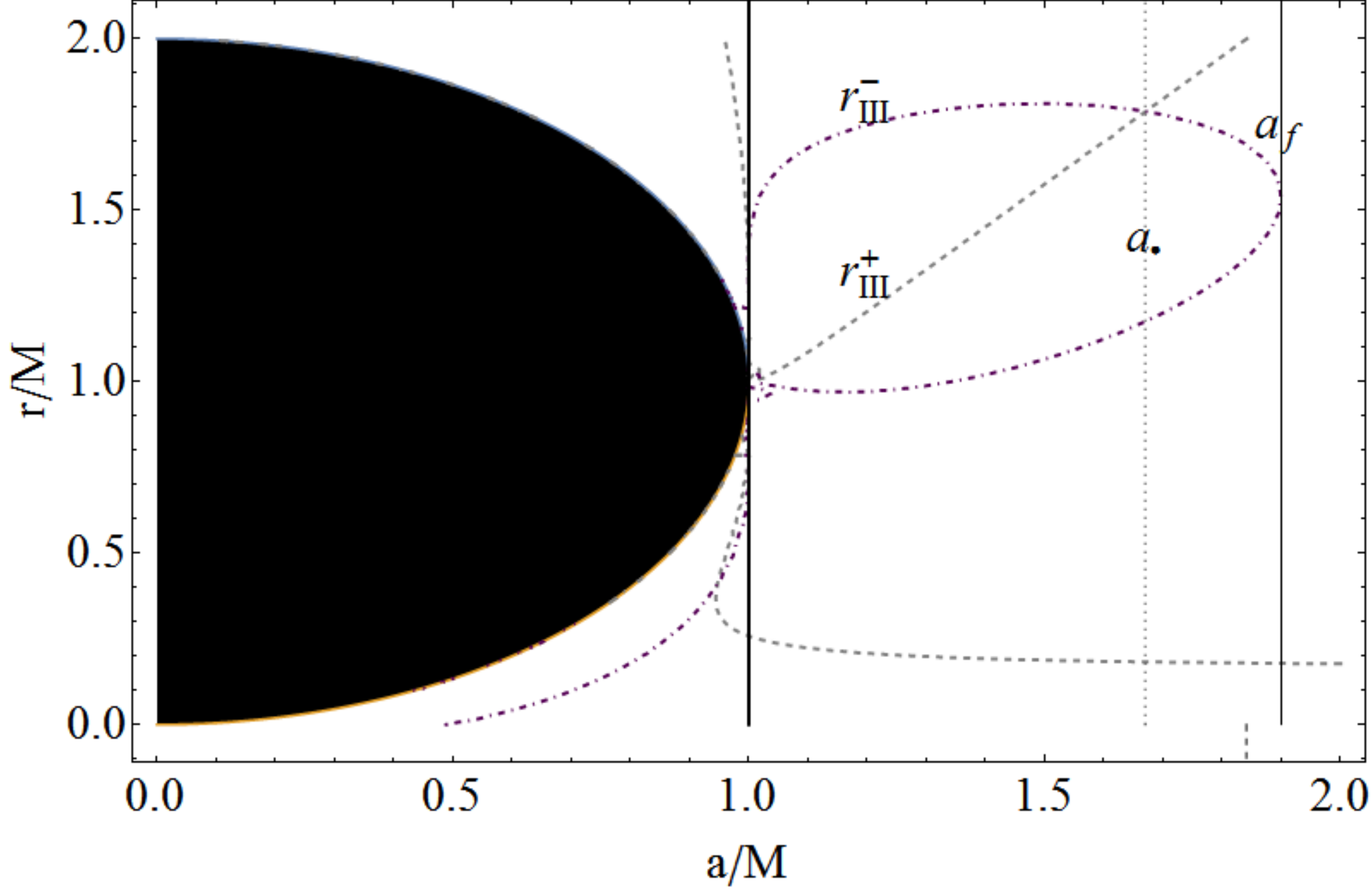}
   \caption{Center-right panels. The radii $r_{II}^{\pm}$ and $r_{III}^{\pm}$, solutions of $\partial_r^2\omega_{\pm}=0$ and
   $\partial_r^3\omega_{\pm}=0$, respectively, are shown as functions of the spin $a/M$. Black regions correspond to  $r<r_{\pm}$.
	Left panel: The frequencies  $\omega_{II}^{\pm}$ and $\omega_{III}^{\pm}$, solutions of $\partial_\omega^2r_s^{\pm}=0$ and
   $\partial_r^3r_s^{\pm}=0$, respectively, are shown as functions of the spin $a/M$.
	Regions bounded by the horizons frequencies $\omega_H^{\pm}$ are black shaded.}
	\label{Fig:MaSuorVIRMscri}
\end{figure}
The radii $r_{III}^{\pm}$ and the frequencies $\omega_{III}^{\pm}$,
as shown in  Fig.\il\ref{Fig:MaSuorVIRMscri}, define closed and limited surfaces.
This implies  that  a Killing throat can always exist,
but a Killing bottleneck  appears only for certain frequencies and values of the dimensionless spin $a/M$.
In fact,  $r_{III}^-$ is defined for   \textbf{NSs}  with spin $a\in]M, a_f]$, where $a_f=1.840M$.
On the other hand,    $r_{III}^+$ and  $\omega_{III}^+$  are not closed and   $\omega_{III}^+$  tends  to the horizon frequency
$\omega_H^->\omega_H^+$  for very small spins. This means that the Killing bottleneck  actually survives  only for  $a\in]M,a_{\bullet}]$, where, $r_{III}^-=r_{III}^+$ at $a_{\bullet}=1.668M<a_f$.
  However, the bottleneck frequencies   satisfy the inequalities
	$\omega_{III}^-<\omega_H^+<\omega_H^-<\omega_{III}^+$. Increasing the spin $a/M$, at constant mass,
	a bending of the frequency $\omega_-$ appears\footnote{Note that the  Killing bottleneck  and Killing  throat  inherit some of the properties of the Killing vectors, particularly, regarding conformal transformations of the metric or vectors.
Consider a Killing throat defined by two Killing  vectors  $(\xi_i,\xi_j)$.
The linear combination $a_\alpha\xi^\alpha$  defines a Killing vector and we can define a  Killing  field up to a conformal transformation.
In other words,
$\mathcal{L}=\xi_t+a_{\phi}\xi_\phi$ identifies a  Killing throat up to a conformal transformation.
The simplest case is when one considers a conformal expanded (or contracted) spacetime  where $\tilde{\mathbf{\xi}}^2\equiv
\tilde{\mathbf{g}}(L,L)=\Xi^2\mathbf{g}(L,L)$.
This holds also   for a conformal expanded  Killing tensor $\tilde{\mathcal{L}}\equiv \Xi \mathcal{L}$.
}.

There are several notable radii associated with the frequencies $\omega_{\pm}$ and light surfaces $r^{\pm}_s$, which are related to the extremes and saddle points  of the curves in the region  $\Sigma_{\epsilon}^+$ (see \cite{observers}).
The function   $\Delta \omega\equiv\Delta^\pm \omega=\omega_+\pm\omega_-$ is considered in  Figs.\il\ref{Fig:appmancokl}
and \ref{Fig:appmancokls}.
The radius $r_{\Delta}$, solution of   $\partial_{r}\Delta \omega=0$, clearly shows the presence of closed surfaces for
$r_{III}^{\pm}$ and  $r_{II}^{\pm}$, and provides a characterization of the Killing bottleneck.
An analysis of $\omega_{\pm}$ and $\Delta \omega$ in  Figs.\il\ref{Fig:appmancokl} and \ref{Fig:appmancokls} shows the
emergence of horizons as the  envelope surface in the plane $r-a$  of the limiting frequencies  $\omega_{\pm}$.
This  important aspect will be  addressed in details in   Sec.\il\ref{Sef:aomega}, revealing the role   played by the Killing horizons  in \textbf{BHs}-\textbf{NSs} connections.
 \begin{figure}[h!]
\centering
\begin{tabular}{lcr}
\includegraphics[scale=.223]{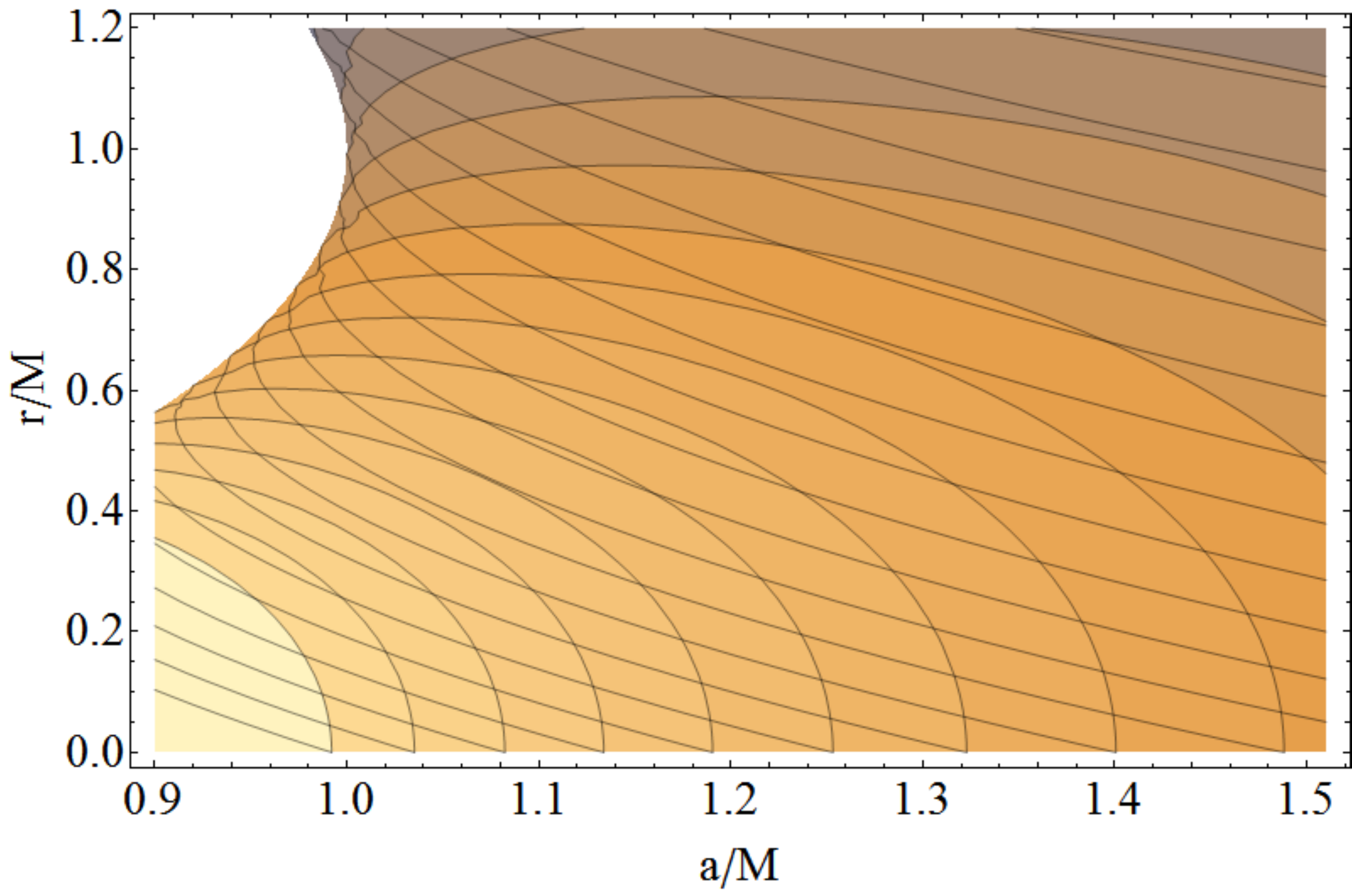}
\includegraphics[scale=.223]{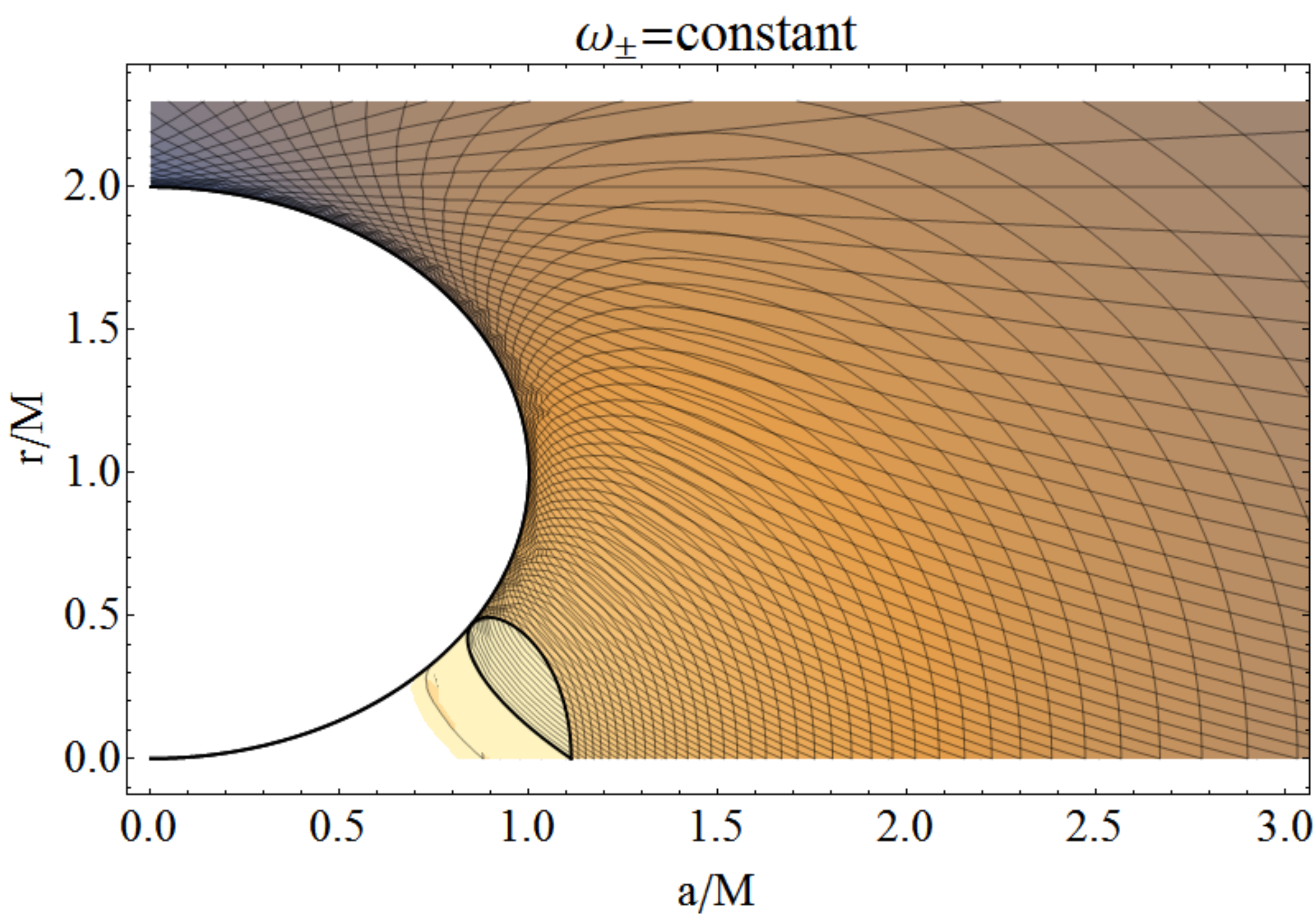}
\includegraphics[scale=.223]{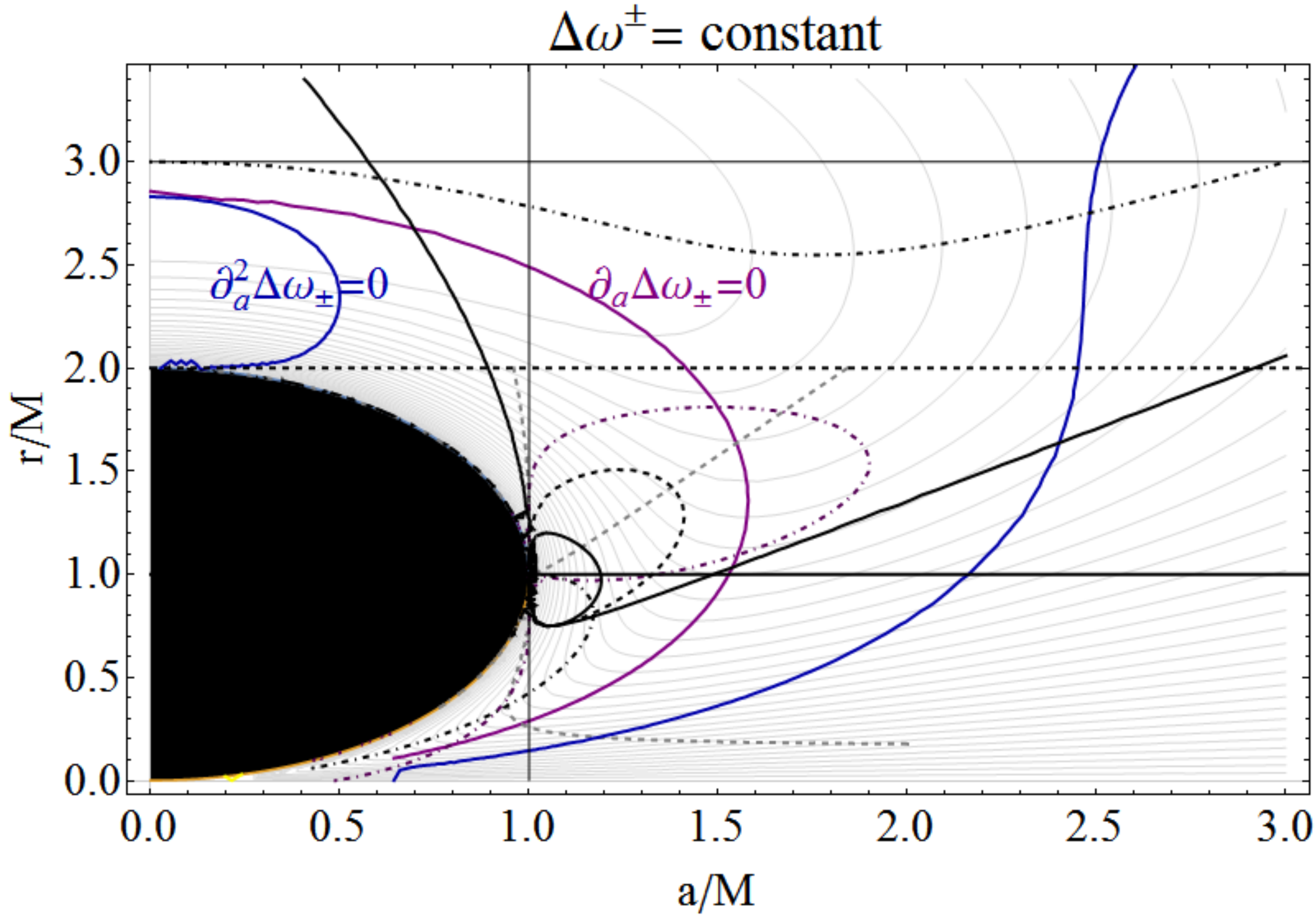}
\end{tabular}
\caption[font={footnotesize,it}]{Left and center panels: Curves $\omega_{\pm}=$constant in different geometries
with spacetime dimensionless spin $a/M$. Horizons  are black thick lines.
Right panel: Curves $\Delta^{-} \omega\equiv \omega^+-\omega^-=$constant and notable radii--see Figs\il\ref{Fig:MaSuorVIRMscri}.}
\label{Fig:appmancokl}
\end{figure}
\begin{figure}[h!]
\centering
\begin{tabular}{c}
\includegraphics[scale=.34]{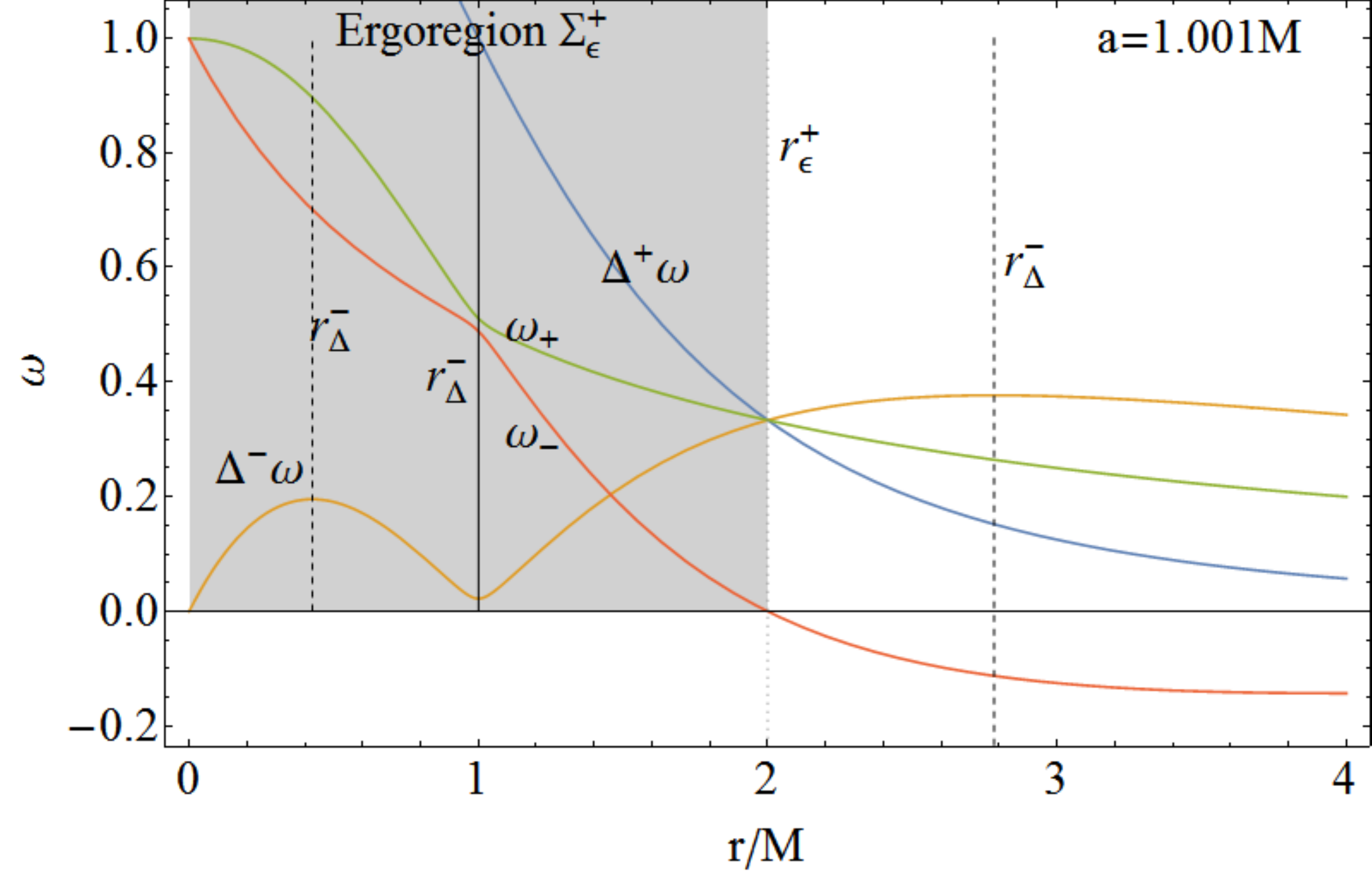}
\includegraphics[scale=.6]{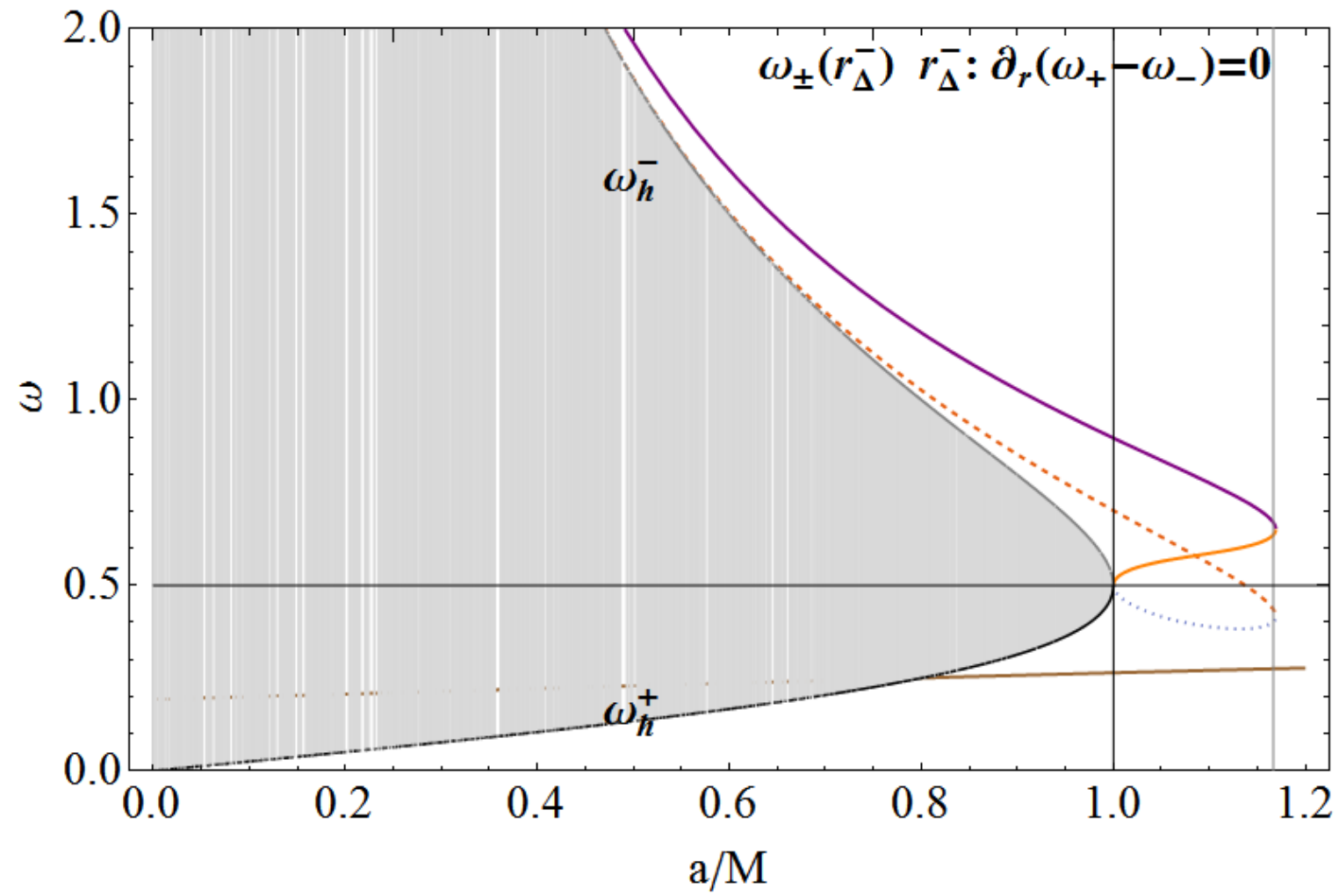}
\end{tabular}
\caption[font={footnotesize,it}]{Right panel:  Plane $\omega-r$ for a \textbf{NS} with spin $a=1.001M$. The plots are for the frequency difference $\Delta^-{\omega}\equiv \omega_+-\omega_-$,
the sum $\Delta^+{\omega}\equiv \omega_++\omega_-$,  and the Killing bottleneck for the surfaces $\omega_{\pm}$.
The radii $r_{\Delta}^-$ and  critical points of the curves $\omega_{\pm}$ are also shown.
Note the role of the static limit $r_{\epsilon}^+$ and the $r_{\Delta}^-$ in the bottleneck definition.
Right  panel: Curves $\omega_{\pm}$ on $r_{\Delta}^-$, i.e., solutions of  $\partial_r \Delta^-\omega=0$.}
\label{Fig:appmancokls}
\end{figure}
\section{Unveiling \textbf{BHs}--\textbf{NSs} connections }\label{Sef:aomega}
In this section, we explore  the entire parameterized family of  Kerr solutions
with $a/M\geq0$. To this end, we introduce the concept of extended plane $\pi_a$, where  the entire collection of metrics of a parametrized  family of solutions can be considered.

We {may} say that a  quantity $\Qa(a)$ in the plane $\pi_a$  induces a  $\Qa-a$ realization of the  extended plane, where $a$ is a parameter that characterizes the entire family of solutions. For the case considered in this work, $a$ is the dimensionless   spin parameter.
A special  {and simple} example of an extended plane  realization is given in Fig.\il\ref{Fig:SoOv},
where we  investigate  constant frequency  surfaces  defining  families of spacetime geometries that we call ``\emph{metric-bundle}s'',
$g^{\pm}_{\omega}$, labeled by a frequency parameter $\omega=$constant. Such definitions are set from  the properties of  stationary observes  and their limiting  frequencies  $\omega_{\pm}$.  In  the  extended plane,    naked singularities and black holes can belong to the  same metric bundle.
In  the extended plane of Fig.\il\ref{Fig:SoOv},  \textbf{BHs} horizons $r_{\pm}$   correspond to the  spin-curve
$a_{\pm}(r)\equiv \sqrt{r(r-2M)}$;    we shall see that in such a plane  the \textbf{BHs}  horizons $a_{\pm}(r)$  define  properties for
{\emph{all}} possible  Kerr geometries, including \textbf{BHs} and \textbf{NSs},  that unveil an interesting connection between
  \textbf{BHs} and \textbf{NSs}.
	
 \medskip

 \textbf{Metric bundles  $g_{\omega}^{\pm}$}

Here, we specify the idea of metric bundles for the Kerr family of solutions. Solving Eq.\il(\ref{Eq:b-y-proc}) for the spin $a$,
we obtain the following two  quantities
\bea\label{Eq.lcospis}
a_{\omega}^{\pm}(r,\omega;M)\equiv\frac{2 M^2 \omega \pm\sqrt{r^2  \omega ^2 \left[M^2-r (r+2M)  \omega ^2\right]}}{(r+2M)  \omega ^2},
\eea
which are plotted  in Figs\il\ref{Fig:SoOv} and \ref{FIG:SPlotooc} as  functions of $r/M$, for different values of the
frequency $\omega$.  Note  that in the region  $r>r_{\epsilon}^+$,   negative orbital frequencies  are possible because  they are  associated  to the retrograde (counterrotating) motion with respect to the central object; this  fact    implies  the possibility of  negative  values of $a_{\omega}^{\pm}$  for $\omega>0$--see  Figs.\il\ref{Fig:SoOv}. However, in this section,  we restrict our  analysis to the ergoregion $\Sigma_{\epsilon}^+$, where $a>0$ and $\omega>0$--see Figs.\il\ref{Fig:SoOv} and  \ref{FIG:SPlotooc}.
Each   spacetime  of the Kerr family  is represented (restricted to the equatorial plane)  in the extended plane of  Fig.\il\ref{Fig:SoOv} by a constant surface
$a_{\omega}^{\pm}/M=$constant (horizontal lines in Fig.\il\ref{Fig:SoOv}).

The metric bundles $g_{\omega}\equiv g_{\omega}^{\pm}$ are defined by the curves $\left.a_{\omega}^{\pm}\right|_{\omega}$  of constant frequency $\omega$ in $\pi_a$-- Fig.\il\ref{Fig:SoOv}.
Each  $g_{\bar{\omega}}=\left.a_{\omega}^{\pm}\right|_{\bar{\omega}}$ for a fixed frequency $\bar{\omega}$ is represented by closed and bounded curves which are continuous   almost everywhere  in $\pi_a$.
Below, we will discuss extensively the properties of these curves.

The bundles $g_{\bar{\omega}}$ contain  an   (almost) continuum and infinite set of metric parameter   values $a/M$. Each value of
$a/M$ sets a specific Kerr geometry. From Fig.\il\ref{Fig:SoOv}, it is clear that eventually some bundles contain both \textbf{BHs} and \textbf{NSs} spacetimes, others define only  \textbf{BHs}, while none of the bundle is constituted by \textbf{NS} only.
In fact,  \emph{all} the metric bundles are tangent at least in one point to  the horizons $a_{\pm}$.
Thus, from the quantities  (\ref{Eq.lcospis}) an alternative definition of \textbf{BHs}  Killing horizons  $r_{\pm}$ emerge.
Indeed, from
 Fig.\il\ref{Fig:SoOv}, it follows that the horizons  $a_{\pm}$ arise as   the envelope surfaces of the curves
$a_{\omega}^{\pm}(r)$, i.e., the metric bundles $g_{\omega}=\left. a_{\omega}^{\pm}\right|_{\omega}$ in $\pi_a$.
This will be a crucial property of the metric bundles with significant consequences that allow us to connect
\textbf{NSs} to \textbf{BHs}  in the extended plane. In fact, since the  curve  $a_{\pm}$ is  tangent to all   metric bundles
$g_{\omega}$, the (inner and outer) horizons   contain \emph{\emph{all}} the frequencies $\omega$ defining  each  metric bundle and, therefore, describing both \textbf{NSs} and \textbf{BHs} in the extended plane.
\begin{figure}[h!]
\centering
\begin{tabular}{lcr}
\includegraphics[scale=.6]{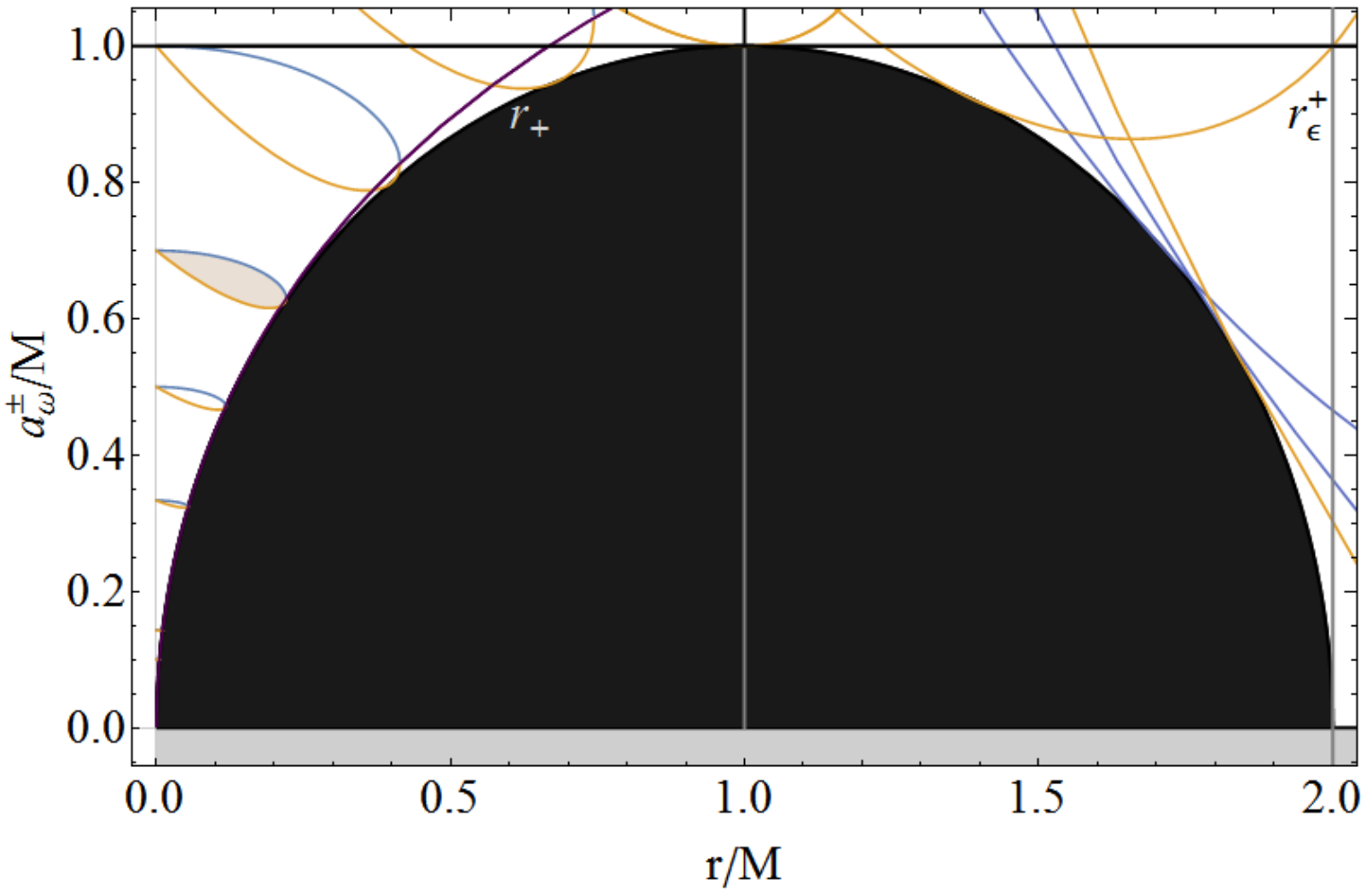}
\includegraphics[scale=.6]{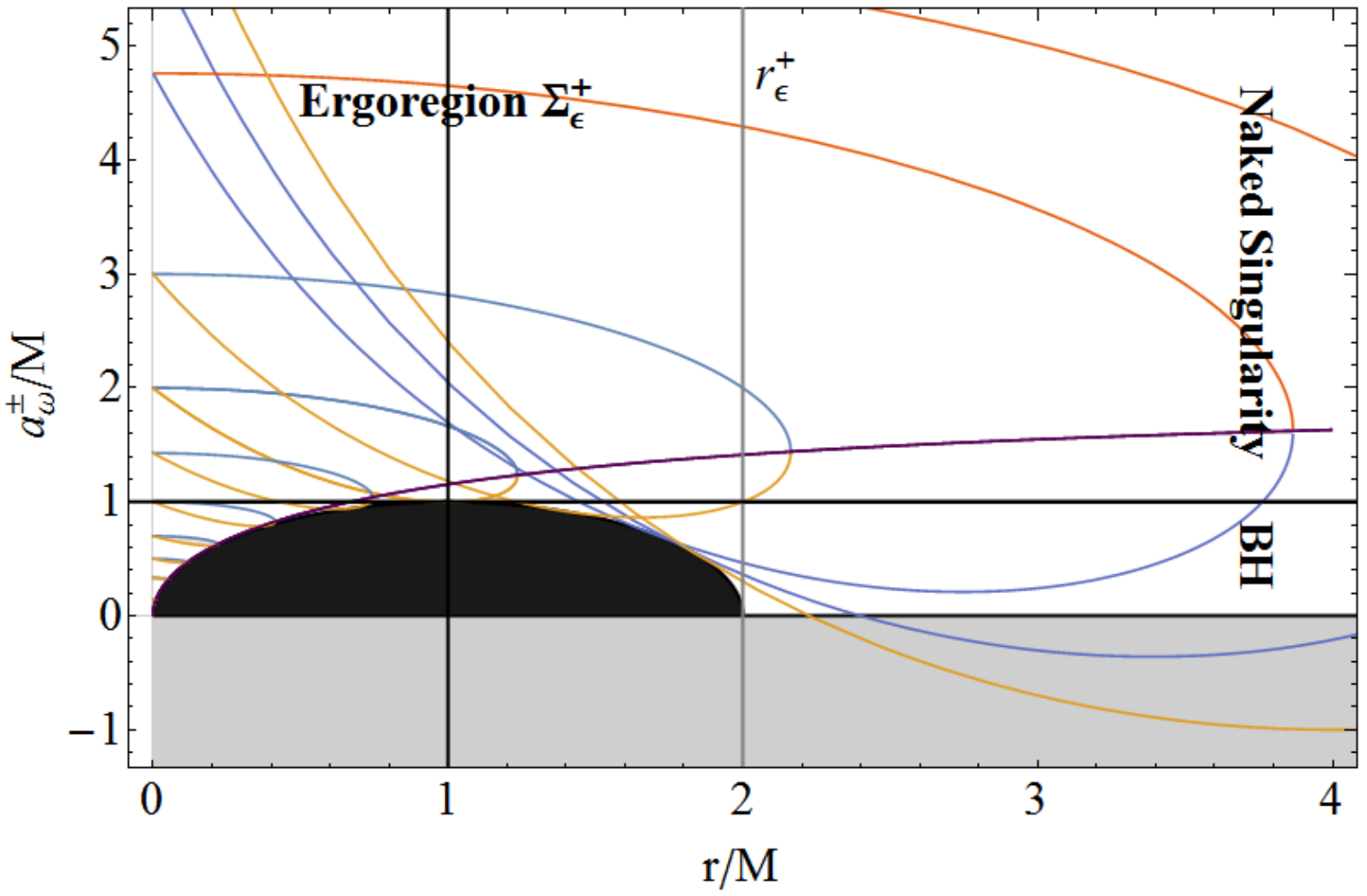}\\
\includegraphics[scale=.67]{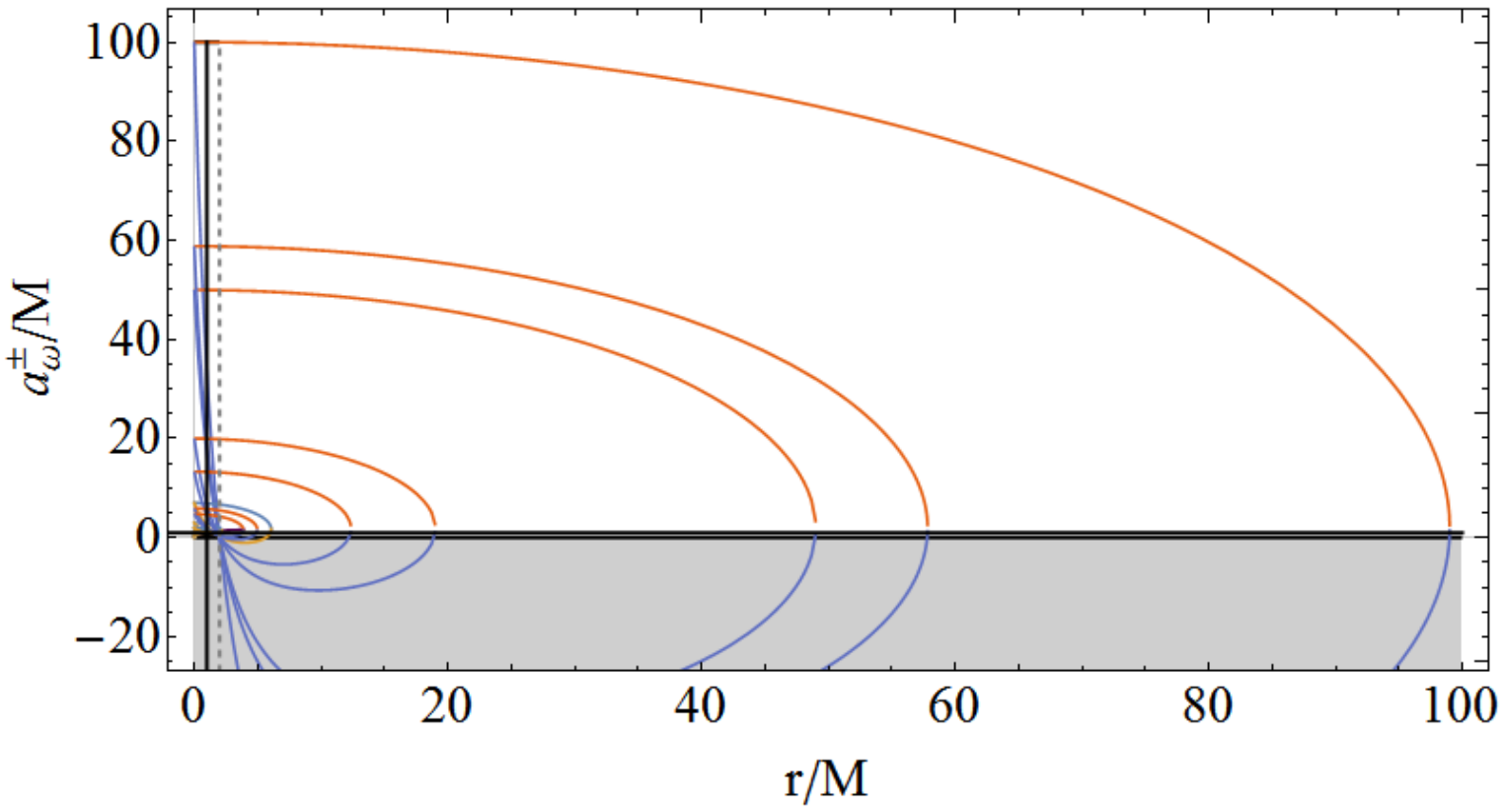}
\end{tabular}
\caption[font={footnotesize,it}]{Spins $a_{\omega}^{\pm}$ versus $r/M$ for different frequencies. The black region is for $r<r_+$ and $r_+$ is the \textbf{BH} outer horizon. See also Fig.\il\ref{FIG:SPlotooc}  for a $3D$ representations of these regions.}
\label{Fig:SoOv}
\end{figure}
\begin{figure}
  \includegraphics[width=3.4cm]{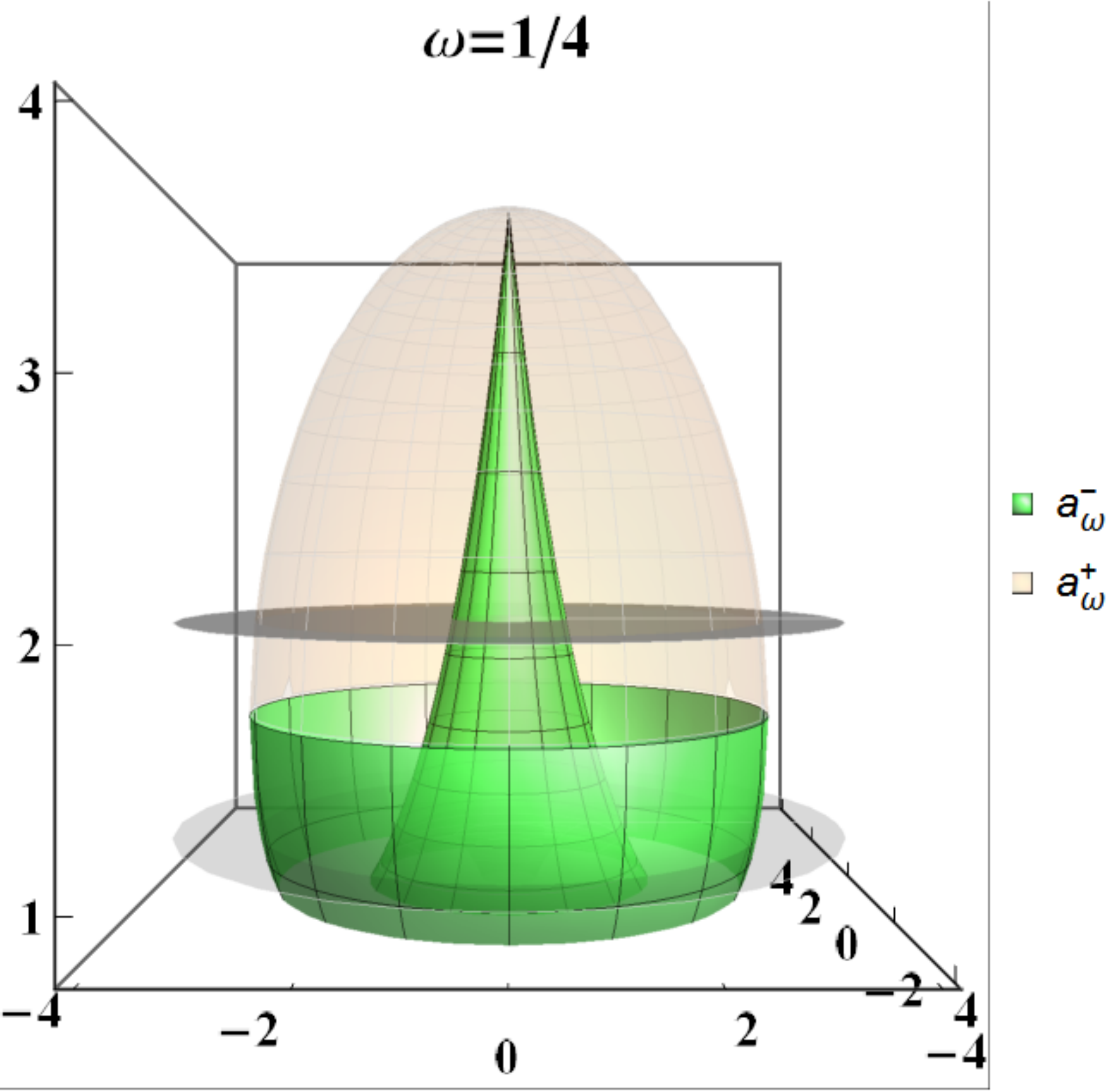}
   \includegraphics[width=3.4cm]{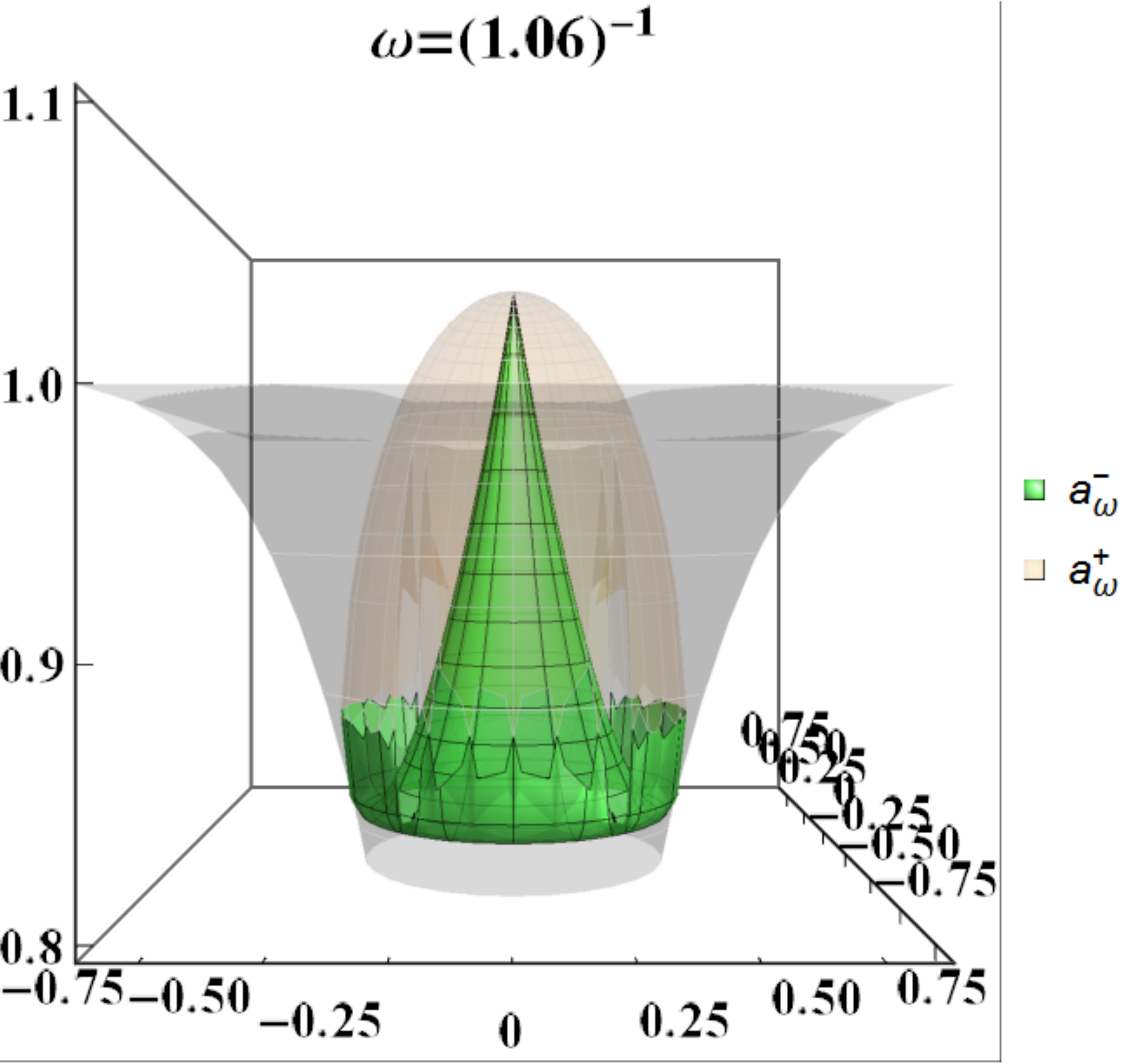}
   \includegraphics[width=3.4cm]{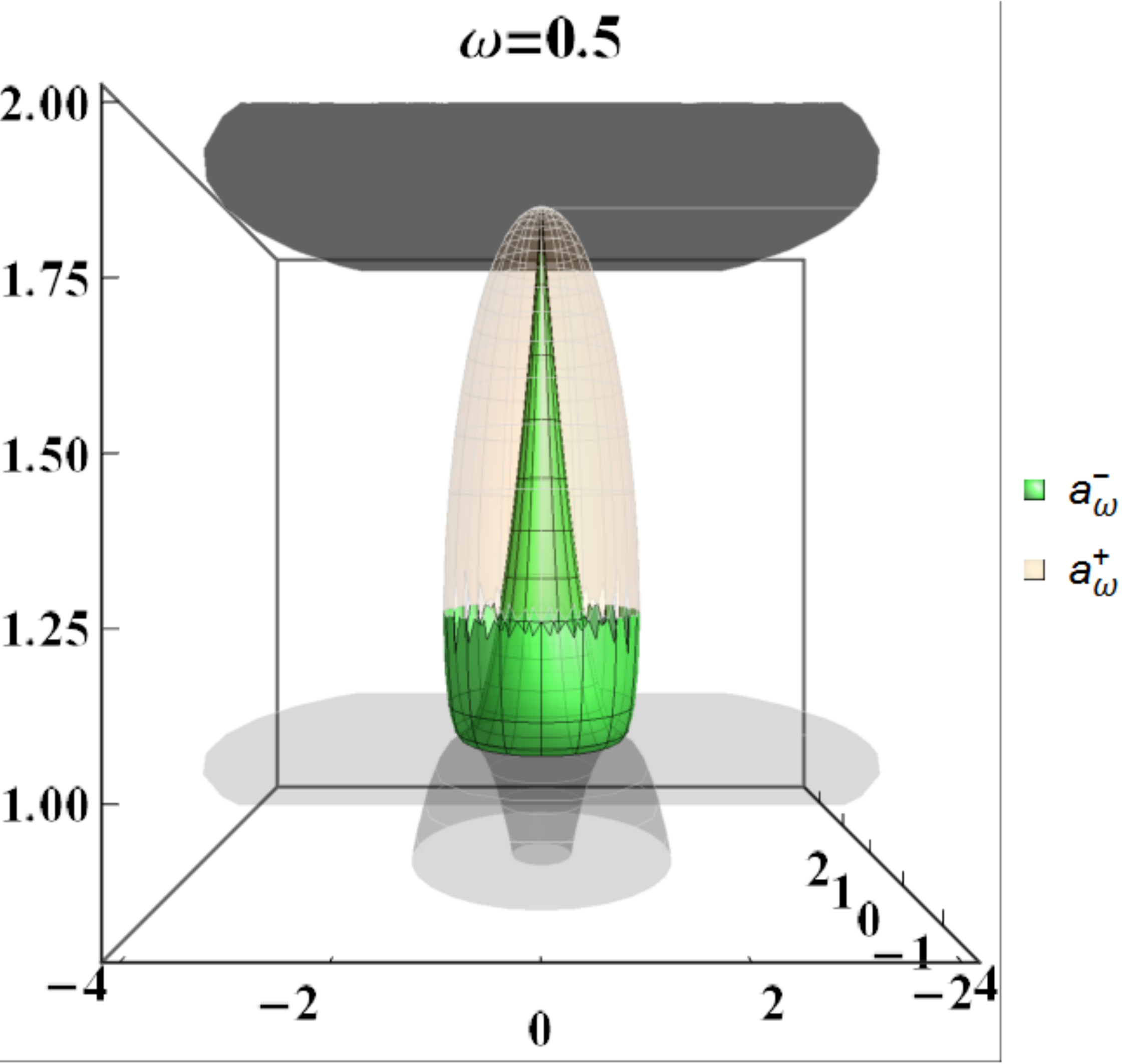}
   \includegraphics[width=3.4cm]{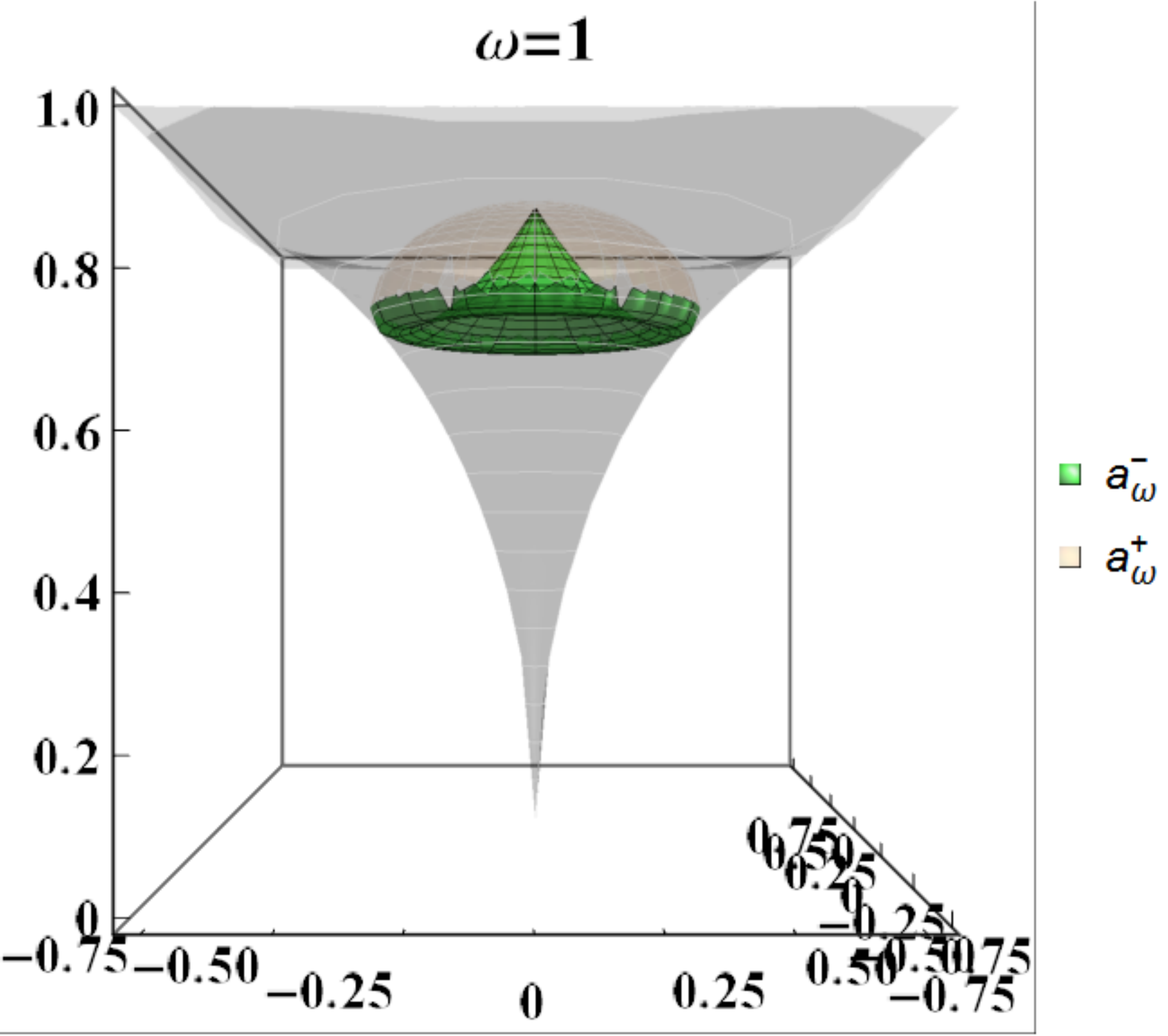}
     \includegraphics[width=3.4cm]{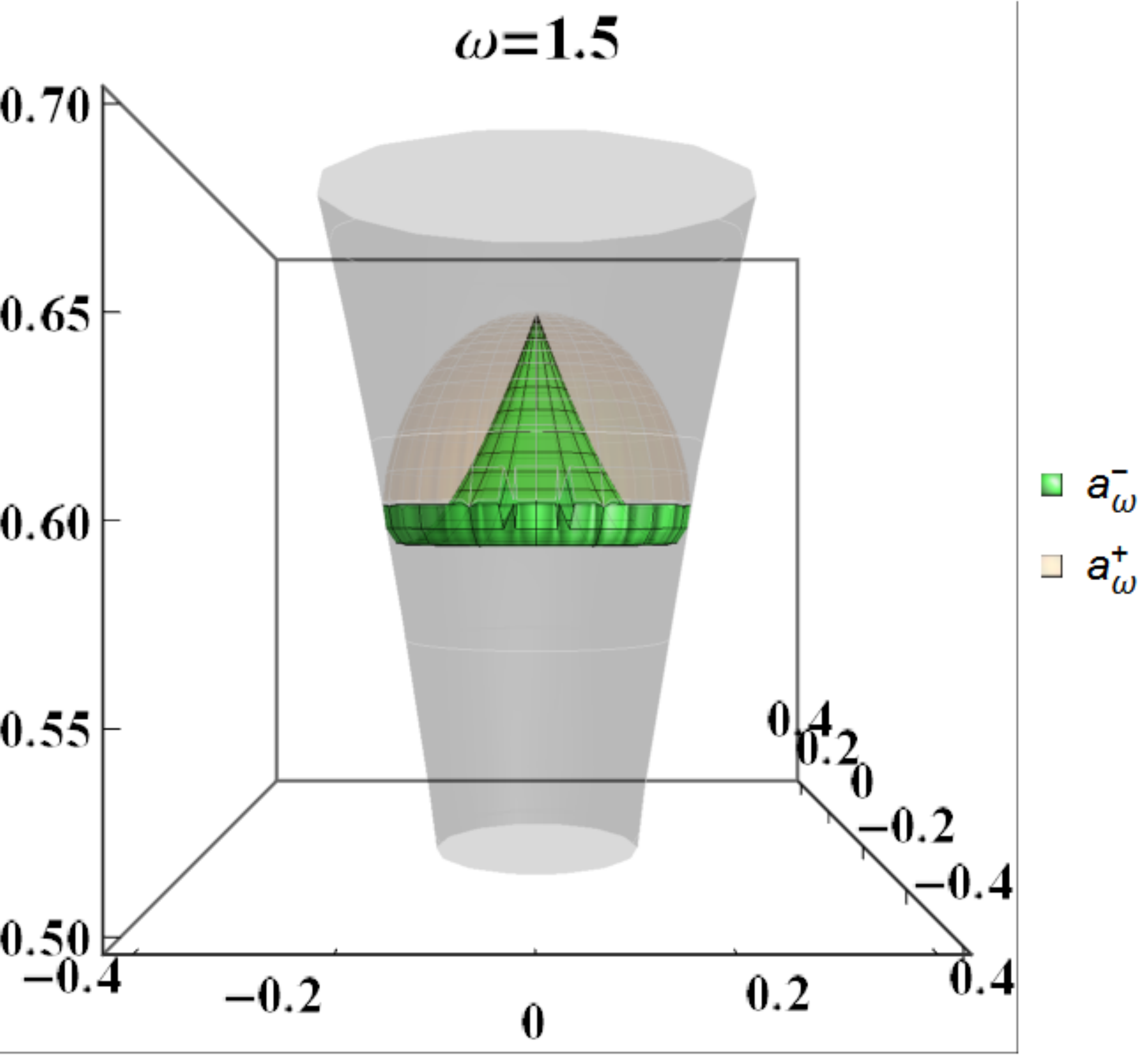}
  \caption{Kerr geometries: Surfaces $a_{\omega}^{\pm}$ (in units of mass) versus the radius $r/M$  for different frequencies
		$\omega$, including \textbf{BH} and \textbf{NS} geometries--see also Eq.\il(\il\ref{Eq.lcospis}).
$3D$ representations of $a_{\omega}^{\pm}$ versus $r/M$ for different frequencies--see also Fig.\il\ref{Fig:SoOv}.
Each surface is a metric bundle $g_{\omega}$ for a fixed value of $\omega_0=$constant. The corresponding metric bundle origin is
$a_0=M/\omega_0$. The surfaces $a_{\omega}^{\pm}$ are represented    as  revolution surfaces   with height
		$a_{\omega}^{\pm}$ (\emph{vertical axes}) and radius $r/M$ (horizontal plane).
		Surfaces are generated by rotating  the two-dimensional curves $a_{\omega}^{\pm}$ around an axis
		(revolution of the  function curves $a_{\omega}^{\pm}$ around the ``z'' axis). Thus,
		$a=$constant with respect to the radius $r/M$ is represented by a circle under this transformation. The disks in the plots are
		either $a=M$  or $r=r_{\epsilon}^+=2M$. The surfaces $a_{\omega}^{\pm}$ are green and pink colored, respectively (as mentioned in  the legend). Horizon surfaces determined by $a_{\pm}=\sqrt{r(2M - r)}$ are gray surfaces. For each fixed frequency $\omega=\overline{\omega}$ there is an associated spin $\bar{a}\equiv \bar{\omega}^{-1}$ (see the discussion in Sec.\il\ref{Sef:aomega}).
		In the \textbf{BH} range, the surfaces $a_{\omega}^{\pm}$ are  in the region $r<r_-$, where $r_-$ is the inner horizon.  }
		\label{FIG:SPlotooc}
\end{figure}
\begin{figure}[h!]
\centering
\begin{tabular}{c}
\includegraphics[scale=.324]{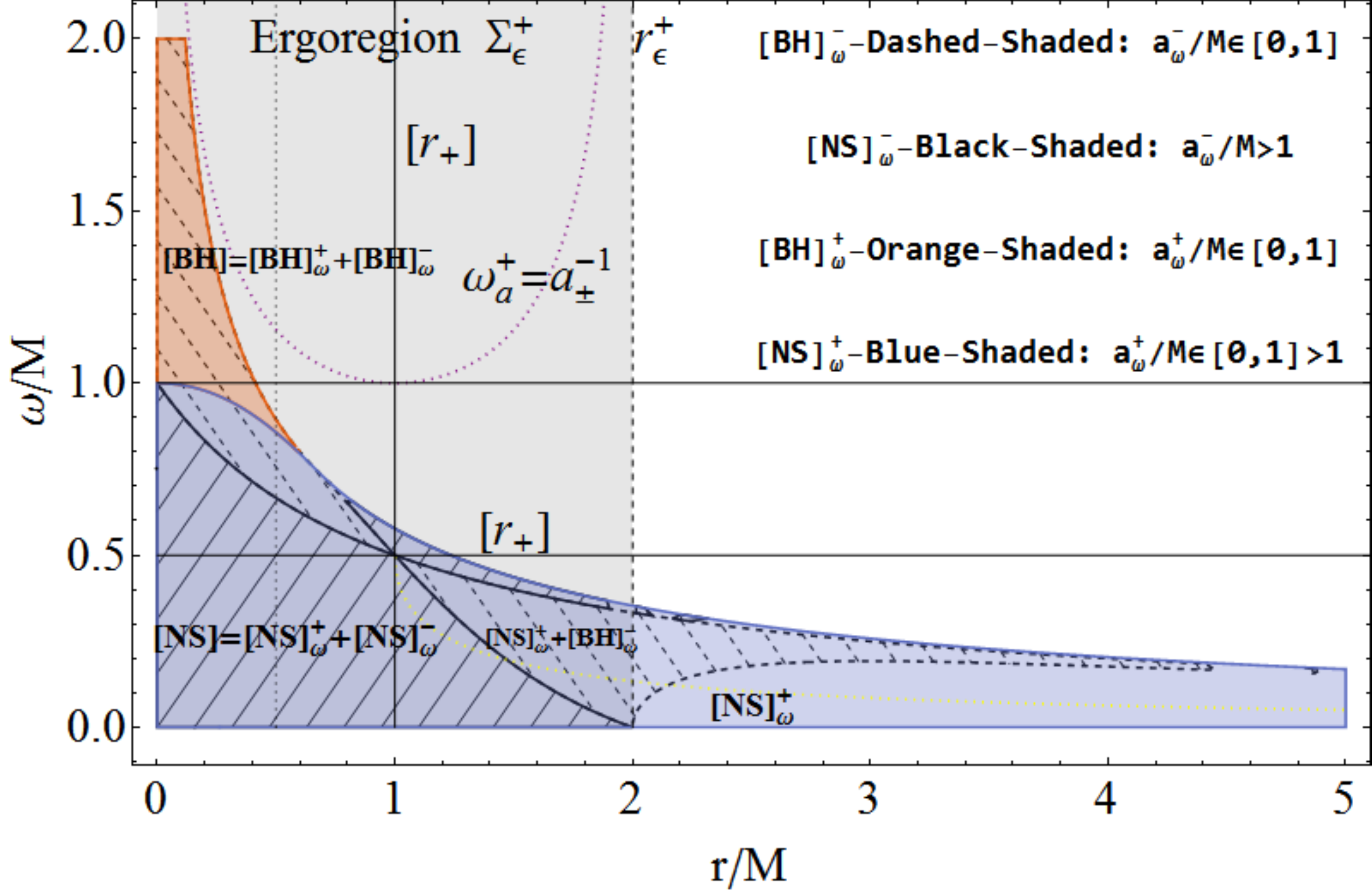}
\includegraphics[scale=.324]{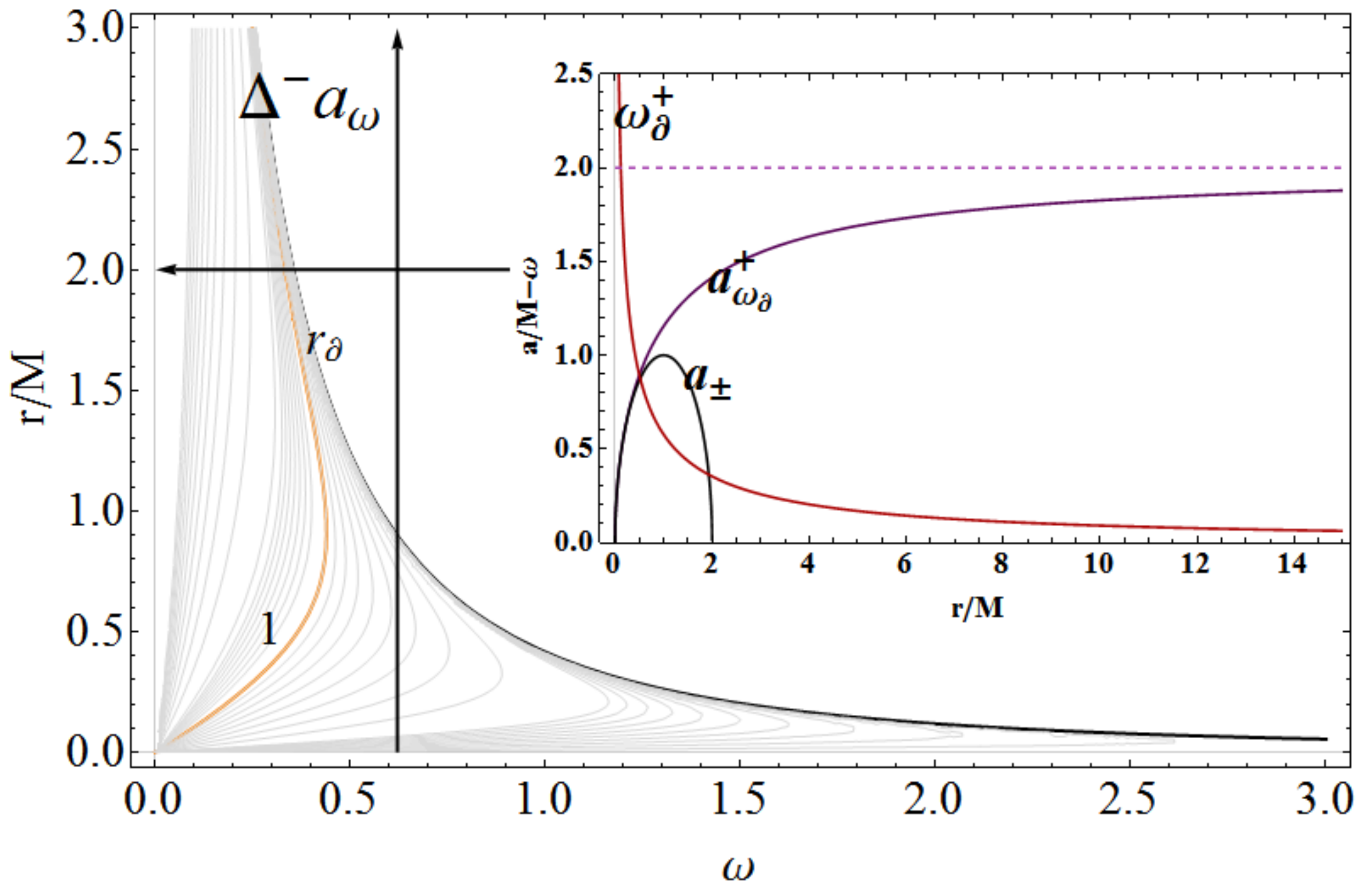}
\end{tabular}
\caption[font={footnotesize,it}]{Left panel: {Regions} $a_{\omega}^{\pm}>M$  (\textbf{NS}) and  $a_{\omega}^{\pm}\in[0,M]$  (\textbf{BH}) versus $r/M$. The ergoregion $\Sigma_{\epsilon}^+$, the \textbf{NS} and the \textbf{BH} regime are shown.
Right panel: $\Delta^- a_{\omega}\equiv a_{\omega}^+-a_{\omega}^-=$constant  in the plane $r-\omega$. The arrows indicate increasing values of $\Delta^- a_{\omega}$. The curve $a=M$ and $r_{\partial}(\omega)$ are shown.
Inside panel: The frequency $\omega_{\partial}^+$, spins $a_{\pm}=\sqrt{r(r-2M)}$, which define the \textbf{BHs} horizons and
$a_{\omega _{\partial }}^+$ defined in Eq.\il(\ref{Eq:rela-a-partialv}) as functions of $r/M$. }
\label{Fig:Tactic}
\end{figure}

From Fig.\il\ref{Fig:SoOv}, it follows that a metric bundle  has its \emph{origin} $a_0$ on the vertical axis $r=0$ of $\pi_a$ and closes on the tangent point to $a_{\pm}$. The  closeness  of the metric bundles  is due to  the spacetime rotation.
To investigate the solutions (\ref{Eq.lcospis}), which define the closed bundles, we solve  the condition
$a_{\omega}^+=a_{\omega}^-$  and find
\bea\label{Eq:rela-a-partialv}
&& a_{\omega}^+=a_{\omega}^-\quad \mbox{on} \quad \frac{r_{\partial}^{\pm}(\omega)}{M}=\pm\frac{\sqrt{\omega ^2+1}}{\omega }-1 \quad\mbox{where}\quad  a_{\omega}^\pm( r_{\partial})\equiv a_{\partial}= 2 \omega r_{\partial}\quad\mbox{and}
 \\
 &&\nonumber a_{\omega_\partial}^+= \frac{2 M \sqrt{r}}{\sqrt{r+2M}}= 2 r \omega_{\partial}^+\quad \mbox{where}\quad
\omega_{\partial}^+= \frac{M}{\sqrt{r(r+2M)}}.
\eea
The frequency $\omega_{\partial}^+$  of  Eq.\il(\ref{Eq:rela-a-partialv}) is a solution of the  equation
$r_{\partial}^+=r$ (see Fig.\il\ref{Fig:Tactic});
the function $\omega_{\partial}^+(r)$  is, therefore, the  frequency associated to  the orbits $r_{\partial}^+$,  where
$a_{\omega}^+=a_{\omega}^-$ defines the metric bundle.
 In fact, $\omega_{\partial}^+(r)$ does not depend on the  spacetime spin because the information on the corresponding geometry can be extracted  from
$\pi_a$   through   $r_{\partial}^+$, i.e.,  from  the condition $a_{\omega}^+=a_{\omega}^-$.
Consequently, $\omega_{\partial}^+$ is a function of the radius  $r=r_{\partial}^+$. The pairs
$(r_{\partial}^+,\omega_{\partial}^+(r))$ identify the corresponding spin  origin $a_0$ (also {origin} of the metric bundle), which is defined by the  frequencies
$\omega_{\partial}^+(r)$  and  $a_{\omega}^+=a_{\omega}^-$  at $r_{\partial}^+$--Fig.\il\ref{Fig:Tactic}.
Asymptotically, for very large values of $r/M$, the value $\omega_{\partial}^+=0$ is approached as shown in Fig.\il\ref{Fig:SoOv}, where
 $a_{\omega}^+=a_{\omega}^-$ is valid  on the line  $a=0$, that is, approaching the limiting geometry of the static and spherically symmetric Schwarzschild spacetime.

 We note that $r_{\partial}^-$ is negative for positive values of the frequency. As we are  restricting our analysis in this work to the case $\omega>0$, we shall not consider $r_{\partial}^-$; nevertheless, an analysis of the case $\omega<0$ would, in fact, provide additional information about the spacetime structure even in the equatorial plane. 
A  very small $\omega_{\partial}^+$,   on the other hand, corresponds to  a very large (origin) spin $a_0=M/\omega$.
Note that the spin $a_0$  corresponds to the frequency $\omega_0=M/a$, which was introduced in Eq.\il(\ref{Eq:b-y-proc})
by considering the behavior of the stationary observer frequencies near the  singularity $r=0$; this is of importance in \textbf{NS} geometries as described in Sec.\il\ref{Sec:bottlem}.
The properties of this special frequency $\omega_0$ have also been extensively discussed in \cite{observers}.
 More generally, as   noted in Secs.\il\ref{Sec:1-st} and \ref{Sec:bottlem},
the dimensionless spin parameter $a/M$ is related to the quantity $M/\omega$ though  the frequencies of the light surfaces
(see \cite{observers}).  Then, the function  $a_{\omega_\partial}^+(r)$  in Eq.\il(\ref{Eq:rela-a-partialv}) is obtained as
$a_{\omega}^{\pm}(r,{\omega_{\partial}^+})$.
As shown in  Fig.\il\ref{Fig:Tactic}, the condition  $a_{\omega_\partial}^+(r)=a_{\pm}$ is valid only at the  spacetime
singularity $r=0$;  otherwise,  $a_{\pm}<a_{\omega_\partial}^+(r)<2M$, while the condition
$ a_{\omega_\partial}^+(r)= 2M$  (a \textbf{NS})  is reached asymptotically for  large  values of $r/M$.

Note that the asymptotic limit of $ a_{\omega_\partial}^+(r)=2M$  is relevant as it corresponds to
a metric bundle $g_{\omega}^{\pm}$ at constant frequency  $\omega=0.5$, which defines the point  of the envelope corresponding to the extreme Kerr  \textbf{BH} solution $a=M$.
Figure \ref{Fig:Slimers} refers to the analysis of  Fig.\il\ref{Fig:SoOv} and  sketch the correspondence between
\textbf{BHs} and \textbf{NSs}  derived from the analysis of  $a_{\omega}^{\pm}$.
The envelope  $a_{\pm}$ of   the  $a_{\omega}^{\pm}$  curves is  defined as the set of points $(a/M,r/M)$ for which
$\partial_{\omega}a_{\omega}^{\pm}=0$, i.e.,  as
the curve tangent to all  $a_{\omega}^{\pm}$ or also as the boundary of the region filled by the curves $a_{\omega}^{\pm}$.
 Then,  small changes of $a$  and  shifts along the orbit radius $r$ leave $\omega$ almost constant as $a_{\omega}^{\pm}$ are continuum functions.

The relation between the  radius  $r_{\partial}^{\pm}$,  the spin $a_{\partial}^{\pm}$ and  frequencies $\omega_{\pm}$ has to be confronted with Eq.\il(\ref{Eq:freq-BH}) for the frequencies
$\omega_{H}^{\pm}$ at the horizons and  radii  $r_\pm$ of the \textbf{BH} Killing horizons.
These quantities play  an  important role  also for the Killing bottleneck emergence  considered in Sec.\il\ref{Sec:bottlem}, as the   surfaces $a_{\omega}^{\pm}(\omega)=$constant  are related  to the solutions $\omega_{\pm}(a)$ of Eq.\il(\ref{Eq:b-y-proc})  and shown in  Fig.\il(\ref{Fig:CORESUNO}). The
relation  $r_{\partial}^{\pm}\omega=a_{\omega}^{\pm}/2$  of Eq.\il(\ref{Eq:rela-a-partialv})   is also used in Fig.\il\ref{Fig:Slimers}
to unveil some \textbf{BHs} and \textbf{NSs}  properties: The  points  on the lines $a_{0}=$constant for $a_0\in]M,a_{\bullet}]$
 lead to the  Killing bottleneck emergence of Fig.\il\ref{Fig:CORESUNO}.

\medskip

\textbf{BHs and NSs in  metric bundles $g_{\omega}$}

As discussed  above, a  metric bundle $g_{\omega}$ can comprise only \textbf{BHs} or \textbf{BHs} \emph{and} \textbf{NSs}. Moreover,
the horizons describe \textbf{BHs} \emph{and} \textbf{NSs} in the extended plane.
In the remaining of this section, we describe this last aspect and the \textbf{BHs}-\textbf{NSs} relation more closely.

Firstly,  Fig. \il\ref{Fig:Tactic}-left represents  \textbf{BHs} and  \textbf{NSs}  in the  $\omega-r$  plane.
 The plane is divided into regions, where the metric bundles  $g_{\omega}$ include \textbf{BHs} or \textbf{NSs};
there are regions with only \textbf{BHs} or  \textbf{NSs} and transition regions that cross different sections and are
connected to  Killing horizons and bottlenecks--see also  Fig.\il\ref{Fig:CORESUNO}.
  A special transition  region is, for example, around the extreme \textbf{BH} horizon
$r=M$  with frequency value $\omega=1/2$, which corresponds to  the extreme Kerr   \textbf{BH}.

We concentrate  on the  restricted  $\pi_a^+\subset\pi_a$ plane determined by  $a\geq 0$  in Figs.\il\ref{Fig:SoOv}.
The restriction of the extended plane  $\pi_a$ to $\pi_a^+$ exploits the symmetry by reflection around the axis  $a=0$,
 where negative origins of metric bundles build up the  horizons $-a_{\pm}<0$.
The plane has the following   remarkable  sections: $\mathcal{P}_\mathcal{S}$, $\mathcal{P}_\mathcal{L}$,
$\mathcal{P}_\mathcal{\odot}$, $\mathcal{P}_\mathcal{H}$, and $\mathcal{P}_
{\mathcal{\otimes}}$.
The line  $\mathcal{P}_\mathcal{S}=(r=0, a=\mbox{constant})$  includes
 \textbf{BHs} and  \textbf{NSs}.
$\mathcal{P}_\mathcal{S}$ represents   the collection of all origins $a_0$.
A variation of the dimensionless spin of the singularity  (at $r=0$) corresponds to a variation of $a_0$ in $\mathcal{P}_\mathcal{S}$.
This line  also represents the singularity  frequencies    $\omega_0=M/a_0$.
Moreover, we define
$\mathcal{P}_\mathcal{L}=(r_{\partial},a_{\partial})$ and  the set   $\mathcal{P}_\mathcal{H}=(a_{\pm}, r)$ for the Killing  horizons. Curves  $a_{\omega}^{\pm}$ at constant $\omega$ closes on $\mathcal{P}_{\mathcal{L}}$ and have
 origins in $\mathcal{P}_{\mathcal{S}}$.
 The line   $\mathcal{P}_\mathcal{\odot}=(a=0, r)$ describes the limiting case of the Schwarzschild solution.
$\mathcal{P}_\mathcal{\odot}$  crosses $\mathcal{P}_\mathcal{H}$ in $r=0$  \emph{and}  $r=2M$. Finally,  the set
$\mathcal{P}_
{\mathcal{\otimes}}=(a=M, r)$ describes the extreme Kerr spacetime and crosses  $\mathcal{P}_\mathcal{H}$ at
$a_{\pm}=M$ \emph{and}  $r=M$.
The collection of all the points  $a_{\omega}=$constant generates the light-curves shown in  Fig.\il\ref{Fig:CORESUNO},  where
 the Killing bottleneck and Killing throat emerge at  $a_{\omega}/M>1$.

\medskip

According to  Fig.\il\ref{Fig:SoOv},  metric bundles can be classified in two classes. \textbf{(1)} The first
class  includes the curves $a_{\omega}^{\pm}$  tangent to
  $r_-$, including those bundles which are  ``entirely''   contained in the \textbf{BH} sector  of $\pi_a^+$, i.e., $a\in[0,M]$
	\emph{and} $r\in [0,r_-]$.
 \textbf{(2)} The second class includes metric bundles    tangent to the outer horizon  $r_+$, containing
\textbf{BHs} \emph{and} \textbf{NSs} in the same bundle.
   These two classes are separated by the  limiting bundle $g_{\omega}^{\pm}$
   with origin $a_0=2M$ and tangent to the maximum of $a_{\pm}$; that is, the point in $\pi_a^+$  with $r=M$, for $a=M$ and frequency
   $\omega=0.5$, describing the extreme Kerr spacetime.
Bundles with origins in the  \textbf{BH} sector  are \emph{completely} contained
       in an ``inaccessible'' region between the singularity,  $r=0$, and the
        inner horizon $r_-$.
Bundles with origin in the the naked singularity sector $a>M$ comprise \textbf{BH} \emph{and} \textbf{NS}  geometries, which are, therefore,  related because they are contained in the same metric bundle.
(Notice that the definition of $g_{\omega}^{\pm}$, through the stationary observes definition, provides a \textbf{BH}-\textbf{NS}  relation). The  case of very strong \textbf{NSs} (very large $a/M$, asymptotic value $a\approx+\infty$) corresponds to the  limiting point
$r_+\approx2M$ and $ a\approx0$--see Fig.\il\ref{Fig:SoOv}.
This suggests that the \textbf{NS} sector is  closely related to the \textbf{BH} sector.
We explore this connection more deeply below.
\begin{figure}[h!]
\centering
\begin{tabular}{c}
\includegraphics[scale=.34]{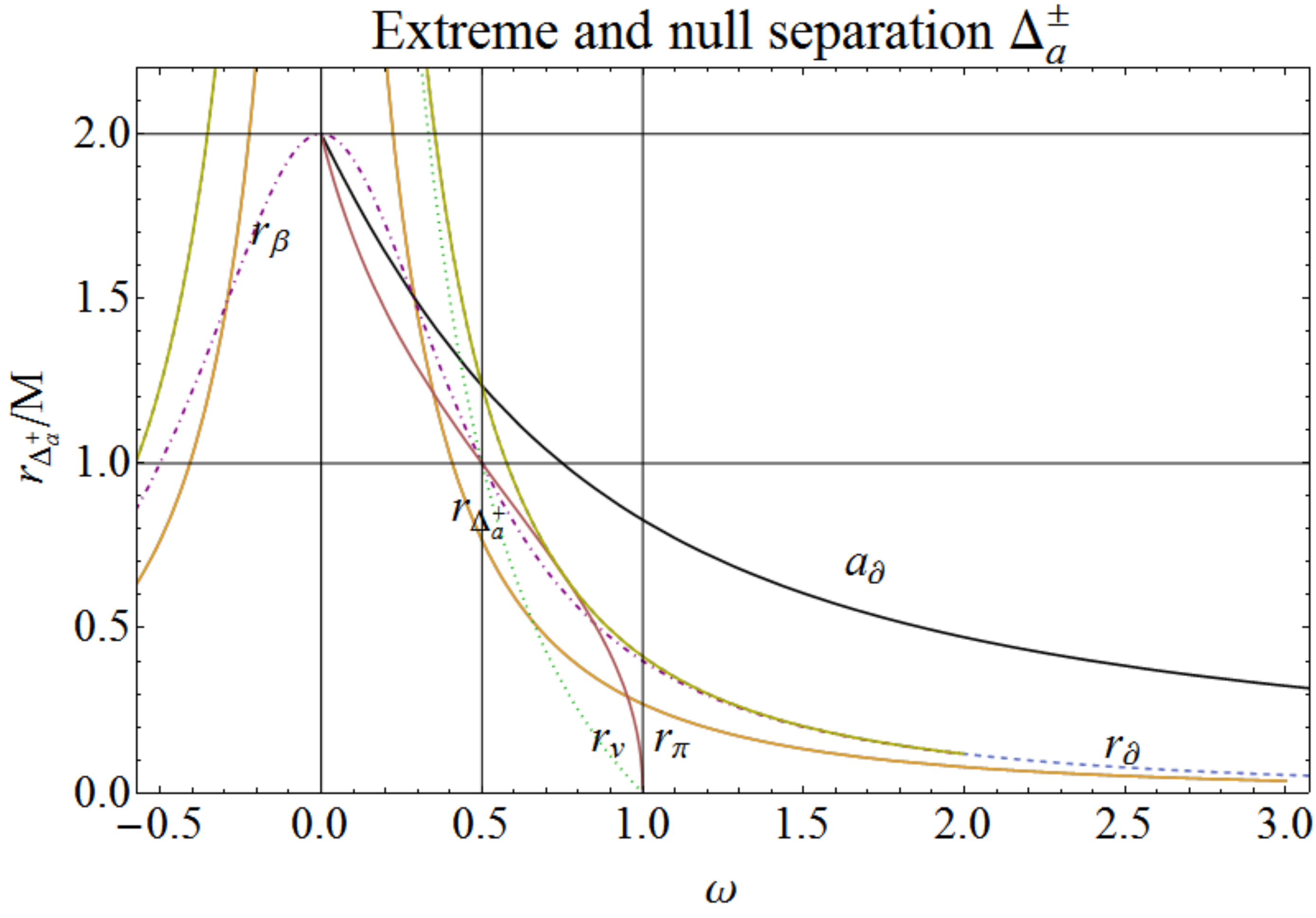}
\includegraphics[scale=.34]{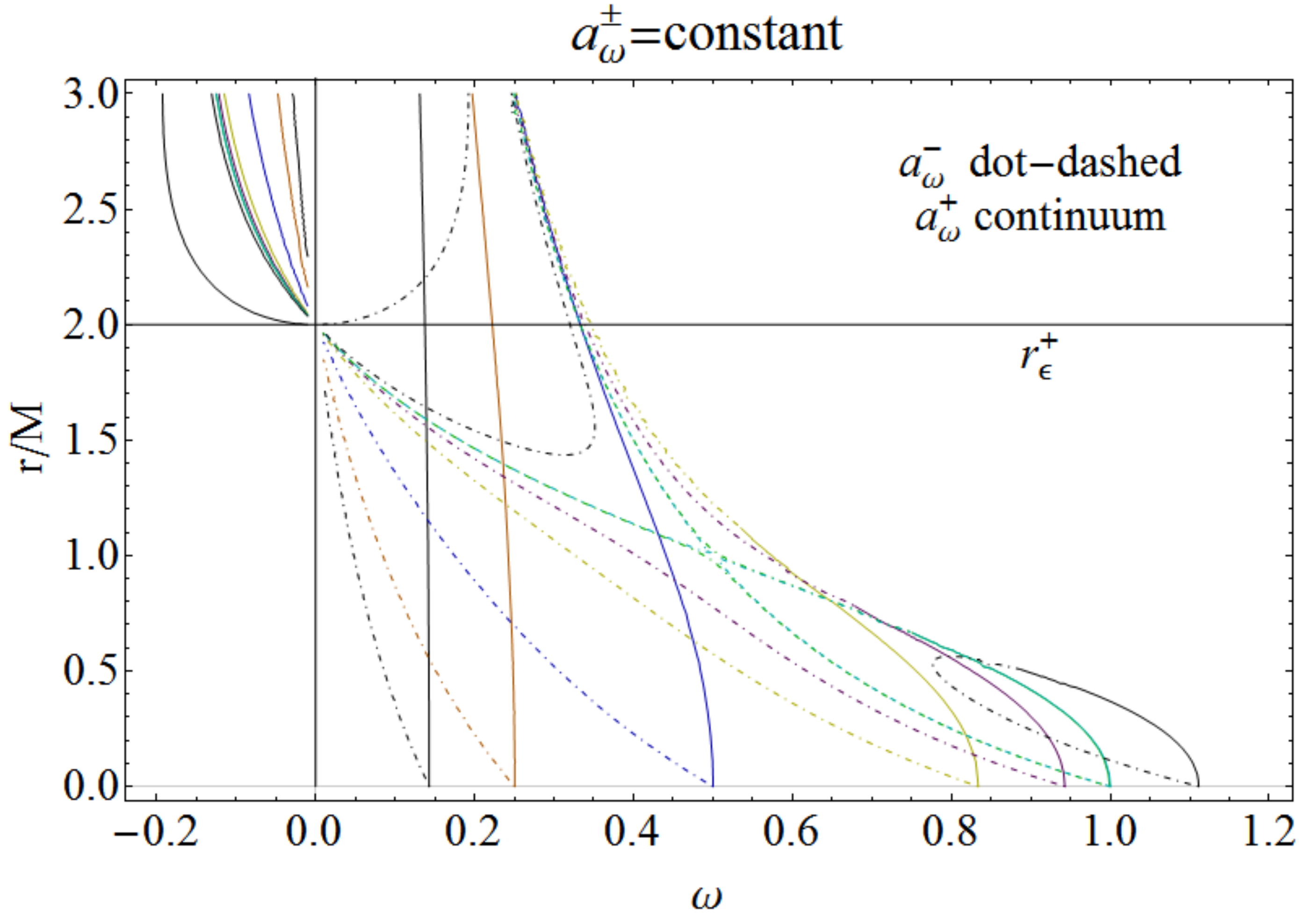}
\end{tabular}
\caption{Left panel:   Radii $\{r_{\Delta},
r_{\beta},
r_{\partial}, r_{\nu}, r_{\pi}\}$
an the spin $a_{\partial}$, according to the Eqs.\il(\ref{Eq:test-cris}), (\ref{Eq:rela-a-partialv}) and (\ref{Eq:rom-grad-mira}), as functions of the  frequency $\omega$. Left panel: Spins $a_{\omega}^{\pm}$= constant given in Eq.\il(\ref{Eq.lcospis})
in the plane $r/M-\omega$. Coalescence of the Killing horizons $r_{+}$  and $r_-$ in the extreme black hole geometries and the emergence of a  Killing throat  and Killing bottleneck in the \textbf{NS} geometries are shown--see also Figs.\il\ref{Fig:cOALESCENzE},\ref{Fig:MaSuorVIRMa},\ref{Fig:possov131},\ref{Fig:possov132} and \ref{Fig:Gpossov4}.
Here $\Delta^{\pm} a=a_{\omega}^+\pm a_{\omega}^-$. The radius $r_{\partial}$ satisfies the equation $\Delta^{-} a=0$. The radii
$r_{\Delta}\equiv r_{\Delta}^{\pm}:\partial_r\Delta^{\pm} a=0 $, and
 $\{r_{\beta},r_{\nu}, r_{\pi}\}$ are limiting  radii.}
\label{Fig:CORESUNO}
\end{figure}

\medskip
\textbf{The horizons as an envelope surface of the metric bundles}

An important consequence of the approach presented here is that  it allows us to establish a relation between \textbf{BHs} and \textbf{NSs}. In fact,
 the outer horizon  in $\pi_a^+$ emerges as an  envelope surface of metric bundles with  only \textbf{NSs} origins. That is, the   \textbf{BH} outer Killing horizon $r_+$ relates  a
   \textbf{BH}  with   $a\in[0,M]$ with a \textbf{NS} with   $a\in]2M,\infty]$. %
\begin{figure}[h!]
\centering
\begin{tabular}{lll}
\includegraphics[scale=.18,angle=90]{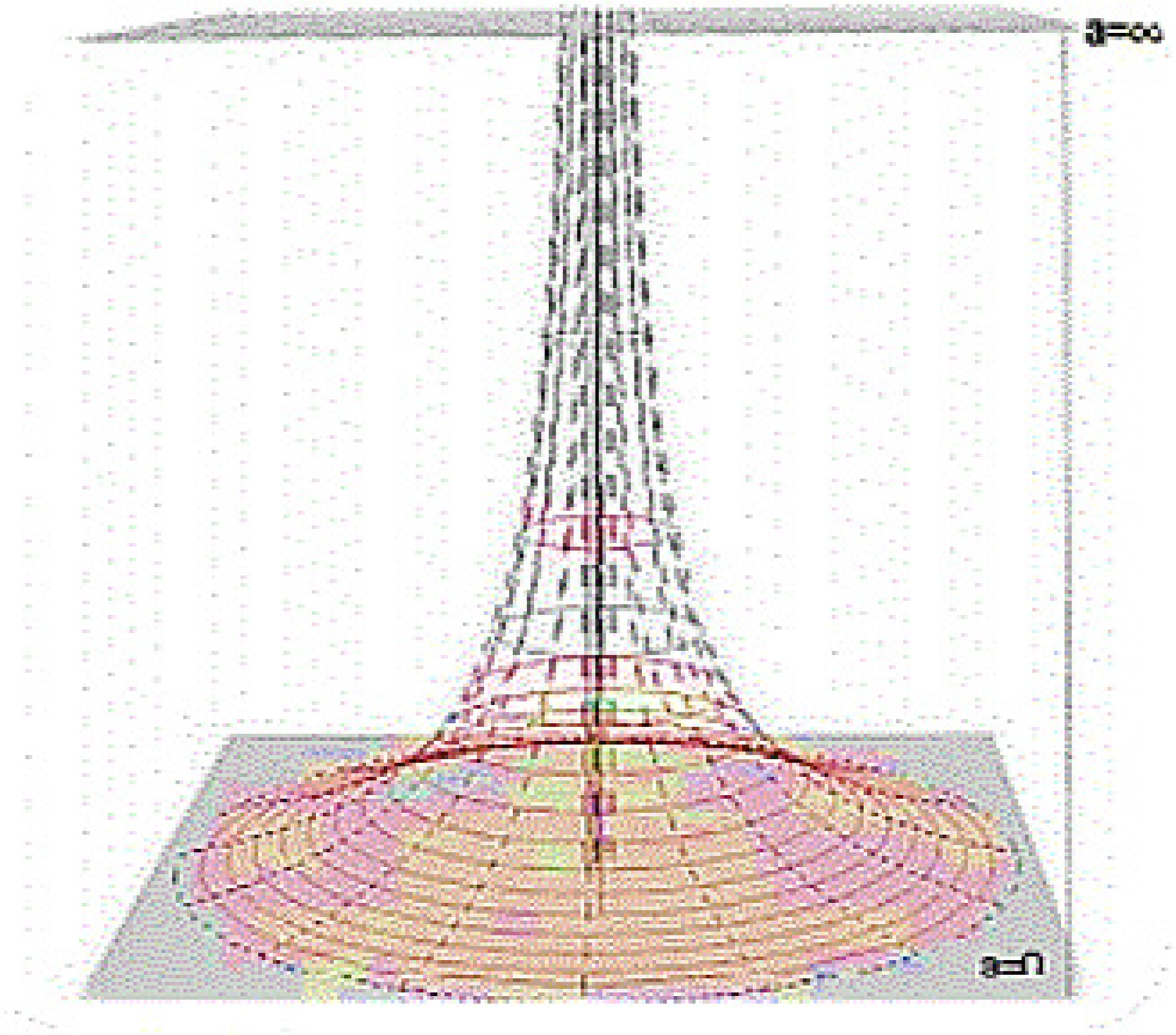}
\includegraphics[scale=.17,angle=90]{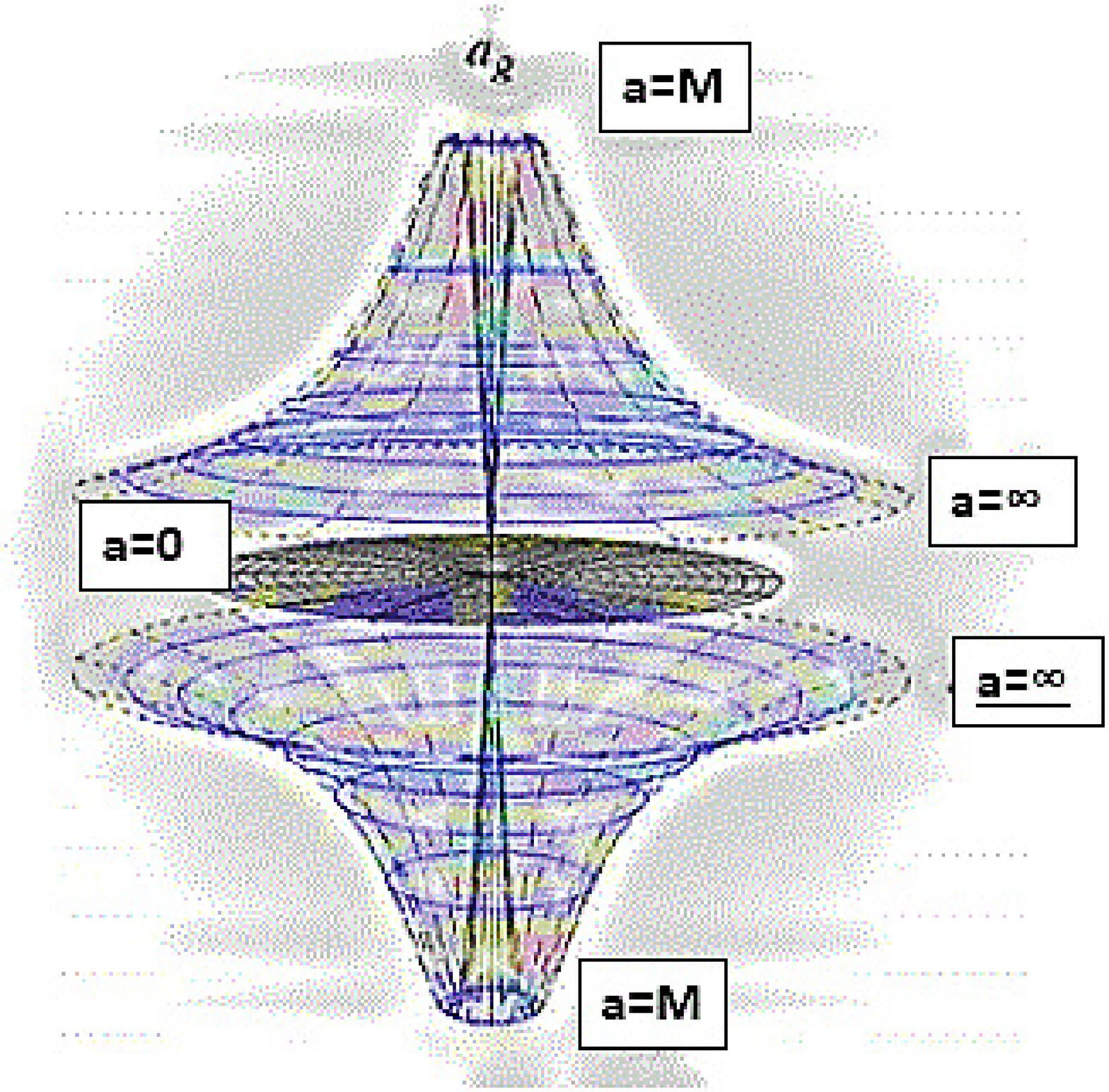}
\includegraphics[scale=.19]{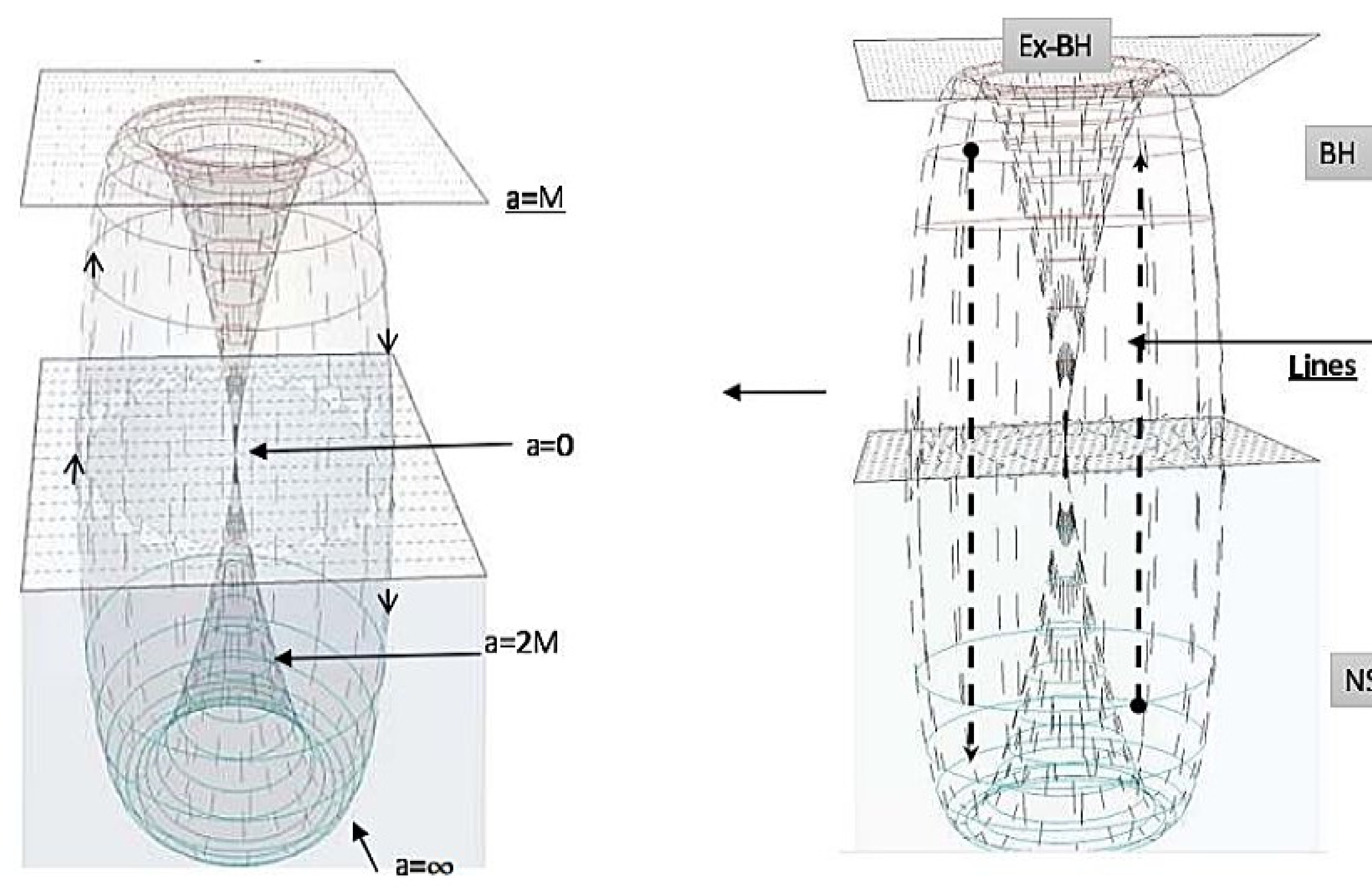}
\end{tabular}
\caption{
Schematic representation of the \textbf{BHs}-\textbf{NSs} correspondence. See also the $a_g(a_0)$ function of Fig.\il\ref{Fig:RoyalS}.
Right first  panel: The spin $a=2M/\omega$ between the planes $a=0$ and $a=\infty$-- cf. Eq.\il(\ref{Eq:freq-BH}).
 Third and fourth panels: Representation of the \textbf{BHs}-\textbf{NSs} correspondence. An own-up-arrow indicates the relation
\textbf{\textbf{NS}}-\textbf{BH}, defining the outer horizon $r_+$. An up-down arrow determines the inner horizons, which are totally included in the \textbf{BH}.
The limiting cases  $a=0$, $a=M$,  $a=2M$ and $a=\infty$ are also shown. }
\label{Fig:Slimers}
\end{figure}
The  inner horizon in $\pi_a^+$  can be constructed by metric bundles with  \textbf{NS} and \textbf{BH}  origins. Viceversa, the \textbf{BH} horizons are tangent to all the metric bundles.
{ All the   \textbf{BHs} \emph{and} \textbf{NSs} frequencies $\omega_{\pm}$ are, therefore, related to
the  horizon frequency $\omega_H$, and all horizon frequencies are the limiting orbital photon frequencies in
\textbf{NS} and \textbf{BH} spacetimes--Figs\il(\ref{Fig:SoOv}).}
{As the horizon is tangent to all the metric bundles,  each  metric  bundle is  defined by one frequency  $\omega$, which coincides with the  horizon frequency  $\omega_H$ on the tangent point in the extended plane.
Frequencies $\omega$ in (\ref{Eq.lcospis}) are actually the limiting frequencies  $\omega_\pm$.
It follows  that  the  metric bundles,
 connected the  singularities frequencies $\omega_0$,  are defined by the origins of the metric bundles
 and the horizon frequencies $\omega_H$.}

Remarkably,  the construction of the \emph{inner} horizon $r_-$ is confined to metric bundles contained   entirely in  the \textbf{BH} sector.
This fact could lead to important consequences, when considering the collapse towards a  \textbf{BH} and the process of formation  of the inner horizon.
These   metric bundles are  all confined in  the $\pi_a^+$ region
$r\leq r_-$ and $a\leq M$.
  This implies  that the inner horizon $r_-(\bar{a})$ of  a \textbf{BH}   with spin $\bar{a}$  is related to a metric bundle with origin $a_0$ in the \textbf{BH} or \textbf{WNS} ($a\in[M,2M]$) regions,
	while the outer horizon $r_+(\bar{a})$ is related to a \textbf{NS} metric bundle.

\medskip

Before continuing, it is convenient to return to the concept of  metric bundle, as shown  in Fig.\il\ref{Fig:SoOv}, and analyze three particular curves in detail:
\textbf{(1)}  the horizontal lines $a=$constant of $\pi_a$;  \textbf{(2)} the vertical  lines $r=$constant   and; \textbf{(3)}  the curves $a_{\pm}$ corresponding to horizons.

\textbf{(1)  The horizontal lines $a$=constant}

For a \textbf{BH}  \emph{or} a \textbf{NS} with $a_0\in \mathcal{P}_\mathcal{S}$, there are  two  curves $a_{\omega}^{\pm}$ of the bundle, which are tangent at a point $p\in \mathcal{P}_\mathcal{H}$  on the horizons.
Each metric bundle $g_{\omega}^{\pm}(a_0)$  is associated to a constant frequency,  $\omega_0=M/a_0$,
defined by the bundle origin $a_0$.
Considering a bundle, there is a pair of points $p_1(a)=(a,r_1(a))$ and  $p_2(a)=(a,r_2(a))$  with $a=$constant and $r_1(a)<r_2(a)$,  which are located respectively on the two curves of the bundle.
{A special case is the pair of points present on the  origin line, $P_{\mathcal{S}}$,  where  $r_1(a_0)=r_2(a_0)=0 $. Also the
horizons $P_{\mathcal{H}}$ for the extreme Kerr spacetime are special. Note that, in general, the condition  $r_1(a)<r_2(a)$ on
the horizon for $a=a_H\in a_{\pm}(r)$ leads to Eq.\il(\ref{Eq:mart-re}).}
 On the orbits $(r_1(a),r_2(a))$,    light-like  orbital frequencies $\omega_{\pm}$  are equal to  $\omega_a(r_1)=\omega_a(r_2)=M/a$, where   $\omega_a\in (\omega_+,\omega_-)$.
There are two special geometries associated to the closure points  $P_{\mathcal{S}}$  and $P_{\mathcal{H}}$:
  $ P_{\mathcal{S}}$ represents the  singularity  $r=0$ with   $\omega_a(0)=\omega_a(0)=M/a$, corresponding to a spacetime with spin $a$.  Moreover,
	the second special geometry is always a \textbf{BH}, whose (inner or outer) horizon  has the  frequency $\omega_H=M/a$, i.e., the frequency of the bundle.
We investigate the spin $a_g$ of this  specific spacetime    below.

 We note that  metric bundles  cross each other  in $\pi_a^+$. This means that,  in a fixed spacetime with a fixed radius, there are two limiting frequency values $\omega_{\pm}$. Therefore,  the maximum number of crossing  points between metric bundles is two.
Consequently,
there are two crossing metric bundles with origins  at $a=1/\omega_{\pm}$,  respectively.

\textbf{(2) Vertical lines $r=$constant.}

Let us now focus  on the vertical lines in $\pi_a^+$ and the intersections on each metric bundle. For a fixed orbit  $r$, there are, in general, two Kerr geometries  corresponding to the spins
$a_1(r)<a_2(r)$ of the same bundle. In addition,  there are the following limiting cases:
\emph{1.} At $r=0$ there is an infinite number of origins,
 where  $a_1(0)=a_2(0)$ on a bundle.
\emph{2.} The point $r_t$ is the
 tangent point of the vertical line  to the bundle, satisfying the condition $a_{\omega}^+(r_t)=a_{\omega}^-(r_t)$-- see Eq.\il(\ref{Eq:rela-a-partialv}) and Fig.\il\ref{Fig:Tactic}.
The condition  $r_{\partial}^+=r_{\pm}$ is satisfied  only in special geometries $a_{R}$.
 In   general, for  $r\in]0,r_{\partial}^+[$,  there are two geometries
   $a_1<a_2$, corresponding to two \textbf{BHs} or one \textbf{BH} and one \textbf{NS}.
 This implies that at a fixed $r$,  there are two geometries  $(a_1,a_2)$  with  frequencies  $\omega=1/a$.
 {The case of the geometries identified in the  extended plane by the vertical lines will  be clarified  at the end of this section because it is necessary to consider the tangency conditions as shown in   Fig.\il\ref{Fig:Bollet}.
With respect to this property, $a_1<a_2$, there are two exceptions represented by the  metric bundles with vertical lines tangent  to
the horizon. There are two asymptotic cases, where $a_1=a_2$; these cases with respect to the horizon points  $(a=0, r=0)$ and  ($a=0,r=2M$) correspond to the limit of the  Schwarzschild spacetime-- Fig.\il\ref{Fig:Bollet}. In general, a vertical line
$r=\bar{r}$  on a metric bundle $g_{\omega_0}$  defines two geometries  with  $\omega_+=\omega_0$ and $\omega_-=\omega_0$.
More details on this point will be addressed at the end of this section.}
 In other words, this property  relates  limiting frequencies of different spacetimes.
 This result is in agreement also with the results presented in Sec.\il\ref{Sec:bottlem}.

 \textbf{(3) The horizon surfaces  $a_{\pm}$}

 All the metric bundles have the frequencies $\omega_{H}=1/a_{\pm}$  at the horizon in the extended plane.
 {As the metric bundle frequency  $\omega_0$ is also a limiting photon orbital frequency $\omega_+$ or $\omega_-$
and as all frequencies  $\omega_0$ represent also the horizon frequency $\omega_{H}$, then
 the limiting photon frequencies  $\omega_{\pm}$ on an  orbit $r$  in all the spacetimes $a\in [0,\infty[$
are the horizons frequencies $\omega_H$ in the extended plane. Viceversa, the set of the horizon frequencies $\omega_H$ in the extended plane is  the collection of all the   limiting orbital  photon frequencies  $\omega_{\pm}$ in any \textbf{BH} or \textbf{NS} spacetime.
This issue will be discussed in detail below. }

\medskip

\textbf{BHs-NSs correspondence}

The relation between \textbf{BHs} and \textbf{NSs} can be formalized by introducing the functions
$a_g(a_0)$ and $a_R(a_0)$ of the origin $a_0$ as follows
\bea\label{Eq:agar}&&
\forall \; a_0>0,\quad a_g\equiv\frac{4 a_0M^2}{a_0^2+4M^2}\quad\mbox{where}\quad a_g\in[0,M]\quad \mbox{and} \quad \lim_{a_0\rightarrow0}a_g=
\lim_{a_0\rightarrow\infty}a_g=0,\quad a_g(a_0=2M)=M,
\\
&&
\mbox{and}\quad
a_R(a_0)\equiv\sqrt{4 \sqrt{(a_0+M)M}M-(a_0+4M)M}  \ .
\eea
The behavior of these functions is plotted in  Figs.\il\ref{Fig:RoyalS}.
For a fixed value of the origin $a_0$, the function $a_g(a_0)$ defines univocally
 the outer  $r_+$ or the inner $r_-$ horizon. %
 This relation includes the Schwarzschild limiting case,
which corresponds to the limit  $a/M\rightarrow+\infty$,
 and the  extreme Kerr spacetime, which is connected with the naked singularity value $a=2M$--see also  Figs.\il\ref{Fig:Slimers}.
 More precisely,  $a_g(a_0)$ is the solution of the equations
 $a_{\omega}^{\pm}(r_{\pm},\omega_0)=a$, where $\omega_0=M/a_0$--while $a_R(a_0)$ is the solution of the equations
$r_{\partial}(a_0)=r_{\pm}$, providing information about the orbits where  the curves $a_{\omega}^{\pm}$ close at the horizons.
The analysis of off-equatorial and charged generalizations considered  in Sec.\il\ref{Sec:RN-KN-cases} reaffirms this result.
\begin{figure}
  \includegraphics[width=9cm]{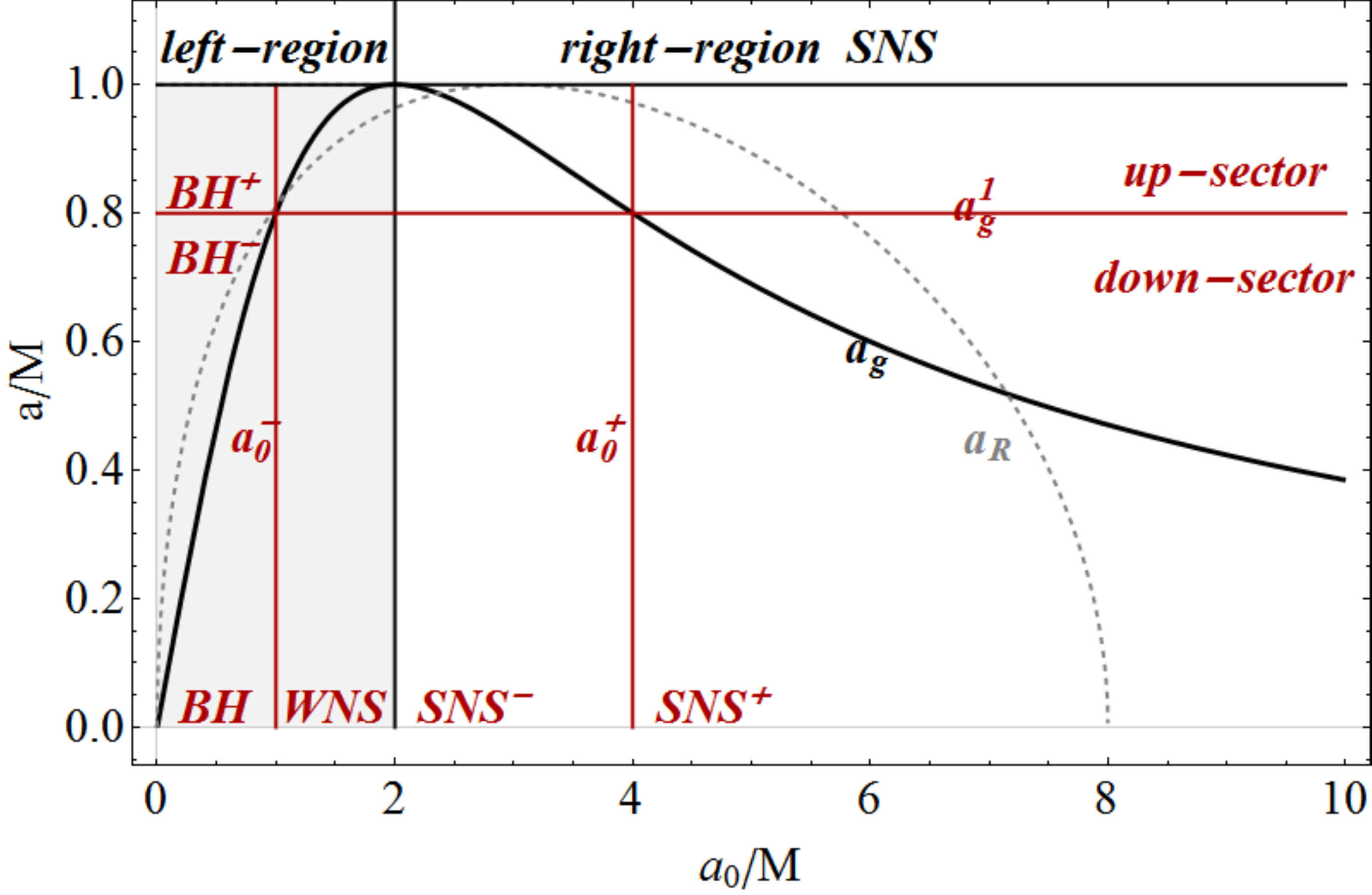}
  \caption{Plots of the spins $a_g$ and $a_R$ defined in Eq.\il(\ref{Eq:agar}) as functions  of the origin $a_0/M$.
	\textbf{BH} and \textbf{NS} regions are gray shaded.  $a_g$ is a representation of the \textbf{BHs}-\textbf{NSs} correspondence
	--see also Figs.\il\ref{Fig:Slimers}. The sectors and regions corresponding to strong naked singularities (\textbf{SNS}) and weak naked singularities (\textbf{WNS}) are explicitly shown (cf. definitions given in Sec.\il\ref{Sef:aomega}).
	Here $a_0^-=M$, $a_0^+=4M: a_g=a_g^+$,  $a_g^1=(3/4)M=a_{g}(a_0^-)$.}
	\label{Fig:RoyalS}
\end{figure}
According to Figs.\il\ref{Fig:SoOv}, each  point $r_H$ on the horizon is  univocally related to a \textbf{NS} or a \textbf{BH} metric
 in  the $\pi_a^+$ plane.
 Each frequency $\omega$  is  in correspondence with a point  $r_H$ on   $r_-$ \emph{or} $r_+$. In this sense,
 we might say that the {information} contained  on the horizon (the frequency $\omega_H$) is
   {extracted} by the functions $a_{\omega}^{\pm}$.

Furthermore, it is immediate to see that from  the expression $a_g(a_0)$ of Eq.\il(\ref{Eq:agar}), we obtain a
relation between the horizon tangent spin (horizon frequency) and the bundle origin (bundle frequency), namely,
$\omega_0^{-1}\equiv a_0^{\pm}/M=\frac{2 r_{\pm}(a_g)}{a_g}\equiv \omega_H^{-1}(a_g)$;
{in particular, from $\omega_0^{-1}\equiv a_0^{\pm}/M=\omega_H^{-1}(a_g)$, it is seen that the bundle frequencies $\omega_0$ represent
all  (and only) the horizon frequencies  on the tangent point,  i.e., $\omega_H(a_g(a_0))=\omega_0$.}
This implicitly relates also the horizons frequencies  with the singularity frequencies; there are then particular
\textbf{BH} spacetimes, where the outer horizon $r_+$ or the inner and outer horizon  $r_{\pm}$ are defined by  bundles
with  \textbf{NSs}. We detail this aspect below. Here we  note that there is  only one fixed point for  the transformation
$a_g(a_0)$, namely $a_0>a_g$,  and   $a_0=a_g$ for $a_0=0$.

We now introduce the concept of couples of related metric bundles, say, \textbf{BD} and \textbf{BC} (see Fig. \ref{Fig:pesoMa}).
 In the first couple, the first bundle has its origin in the \textbf{NS} region (tangent to the outer horizon) and the  second bundle is
completely or partially contained in the  \textbf{BH}  region (tangent to the inner horizon).
The two bundles with origins  ($a_0,a_0^{\prime}$)  share  equal tangent spin $a_g$. That  is, if  $a_0>a_0^{\prime}$, then the bundle with origin  $a_0$, frequency   $\omega_0=Ma_0^{-1}$, has the contact spin  $a_g$ in $r_g\in r_+$ with horizon frequency  $\omega_H^+(r_g,a_g)=\omega_0=Ma_0^{-1}$. The second  bundle of the couple
has its origin at $a^{\prime}_0$, frequency  $\omega^{\prime}_0=M/a_0^{\prime}$,  tangent  spin  $a_g$ in $r^{\prime}_g\in r_-$ and  horizon frequency  $\omega_H^-(r^{\prime}_g,a_g)=\omega^{\prime}_0=M/a_0^{\prime}$.
Bundles $g_{\omega_0}$ and $g_{\omega_0^{\prime}}$ determine, in the sense of the envelope surface, respectively,
the  outer horizon $r_+=r_g$ and the inner horizon $r_-=r_g^{\prime}$ of the  \textbf{BH} spacetime with spin $a_g$.
The relation between the tangent points  $(a_g, r_g)$ and the origin $a_0$ and the relations between  \textbf{BHs} and  \textbf{NSs}  through the   bundles will be addressed in full details below.
Importantly, the condition of the bundle correspondence, i.e., $a_g(a_0)=a_g(a_0^{\prime})$, leads to the non-trivial solution
$a^{\prime}_0=a_p\equiv 4M^2/a_0$ (see \textbf{DD}  and \textbf{BB}  model bundles of Figs.\il\ref{Fig:pesoMa}, \ref{Fig:PostMerge},
and \ref{Fig:manet} and Tables \ref{Table:ABCDmodels} and \ref{Table:number-cicle}).
In terms of the horizon frequency  (equal to the  bundles frequencies), there is
$\omega^{\prime}_0=\frac{1}{4 \omega_0}$; in fact, using  Eq.\il(\ref{Eq:freq-BH}), this relation can be written in compact form as
$ \omega_H^+\omega_H^-=\frac{1}{4}$ (or we can write $a_0^+(a_g)a_0^-(a_g)=4M^2$), which is independent of the spin $a$, in general.
In Fig.\il\ref{Fig:pesoMa}, the solutions $a_0^{\pm}/M=\frac{2 r_{\pm}(a_g)}{a_g}$  correspond to  $a=a_0$ and  $a=a_p$.

In the second  couple, \textbf{BC}, the origin spin of one bundle $a_0^{\prime}$ is the tangent point $a_g(a_0)$ of the second bundle with origin $a_0$ (therefore, $a_0^{\prime}$ is always a \textbf{BH} and the other bundles are all \textbf{BHs}).
An example of these couples are the models \textbf{BB} and \textbf{CC} of Figs.\il\ref{Fig:pesoMa}, \ref{Fig:PostMerge},
and \ref{Fig:manet} and Tables \ref{Table:ABCDmodels} and \ref{Table:number-cicle}.

To enlighten some properties of the  \textbf{BC} and  \textbf{BD} bundles, we consider the difference $G[a_0]\equiv(a_0-a_g)$ as a
function of  $a_0$ and the    recurrence relation  for  $a_g$, i.e.,   $a_g[n+1]=\frac{4 a[n]M^2}{a[n]^2+4M^2}$ where $a[0]=a_0$.
Then, $a_g[n+1]$ decreases  with the cycle order $n$  (see Fig.\il\ref{Fig:pesoMa}). In fact,
in \textbf{BC} bundles, the cycles are confined to the  inner horizon in the extended plane
(apart from the starting point $a_0$).
Naturally, the only fixed point, $a_p=a_0$, of this relation is in
  $a_0=2M$, corresponding to the extreme \textbf{BH}.

Bundles  \textbf{BD}  have coincident cycles and  bundles  \textbf{BC}  have partially coincident cycles; therefore,
 \textbf{BD} and \textbf{BC} bundles are related by partially coincident  cycles (see Table\il\ref{Table:number-cicle} and
Fig.\il\ref{Fig:pesoMa}).

  \begin{table}[h!]
\caption{\label{Table:number-cicle}Recurrence relation $a_g[n]$ for different starting points, see also Fig.\il\ref{Fig:pesoMa}.
\textbf{BD} bundles with coincident cycles (\textbf{BB} and \textbf{DD}) are clearly denoted
(excluding the initial point) and partially coincident cycles in \textbf{BC} bundles  are also shown.
Cycles (with the exclusion of the initial point $a_0$) are entirely confined in the \textbf{BH} region.
Models $\mathbf{XX}$
where $\mathbf{X}=\{\mathbf{A,B,C,D}\}$ are defined  in Fig.\il\ref{Fig:PostMerge}. }
\centering
 \begin{tabular}{|l|l|l|l|l|l|l|l|}
  \hline
 $a_0 $ & $a_g[1]$      &$a_g[2] $ &       $  a_g[3]$     &  $  a_g[4]  $ &      $  a_g[5]$    &     $a_g[6]  $   &   $  a_g[7]$\\\hline
1 ({\tiny{\textbf{BB}}})&0.8 & 0.689655 & 0.616366 & 0.562903 & 0.521586 & 0.48837 &0.460889\\
2 ({\tiny{\textbf{CC}}})&1.& 0.8& 0.689655& 0.616366& 0.562903& 0.521586& 0.48837\\
4 ({\tiny{\textbf{DD}}})&0.8& 0.689655& 0.616366& 0.562903& 0.521586& 0.48837& 0.460889\\
1/2 ({\tiny{\textbf{AA}}})&0.470588& 0.445902& 0.424787& 0.406451& 0.39033& 0.376008& 0.363172\\
$a_p$(1/2)=8&0.470588& 0.445902& 0.424787& 0.406451& 0.39033& 0.376008& 0.363172\\
$a_g$(1/2)=0.470588&0.445902& 0.424787& 0.406451& 0.39033& 0.376008& 0.363172& 0.351579\\
\hline
\end{tabular}
\end{table}

\begin{figure}[h!]
\centering
\begin{tabular}{lcr}
\includegraphics[scale=.6]{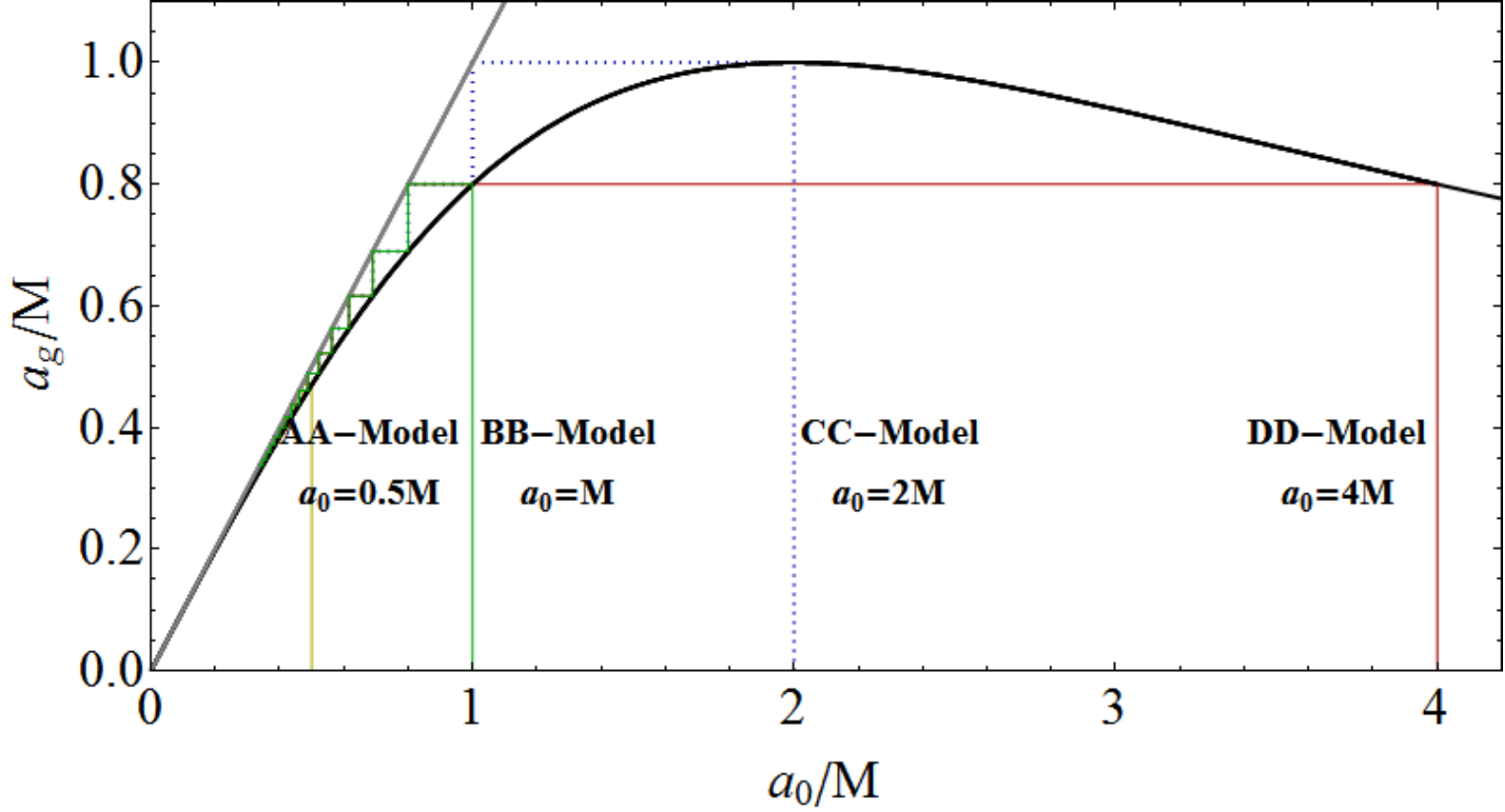}
\includegraphics[scale=.6]{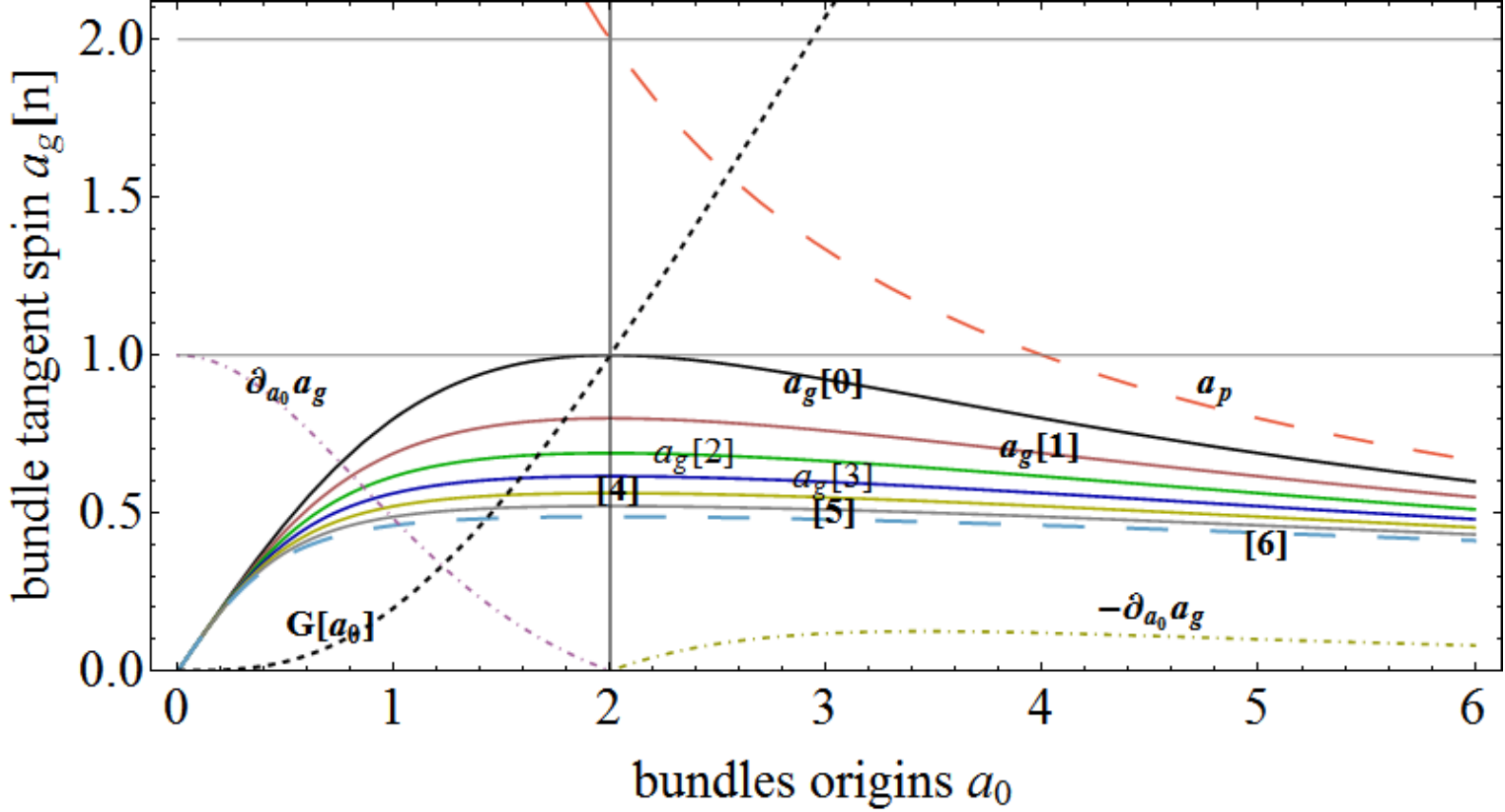}\\
\includegraphics[scale=.6]{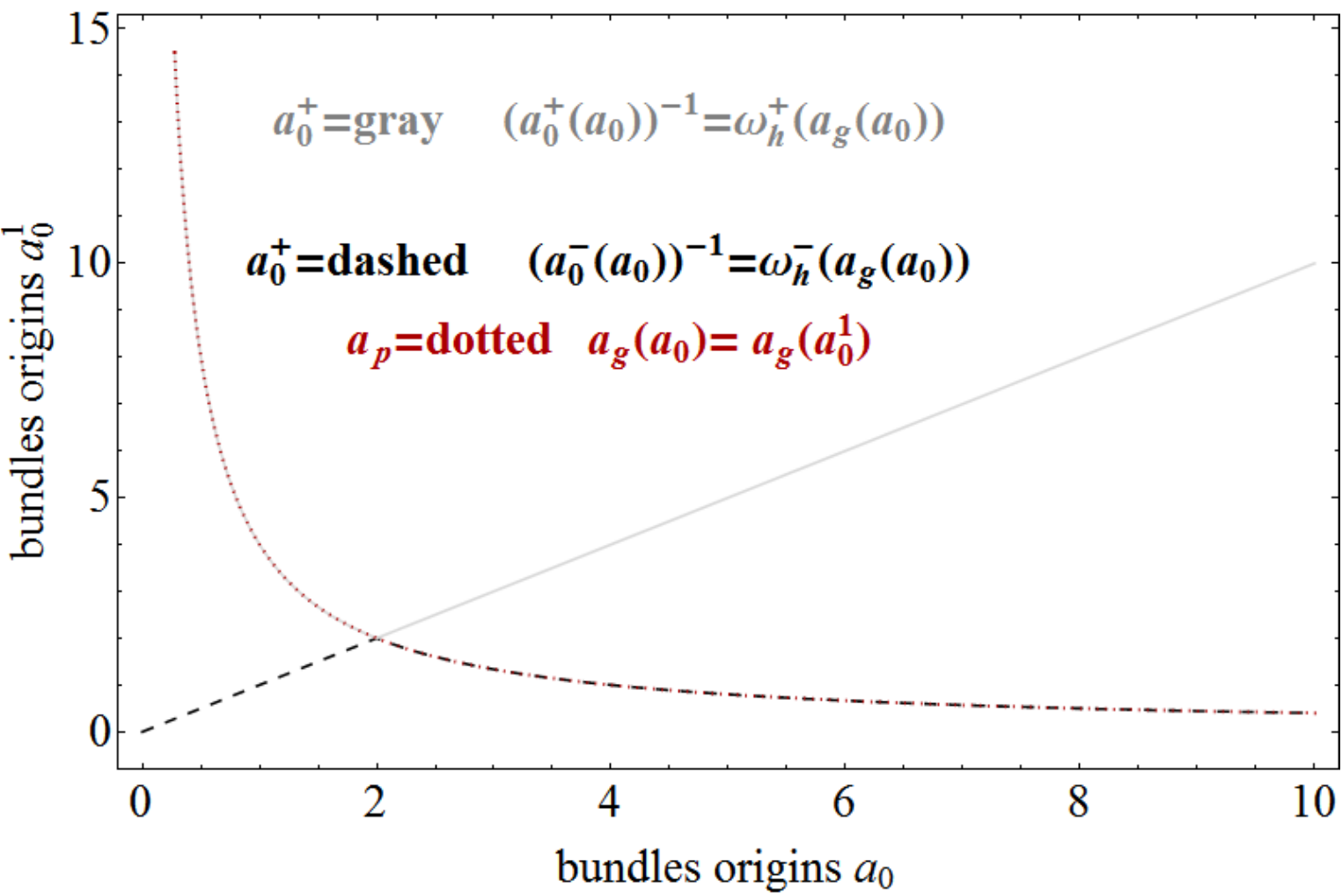}
\includegraphics[scale=.6]{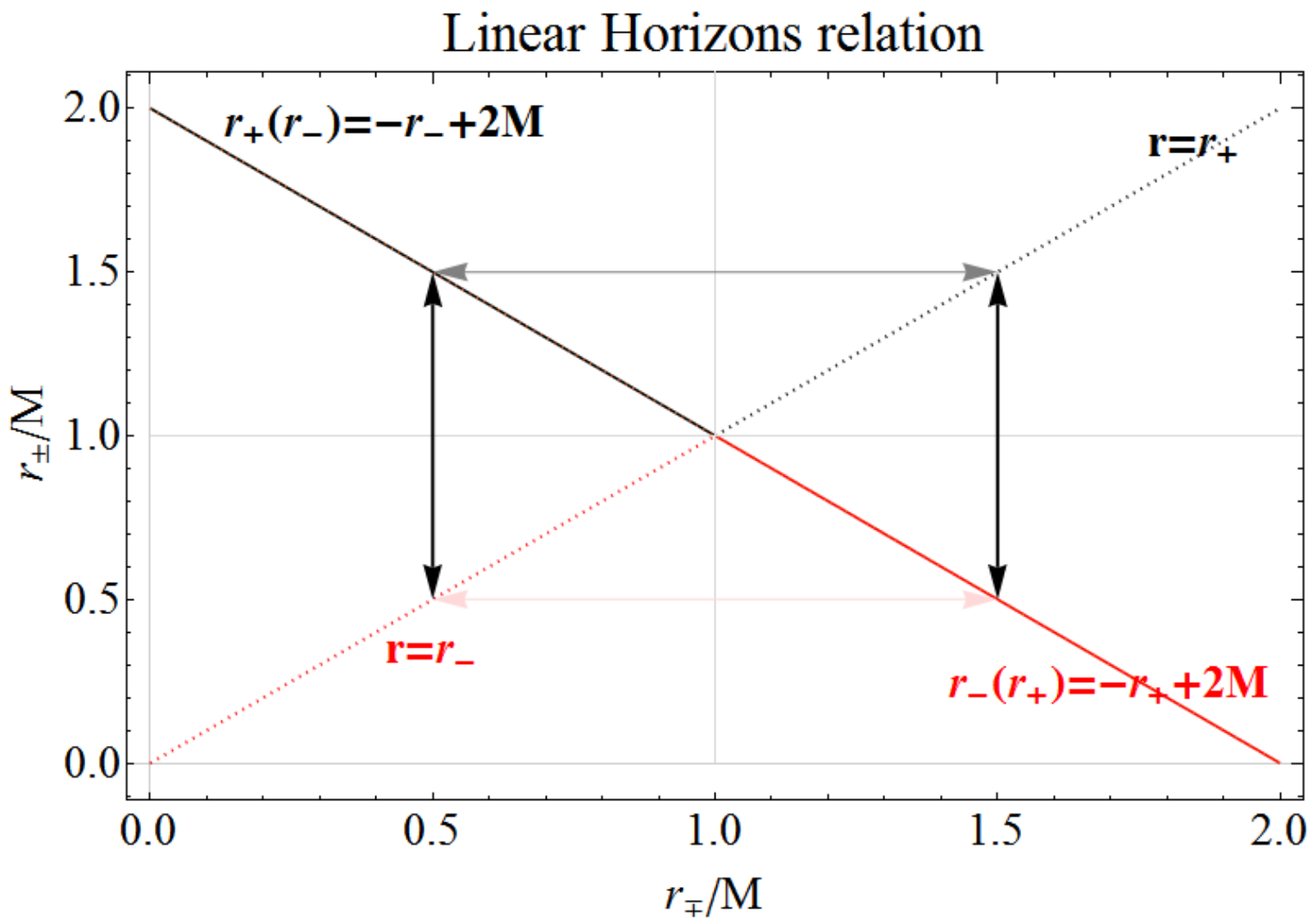}
\end{tabular}
\caption[font={footnotesize,it}]{{Below right panel: The functions
$r_-(r_+)$ (red line) and  $r_+(r_-)$ (black line)  (on the lines $a$=constant of the extended plane).
The region $r_\pm\times r_{\mp}\in [0,2M]\times[0,2M]$ describes  \textbf{BH} horizons in the extended plane
along the  red and black lines or,
equivalently, the dotted lines, according to the symmetries
enlightened in the plot}.
Below left panel: The solutions $a_0^{\pm}/M=\frac{2 r_{\pm}(a_g)}{a_g}$ as functions of bundle origin  $a_0$;
they correspond to  $a=a_0$ and  $a=a_p$, i.e.,
$a_g(a_0)=a_g(a_p)$ and $\omega_0^{-1}\equiv a_0^{\pm}/M=\frac{2 r_{\pm}(a_g)}{a_g}\equiv \omega_H^{-1}(a_g)$, determining
the related  \textbf{BC} or \textbf{BD}  bundles--Table \ref{Table:number-cicle}.
Upper panels: Recurrence relation for the point $a_g[n]$
as function of the starting points $a_0$.
\textbf{BD} bundles have  coincident cycles (\textbf{BB} and \textbf{DD}) with the exception of the initial point;
\textbf{BB-CC} bundles have   partially coincident cycles.
Cycles (with the exclusion of the initial point $a_0$) are entirely confined in the \textbf{BH} region.
Models $\mathbf{XX}$
with $\mathbf{X}=\{\mathbf{A,B,C,D}\}$ are defined  in Fig.\il\ref{Fig:PostMerge}.
For a fixed $a_0$, $a_g[n]$ decreases with the number of cycles.
The importance of the bundle with origin $a_0=2M$ is also shown; this is related to the extreme Kerr  BH and there
is  a maximum for each  cycle; $a_0=2M$ is the fixed point of the transformation $a_0=a_p$.
The difference  $G[a_0]\equiv a_0-a_g$ increases reaching $G[2M]= 1M=2M-a_g$ only for the first cycle-- Table\il \ref{Table:number-cicle}.}
\label{Fig:pesoMa}
\end{figure}
{To conclude this analysis, we note that it is possible to find a linear relation between the horizons curves
(in the extended plane) by  re-expressing  the curve of
the inner  horizon as a function of   the outer horizon  (on the lines  $a=$costant)  and, viceversa, i.e.,
$r_-(r_+)$ and  $r_+(r_-)$--Fig.\il\ref{Fig:pesoMa}.}

\textbf{BH-NS correspondence: one-to-one relation}

There is a one-to-one correspondence between the points of the horizons, the horizon frequencies,
 and the spins $a\in[0,\infty[$ for \textbf{NSs} or \textbf{BHs}.
\textbf{BHs} are  related to  a part of  the $r_-$ curve,
  \textbf{SNSs}  ($a>2M$)  correspond to  $r_+$,  and
	\textbf{WNSs} ($a\in[M,2M]$)  are in correspondence with the envelope surfaces
 of the  inner horizon, i.e.,  $a_-\in \pi_a^+$ for $r\in[0.8M,M]$.
 The limiting cases of  Schwarzschild, $a=0$, and  extreme  Kerr \textbf{BH}, $a=M$, are
 connected with the  limit $a=\infty$ and the double point $a_0=2M$, respectively.

In fact, we can say that $\forall \bar{a}\in]0,+\infty[$, $\exists!\, \bar{\omega}_0\equiv M/\bar{a}$ and, \emph{viceversa},
 $ \forall \bar{\omega}_0$ there is  \emph{one and only one} $a_0\in]0,M]: \bar{\omega}_0=\omega_H$, where $\omega_H$ is the horizon frequency, i.e., we connect points of
$\mathcal{P}_\mathcal{S}$ to points
 of $\mathcal{P}_\mathcal{H}$ of the horizons by considering that each metric bundle  relates univocally an origin $a_0$ of $\mathcal{P}_\mathcal{S}$   with a tangent point of  $\mathcal{P}_\mathcal{H}$.
 ($a_g$ is solution of $a_0=\omega_H^{-1}$).
Any origin of the metric bundle $g_{\omega_0}^{\pm}$, associated to a frequency $\omega_0 =$constant, is associated to one and only one point $p_{\pm}:\; p_-\in r_-$ or the outer  $p_+\in r_+ $, according to  the origin  $a_0$.

A special case is  the extreme  Kerr \textbf{BH} solution,  where the origin spin $a_0=2M$, associated to the metric bundle  $g_{\omega_0}$  of constant frequency  $\omega_0=1/2$, crosses the horizons at   $r_\pm=M$ and $a_\pm=M$.
 In general,  for a fixed origin  $a_0$,  there are two  frequencies  $\omega_H$ and $\omega_0\equiv M/a_0\neq\omega_H$, respectively, where $\omega_0=M/a_0$ is the frequency of the metric bundle
$g_{\omega_0}$, and $\omega_H\neq \omega_0$ is the horizon frequency defined by   the spin $a_{\pm}$ defined by the tangent point.  $p_{\pm}\equiv(a_{\pm},r_{\pm})\in \pi_a^{+}$.

The one-to-one \textbf{BHs}-\textbf{NSs} correspondence  is   described  by the function $a_g$  as given in Eq.\il(\ref{Eq:agar}) and illustrated in Fig.\il(\ref{Fig:RoyalS}).
We can say that each \textbf{BH} solution is connected to one and only one   $\mathbf{NS}$ in  $\pi_{a}^+$, as it emerges from the   analysis  of the  Killing horizons and  light frequencies on the  equatorial plane and, viceversa, each \textbf{NS} is related to one  \textbf{BH}.
 These considerations include the  limiting   cases of the Kerr extreme spacetime,  where the associated metric bundle has origin
$a_0=2M$,  and the  limiting case of the static Schwarzschild \textbf{BH}, which is connected to the   \textbf{NS} with $a=+\infty$.
The  \textbf{BH}-\textbf{NS} relation  allows us to consider a spin  shift from an initial $\mathbf{BH}_1$ ($\mathbf{NS}_1$) source as corresponding to  a shift of the respective $\mathbf{NS}_1$
 ($\mathbf{BH}_1$); therefore,    the pair $\mathbf{BH}_1-\mathbf{NS}_1$  shifts to the pair $\mathbf{BH}_2-\mathbf{NS}_2$.
The  segment $a_{\mathbf{BH}_1}-a_{\mathbf{NS}_1}$ of  $\mathcal{P_S}\in \pi_a^+$ transforms into
     $a_{\mathbf{BH}_2}-a_{\mathbf{NS}_2}$,  becoming  larger or smaller depending on the  curve $a_g$ as shown in Fig.\ref{Fig:RoyalS}.

{To conclude this section, we analyze the relation between the tangent radius $r_g$ on the horizon and the  bundle origin $a_0$.
We start with a description of Fig.\il\ref{Fig:RoyalS}.}

\textbf{Analysis of Figure \ref{Fig:RoyalS}}

 Figure \il\ref{Fig:RoyalS} represents the spin  $a\in  \textbf{BH}$ versus   $a_0$.
Spins  $a\in  \textbf{\textbf{BH}}$  and   $a_0$  are connected through  $a_g$.
The sectors and regions in this figure  are determined by the  following special boundary spins, considering that $a_0\in [0,+\infty]$ and $a_g\in[0,M]$:

\medskip

\textbf{$\bullet$} the origin $a_0^-=M$ distinguishes  \textbf{NSs} from  \textbf{BHs};

\textbf{$\bullet$} the  spin $a_0=2M$ defines the  \emph{left region}  where $a_0\in[0,2M]$ (for \textbf{BH} and \textbf{WNS}) and \emph{right region}  where $a_0>2M$ (for $\mathbf{SNSs}=\mathbf{SNS}^-\cup\mathbf{SNS}^+$);

\textbf{$\bullet$} {the spin $a_g^1\equiv a_{g}(a_0^-)=0.8M$ defines the \emph{up-sector}   where $a\in[a_g^1,M]$ (for $\mathbf{BH}^+$) and the  \emph{down sector}  where  $a\in[0,a_g^1[$ (for $\mathbf{BH}^-$)};

\textbf{$\bullet$} the spin
$a_0^+=4M: \;a_g=a_g^+$ distinguishes strong naked singularities $\mathbf{SNS}^+$ and $\mathbf{SNS}^-$.

\medskip

Consequently,

\textbf{$\bullet$} a bundle origin $a_0$ in  the \textbf{BH}-region corresponds through the horizon curve to
$\mathbf{BH}^-$ singularities ($a_g:\quad a_0\in \textbf{BH}\mapsto a_g \in \textbf{BH}^-$);

\textbf{$\bullet$} origins  $a_0$ in the \textbf{WNS}-region correspond to  $a_g \in\mathbf{BH}^+$.

Therefore,   $a_g\in \mathbf{BH}=\mathbf{BH}^-\cup\mathbf{BH}^+$  correspond, through the tangency with the inner horizon curve
$a_-$, to  $a_o\in$\textbf{BH}$\cup$\textbf{WNS};

\textbf{$\bullet$}  the origin $a_0$ in $\mathbf{SNS}^-$ corresponds to  $a_g\in\mathbf{BH}^+$;

\textbf{$\bullet$} the origin  $a_0$ in $\mathbf{SNS}^+$ corresponds to  $a_g\in\mathbf{BH}^-$.

Therefore,   $a_g\in\mathbf{BH}=\mathbf{BH}^-\cup\mathbf{BH}^+$ is related to $a_0\in \mathbf{SNS}^-\cup\mathbf{SNS}^+$.

\medskip

%
%
%

Because $\partial_{a_0}a_{g}\geq0$,  increasing the origin spin $a_0\in \textbf{BH}\cup \textbf{WNS}$ and
the tangent spin $a_g\in \textbf{BH}^-\cup \textbf{BH}^+$ increases the \emph{left region}.  %
{
Let us consider the  \textbf{BHs} and \textbf{NSs}  correspondence   determined by  the tangent point spin $a_g$. Note that  a fixed origin   $a_0\in\mathbf{SNS}^{+}$  (down-right region) corresponds to the outer horizon tangent point (we identify
$a_g(a_0)\in \mathbf{BH}$). This is sufficient for the determination of
the  spin   $a_0^-\in \textbf{BH}^{-}$ and  the  origin of the  bundle metric   tangent to the inner horizon in $(a_g,r^-_g)$; therefore, this defines the inner horizon $r^-_g$ of the $\textbf{BH}^-$ spacetime with  spin $a_g$.}

Similarly,
 in the \emph{right-up} region  there are couples  following related geometries $(a_g,a_0)$:
\textbf{(1)}
 $\mathbf{BH^+}-\mathbf{WNS}$ and  \textbf{(2)} $\mathbf{BH^+}-\mathbf{SNS}^{-}$.

We note that the origins $\mathbf{SNS}^{\pm}$ ($a_0$)  are in correspondence with a  $a_g\in \textbf{BH}$;
also the $a_0\in \mathbf{WNS}$ and $a_0 \in \mathbf{BH}$  are in correspondence   with a  $a_g\in \textbf{BH}$.

Therefore, in general there is a correspondence
 $a_0\mapsto a_g\in \mathbf{BH}\mapsto a^{\prime}_0$, i.e, the origin of the  bundle defines a tangent point to the horizon $a_g$ and
$r_+$ or $r_-$ in the spacetime with $a_g$; correspondingly, there is the bundle   with origin $a^{\prime}_0$ whose  tangent point to the horizon  is $a_g $and $r_-$ or $r_+$, respectively.

  Therefore, the following triple relations hold
  \bea\label{Eq:aptriplet}
\underset{\textbf{(a)}}{ \underbrace{[a_0\in \mathbf{SNS}^+\mapsto a_g\in \mathbf{BH}^-\mapsto a^{\prime}_0\in \mathbf{BH}]}},&&\quad
\underset{\textbf{(b)}}{ \underbrace{[a_0\in \mathbf{SNS}^-\mapsto a_g\in \mathbf{BH}^+\mapsto a^{\prime}_0\in \mathbf{WNS}]}};
\\\nonumber
&{\LARGE{\mathrlap{\mathrlap{\searrow}{\nearrow}}
{\mathrlap{\nwarrow}{\swarrow}}}}&
\\\label{Eq:aptrip-1}
\underset{\textbf{(c)}}{ \underbrace{\left[
a_0
\in
\mathbf{WNS}
 \mapsto
a_g \in \mathbf{BH}^+ \mapsto a^{\prime}_0 \in \mathbf{SNS}^-\right]}},
&&\quad\underset{\textbf{(d)}}{ \underbrace{[a_0\in \mathbf{BH}\mapsto a_g\in \mathbf{BH}^-\mapsto a^{\prime}_0\in \mathbf{SNS}^+]}}
 \eea
{The geometries of the  triple relations given  in Eq.\il(\ref{Eq:aptriplet}) are bounded together by $a_0\in a_+ $ and the
inner horizon $r_-$; on the other hand, the geometries of the  two    relations  in Eq.\il(\ref{Eq:aptrip-1}) are
bounded together by $a_0\in a_-$.
The geometries of the last  two  triple relations are bounded  by  $a_g$.
  The sets (\textbf{(b)} and \textbf{(c)}) and (\textbf{(a)} and \textbf{(d)})  are  related by an exchange of spins $a_0$, $a_0^{'}$. Bundles corresponding to the tangent   $a=a_g$ (same tangent point spin) belong to triplets connected by arrows.
	We shall see this also below considering some examples--see also Table \ref{Table:number-cicle}.}

\medskip

{The black hole area, delimited by the (outer) horizon, is a crucial quantity, determining  the  thermodynamic properties
of \textbf{BHs}. Given the relevance of this concept, we study in Sec.\il(\ref{Sec:appendix-area}) some properties of the areas
of the regions delimited  by  metric bundles  and compare them with the horizon area $\mathcal{A}_{r_{\pm}}^{+}=\pi/2$ in the extended plane $\pi_a^+$, i.e., the region in the plane $\pi_a^+$ bounded by the horizon curve $a_{\pm}$.}

\medskip

To conclude, we show that the horizons frequency $\omega_H$   defines the bundles frequencies. {We will resume part of the discussion carried out in relation with   Eq.\il(\ref{Eq:agar}).}
We  also  investigate the limiting  frequencies obtained  from the bundles crossing and some properties of the tangents to the horizons in the extended plane.

\textbf{On the frequencies}

The bundle frequency is  a  limiting photon-like frequency,
$\omega_+$ \emph{or} $\omega_-$ as introduced  in Sec.\il(\ref{Sec:1-st}), at the point $(a,r)$ of the bundle.
The second, limiting photon frequency at the point $(a,r)$, $\omega_-$ \emph{or} $\omega_+$, respectively,  is determined
  by the bundles crossing. As previously discussed, there is a maximum of two bundles at the intersection.
This  second  frequency is, therefore, the solution of $a_{\omega_1}^{\pm}(\bar{r})=a_{\omega_2}^{\pm}(\bar{r})$, where $\omega_1$ is the known frequency of  the bundle $g_{\omega_1}$, $\bar{r}$ is a fixed point of the bundle. The second photon  frequency $\omega_2$, identifying the related bundle $g_{\omega_2}$, is a function of $\omega_1$ and $\bar{r}$:
\bea\label{Eq:gen1-dat}
a_{\omega_1}^+&=& a_{\omega_2}^\pm:\quad\mbox{for}\quad
\omega^{(+,\pm)}_1=\varpi_-
\\\label{Eq:gen2-dat}
a_{\omega_1}^-&=& a_{\omega_2}^-:\quad\mbox{for}\quad
\omega^{(-,-)}_1=\varpi_+\quad \varpi_\pm\equiv-\frac{(r-2) \left[\left(r^2+4\right) \omega_2\pm4 \sqrt{r^2 \omega_2^2 \left[1-r (r+2) \omega_2^2\right]}\right]}{16 r^3 \omega_2^2+(r-2)^2 (r+2)}.
\eea
The solutions $a_{\omega_1}^\pm$ for $\omega_2=$constant are shown in  Fig.\il\ref{Fig:PostMerge}.
Note that  the limiting  curve for the constant frequencies curves    $\omega^{(+,\pm)}_1,\omega^{(-,-)}_1$ is $r_{\partial}^+$ of
Eq.\il(\ref{Eq:rela-a-partialv})--see also  Fig.\ref{Fig:Tactic},\ref{Fig:CORESUNO}.

By definition, the  frequency $\omega$ is constant along the  bundle; thus, the bundle frequency $\omega$ is  the bundle origin frequency
$\omega_0=M/a_0$ and,  particularly, the frequency at   $(a_g,r_g)$, where $a_g$ is  the bundle spin  at the tangent point  $r_g$ on the horizon curve.
 We can show that the bundle frequency coincides with the horizon frequency  at the tangent point $r_g$,
that is, $\omega_0=\omega_H(a_g,r_g)$.

First, note that the bundle frequency $\omega=\omega_0$ is defined in   $[0,\infty]$.
The analysis of the frequency variation domain gives us indications  that both the inner $\omega_H^-$ and outer horizon frequency $\omega_H^+$ define the bundle frequency,   as there is  $\omega_H^+\in[0,1/2]$ and  $\omega_H^-\in[1/2,+\infty[$  (for $a\in[0,M]$)--see also Fig.\il(\ref{Fig:MaSuorVIRMa}).
Then,   condition
$\omega_0=\omega_H$,
where $\omega_H$ is the union $\omega_H^+\cup\omega_H^-$,
leads to   $a_g(a_0)$, relating the tangent-point-spin $a_g$ and the bundle origin  $a_0=\omega_H^{-1}(a_g)$.

Fig.\il\ref{Table:ABCDmodels} and Table \ref{Fig:PostMerge} show  some notable numerical examples, proving that the bundle frequencies are, in fact, the horizon frequency at the tangent point to the horizons. Connections between the two bundles $g_{\omega_1}$ and
$g_{\omega_2}$ at equal $a_g$, which are related to the horizons points $r_{\pm}(a_g)$, are also shown in Table \ref{Fig:PostMerge}.
The tangent spin $a_g$ can be also obtained  from the relation  $a_{\omega}(r_\pm,\omega_0)=a$ by solving for     $a_0(a)$ and representing the result as $a_g(a_0)$ in Eq.\il(\ref{Eq:agar}).

Then, we can  consider the bundle on the horizon  ($r=r_g$) with the bundle frequency $\omega$ expressed as the
origin frequency $\omega=\omega_0$.
In terms of frequencies,
 from
$\omega_H=\omega$,
we obtain
$a={4 \omega_*}/({4 \omega_*^2+1})$, where $ \omega_*\in[0,1/2]$
for  $\omega_H=\omega_H^+ (r_+)=\omega_*$, and $\omega_*\geq1/2$
for  $\omega_H=\omega_H^- (r_-)$. This shows also the role of
the  inner horizon frequency.
This relation can be also found from  the spin $a_g(a_0)$, where  $a_0=\omega_0^{-1}$.
Note that we can eliminate the frequency $\omega$ from  $a_{\omega}^{\pm}$ of  Eq.\il(\ref{Eq.lcospis}) and
parametrize the metric bundles  in terms of  $a$ and  $ r$, using the  condition
 $\omega=\omega_H$ in
$a_{\omega}^{\pm}$.
In this case, there is
$a\in[0,M]$  as the spin is the horizon  tangent-point spin $a_g$:
\bea\label{Eq:UE-dif}
a_{\omega}^{\mathbf{(\natural)}}(\omega_H^{\mathbf{\flat}})\equiv\mathbf{(\natural)}\frac{ \sqrt{a^2 r^2 \left[8 r_\flat-a^2 (r (r+2)+4)\right]}+\mathbf{(\natural)} ar_\flat}{a^2 (r+2)},\quad \mathbf{(\natural)}=\pm; \quad \flat=\mp.
  \eea
  The bundles $a_{\omega}^{\mathbf{(\natural)}}(\omega_H^{\mathbf{\flat}})$ are, therefore,
parameterized for the tangent point $a=a_g$--see Fig.\il\ref{Fig:PostMerge}.
From the  condition of  coincidence between bundle and horizon   in the extended plane (i.e., $a_{\omega}^{\pm}=a_{\pm}$),  we obtain
$r=\frac{2}{4  \omega^2+1}$ and
assuming  $r=r_\pm $, then  $\omega=\omega_H$.
\begin{figure}[h!]
\centering
\begin{tabular}{lcl}
\includegraphics[scale=.39]{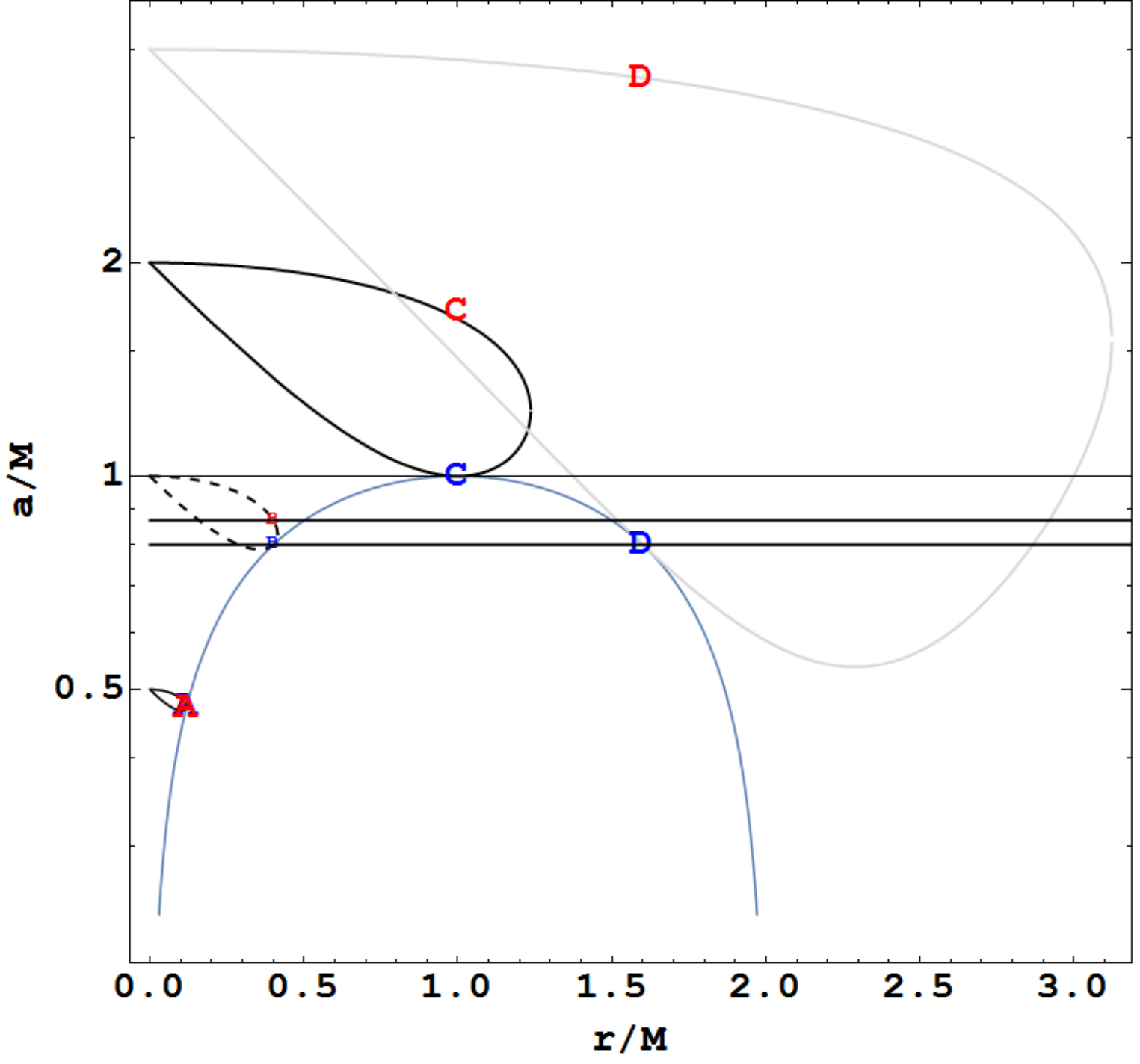}
\includegraphics[scale=.39]{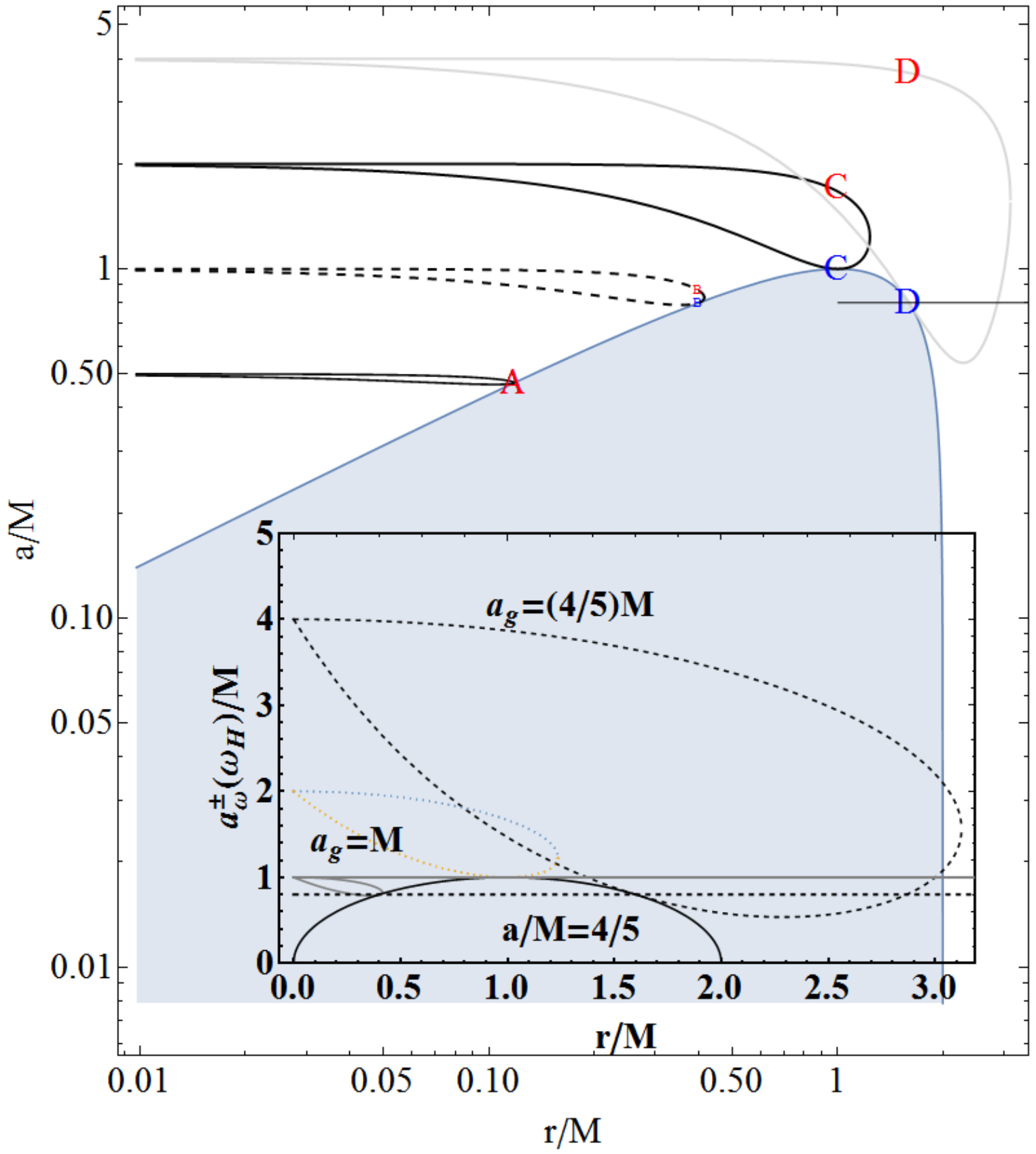}
\includegraphics[scale=.43]{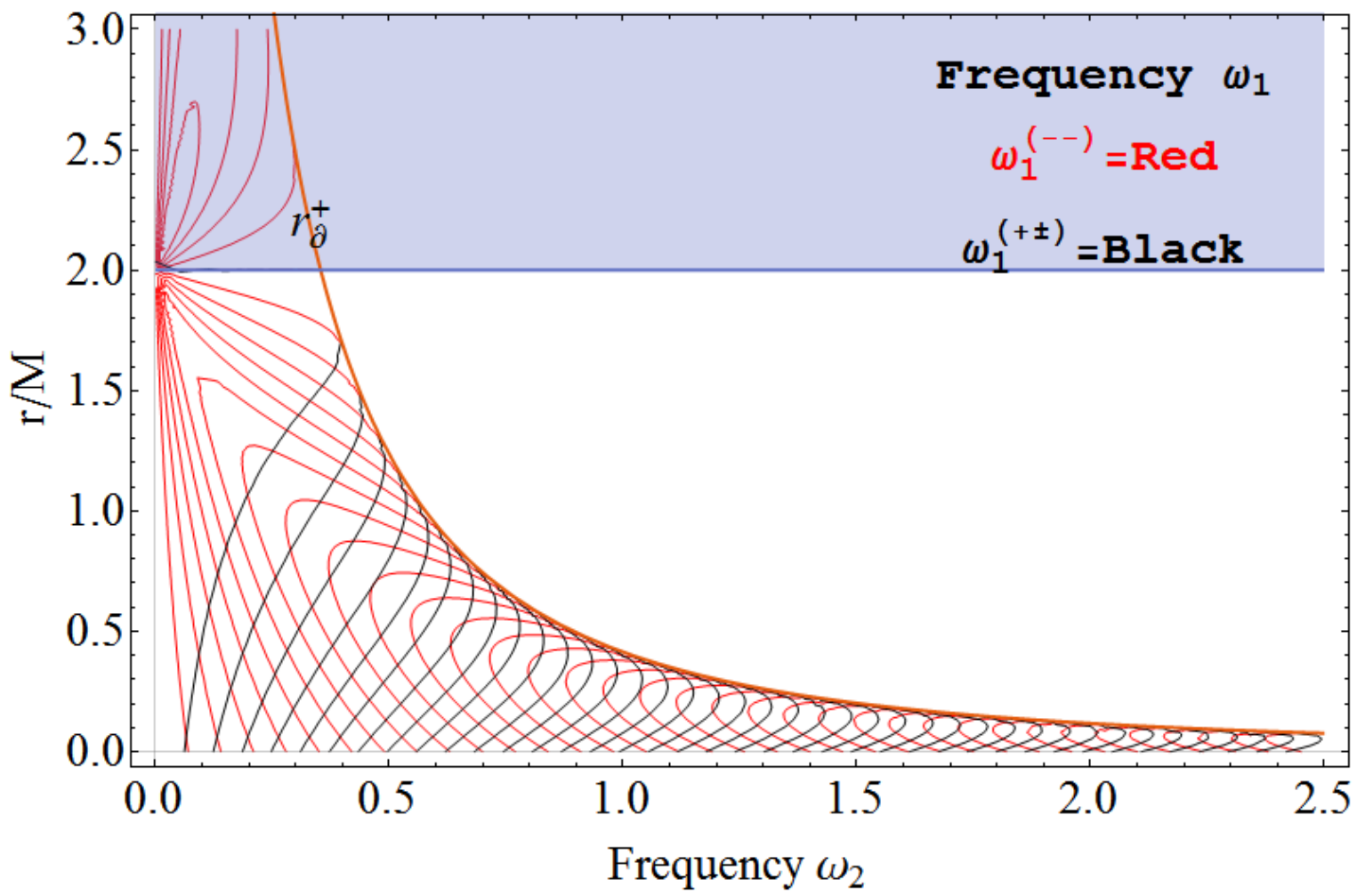}
\end{tabular}
\caption[font={footnotesize,it}]{\emph{Left} and  \emph{enter  panels}: Models $\mathbf{XX}$
where $\mathbf{X}=\{\mathbf{A,B,C,D}\}$  considered  in Table\il \ref{Table:ABCDmodels}.
The spin $a_{\omega}$ as a function of   $r/M$
for different origins $a_0$ with tangent point $(r_g,a_g)$ on the horizon curve.
 The spins $a_g <a_g^o$ are represented on  $\mathbf{X}^i<\mathbf{X}^o$, where  $a_g^o$ ($\mathbf{X}^o$) corresponds to the outer spin of the bundle for  $r=r_g$.
$\mathbf{X=A}$: origin  $a_0=1/2$, frequency  $\omega=2$;
$\mathbf{X=B}$: $a_0=1$, $\omega=2$;
$\mathbf{X=C}$ $a_0=2$, $\omega=1/2$;
 $\mathbf{X=D}$ $a_0=4 $, $\omega=0.25$.
The bundle  $D$ is related to  $\mathbf{C}$ as $a_g(\mathbf{D})=a_g(\mathbf{B})$. The model $\mathbf{D}$ ($\mathbf{B}$) defines
$r_+(a_g)$ ($r_-(a_g)$) for the \textbf{BH} spacetime with spin $a=a_g$.
(\emph{Center  panel:}) Shaded  region is delimited by the horizon curve.
Models  $\mathbf{XX}$ are shown.
Inside panel:  bundles parameterized  in terms of the horizon frequency  $\omega_H$ of Eq.\il(\ref{Eq:UE-dif}) are shown with $a_g$ values.
\emph{Right panel:} The shaded region is $r>r_\epsilon^+=2M$.
Frequencies  $\omega_1$=constant of Eqs.\il(\ref{Eq:gen1-dat})-(\ref{Eq:gen2-dat}) in terms of the frequencies $\omega_2$ and  $r$.}
\label{Fig:PostMerge}
\end{figure}
\begin{table*}[h!]
\caption{\label{Table:ABCDmodels} Models $\mathbf{XX}$,
where $\mathbf{X}=\{\mathbf{A,B,C,D}\}$ as defined  in Fig.\il\ref{Fig:PostMerge}. }
\centering
\begin{tabular}{|l|l|l|}
  \hline
  \hline
 Models:  & \textbf{AA-Model} & \textbf{BB-Model} 
 \\
  Bundle frequency: &$\mathbf{\omega_0=2}$ &  $\mathbf{\omega_0=1}$ 
  \\
  Tangent point $r_g$: & $r_g = 2/17$& $r_g = 2/5\in r_ -$ 
  \\
  Bundle spin $a_g$ at $r_g$:& $a_g=0.470$ & $a_g=\underline{a_ {\omega}^-} (r_g) = 4/5$ \textbf{{(D)}}
  \\
  &${a_ {\omega}^+} (r_g) =0.474$ ($A^{o}>A^i$)& ${a_ {\omega}^+} (r_g) = 13/15$ ($B^{o}>B^i$)
  \\
  Horizon frequency $\omega_H$ at  $r_g$: &$\mathbf{\underline{\omega_H^-(r_g, a_g)} = 2\equiv \omega_0}$ & $\mathbf{\underline{\omega_H^-(r_g, a_g)} = 1\equiv \omega_0}$ 
 \\
Bundle frequency for $r_ +$ or $r_-$: &${\omega_H^+(r_g, a_g)} =
 1/8$ &${\omega_H^+(r_g, a_g)} =
 1/4$ ($D_i$) 
 \\
 Horizon frequency ($X^{o}$): &$\omega_H^-(r_g,  {a_ {\omega}^+}(r_g)) =1.984$
$\omega_H^+(r_g, a_ {\omega}^+(r_g)) =0.126$ &$\omega_H^-(r_g,  {a_ {\omega}^+}(r_g)) =0.865$
$\omega_H^+(r_g, a_ {\omega}^+(r_g)) =0.289$
\\
  \hline
  \hline
  \hline
 Models:   & \textbf{CC-Model} &\textbf{DD-Model}\\
  Bundle frequency: & $\mathbf{\omega_0=1/2} $ &$\mathbf{\omega_0=1/4} $
  \\
  Tangent point $r_g$: & $ r_g =1 $& $ r_g =8/5 $
  \\
  Bundle spin $a_g$ at $r_g$:&$a_g=1 $ & $a_g=4/5$ \tb{\textbf{(B)}}
  \\
  &${a_ {\omega}^+} (r_g) =5/3$ $(C^o>C^i)$&${a_ {\omega}^+} (r_g)=3.644$
  \\
  Horizon frequency $\omega_H$ at  $r_g$: & $\mathbf{\underline{\omega_H^-(r_g, a_g)} = 1/2\equiv \omega_0}$  &
 $\mathbf{ \underline{\omega_H^+(r_g, a_g)} = 1/4\equiv \omega_0}$ ($B_i$)
 \\
Bundle frequency for $r_ +$ or $r_-$:&${\omega_H^-(r_g, a_g)} =
 1$ \textbf{\texttt{(B)}}
 \\
 Horizon frequency ($X^{o}$):&$\emptyset$&$\emptyset$\\
  \hline
  \end{tabular}
\end{table*}

\textbf{On the tangent lines}

A relevant aspect of the metric bundle is that it is tangent to the horizon the extended plane.
The  tangent line  with respect to the  horizon is horizontal  only for extreme Kerr \textbf{BH} (where the line $a =$constant has a double contact point on the tangent  bundle) and asymptotically vertical in the static case.
To  study the tangents at the horizon,  we consider the variations\footnote{The definition of metric bundle, tangent to the horizon in the extended plane presented in this work, should not  be confused with the definition  of \emph{bundle metric} and of the \emph{tangent bundle} of a differentiable manifold
in differential geometry.
A \emph{metric} on a \emph{vector bundle} is a choice of smoothly varying inner products
on the fibers.
While the a \emph{tangent bundle} \textbf{TM} could be defined  as the  disjoint union of tangent vectors in $M$.
Although different  in their definitions, it is clear that the concepts introduced in this analysis can be read in terms of \emph{tangent bundles} in a  differential manifold, providing a deep insight on the properties considered here. While it is not  the goal of this work,
 it is worth noting that  we may consider the horizon as a one-dimensional surface  embedded in the  extended plane
considered as  $R^2$ (including the reflection $\pi_a^-$).
 For a general and smooth  curve $c(r)$ in $R^2$,  the associated tangent bundle  may be seen as a regular surface  in  $R^4$,
written as
$T(r, \epsilon)=[c(r), \epsilon c'(r)]$ with
$ \epsilon\in [-1,1]$. The  metric bundles  identify at every point on the horizon
a tangent vector.  We study the tangent to the horizons and the metric bundles  ending Sec.\il(\ref{Sef:aomega}).
}
$\partial_r a_{\pm}=({M-r})/{\sqrt{-(r-2M) r}}$, with  $a_{\omega}^{\pm}$ in the form of Eq.(\ref{Eq.lcospis})
(or, alternatively,  Eq.\il(\ref{Eq:UE-dif})).
The tangent curve is:
\bea\label{Eq:a-tangent}
a_{tangent} (r)\equiv \frac{r (M-r_g)+M r_g}{\sqrt{-(r_g-2M) r_g}}\quad \mbox{where}\quad r_g\in[0,2M],
 \eea
and the tangent point $r_g$ is a parameter.

The  bundle-horizon tangent line at the point $(r_g,a_g)$ is provided by the relation
$\partial_r a_{\omega}^{\pm}=\partial_r a_{\pm} $, where  $a_{\pm}$ is the horizon curve in the extended plane.
These solutions are  shown in  Fig.\il\ref{Fig:manet}, where  some properties of the tangents are highlighted.

The solution of the tangency condition
$a_g=a_{\pm}$
 leads to the functions $r^{real}_g$ and $r^{\checkmark}_g$:
\bea\label{Eq:rgpmrg}
a_g=a_{\pm}:\quad \frac{r^{real}_g}{M}\equiv\frac{2 a_0^2}{a_0^2+4M^2}\quad \mbox{and}\quad \frac{r^{\checkmark}_g}{M}\equiv \frac{8M^2}{a_0^2+4M^2}.
\\
\mbox{Viceversa: }\quad
\frac{r_g^\mp}{M}\equiv \frac{r_\mp(a_g)}{M}=
1\mp\sqrt{\frac{\left(a_0^2-4M^2\right)^2}{\left(a_0^2+4M^2\right)^2}}.
\eea
Figure \ref{Fig:Bollet} shows that $(r^{real}_g,r^{\checkmark}_g)$ coincide with  $(r_g^-,r_g^+)$.
\begin{figure}[h!]
\centering
\begin{tabular}{ll}
\includegraphics[scale=.43]{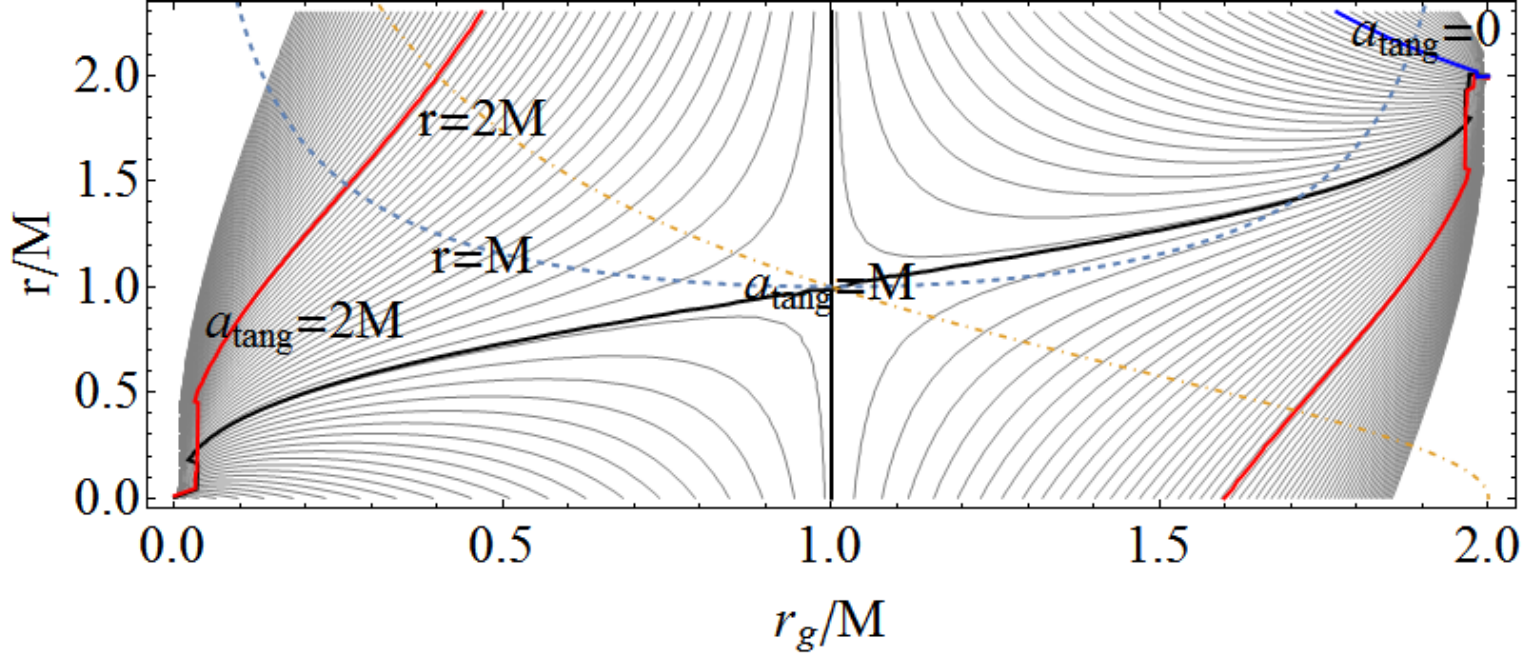}
\includegraphics[scale=.43]{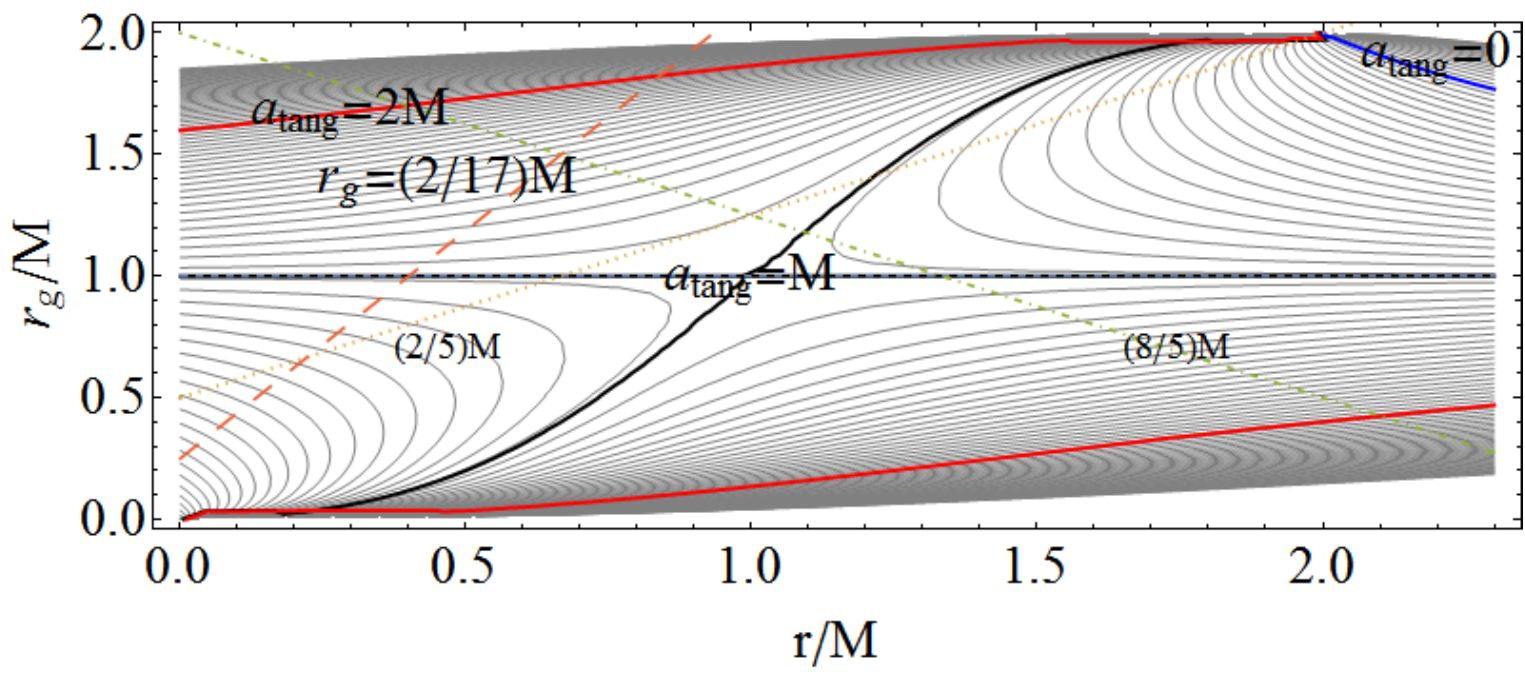}
\\
\includegraphics[scale=.7]{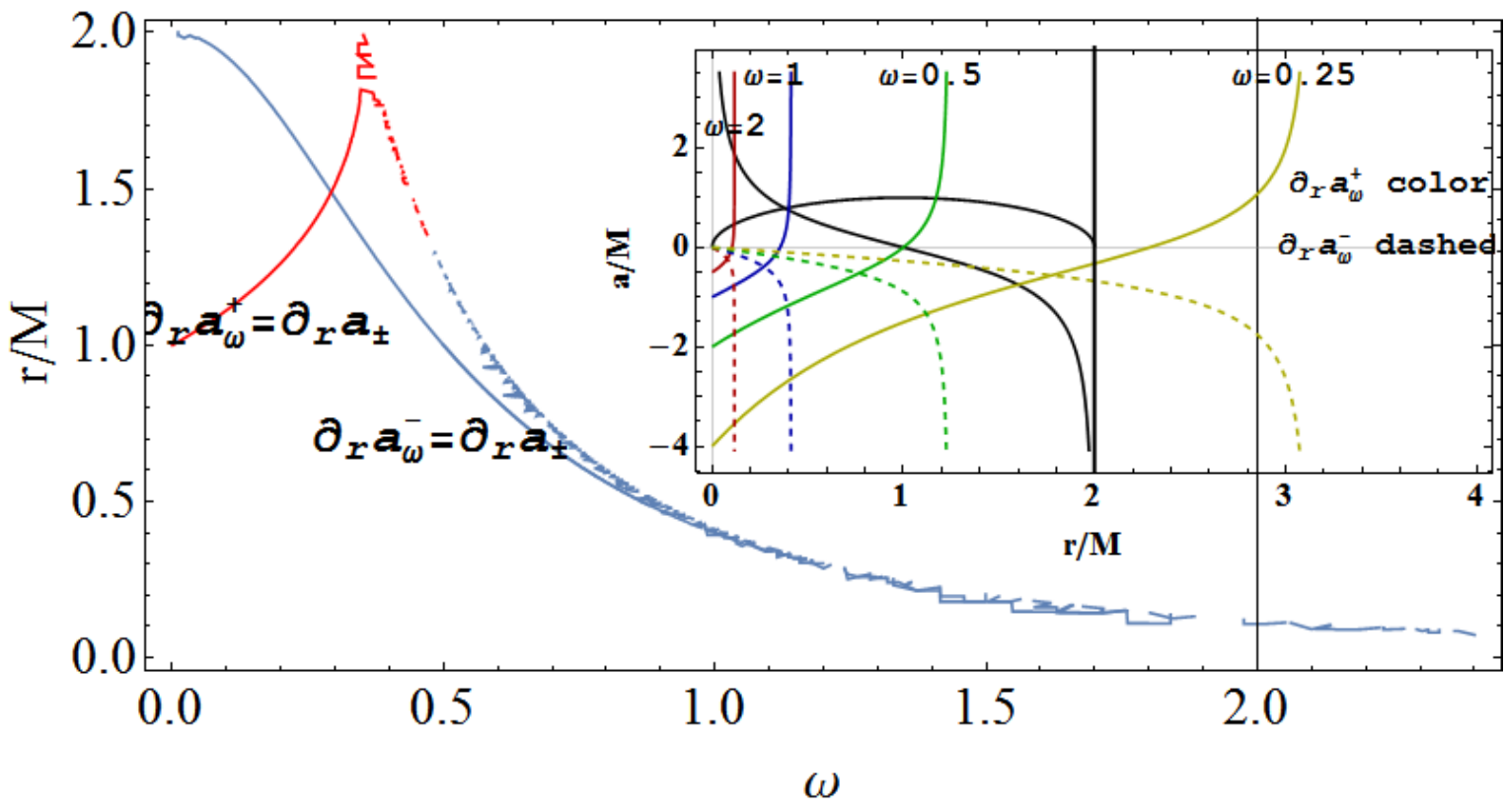}
\end{tabular}
\caption[font={footnotesize,it}]{Below panel:
 Solutions  $\partial_r a_{\omega}^{\pm}=\partial_r a_{\pm}$ in the plane $r/M-\omega$.
Bundles frequencies and  the contact point to the horizons are represented for
$\omega=0.5$. The \textbf{CC} model  of  Fig.\il\ref{Fig:PostMerge} and Table \ref{Table:ABCDmodels}, where $r_g=M$ and $a_g=M$.
Inside  panel: The black curve is the  horizon $a_{\pm}$  and the function $\partial_r a_{\pm}$; there are  the two vertical asymptotes
for the point
$(a=0,r=0)$ and  (a=0,r=2M) corresponding to the limiting static geometry.
Models  $\mathbf{XX}$ are shown with bundle frequencies for the evaluation of $(\partial_r a_{\omega}^{\pm}(X))$,
where $\mathbf{X}=\{\mathbf{A,B,C,D}\}$ is considered  in Table\il\ref{Table:ABCDmodels} and Fig.\il\ref{Fig:PostMerge}.
Considering the models curves $(\partial_r a_{\omega}^{\pm}(X))$, their intersections
 with $\partial_r a_{\pm}$  provide the tangent point $r_g$  and the tangent spin point  $a_g$.  Moreover,
they indicate the bundle curve $a^{+}_\omega$ or $a^{-}_\omega$ tangent to the horizon, as well as the tangent line inclination.
\emph{Upper panels:} $a_{tan}=$constant in terms of $(r,r_g)$; the black thick curve is $a_{tan}=M$;
curves for some special values of the $r$ and $r_g$  are also shown. These are related to the models
$\mathbf{XX}$,
where $\mathbf{X}=\{\mathbf{A,B,C,D}\}$ is considered  in Table\il\ref{Table:ABCDmodels} and Fig. \il\ref{Fig:PostMerge}.}
\label{Fig:manet}
\end{figure}
The functions  $(r^{real}_g,r^{\checkmark}_g)$ and   $(r_g^-,r_g^+)$ reveal  some properties enlightened in  Table\il\ref{Table:ABCDmodels} and  further interesting  symmetry properties of the \textbf{NS-BH} correspondence.
Fig.\il\ref{Fig:Bollet} shows the relation between the tangent point $r_g$  on the horizon and  the bundle origin $a_0$. Moreover,
 it points out  the correspondence between the  two metric bundles  $g_{\omega_0}$ and $g_{\omega_1}$,  with equal tangent spin $a_g$  and with different origin in \textbf{BH} and in \textbf{NS}. As made  explicit in Table\il\ref{Table:ABCDmodels},
such bundles are related to the construction of the inner and outer horizon  $r_{\pm}(a_g)$  of  the \textbf{BH} spacetime with $a=a_g$.

It is clear that  $r_g^{real}$  is a combination of  $r_g^{\pm}$  and provides the
tangent point $r_g(a_0)$ for the origin $a_0$.  In fact, by fixing $a_0$  there is only one tangent point  in $r_g^{real}$ equal to $r_g^-$ for $a_0<2M$, or else  equal to $r_g^+$ for $a_0>2M$; however, the second point at  $a_0$ on the curve $r_g^{\checkmark}$ (shaded region)  has no immediate meaning with respect to the bundle $g_{\omega_0}$; on the other hand, the point $r_g^{\checkmark}(a_0)$  provides    the second horizon ($r_-$ or $r_+$) in the spacetime with  $a=a_g$. Therefore, the connecting bundle $g_{\omega_0}$ with $g_{\omega_1}$ is
tangent to the horizon at $r_g^{\checkmark}(a_g)$. This case has been also represented in Table\il\ref{Table:ABCDmodels}
with respect to the $\mathbf{BB}$ and \textbf{DD} models.
Note that the \textbf{CC} model, extreme Kerr spacetime, corresponds to $a_0=2M$ and $r_{g}^{real}=r_g^{\checkmark}=M$ ($r_g^{+}=r_{g}^-=M$).

We return to the analysis of $(r_{g}^{real},r_g^{\checkmark})$ ($r_g^{\pm}$)  in Fig.\il\ref{Fig:Bollet}.
Let us consider as an example the \textbf{BB} and \textbf{DD} models.
For  $a_0<2M$ the bundles are tangent to the  inner horizon (note also the saddle point   at $a_0/M={2}/{\sqrt{3}}\approx1.1547$).
According to the \textbf{DD} model $a_0=4M$,  the correct tangent point is on $r_g^{real}=r_g^+$. The  second point, for $a_0=4M$,
on the $r_g^{\checkmark}=r_g^-$ is the tangent point  $r_g(B)$ in the  \textbf{BB} model with $a_0=M$. The
\textbf{BB} and \textbf{DD} models share same tangent spin $a_g$.

In conclusion, in this \textbf{BH} spacetime:
\bea\nonumber
\mbox{the outer horizon is} \quad r_+=r_g(D)=
r_g^{real}(D)=r_g^+(D)=r_g^{\checkmark}(B)=r_g^+(B),
\\ \label{Eq:fast-can}\mbox{the inner horizon is}\quad
r_-=r_g(B)=
r_g^{real}(B)=r_g^-(B)=r_g^{\checkmark}(D)=r_g^-(D).
\eea
 Consequently, we could say that  $r_g^{real}(a_0)$ represents the horizon curve $a_{\pm}$ as the envelope surface in the  extended plane (note that asymptotically, for large values of  $a_0$, $r_g^{real}(a_0)$ approaches $2M$ from left).
On the other hand,  $r_g^{\checkmark}(a_0)$ provides information on the corresponding metric bundle  and  the second horizon for $a=a_0$.
By using the
couple $(r_g^{real},r_g^{\checkmark})$, it is sufficient the knowledge of the  bundle origin to characterize the  \textbf{BH} spacetime defined by the tangent bundle.
\begin{figure}[h!]
\centering
\begin{tabular}{ll}
\includegraphics[scale=.7]{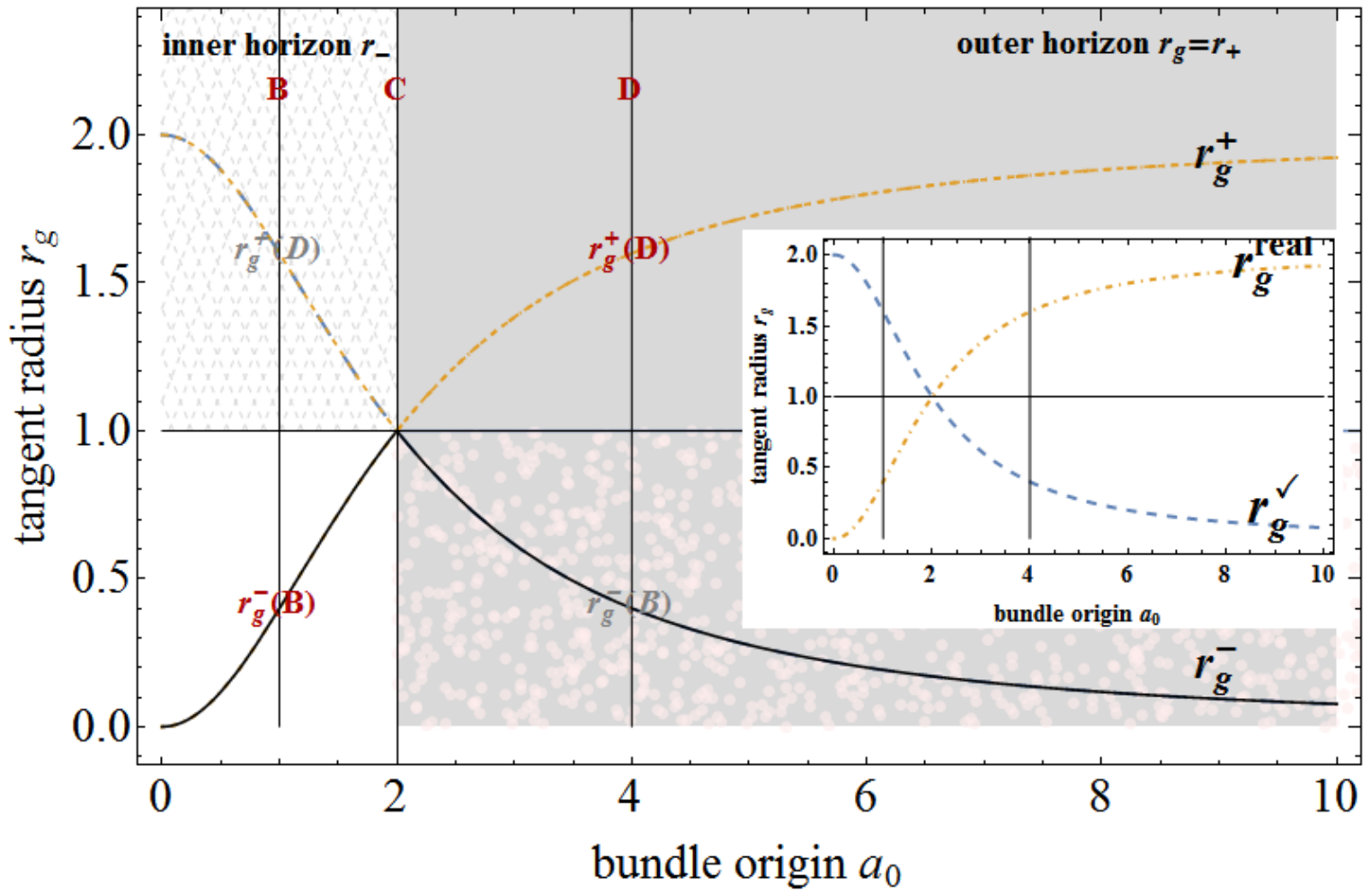}
\end{tabular}
\caption[font={footnotesize,it}]{
Tangent radius $r^{\pm}_g$ of Eq.\il(\ref{Eq:rgpmrg}) as a function of the tangent spin $a_g$. Inside panel:
 $(r_g^{real},r_g^{\checkmark})$.
The gray region represents \textbf{SNS} with $a_0>2M$ and the shaded regions are values of
$r_g^{\checkmark}$. The models  $\mathbf{XX}$ are also shown,
where $\mathbf{X}=\{\mathbf{A,B,C,D}\}$ is  considered  in Table\il\ref{Table:ABCDmodels} and Fig.\il\ref{Fig:PostMerge}.}
\label{Fig:Bollet}
\end{figure}
\section{The Kerr-Newman geometries}
\label{Sec:RN-KN-cases}
The investigation of Sec.\il(\ref{Sec:bottlem}) and Sec.\il(\ref{Sef:aomega}) is performed  here  for the case of
Kerr-Newman (\textbf{KN}) and Reissner-Nordstr\"om (\textbf{RN})   spacetimes and in the region outside the equatorial plane of the
Kerr spacetime. This further analysis  will allow us to better evaluate  the role of  the  frame-dragging. The analysis of
Fig.\il\ref{Fig:SoOv}  is presented  in  Fig.\il\ref{Fig:SmallPar}    for electrically charged  geometries with $a=0$.
We prove that  the closure   of the metric bundles   is a consequence of the rotation of the  singularity:
the  correspondent curves, defining the \textbf{BHs} horizons for the static \textbf{RN} case, are open; the  analysis of the
\textbf{KN} case represented in Fig.\il\ref{Fig:SmallPar} better shows  the influence of the spin in the bending and separation into the two  families of closed curves on the equatorial plane.

 The Kerr-Newman geometry corresponds to  an electro-vacuum  axially symmetric  solution
  with a net electric charge $Q$,  described by
metric  (\ref{alai}) with $\Delta_{KN}\equiv r^2+ a^2+ Q^2-2M r$.
The solution  $a=0$ and $Q\neq0$ constitutes the static case of the spherically symmetric and charged  Reissner-Nordstr\"om
spacetimes.
The horizons and the outer and inner static limits for the \textbf{KN} geometry  are respectively
\bea\label{Eq:KN.RN.shown}
r_{\mp}=M\mp\sqrt{M^2-(a^2+Q^2)},\quad r_{\epsilon}^{\mp}=M\mp\sqrt{M^2-a^2 \cos ^2\theta-Q^2},
\eea
are depicted in Fig.\il\ref{Fig:Clasxcson}. \textbf{KN} naked singularities  are defined for
$\mathcal{Q_T}^2/M^2>1$, where $\mathcal{Q_T}^2\equiv (a^2+Q^2)$ is  the \emph{total \textbf{KN} charge}.
This condition implies that either $Q^2>M^2$ \emph{or} $a^2>M^2$ give rise to a \textbf{NS}--\cite{Pu:Neutral,Pu:Charged,Pu:KN}.
\begin{figure}
  \includegraphics[width=.57\hsize]{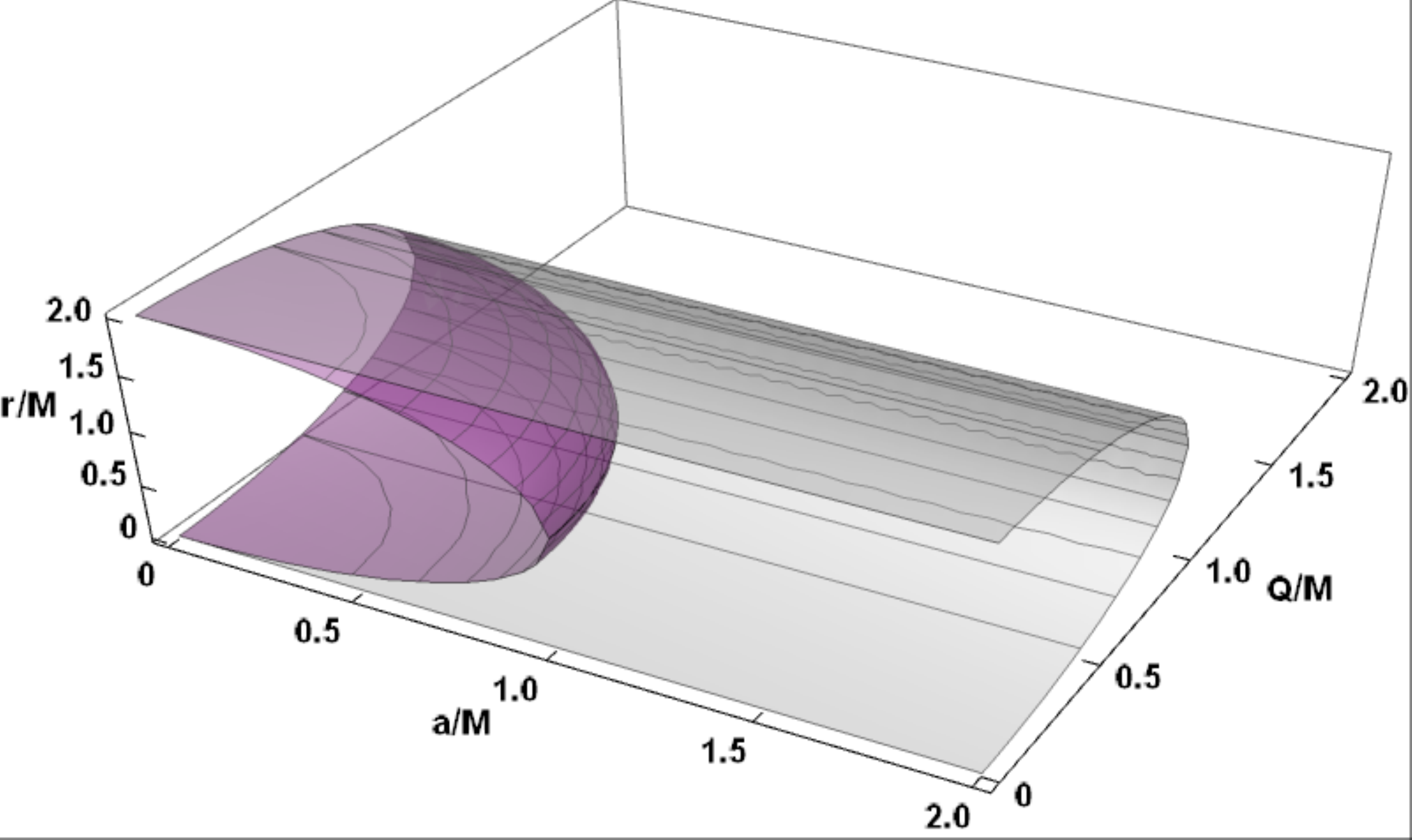}
  \caption{Kerr-Newman solution--the equatorial plane. Plot of the horizons $r_{\pm}/M$ (purple) and static limit
	$r_{\epsilon}^+/M$ (gray) as functions of $a/M$ and $Q/M$--Eqs\il(\ref{Eq:KN.RN.shown}).}
	\label{Fig:Clasxcson}
\end{figure}
Following Sec.\il(\ref{Sec:bottlem}),  the   frequencies  $\omega^{\pm}_H\equiv \omega_{\pm}(r_{\pm})$ at the horizons $r_{\pm}$ are given by
\bea\label{Eq:spoksingle}
\omega_H^-=\frac{a M\left(2 M\sqrt{M^2-(a^2+Q^2)}-Q^2+2M^2\right)}{4M^2 a^2+Q^4},\quad \omega_H^+=
\frac{aM}{2M \sqrt{M^2-(a^2+Q^2)}-Q^2+2M^2}
\eea
and are represented in Fig.\il\ref{Fig:Clasxcsonfatto}.
\begin{figure}
  \includegraphics[width=7cm]{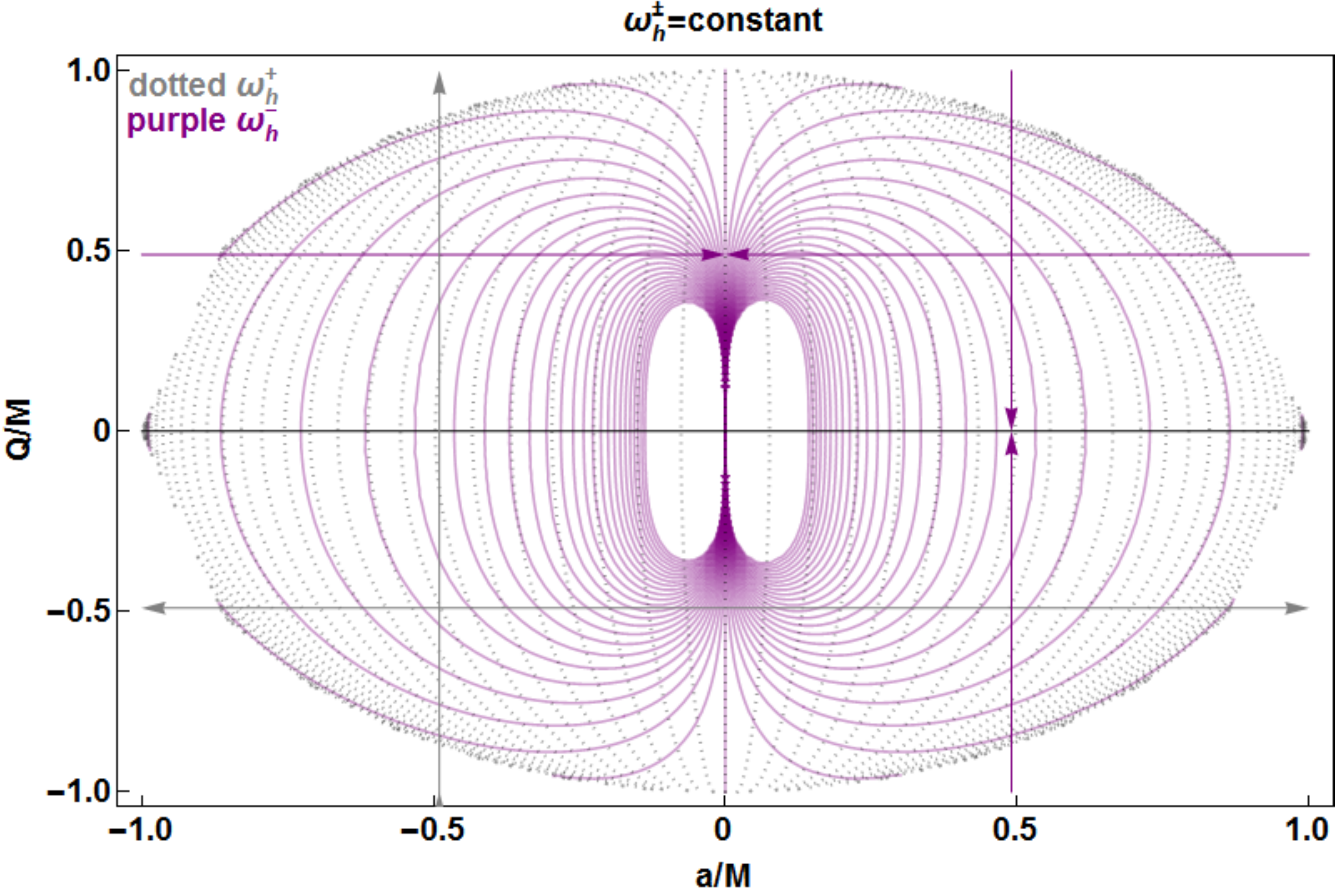}
   \includegraphics[width=5cm,angle=90]{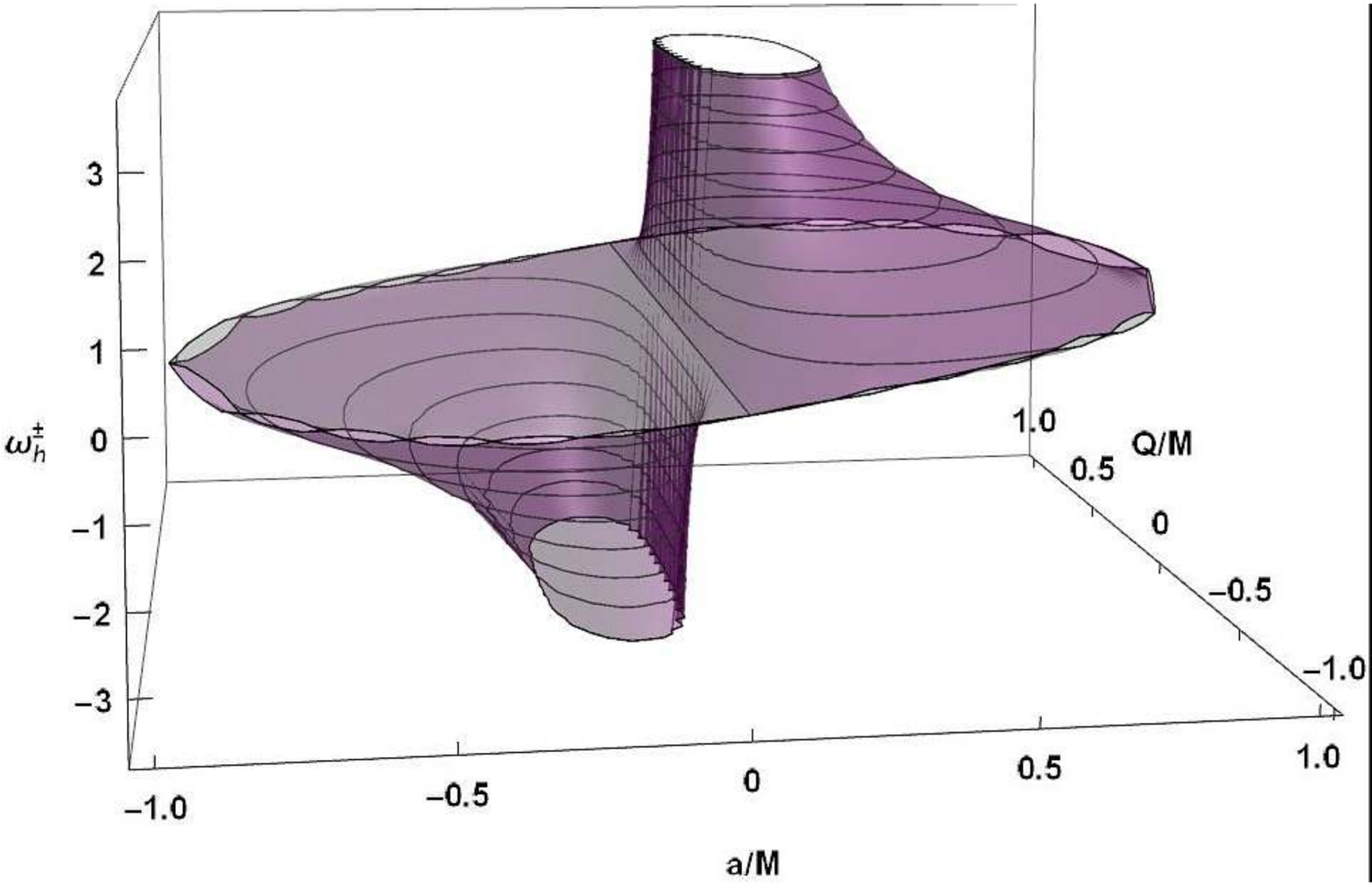}
  \caption{Kerr-Newman spacetimes. The frequencies $\omega_H=\omega_{\pm}(r_{\pm})$ at the horizons $r_{\pm}$--see Eq.\il(\ref{Eq:spoksingle}). \emph{Left panel:}  Curves $\omega=$constant. Arrows indicate the increasing magnitude of the frequencies
	$\omega_H^-$ (purple), $\omega_H^+$ (gray). \emph{Right panel:}
$3D$ plot of the frequencies  $\omega_H^{\pm}$ as functions of  the dimensionless spin $a/M$  and charge $Q/M$.  }\label{Fig:Clasxcsonfatto}
\end{figure}
In this section, we consider the problem faced in the Secs.\il(\ref{Sec:bottlem}) and  (\ref{Sef:aomega}) and analyze
the entire range   $Q^2\geq0$ and $a^2\geq0$. We  test the  conjectures  presented in  Sec.\il(\ref{Sef:aomega}) and reproduce
 the analysis of Sec.\il(\ref{Sec:bottlem}),  in particular, for the case of spherical symmetry,
  when the frame-dragging due to the source's spin is absent, isolating the effects of the electric charge  from the
	rotation component of the total charge $\mathcal{Q}_T$.
In doing so, we  generalize the extended plane $\pi_a^+$  used  in Secs.\il(\ref{Sec:bottlem}) and  (\ref{Sef:aomega}),  considering a
two-parameter family of solutions and passing
from the $\textbf{(1+1)}$ dimensional problem of the Kerr spacetimes to a $\textbf{(1+2)}$ problem of the \textbf{KN}  solutions.
Fixing one of the two components of the  total charge, we can  obtain an entire parametrized family of Einstein solutions.
The  off-equatorial case will also  be  briefly addressed.

In order to  understand the effects of the  two charge parameters $a$ and $Q$,   it is  useful to look at the solutions
(\ref{Eq:KN.RN.shown}) in the extreme cases. The axial symmetry of the metric is due to the presence of the spin of the central singularity
. The presence of the electric charge actually  ``balances'' the effects of the spin {in several ways, as we shall see below}. In fact,  we consider on the equatorial plane  the static limits and the horizons  of Eqs.\il(\ref{Eq:KN.RN.shown}) as follows
\bea\label{Eq:otov-limit}
&&\mbox{static limits:}\quad\left.r_{\epsilon}^{\pm}\right|_{\theta=\pi/2}=M\pm \sqrt{M^2-Q^2}, \quad \mbox{implying}\quad Q/M<1\quad \mbox{or}\quad Q_{\pm}^{\epsilon}=\sqrt{-(r-2M) r};\\
&&\mbox{and}\quad
\left.r_{\epsilon}^{\pm}\right|_{\theta=\pi/2}=M \quad \mbox{for}\quad  Q=M,
\\\label{Eq:horiKK-a-Q}
&&
\mbox{\textbf{BHs} horizons:}\quad
a_{\pm}=\sqrt{-Q^2-r(r-2M)} \quad \mbox{or also }\quad Q_{\pm}=\sqrt{-a^2-r(r-2M) } .
\eea
Here, $a_{\pm}$ and $Q_{\pm}$ are solutions of $r_{\pm}=r$.
On the equatorial plane, there are  two static limits, independently of the spin,  only for \textbf{KN BHs} or \textbf{NSs} having   $Q<M$. In other words, the charge component of the \textbf{KN-NS } (and only for  \textbf{NSs})  is not  ``predominant'' with respect to the spin, i.e., for \textbf{KN-NSs} with  $a\geq M$ but $Q<M$.
On  the equatorial plane, the static limits $r_{\epsilon}^{\pm}$ can be compared with   the event horizons of  the \textbf{RN}
\textbf{BH} geometry, as in the  Kerr geometry the static limit $r_{\epsilon}^+$ coincides the Schwarzschild horizon $r=2M$.
Conversely, this  similarity appears even more clearly in the definition of
 $Q_{\pm}^{\epsilon}$ in Eq.\il(\ref{Eq:otov-limit}), which is equal to  the horizons $a_{\pm}$ in the $a-r$ plane  of the  Kerr geometry.
When  $Q=M$,  there is  one static limit radius only,   independently of the spin  $a/M$.
In this sense, the  spacetime dragging is totally balanced, on the equatorial plane,  by the  electric charge   $Q^2>M^2$.
For ($\theta\neq\pi/2$, $\theta\neq0$), a static limit exists provided the charge satisfies the condition
 $Q^2\leq M^2-a^2\cos^2(\theta)$.
On the other hand, for $Q=M$ ($a=M$) the Killing horizon is defined for
 $a=0$ ($Q=0$) only.
The photon orbital frequencies in the \textbf{KN} geometry  are
\bea\label{Eq:esa.tt}
&&
\omega_{\mp}=\frac{a M\left(Q^2-2 Mr\right)\mp M\sqrt{r^4 \left[a^2+Q^2+(r-2M) r\right]}}{a^2 \left[r (r+2M)-Q^2\right]+r^4}.
\eea
Analogously to Eq.\il(\ref{Eq.lcospis}), we can define the functions
\bea
&&\label{Eq:Muss.churnc-parl}
a_{\omega}^{\mp}=\frac{\mp\sqrt{r^4 \omega ^2 \left\{\omega^2 \left[Q^2-r (r+2M)\right]+M^2\right\}}+\omega M (Q^2 -2 r M) }{\omega ^2 \left[Q^2-r (r+2M)\right]},
\quad
\mbox{and }
\\
&&\label{Eq:Q-para-metric}
(Q_{\omega}^{\pm})^2\equiv \frac{r \left\{\omega^2 \left[a^2 (r+2M)+r^3\right]-4 a M^2\omega -rM^2+2M^3\right\}}{(a \omega -M)^2} ,
\eea
where, in particular,  for the \textbf{RN} static case ($a=0$), we find
\bea
&&
\textbf{RN}:\quad (Q_{\omega}^{\pm})^2= r \left(\frac{r^3}{M^2} \omega ^2-r +2M\right)\quad\mbox{or}\quad
\omega_{\pm} =\pm \frac{M\sqrt{Q^2+(r-2M) r}}{r^2}.
\eea
Note that while the horizons $r_{\pm}$ can be re-parametrized for the total charge $\mathcal{Q}_T$ and its variation with respect to the  parameter $\mathcal{Q}_T$ is exactly the same as for  the corresponding radii  $r_{\pm}$ in the \textbf{RN} or Kerr solution,
the surfaces $\omega_{\pm}$ do not depend directly on  $\mathcal{Q}_T$; this means that the two parameters play a different role in the solutions $\omega_{\pm}$=constant,
although the envelope surfaces depend  on $\mathcal{Q}_T$ alone.
For  $Q=0$, the surfaces $a_{\omega}^{\pm}$ are shown in Fig.\il(\ref{FIG:fortCarl8}).
The surfaces $\omega_{\pm}=$constant of Eqs.\il(\ref{Eq:Muss.churnc-parl}) are shown in Figs.\il\ref{Fig:SmallPar} and \ref{Fig:fintyt}.
This is a generalization of  the case
$ Q\neq0 $ of the analysis shown in  Fig.\il\ref{Fig:SoOv}.
\begin{figure}
  \includegraphics[width=7cm]{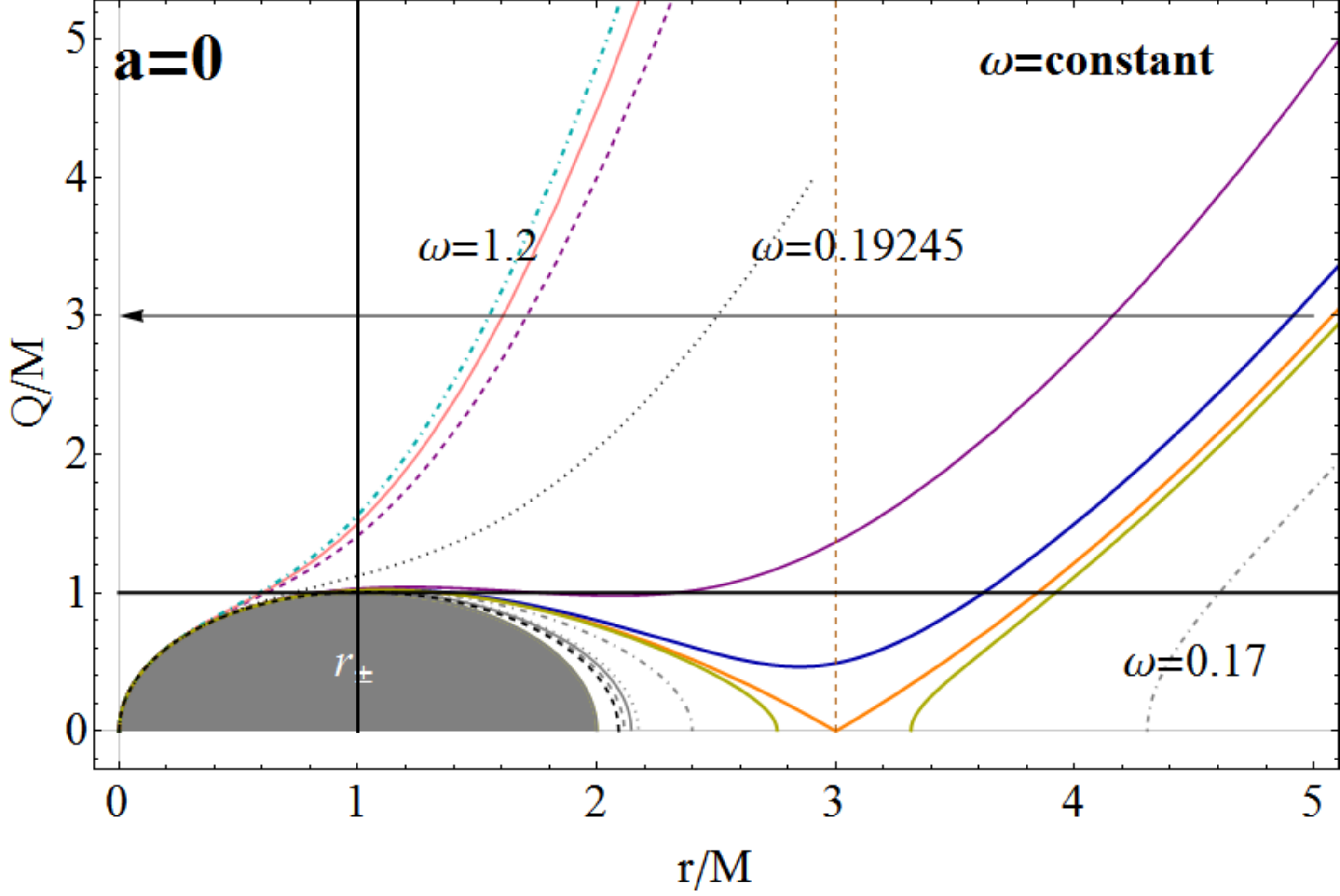}
  \includegraphics[width=7cm]{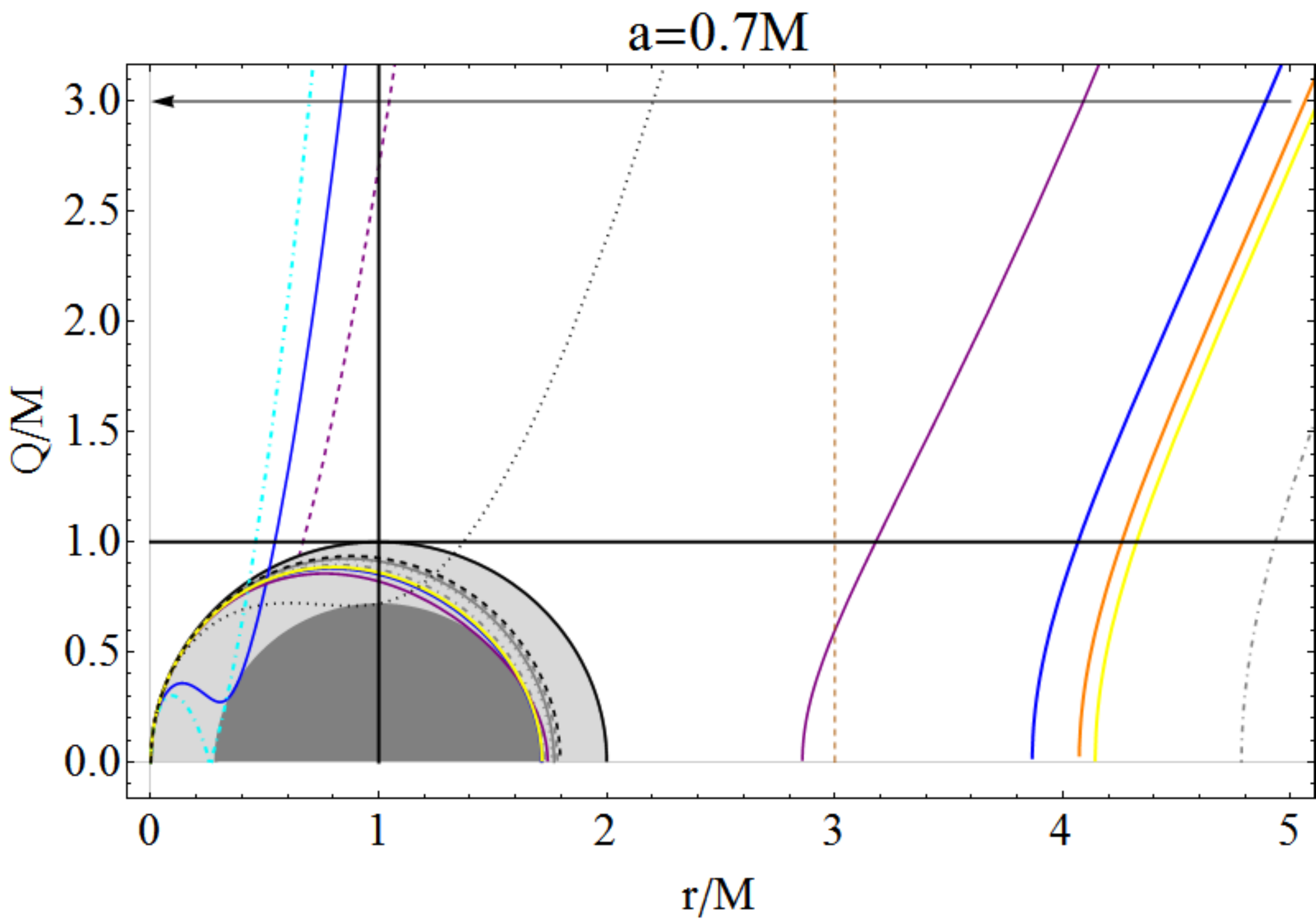}
   \includegraphics[width=7cm]{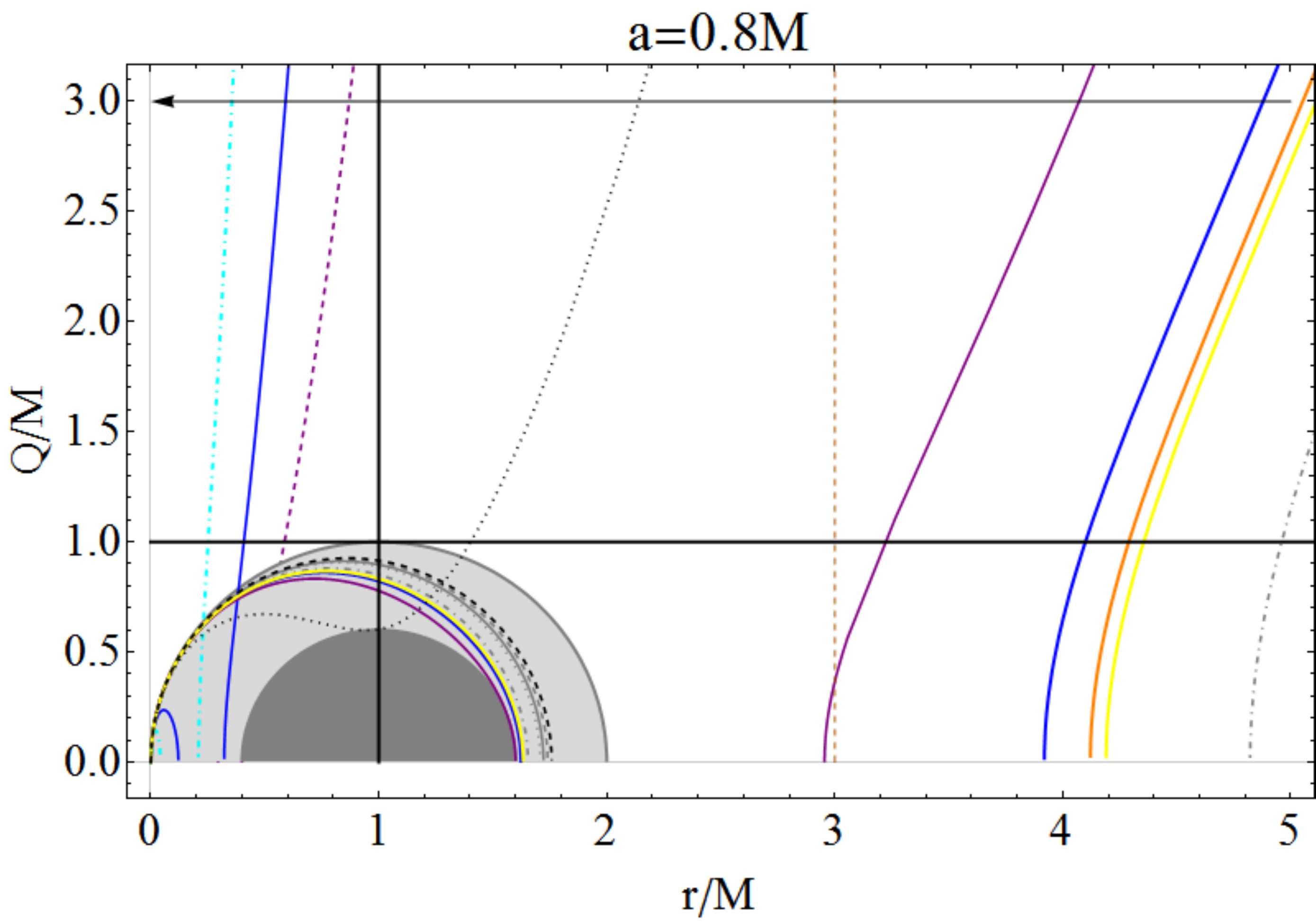}
     \includegraphics[width=7cm]{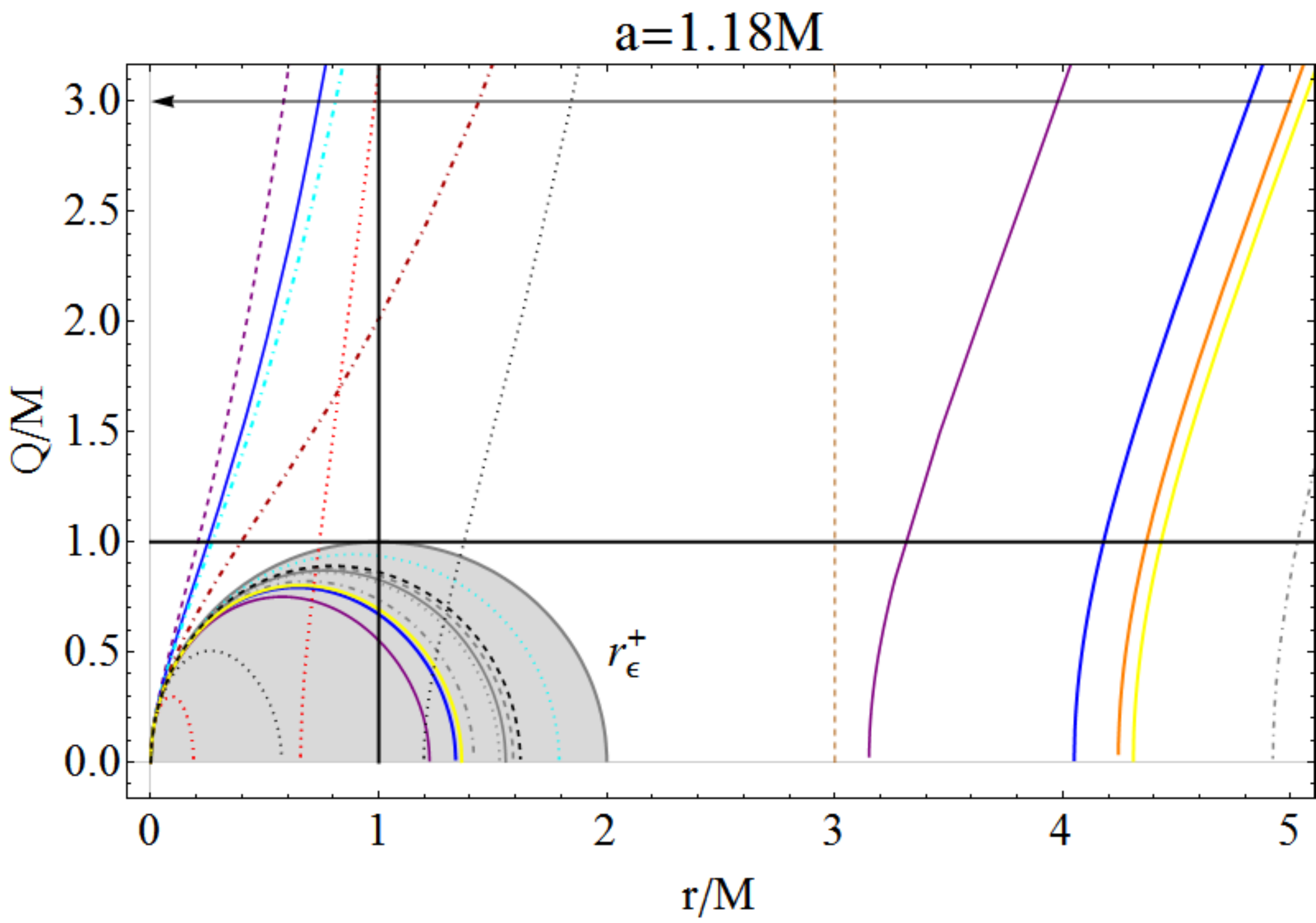}
  \caption{Equatorial plane of the Kerr-Newman spacetimes: The surfaces $\omega_{\pm}=$constant as functions of the radius $r/M$ and the electrical charge $Q/M$--Eq.\il(\ref{Eq:Muss.churnc-parl}),
 for different spins.
  This is a generalization of the analysis shown in  Fig.\il\ref{Fig:SoOv} to the case
$ Q\neq0 $.
The black region is for $r<r_{\pm}$ and the gray region is the ergosurface $\Sigma_{\epsilon}^+$. See also Fig.\il\ref{Fig:fintyt}.
}\label{Fig:SmallPar}
\end{figure}
Also in this case, we  consider some  limiting solutions to fix the different contributions of the two charge components:
\bea\label{Eq:egem-genti-fas}
\lim_{r\rightarrow0}{Q_{\omega}^{\pm}}=0,\quad \lim_{r\rightarrow0}{a_{\omega}^{\pm}}=\frac{M}{\omega},\quad
\lim_{r\rightarrow0}{{\omega}_{\pm}}=\frac{M}{a}.
\eea
An analysis of the spins $a_{\omega}^{\pm}=$constant for the static limits  $r_{\epsilon}^{\pm}$ on the equatorial plane  is shown in
Fig.\il\ref{Fig:policc}.

Equations (\ref{Eq:egem-genti-fas}) show that  in the limits considered the frequency is related to the spin source independently of the electric charge --Fig.\il\ref{Fig:ciaghiacc}.
This suggests that  we should  define  the origin of the  \textbf{KN} metric bundles $g_{\omega}^{\pm}$
as dependent on the spin $a/M$ only.
\begin{figure}[h!]
\centering
\begin{tabular}{lcr}
\includegraphics[scale=.3]{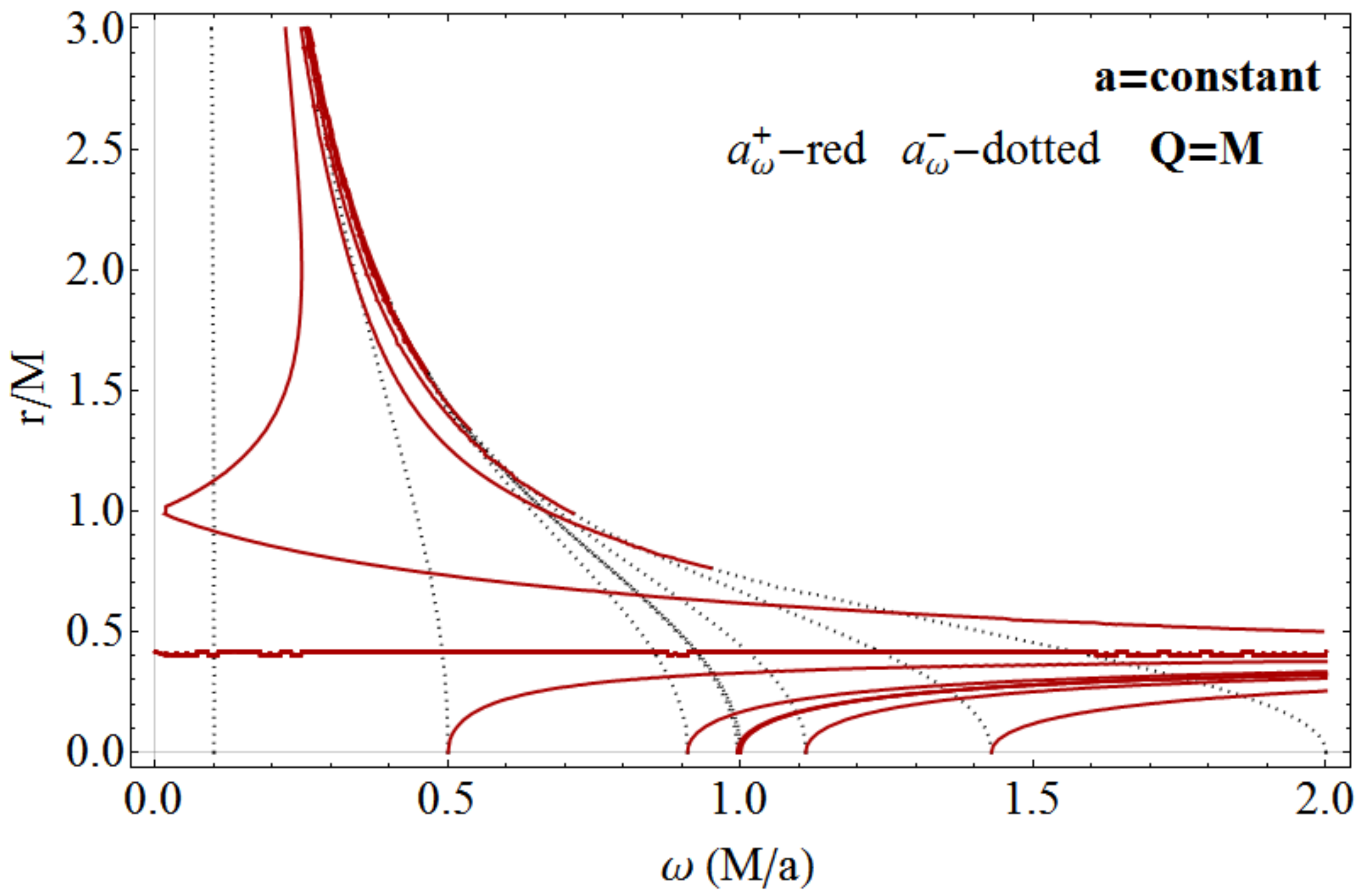}
\includegraphics[scale=.3]{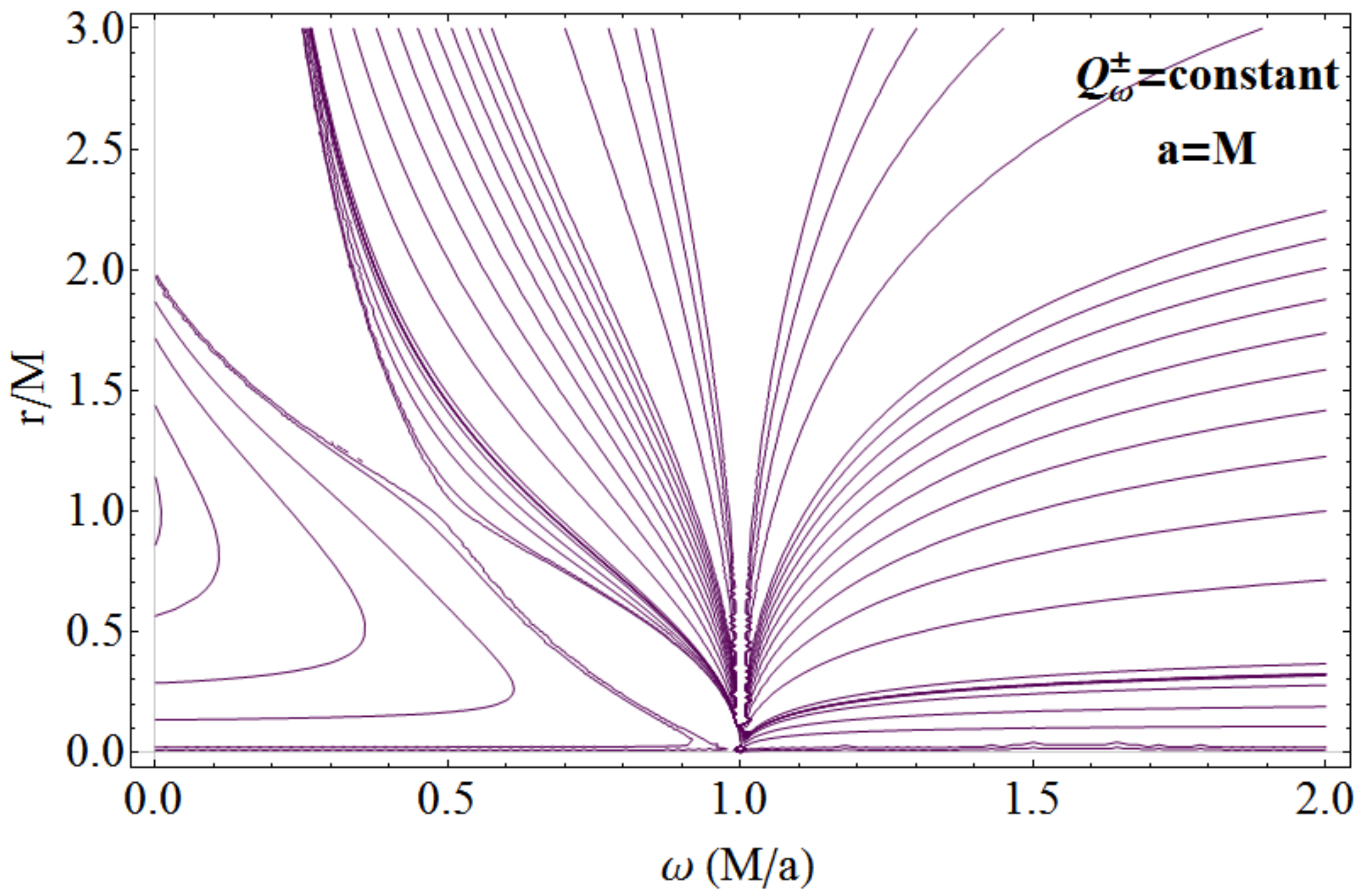}\\
\includegraphics[scale=.3]{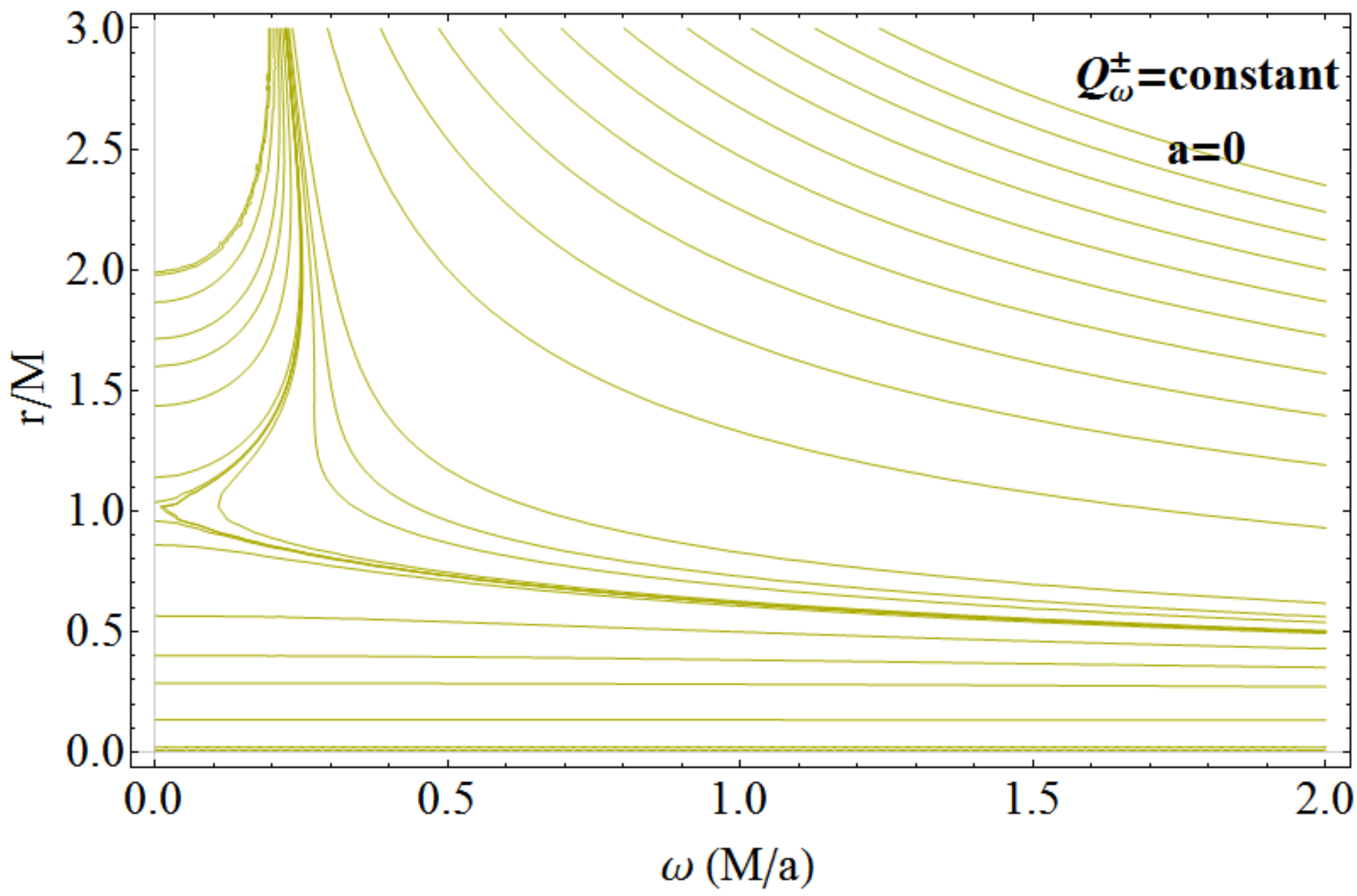}
\end{tabular}
\caption[font={footnotesize,it}]{
Equatorial plane of the Kerr-Newman spacetimes. The frequencies $\omega_{\pm}=$constant, charge
$Q_{\omega}^{\pm}=$constant, and spins $a_{\omega}^{\pm}=$constant for some limiting cases--
Eq.\il(\ref{Eq:egem-genti-fas}).}
\label{Fig:ciaghiacc}
\end{figure}
It should be considered that the Killing horizons are characterized by the ``rotation charge'', but $\omega_H^{\pm}$  does not ``carry'' any electric charge;
that is, we can always define metric  bundles considering  $\omega_0=M/a$ and the horizon frequencies.
Explicitly, we can  generalize the analysis of  Sec.\il\ref{Sef:aomega}, considering a surface $a_g(a_0;Q)$ in the case $a_0\neq0$,
 where $Q$ is a parameter, and we obtain
\bea&&\label{Eq:gresti}
a_0= \frac{2M^2-Q^2\mp2 M\sqrt{M^2-(a^2+Q^2)}}{a}\quad (r_\mp),\quad a^\mp_g(a_0)=\frac{a_0 \left(2M^2-Q^2\right)\mp2 M\sqrt{a_0^2 \left(M^2-Q^2\right)-Q^4}}{a_0^2+4M^2} %
\\
 &&\mbox{where }\quad a_0>a_L(Q)\equiv\sqrt{-\frac{Q^4}{Q^2-M^2}}\quad\mbox{implying }\quad
Q^2\in]0,M^2[,
\eea
--Fig.\il\ref{Fig:Boldnessplot}.
Adopting a notation analogue to the one used in  Sec.\il\ref{Sef:aomega},
 we solve the equation $a_{\omega}^{\pm}(r_{\pm},\omega_0)=a$
 (similarly, we could have used $Q_{\omega}^{\pm}$) and introduce the
  two functions  $a_{g}^{\pm}$.
 However, we can exploit the fact that all the curves in Figs.\il\ref{Fig:SmallPar} and \ref{Fig:fintyt} tend to
the point  ($r=0$, $Q=0$),  that is, to the  Kerr singularity. Approaching the static limit in the extended plane,
we consider the solutions of
 $Q_{\omega}^{\pm}(r_{\epsilon}^{\pm},\omega_0)=Q$:
 \bea\label{Eq:MaesMirtilo}
 Q_{\omega}^{\epsilon}\equiv \sqrt{2} \sqrt{a^2-a a_0+\sqrt{2} M\sqrt{a (a_0-a)}}
 \eea
as shown in Fig.\il\ref{Fig:MaesMirtilo}.
\begin{figure}
  \includegraphics[width=7cm]{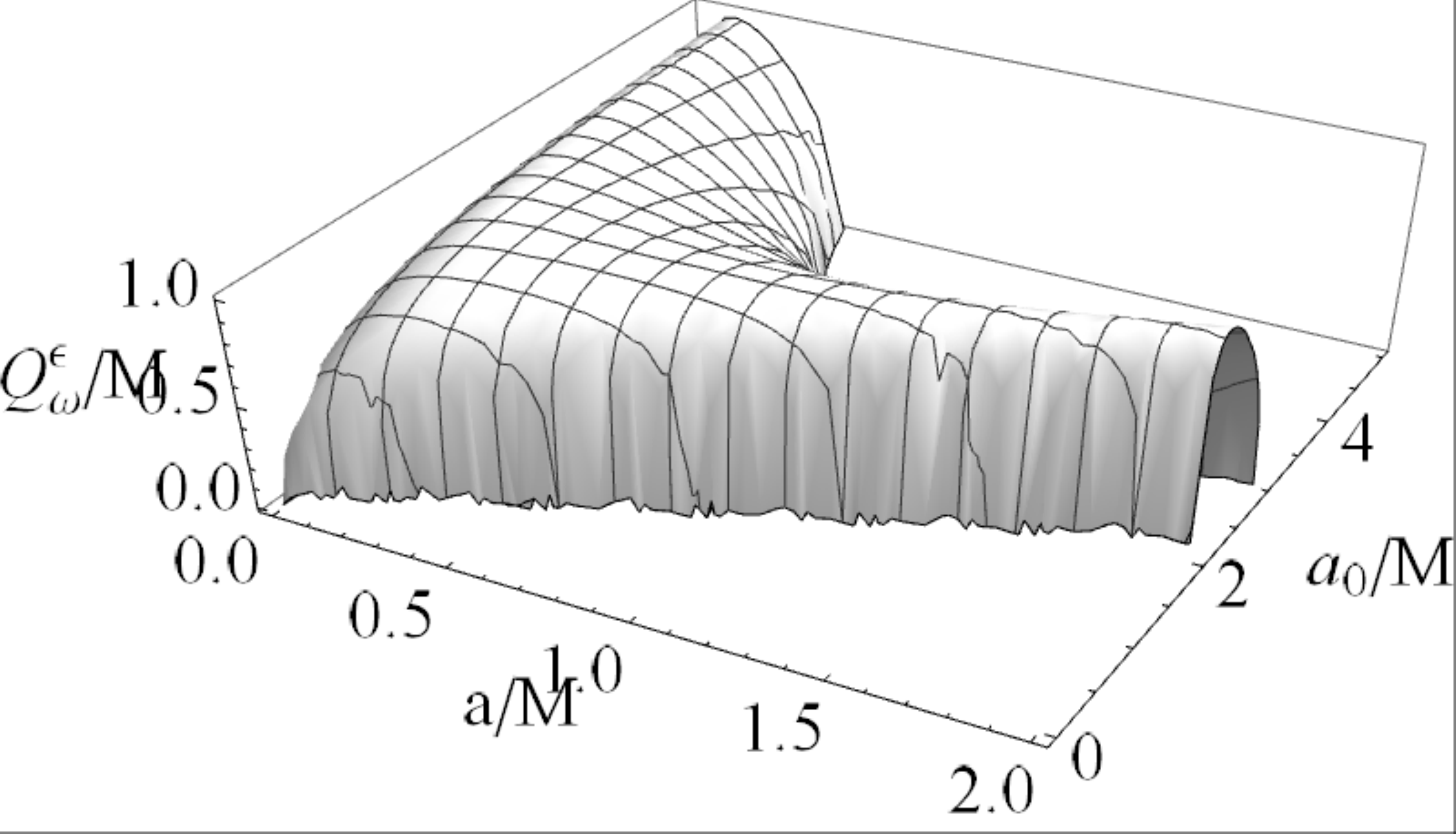}
  \caption{Charge  $Q_{\omega}^{\epsilon}$, solution of $Q_{\omega}^{\pm}(r_{\epsilon}^{\pm},\omega_0)=Q$  defined in Eq.\il(\ref{Eq:MaesMirtilo}).}\label{Fig:MaesMirtilo}
\end{figure}

\begin{figure}
  \includegraphics[width=7cm]{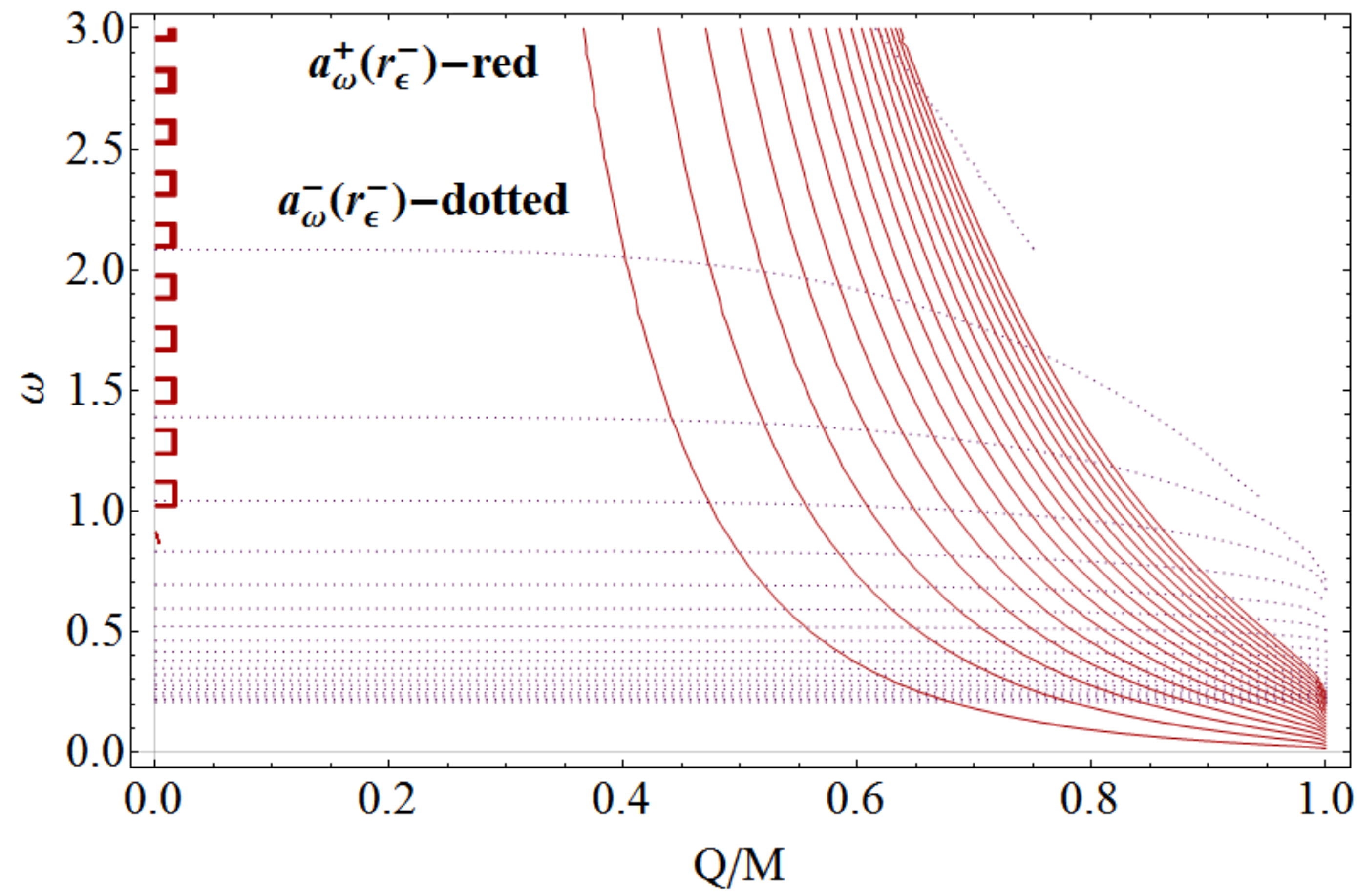}
  \includegraphics[width=7cm]{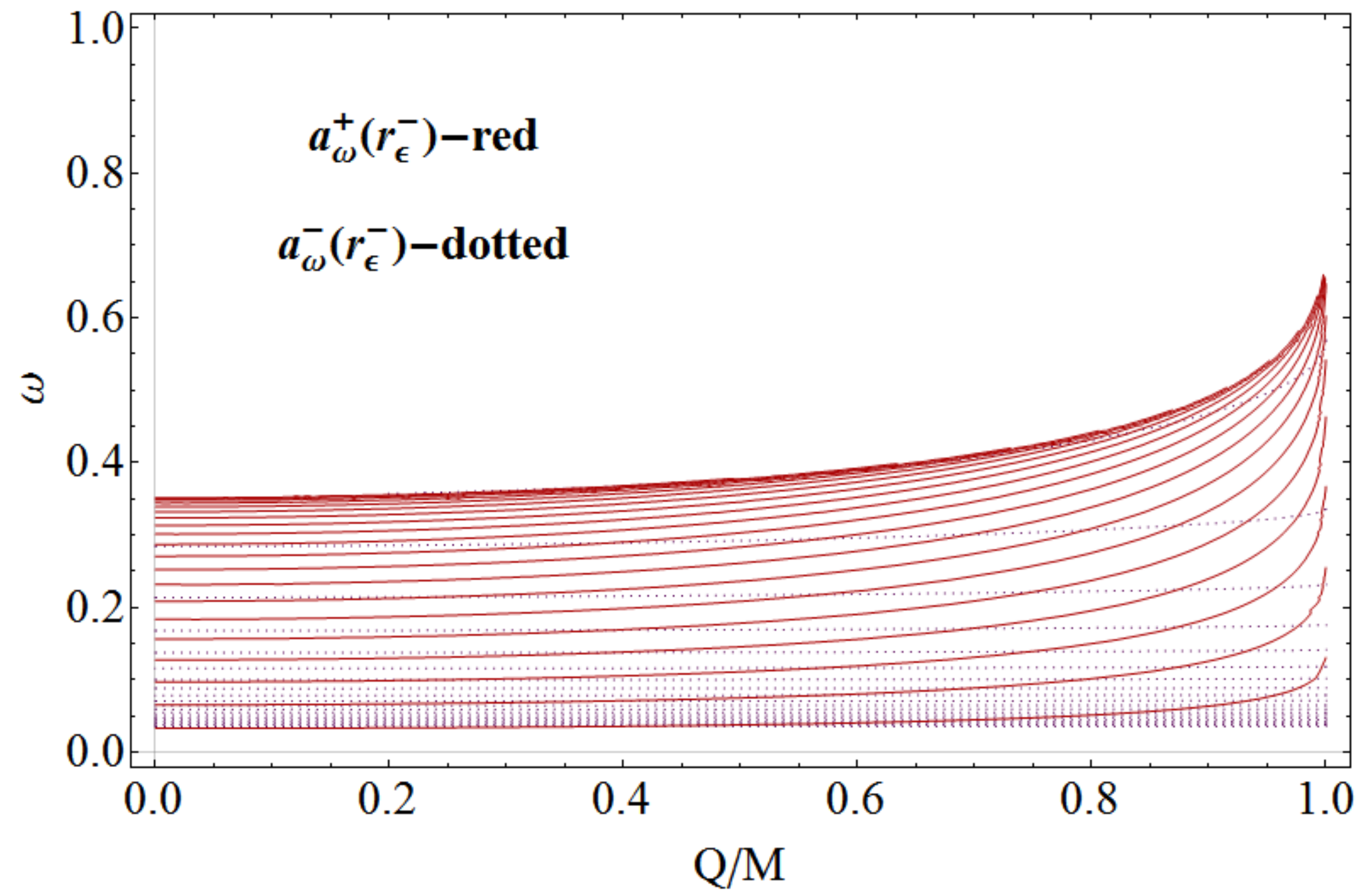}
  \caption{Spins $a_{\omega}^{\pm}=$constant on the static limits  $r_{\epsilon}^{\pm}$ on the equatorial plane }\label{Fig:policc}
\end{figure}
\begin{figure}
  \includegraphics[width=7cm]{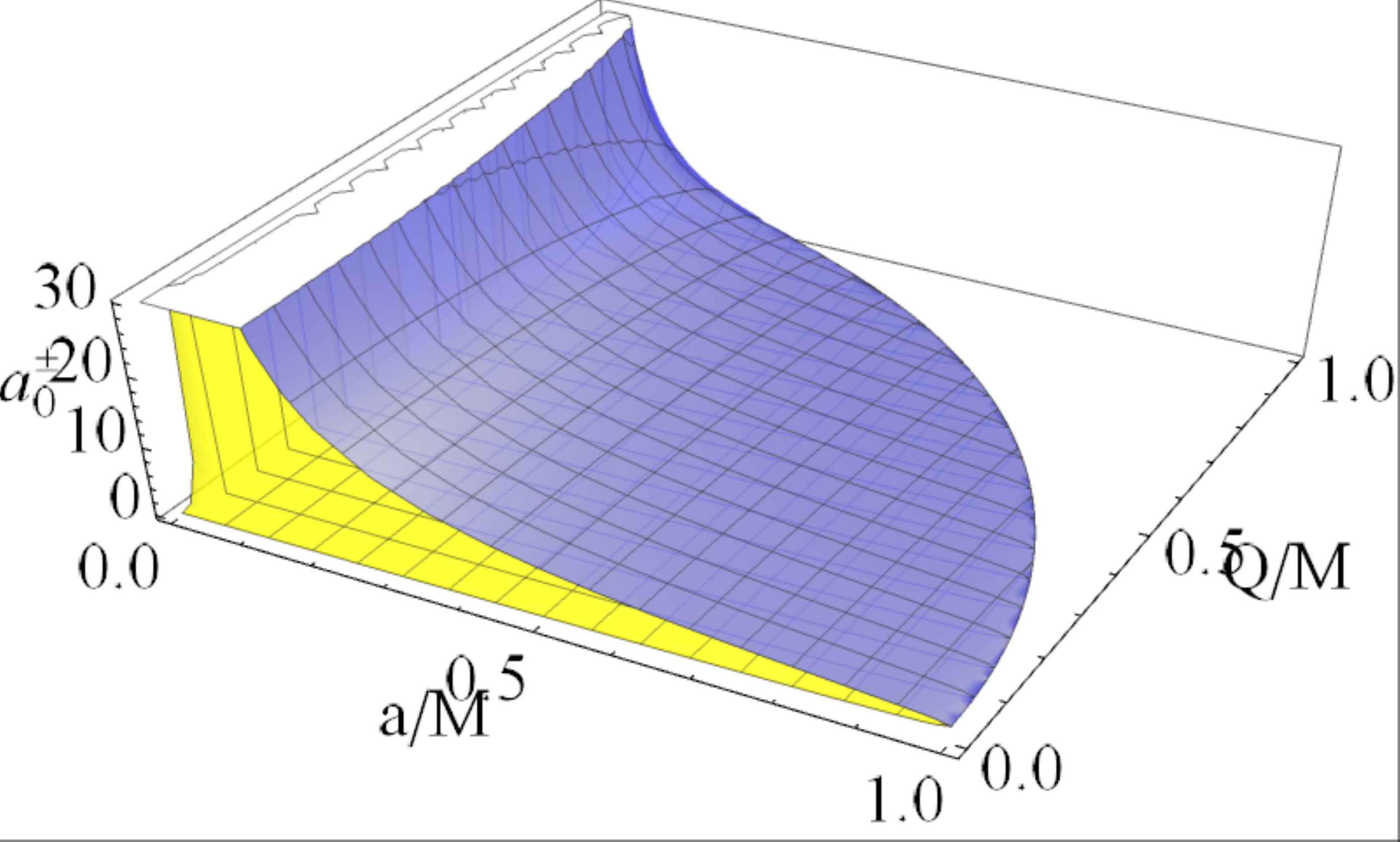}
  \includegraphics[width=7cm]{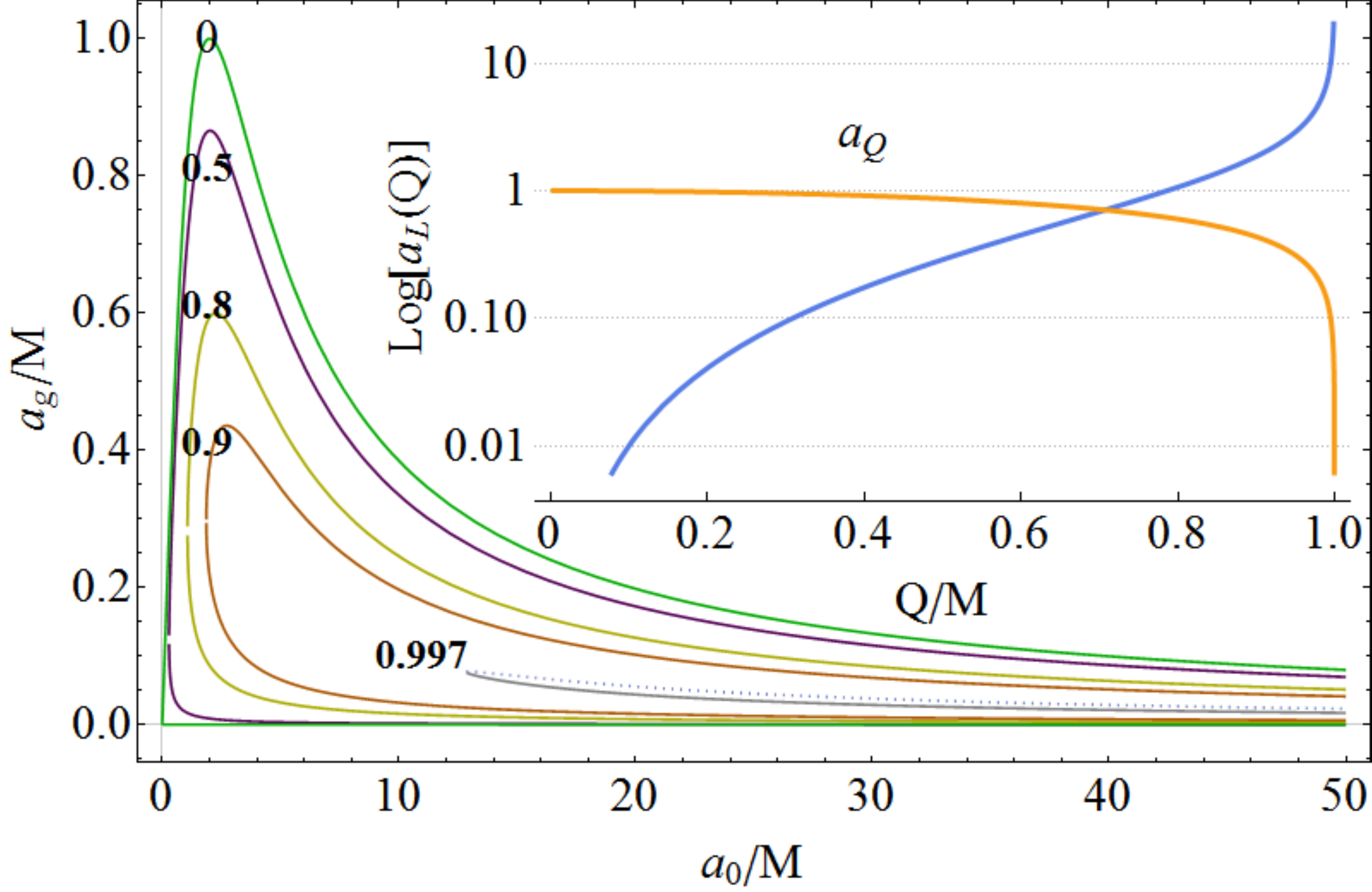}
  \caption{\emph{Left panel}: Three dimensional plot of $a_0/M$, Eq.\il(\ref{Eq:gresti}), as a function of $a/M$ and the charge $Q/M$.
	\emph{Right panel:}  Spin $a_g/M$ as a function of $a_0/M$ for different charge to mass ratios $Q/M$.
	\emph{Inside panel}: Plot of $a_L(Q)$, where  $a_Q\equiv\sqrt{M^2-Q^2}$, $a_g$ is defined for $Q\in]0,M[$ and  $a_g\geq a_L(Q)$--see Eq.\il(\ref{Eq:gresti}).  \textbf{BH} Killing horizons  are defined by the condition $a\leq a_Q$ (for $a>0$).}\label{Fig:Boldnessplot}
\end{figure}

We can see that in the extended plane it is necessary to consider the entire range of parameter values $(a/M,Q/M)$, including
the \textbf{BH} case to describe the \textbf{NSs}. For the equatorial plane, the analysis carried out for the case $Q = 0$ is confirmed also in presence of an electric charge. For predominant spins, any curve $\omega=$constant crosses the horizons at some points.
We also see the  bending of the curves limited above from the inner horizon $ r _- $, confirming   the results of Sec.\il\ref{Sef:aomega}, although with  differences which are   evident  as the electric charge increases.
\begin{figure}
  \includegraphics[width=7cm]{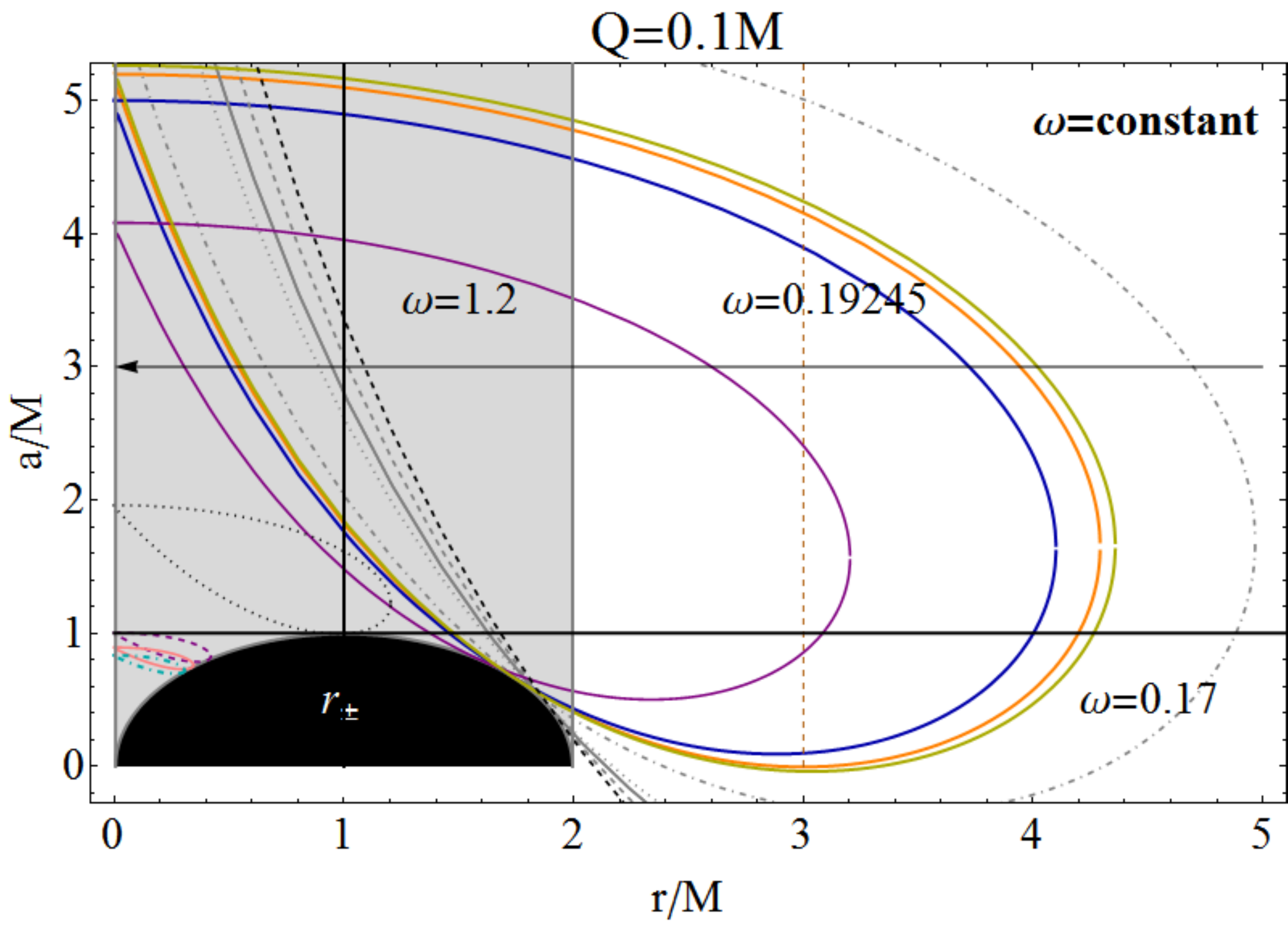}
    \includegraphics[width=7cm]{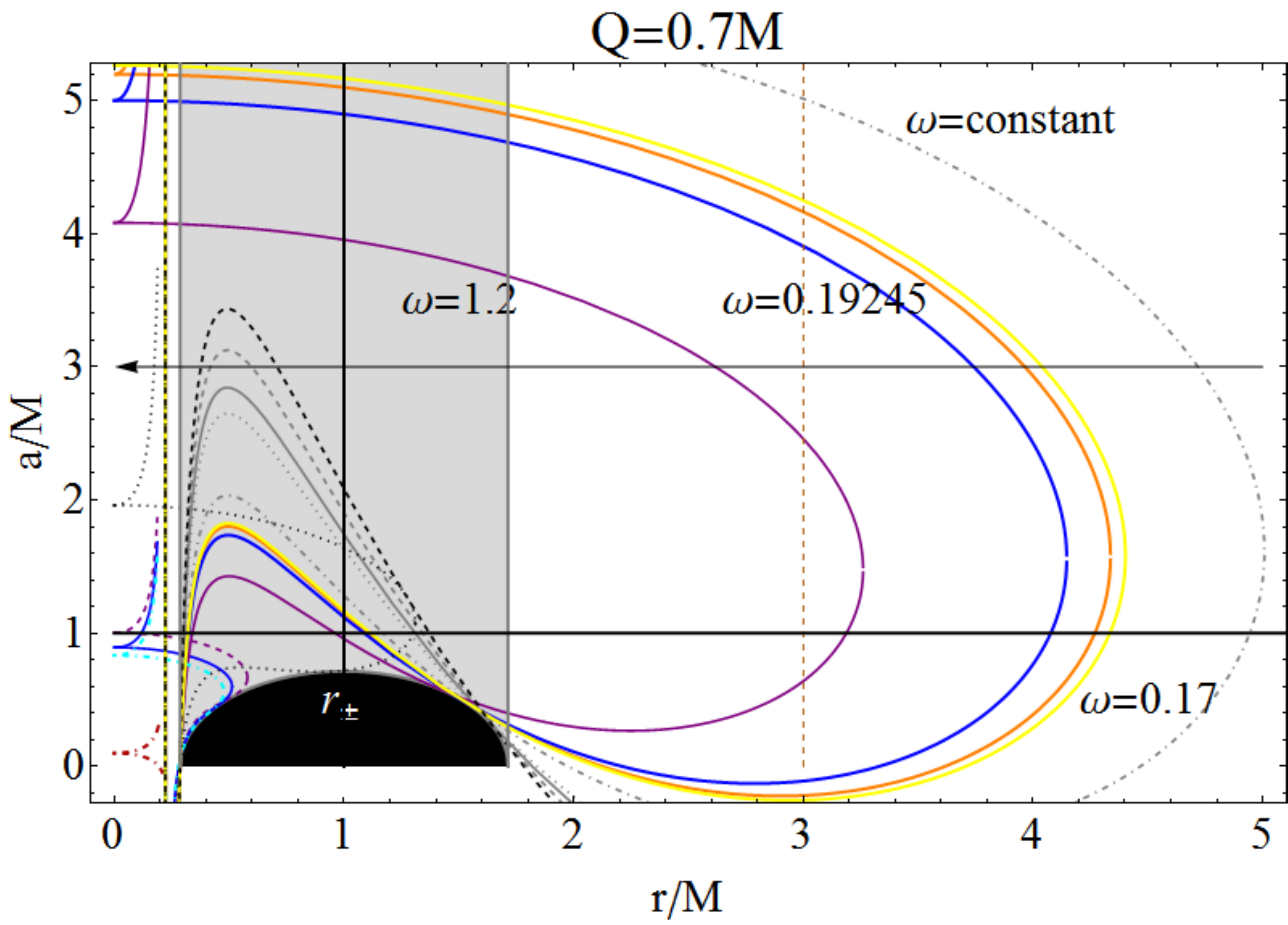}
  \includegraphics[width=7cm]{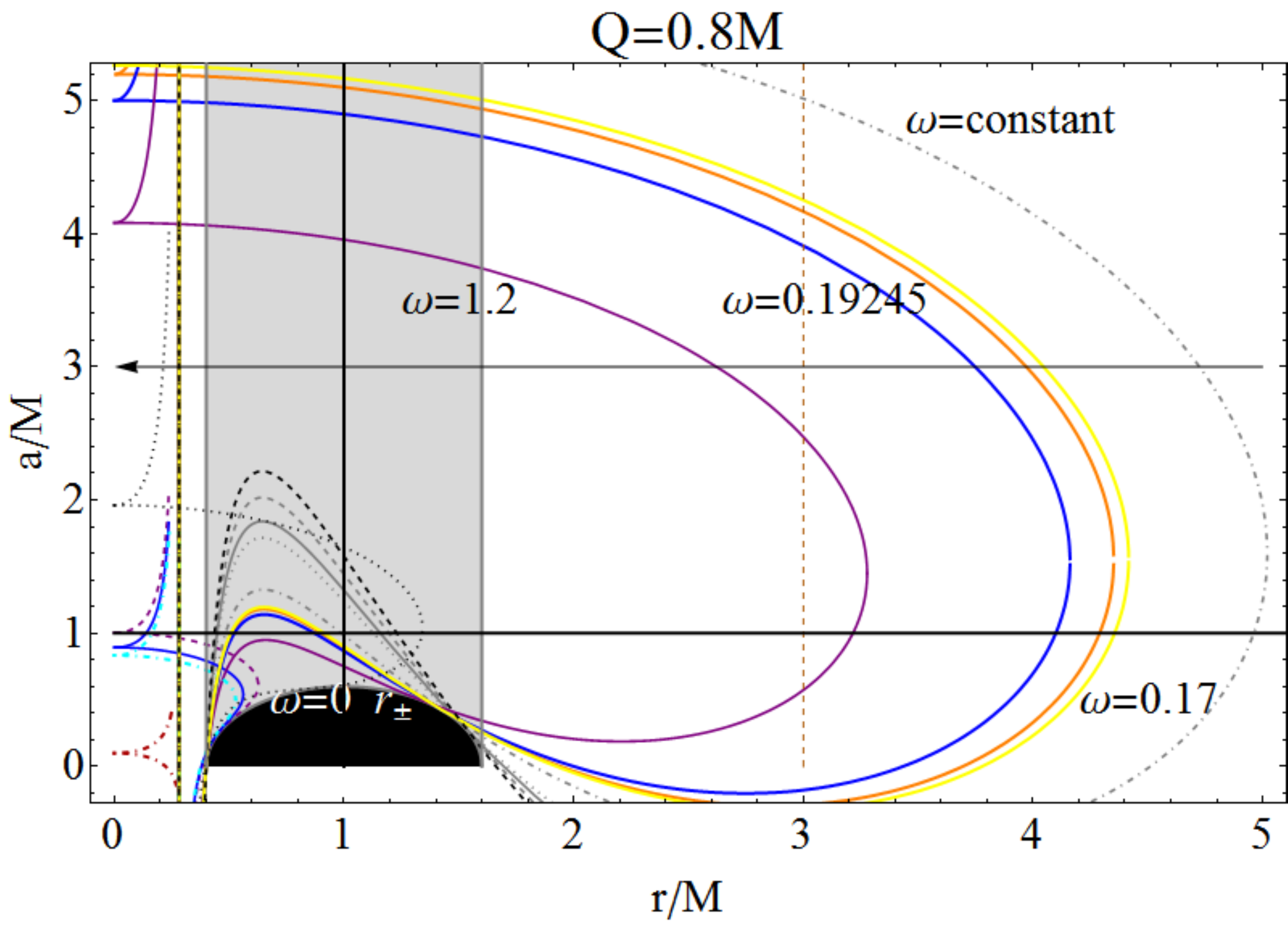}
  \includegraphics[width=7cm]{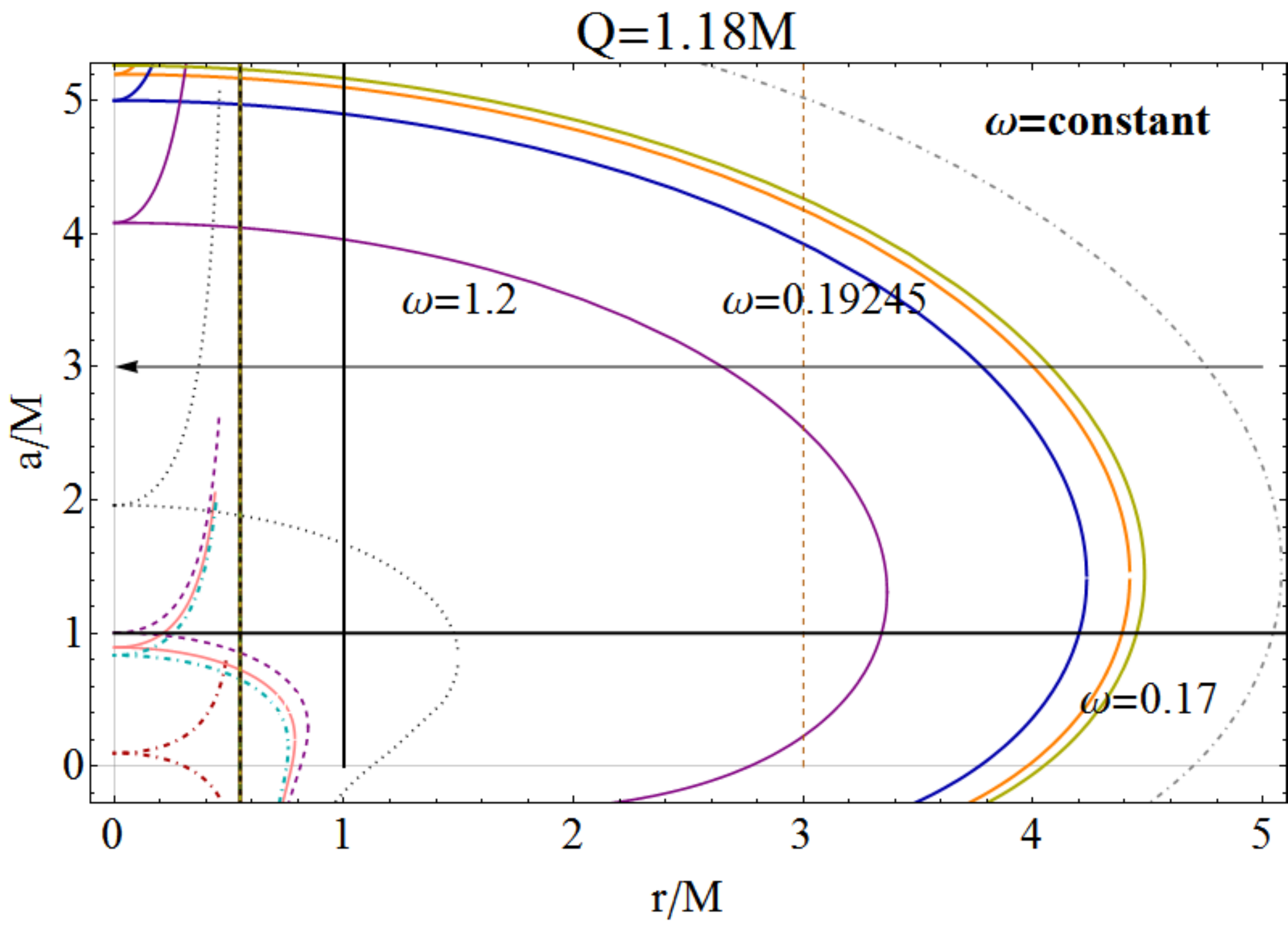}
  \caption{Equatorial plane of Kerr-Newman spacetimes: The surfaces $\omega_{\pm}=$constant as functions of the radius $r/M$ and the
	spin $a/M$--Eqs\il(\ref{Eq:Muss.churnc-parl}),
 for different charges $Q$.
  This is a generalization to the case
$ Q\neq0 $ of the analysis shown in  Fig.\il\ref{Fig:SoOv}.
The black region is for $r<r_{\pm}$ and the gray region represents the ergosurface $\Sigma_{\epsilon}^+$.
See also Fig.\il\ref{Fig:SmallPar}.}\label{Fig:fintyt}
\end{figure}
The generalization of the analysis of Fig.\il(\ref{Fig:Tactic}) is  presented in  Figs.\il(\ref{Fig:possov131}) and (\ref{Fig:possov132}).
We will not  enter into the details of this analysis; instead, we only mention that in the \textbf{KN} case a more articulated situation
for the Killing throat  and bottleneck appears, when the effects of the electric charge are combined with those of the frame-dragging.
The frequencies $\omega_{\pm}$ are plotted as functions of $r/M$ for different values of the charge $Q/M$ and spin $a/M$ in Fig.\il(\ref{Fig:Gpossov4}). In these plots, the static and  axisymmetric cases are compared and the contribution of an electric charge are confronted with those of weak \textbf{NS} geometries.
In the static geometries,  the Killing throat and bottleneck  still appear,  while the effects of the rotation emerge as a disruption of  the symmetry around the axis  $\omega=0$  of the static case and become evident in the coalescence phases of the horizons.
\begin{figure}[h!]
\centering
\begin{tabular}{lccr}
\includegraphics[scale=.45]{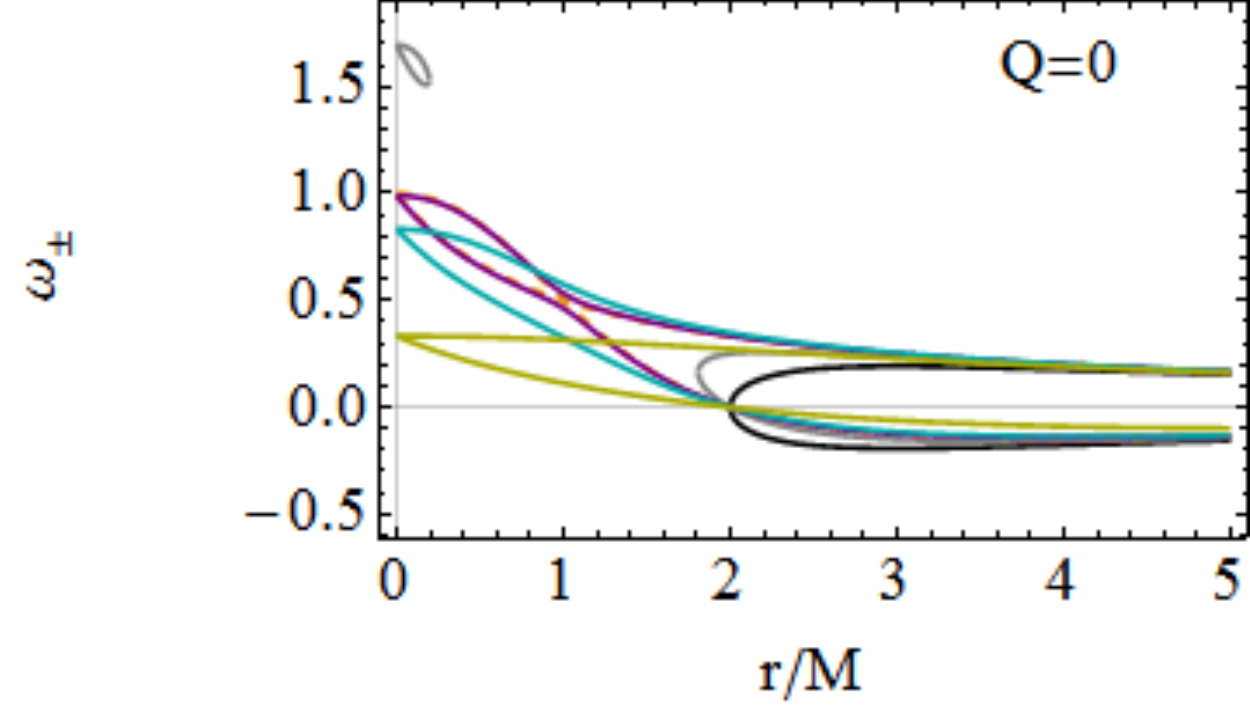}
\includegraphics[scale=.45]{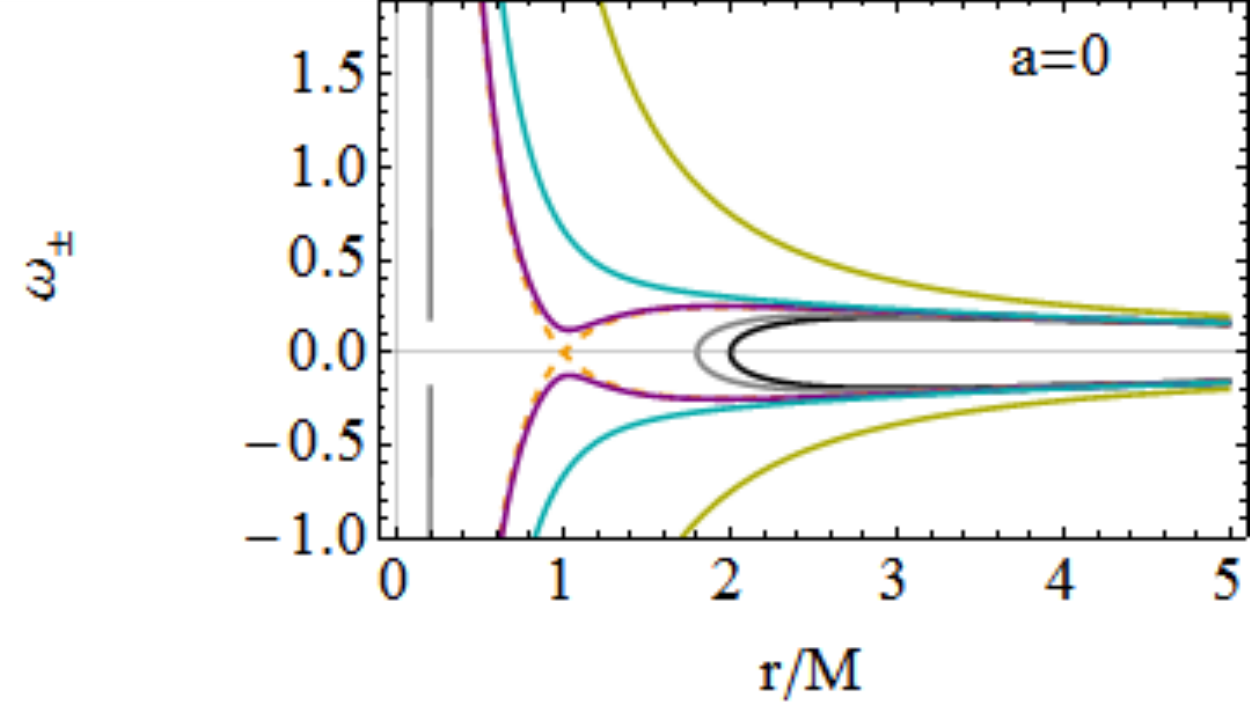}
\includegraphics[scale=.3]{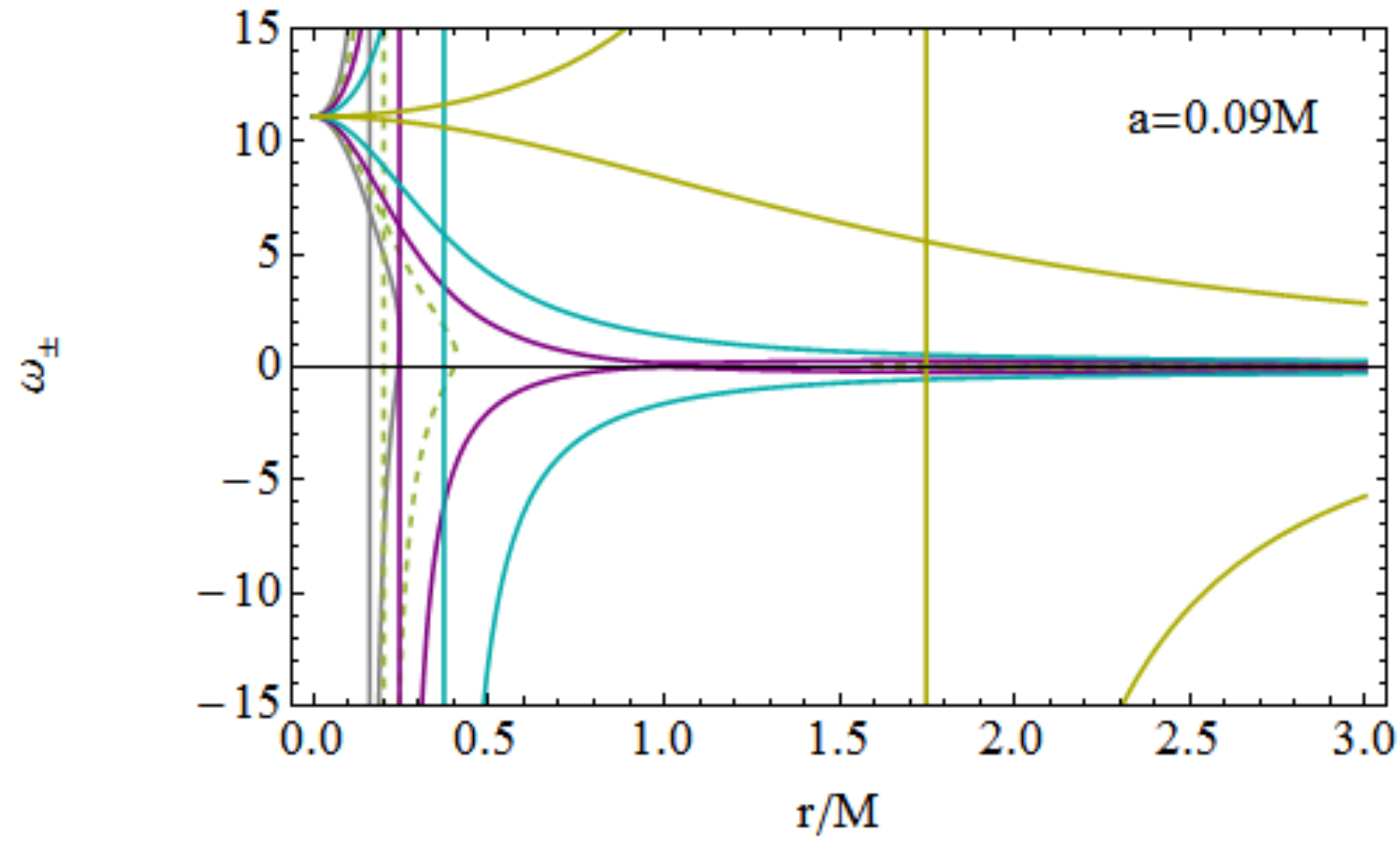}\\
\includegraphics[scale=.3]{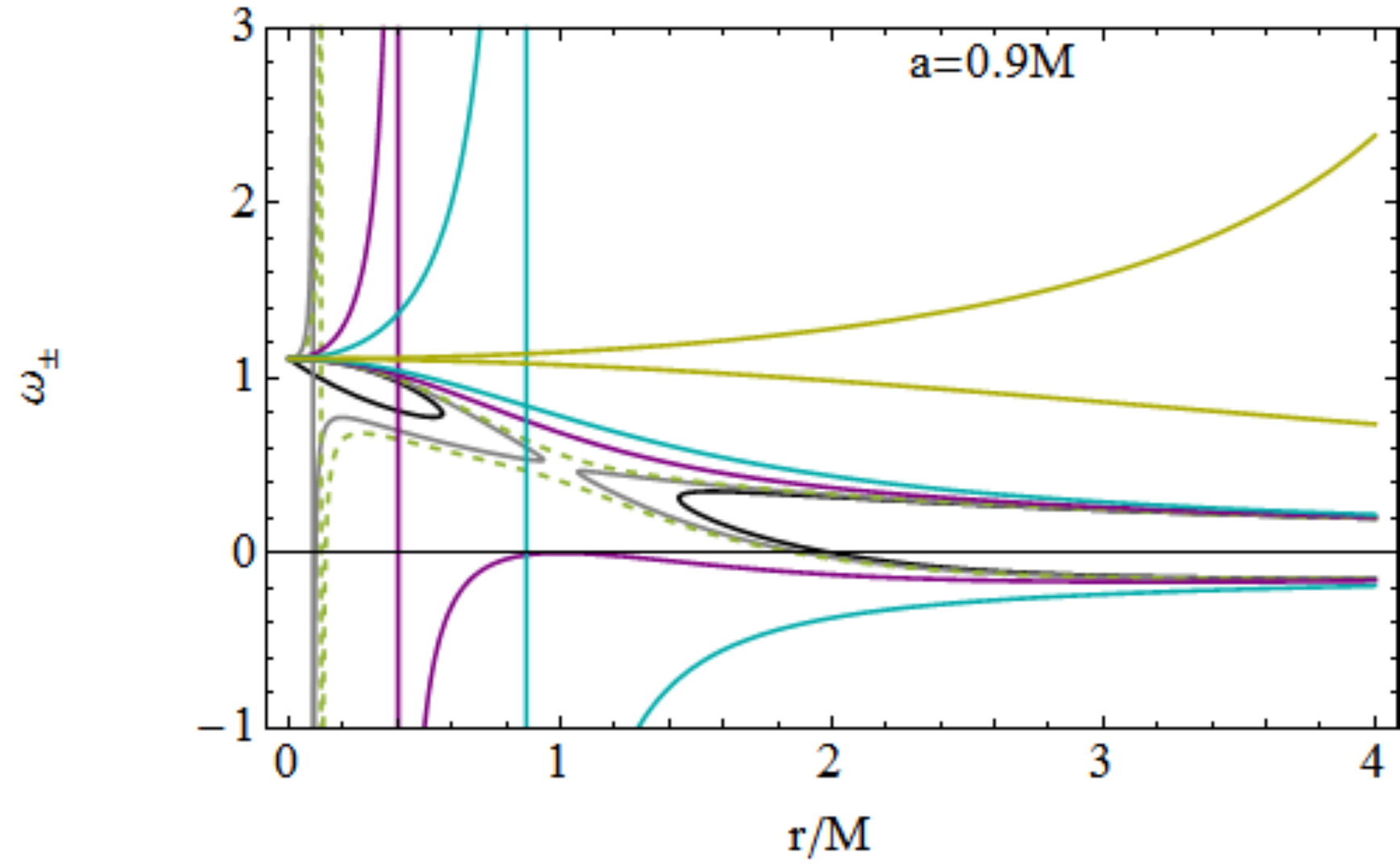}
\includegraphics[scale=.3]{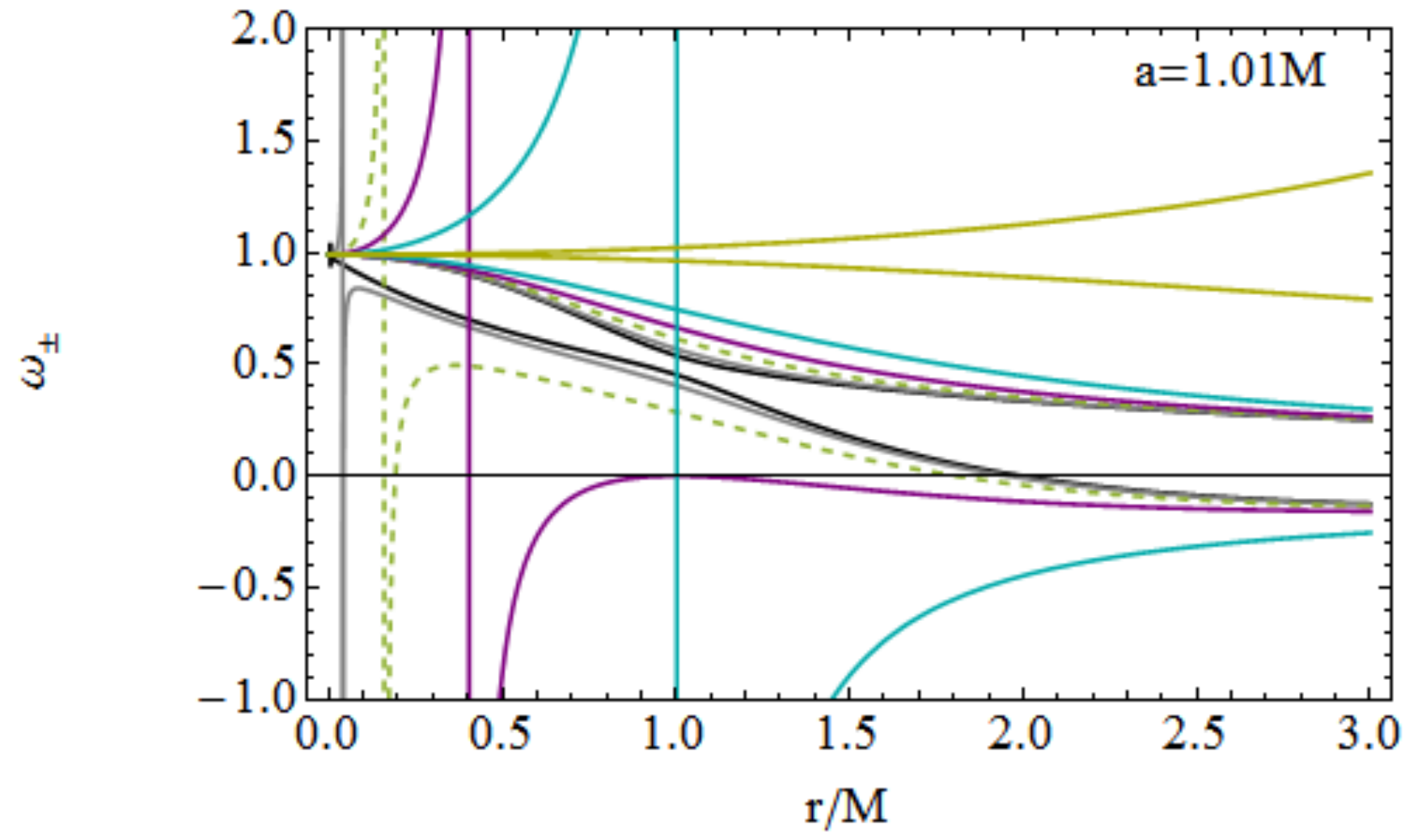}
\includegraphics[scale=.3]{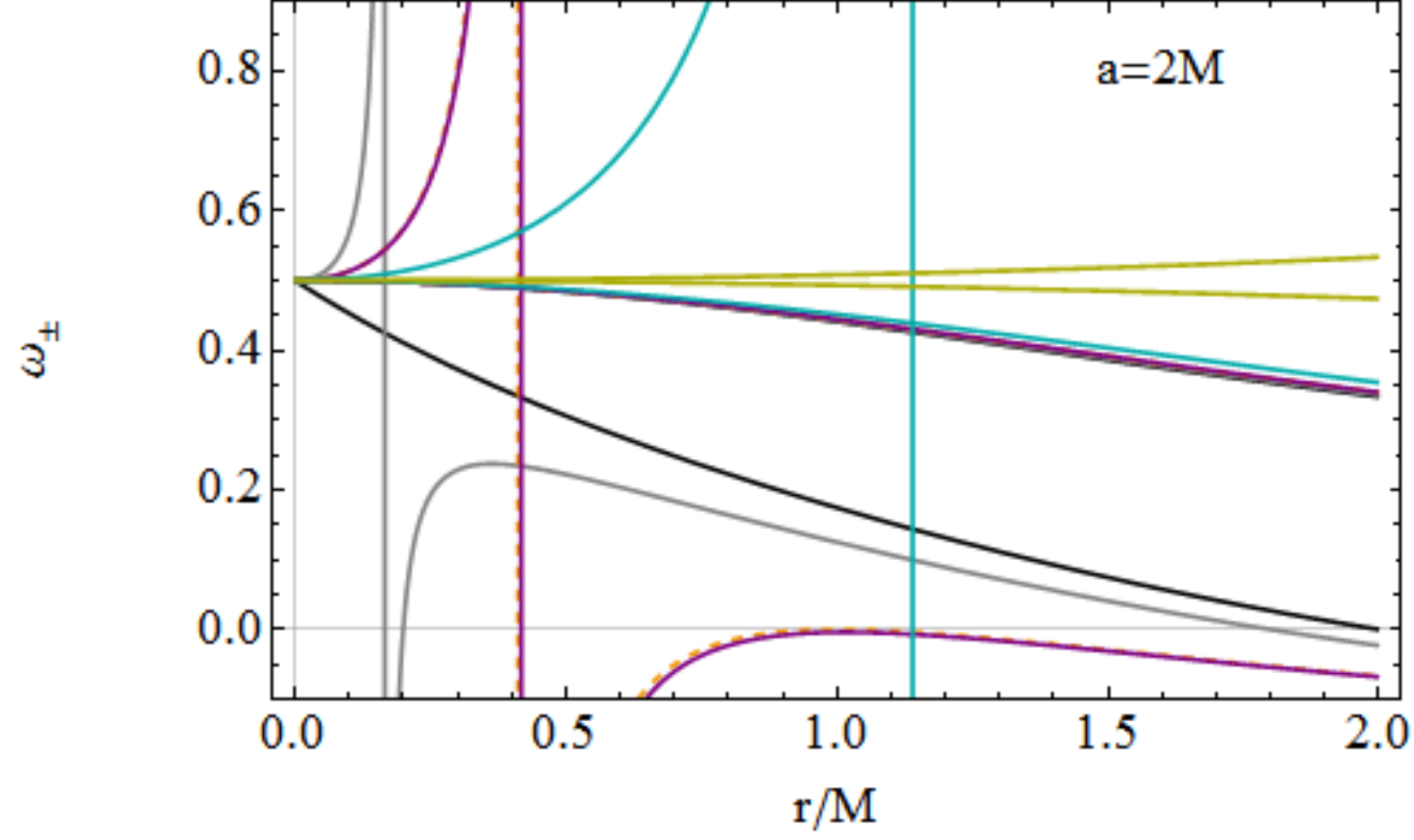}
\end{tabular}
\caption[font={footnotesize,it}]{
Equatorial plane of the Kerr-Newman spacetime: The frequencies $\omega_{\pm}$ given in  Eq.\il(\ref{Eq:Muss.churnc-parl}) as functions
of $r/M$ for different values of the charge $Q/M$ and spin $a/M$.
The limiting cases of the Kerr spacetime ($Q=0$) is shown in the first upper left panel and the static   Reissner-Nordstr\"om geometry
  with $a=0$ is the second left upper panel. The analysis shows the emergence of coalescence of the Killing horizons $r_{+}$  and $r_-$ in the extreme black hole geometries and the emergence of a  Killing throat  and Killing bottleneck in the \textbf{NS} geometries
	--see also Figs.\il(\ref{Fig:cOALESCENzE},\ref{Fig:MaSuorVIRMa},\ref{Fig:CORESUNO},\ref{Fig:possov131} and \ref{Fig:possov132}).}
\label{Fig:Gpossov4}
\end{figure}
Figures\il(\ref{Fig:AGpossov15}), (\ref{Fig:spossov10}) and (\ref{FIG:fortCarl8}), on the other hand, show the solutions  of
$\partial^2_r\omega_{\pm}=0$  and  $\partial^3_r\omega_{\pm}=0$ defining the Killing bottlenecks of  naked singularities, which are  generalizations of the analysis presented in
  Fig.\il(\ref{Fig:MaSuorVIRMscri}) for the case  $Q=0$ (Kerr spacetime).
The surfaces $a_{\omega}^{\pm}(r,\theta)$ are shown in Fig.\il(\ref{FIG:fortCarl8}), giving a view of the solutions for the light surfaces in the off-equatorial case.

\begin{figure}[h!]
\centering
\begin{tabular}{lccr}
\includegraphics[scale=.3]{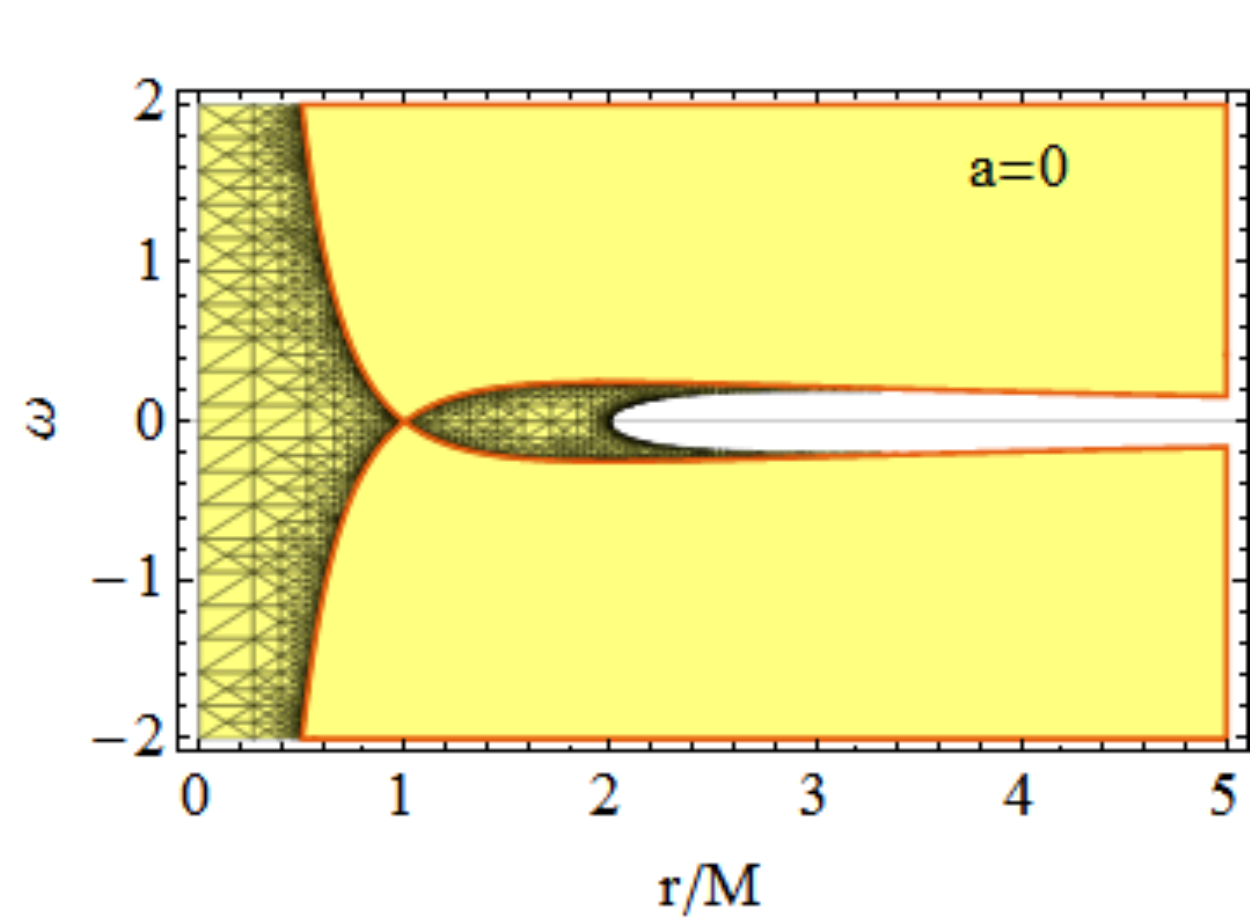}
\includegraphics[scale=.3]{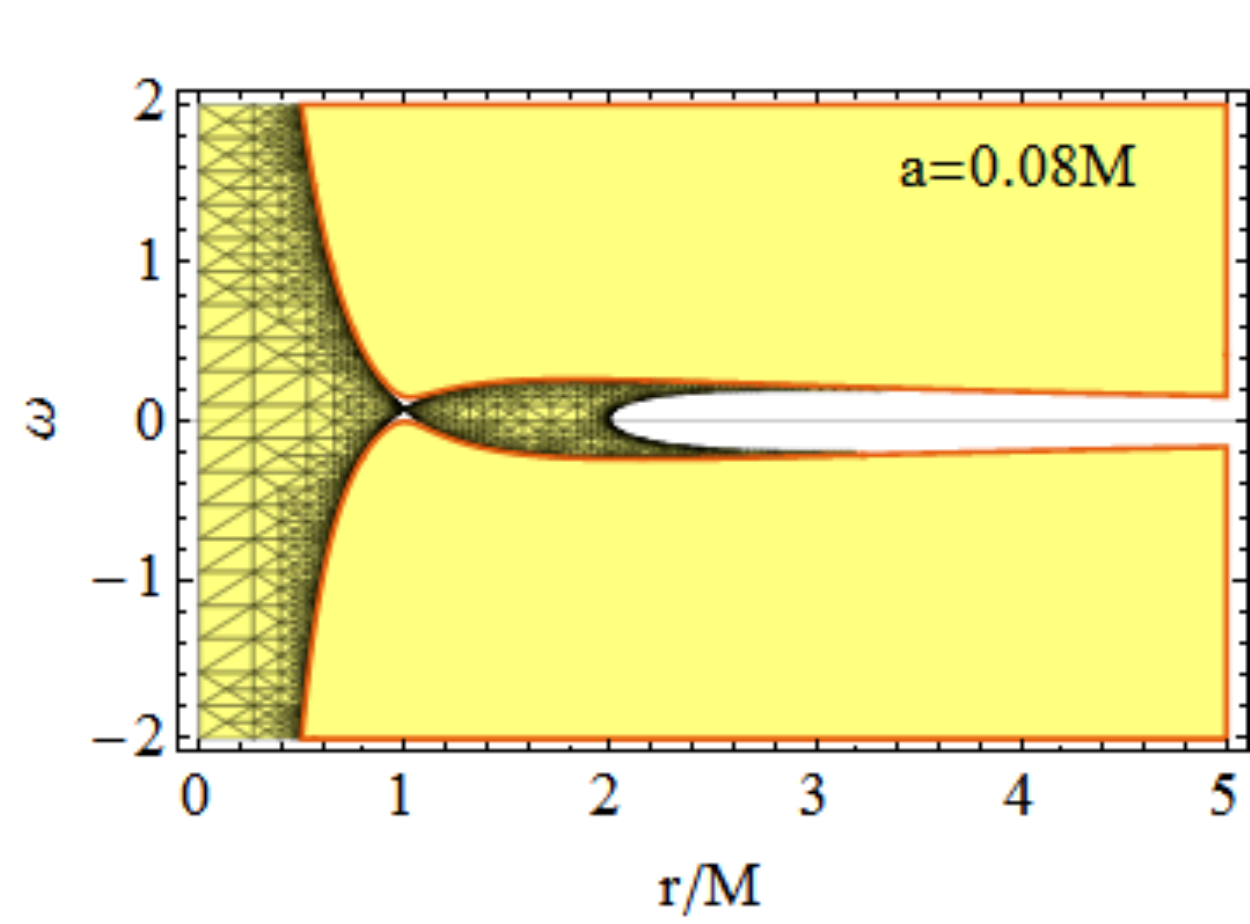}
\includegraphics[scale=.3]{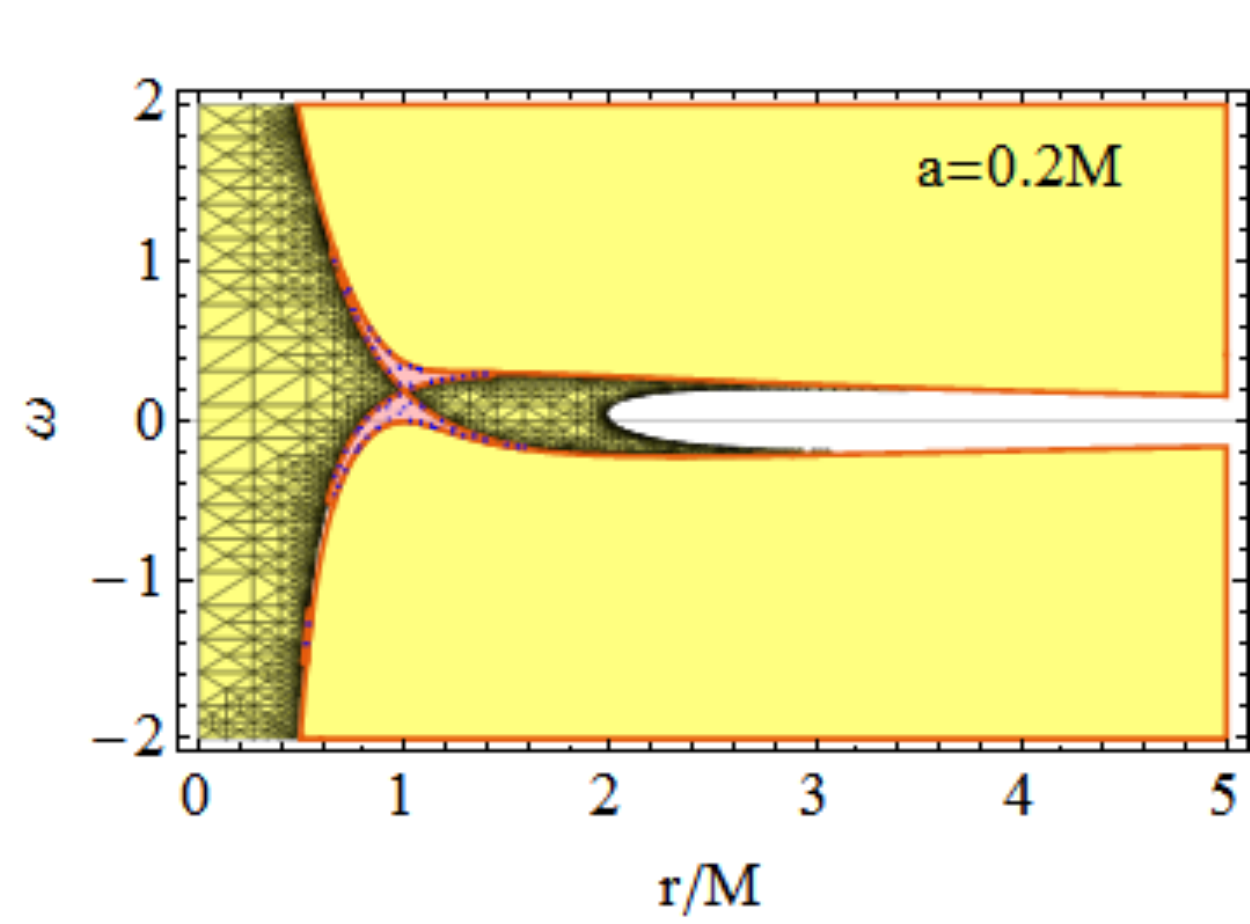}
\includegraphics[scale=.3]{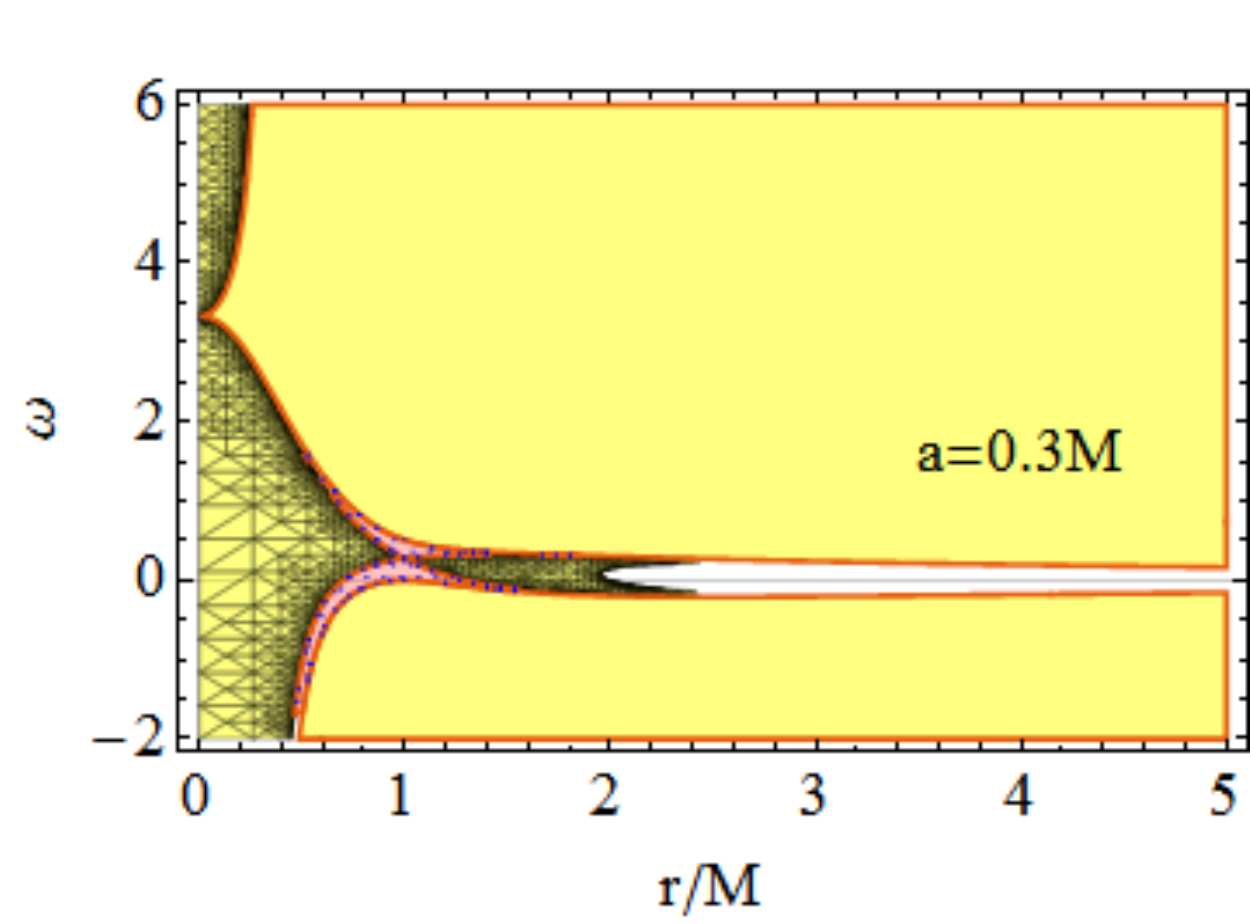}\\
\includegraphics[scale=.3]{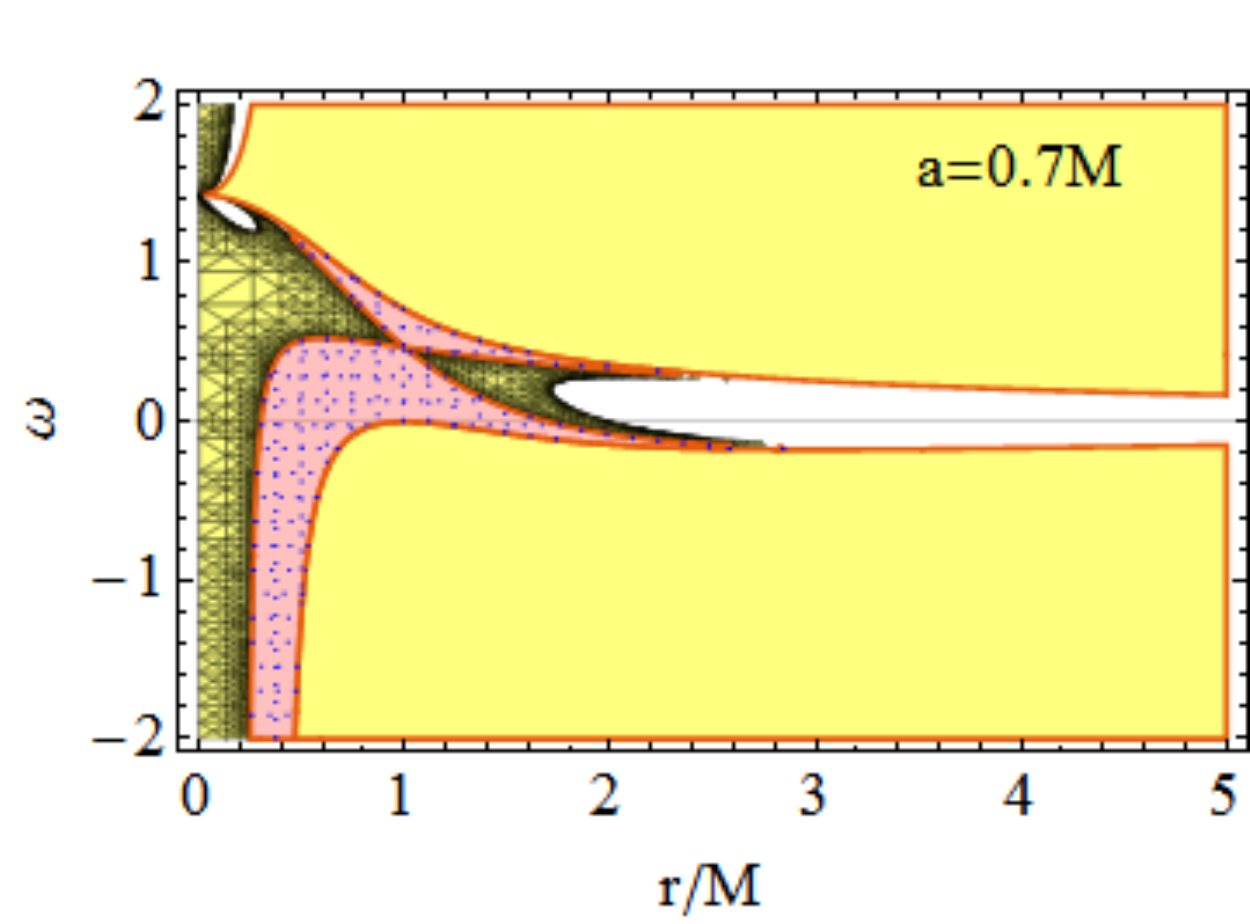}
\includegraphics[scale=.3]{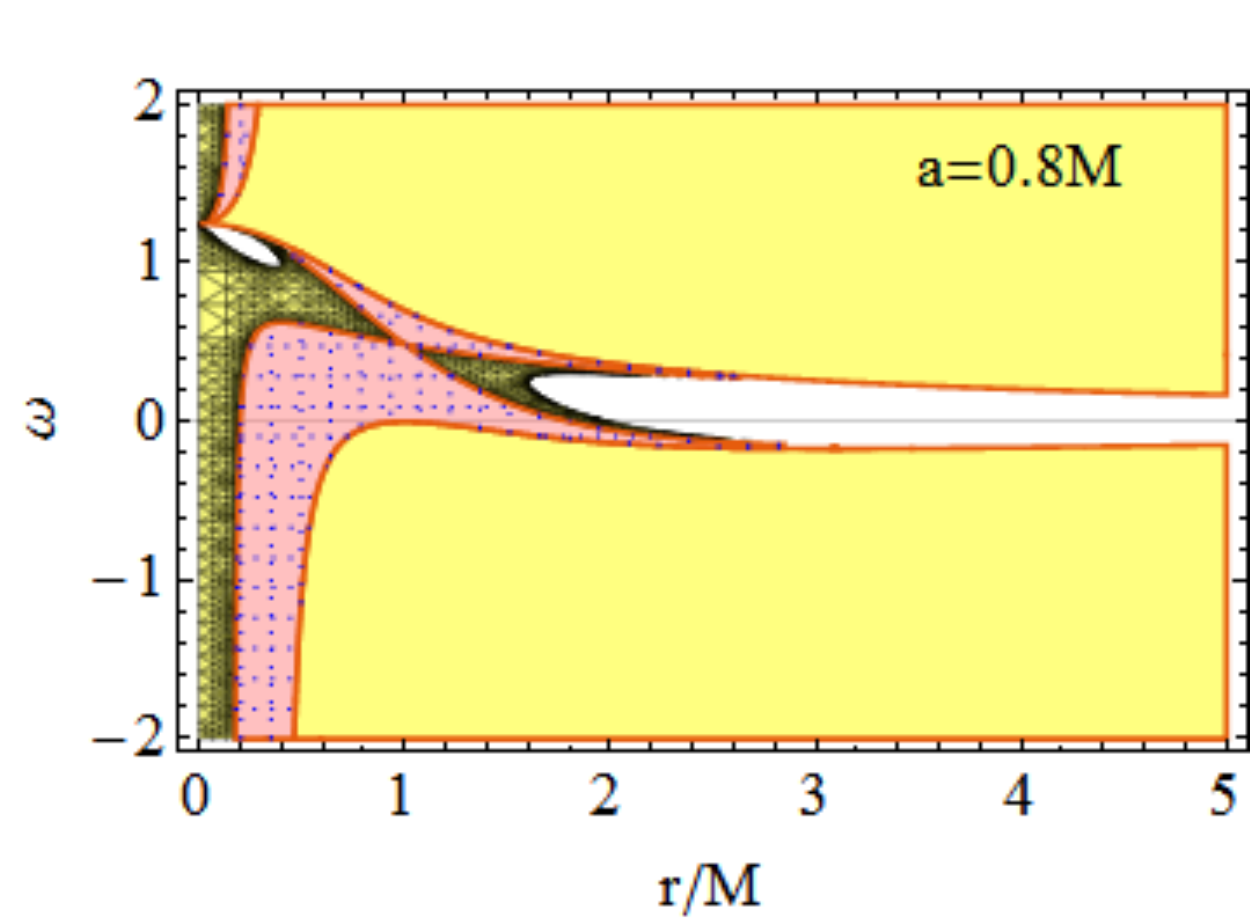}
\includegraphics[scale=.3]{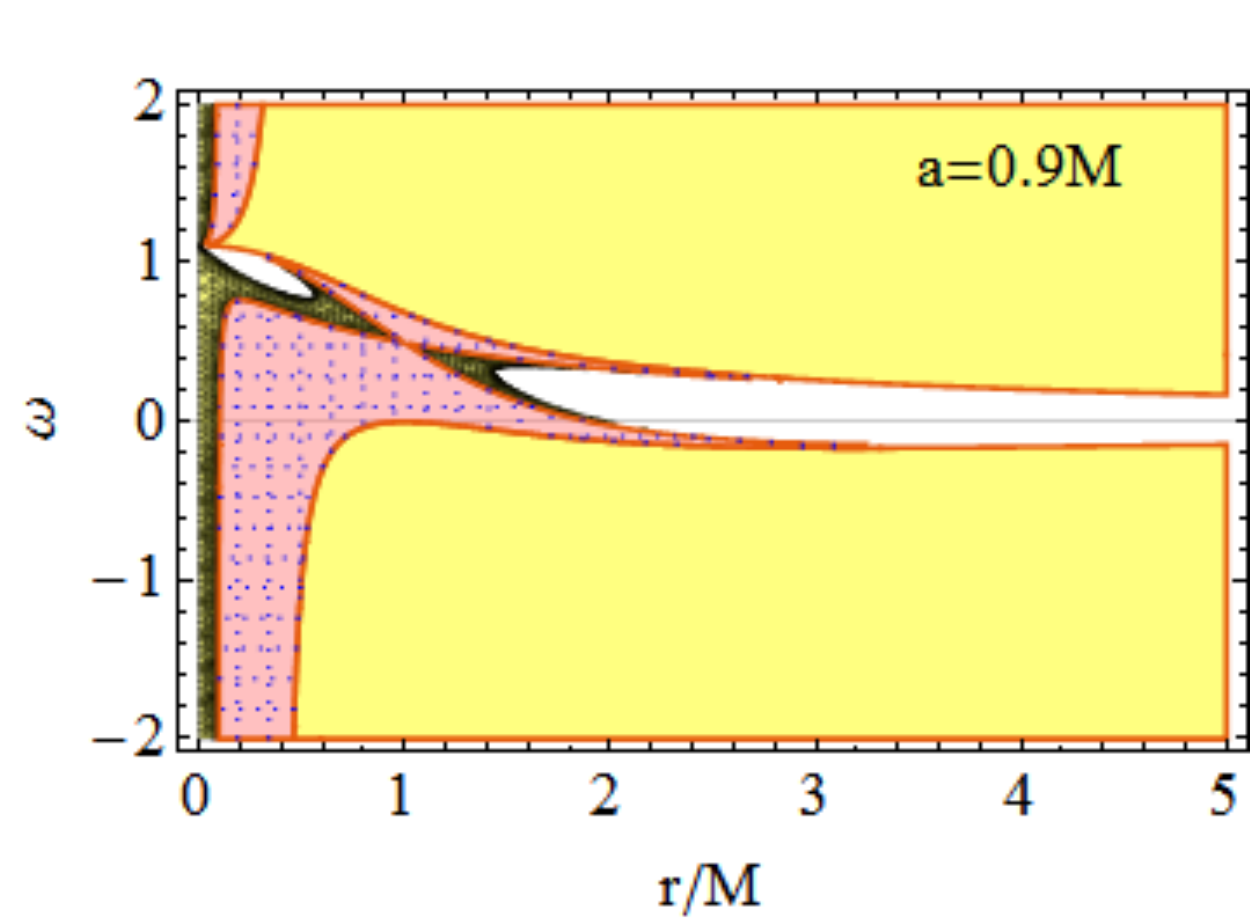}
\includegraphics[scale=.3]{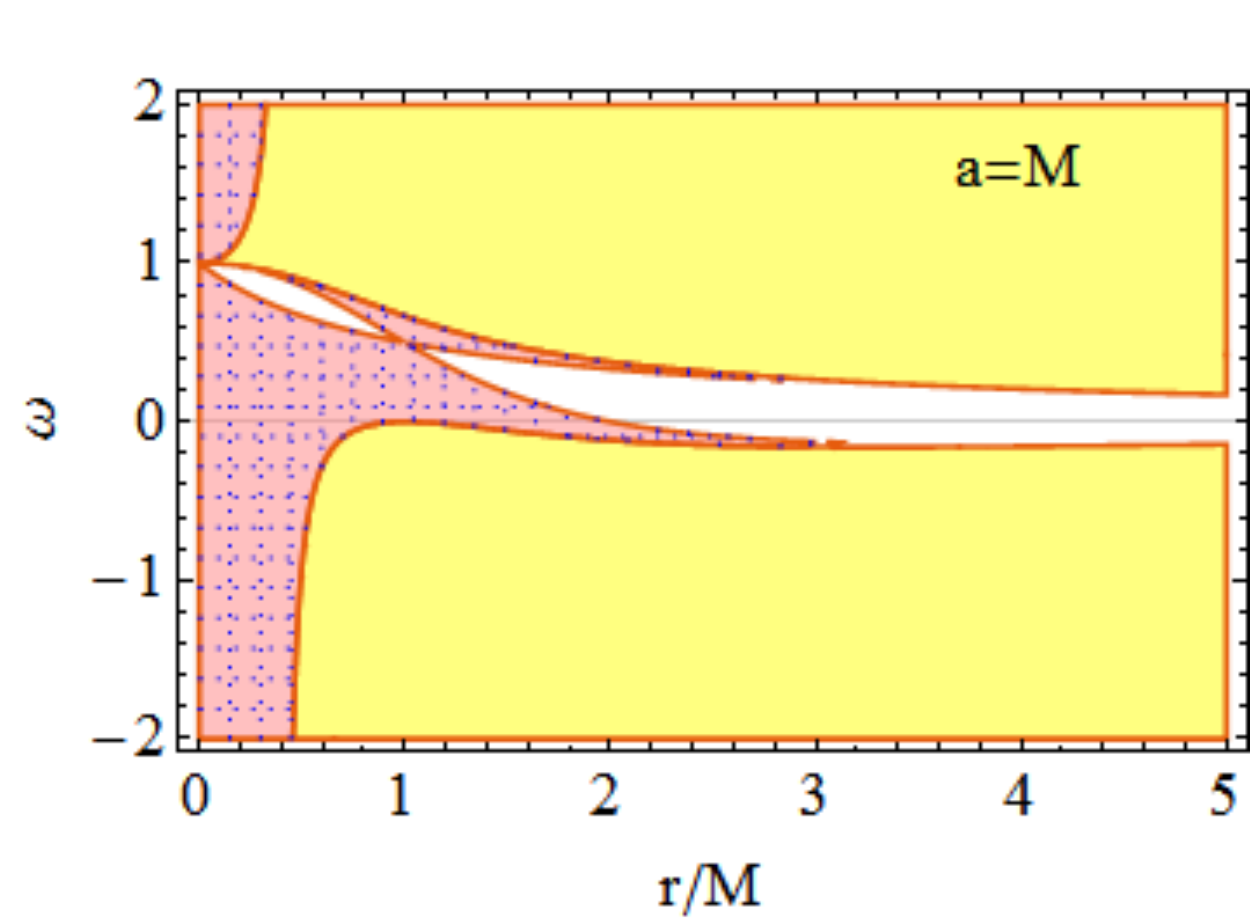}\\
\includegraphics[scale=.3]{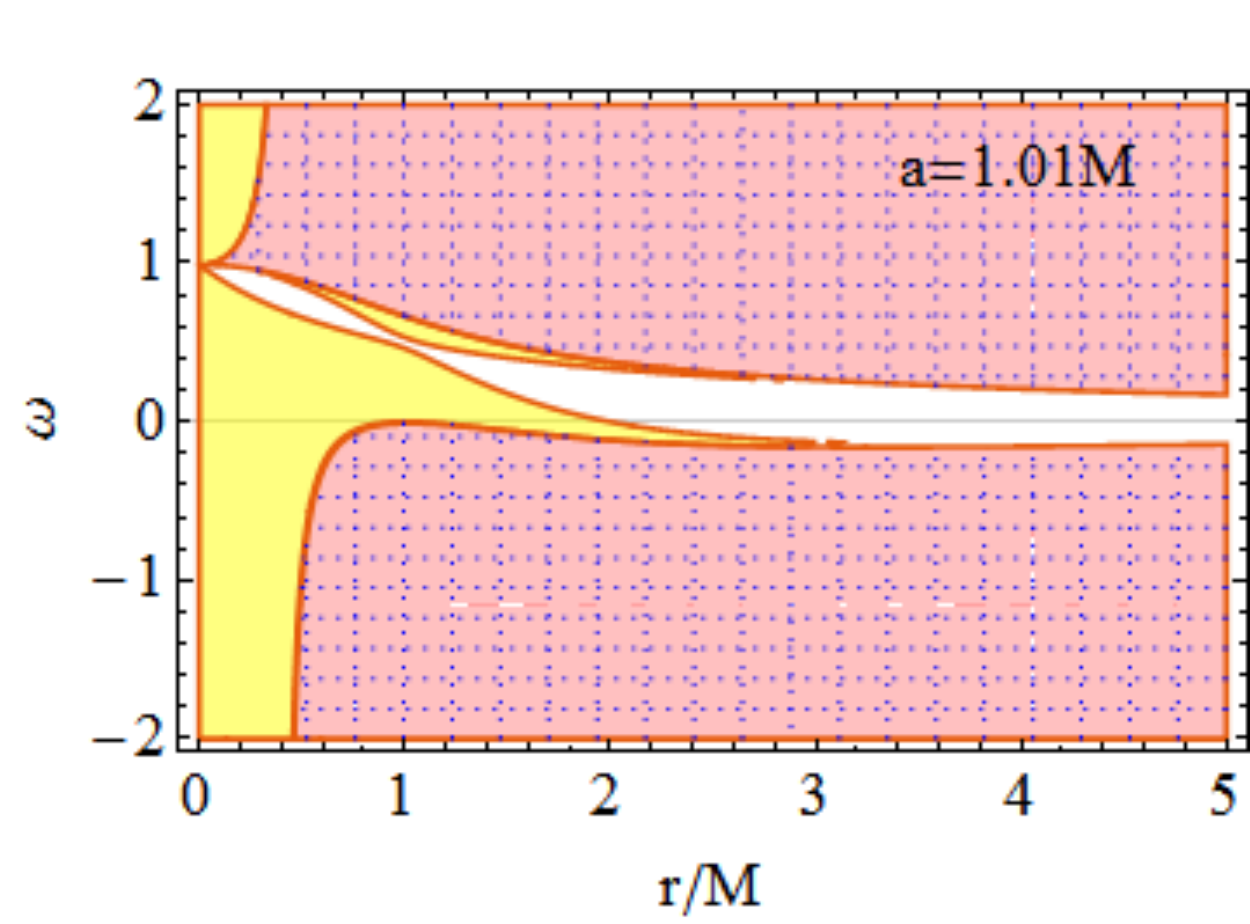}
\includegraphics[scale=.3]{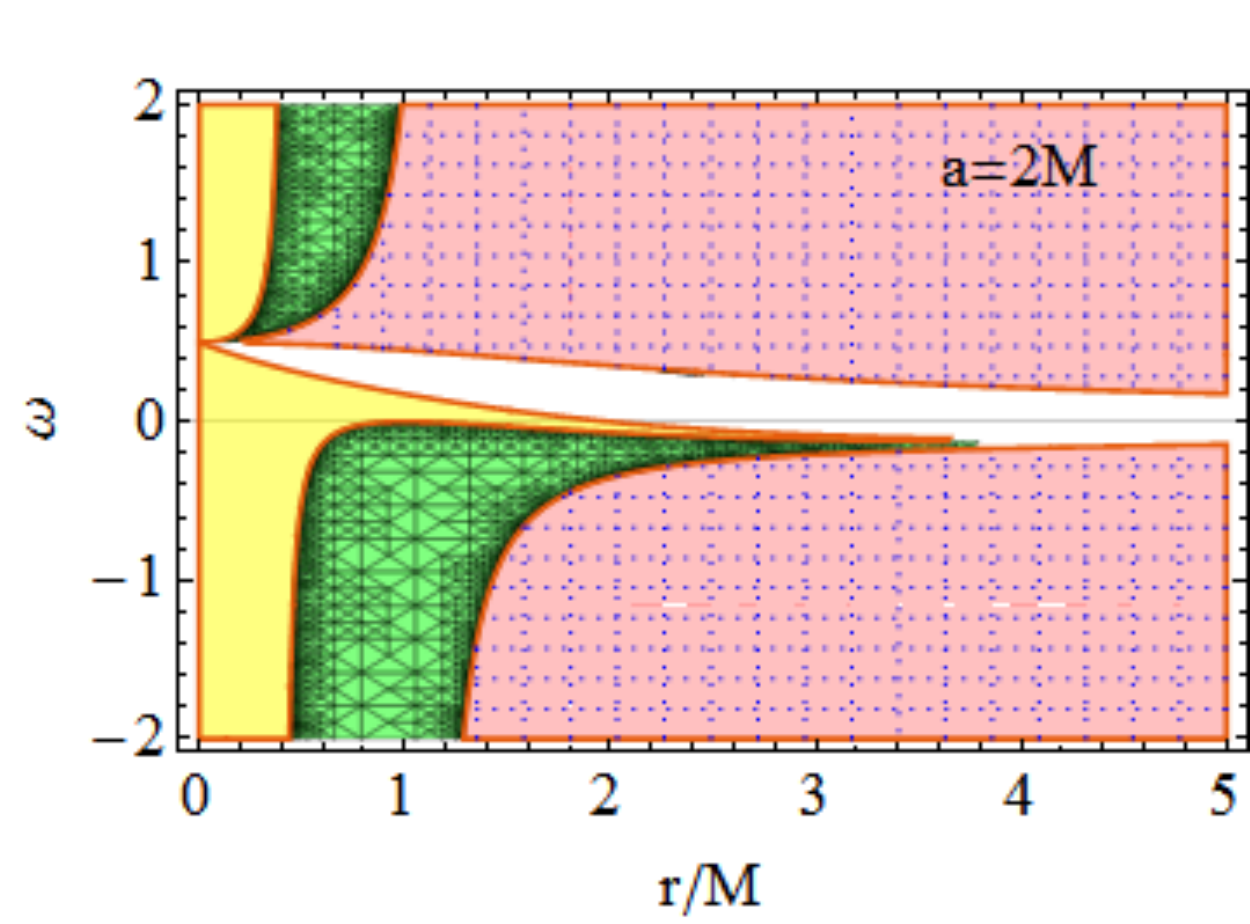}
\end{tabular}
\caption[font={footnotesize,it}]{Equatorial plane of the Reissner-Nordstr\"om  spacetime:
 generalization of the analysis of Fig.\il\ref{Fig:Tactic} and  \ref{Fig:possov132}  for the case $a=0$ and  $Q\neq0$.}
\label{Fig:possov131}
\end{figure}
\begin{figure}[h!]
\centering
\begin{tabular}{llll}
\includegraphics[scale=.3]{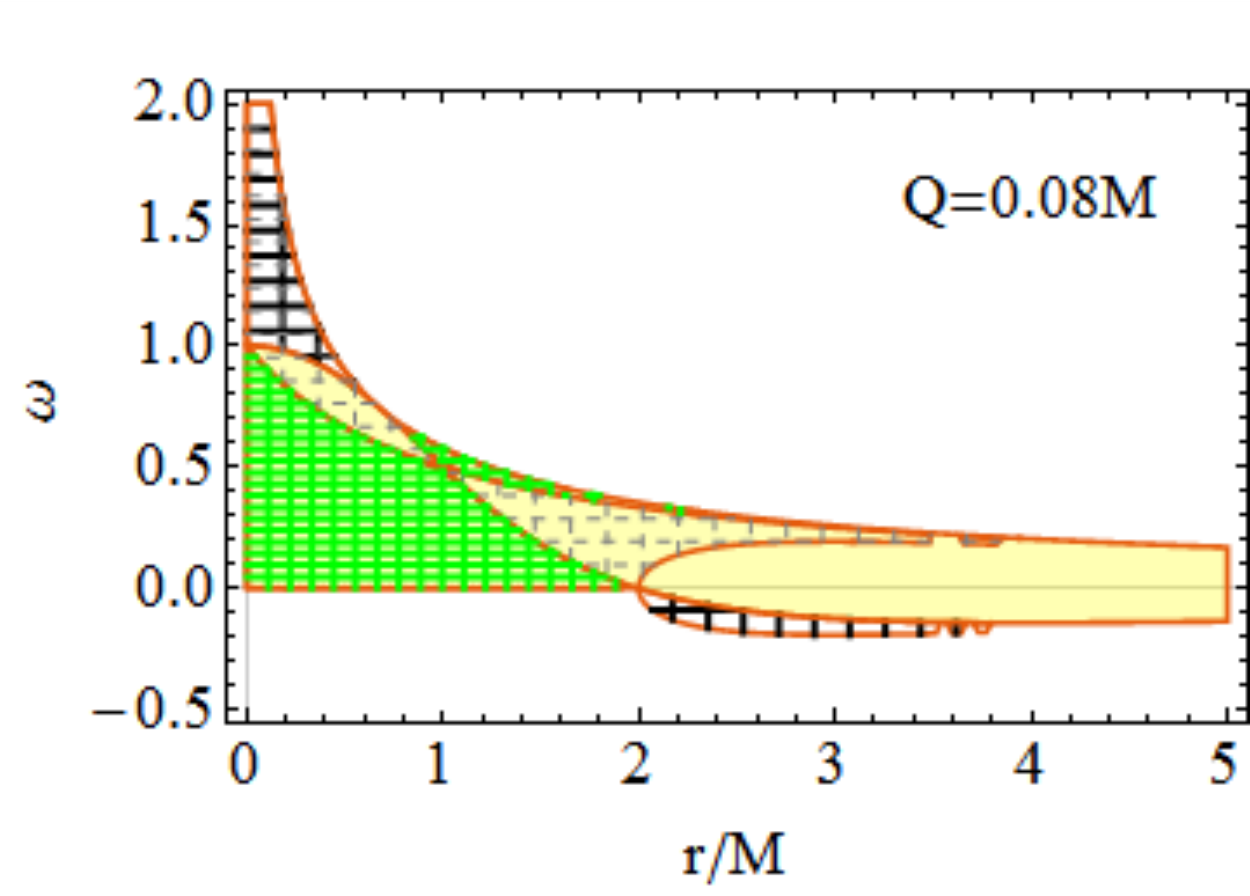}
\includegraphics[scale=.3]{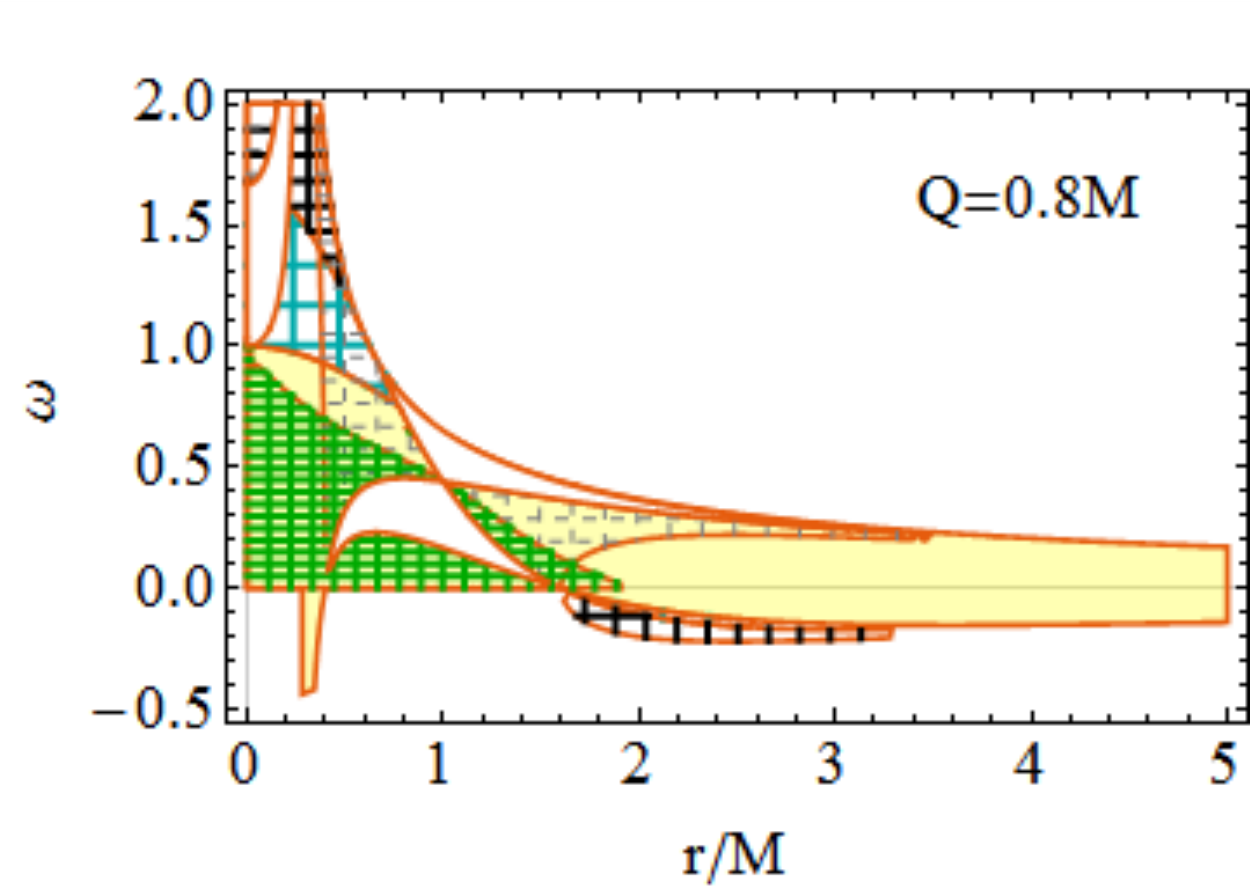}
\includegraphics[scale=.3]{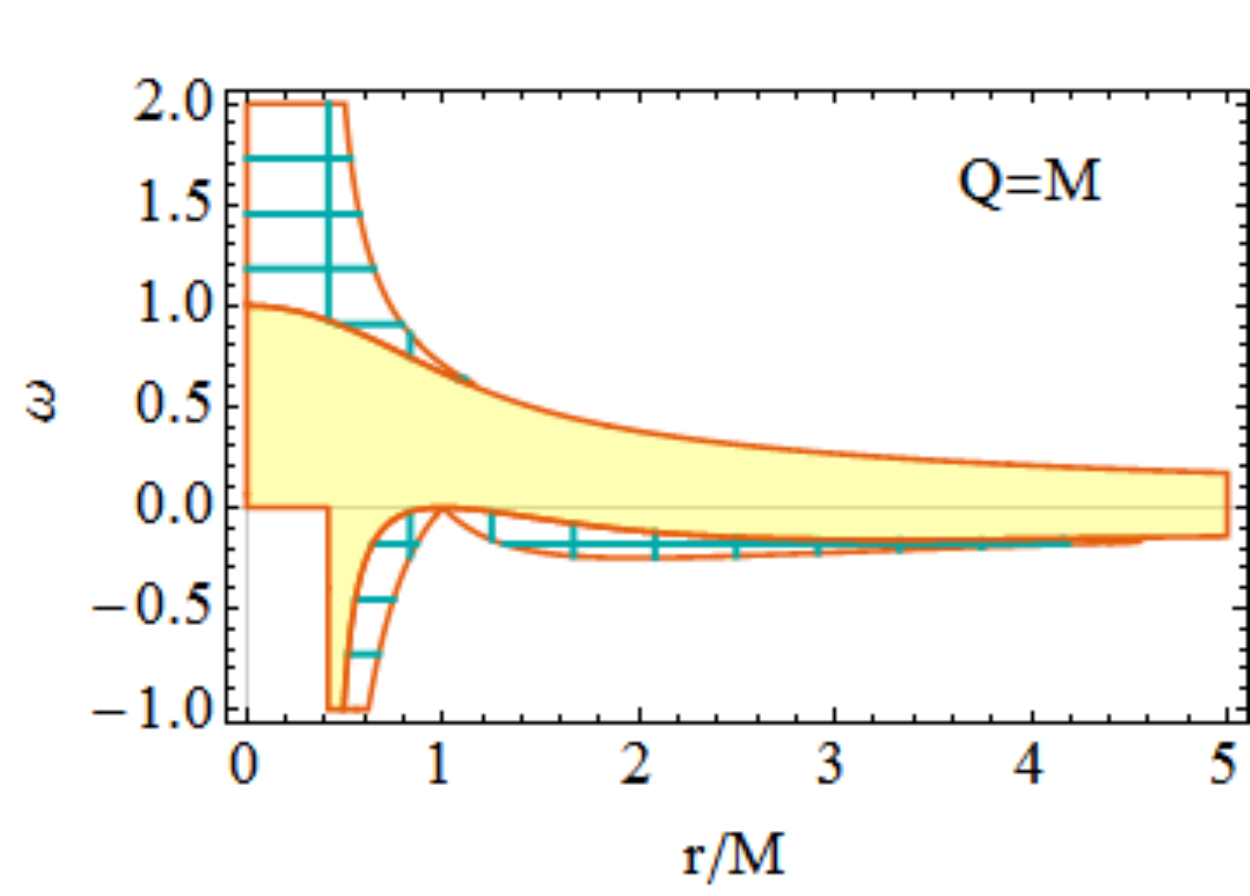}
\includegraphics[scale=.3]{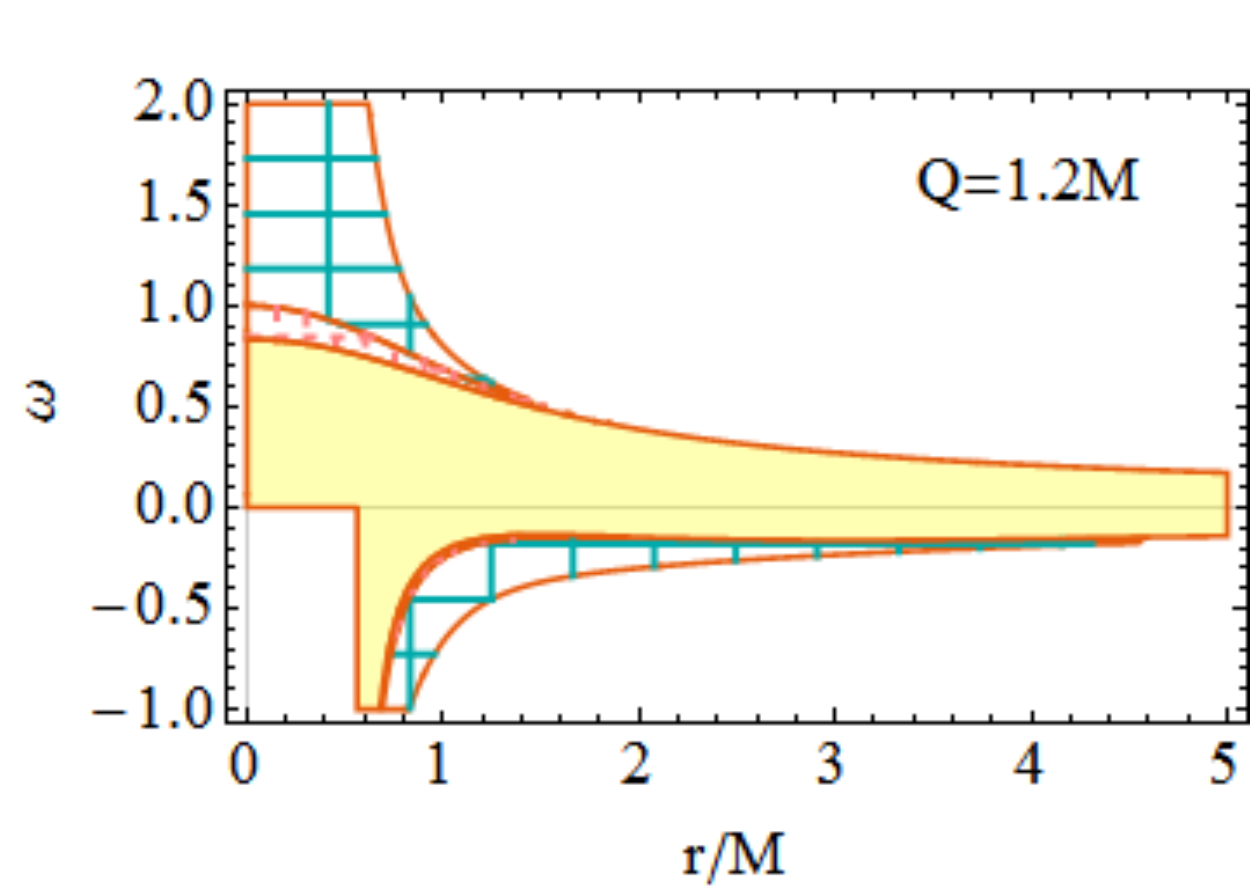}
\includegraphics[scale=.3]{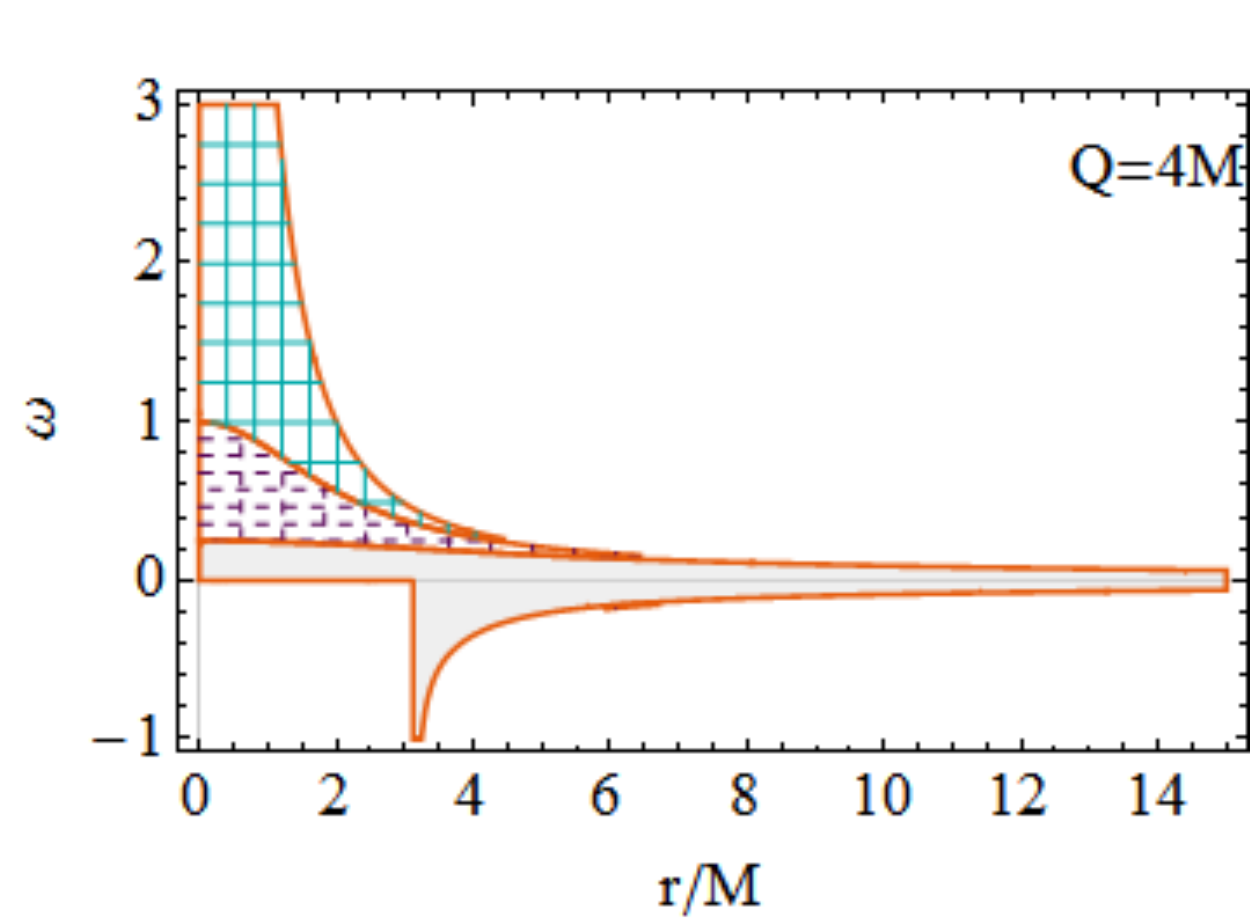}
\end{tabular}
\caption[font={footnotesize,it}]{Equatorial plane of the Kerr-Newman spacetime: generalization  of the analysis
of Fig.\il\ref{Fig:Tactic} to the case $Q\neq0$--see also Fig.\il\ref{Fig:possov131} }
\label{Fig:possov132}
\end{figure}
This analysis  confirms the results  of  Secs.\il\ref{Sec:bottlem} and \ref{Sef:aomega}  and the role of the frame-dragging.

\section{Concluding remarks  and future perspectives}\label{Sec:con-RAEM-FP}
In this work, specially in Sec.\il\ref{Sec:bottlem},  we explored the Killing throats  and bottlenecks,  arising  from  the properties of stationary observers in the Kerr geometries.
In the case of \textbf{WNSs} ($a/M\in ]1,2]$), the Killing throats show   ``restrictions'', which we   identify as Killing bottlenecks. To explore the properties of the bottlenecks, we introduced  in Sec.\il(\ref{Sef:aomega}) the concept of extended plane, which is a graph relating a particular characteristic of a spacetime in terms of the parameters entering the corresponding spacetime metrics.
More precisely,   the analysis of some peculiar characteristics of the  bottlenecks, defined as  horizons remnants, indicated some links  between \textbf{BHs} and \textbf{NSs} (and in particular \textbf{WNSs}). To  compare the \textbf{BH} and \textbf{NS} characteristics, it was convenient    to  introduce in Sec.\il\ref{Sef:aomega}  the concept of metric bundles, i.e.,  curves in the extended plane $\pi_a^+$, representing a collection of metrics  defined by a particular photon  orbital frequency, named metric bundle frequency.
The bundle frequencies (and the orbital limiting frequencies in any point on the equatorial plane of any spacetime of the family)    are all and only the frequencies of the horizon defined in the extended plane. This analysis has been done mainly  on the equatorial plane
of axisymmetric geometries.  Metric bundles   show, in fact, the  remarkable property to be tangent to the horizon curve in the   extended plane, the space where the curves representing the metric bundles are defined. In the case of the the equatorial plane of the Kerr metric,
 the extended plane is essentially equivalent to the function that relates the frequency with the spin.  Notably,
the metric bundle associated to the extreme Kerr \textbf{BH} corresponds to a regular curve tangent to the horizon with
bundle origin $a_0=2M$. As proved in   Sec.\il\ref{Sef:aomega},
\emph{all}  the metric bundles are tangent to the horizon in the extended plane $\pi_a^+$ and the horizon emerges as the envelope surface of the metric bundles in $\pi_a^+$. A bundle frequency corresponds to  one  limiting photon orbital frequency for all   \textbf{BHs}
 or some \textbf{BH}s and \textbf{NS}s geometries. The bundle frequency coincides with the frequency at the (inner or outer)
horizon curve, which is tangent to the metric bundle.
On the other hand, the metric bundles are all defined by all and only the frequency     of the horizon in $\pi_a^+$.
Viceversa, all the horizon frequencies in the extended plane are metric bundle frequencies.
In Sec.\il\ref{Sec:RN-KN-cases}, we consider the  static and  electrically charged Reissner-Nordstr\"om spacetime
and the Kerr-Newman axisymmetric  electrovacuum solution and show that the bending (closing) of the curves of the  metric bundles
is due to the rotation of the gravitational source.
Therefore, we can say that the horizons frequencies  determine the \textbf{BHs} \emph{and} \textbf{NSs}   limiting photon orbital frequencies. This fact establishes a connection between \textbf{BHs} and \textbf{NSs}:
  the  inner \textbf{BH} horizon is connected to  \textbf{BH} and \textbf{WNS} origin bundles, whereas
 the outer horizon  sets the \textbf{BHs-SNSs} correspondence.   In the extended plane,
\textbf{NS}s are associated with  portions of the horizon.
  In this sense, the inner horizon  is partially constructed by \textbf{BH} metric bundles.
	The inner horizon is associated with  \textbf{WNS} origins.
	This last   property turns out to be related to the Killing bottlenecks appearing in the light surfaces.
	Interestingly, the outer horizon in $\pi_a^+$ is  generated  by \textbf{SNSs} metric bundles only.
	This fact has the interesting consequence that only  the horizon frequencies determine   the frequencies $\omega_{\pm}$
	at each point, $r$, on the equatorial plane  of a Kerr \textbf{BH} or \textbf{NS} geometry:
	all the frequencies $\omega_{\pm}(r)$ on the equatorial plane are only those of the horizon in $\pi_a^+$,
	the horizon in $\pi_a^+$ contains information about all limiting photon orbits also in \textbf{NS} spacetimes.
	In Secs.\il\ref{Sec:bottlem} and \ref{Sef:aomega}, we have also introduced the concept of  inner horizon confinement.
   In this sense, \textbf{NS}s are ``necessary'' for the  construction of horizons.
	The outer horizon is associated with  a \textbf{NS} (the bundle origin) in the extended plane and the inner horizon to a
	\textbf{WNS} or a \textbf{BH}.
 Therefore, we believe that this result could be
of interest for the investigation of gravitational collapses in which
connections between \textbf{NS} and \textbf{BH} solutions are expected and emerge \cite{Goswami:2006ph,CS181,Manko:2018yrc}.

Some further aspects of these properties are currently  under investigation.
Firstly, it would be necessary to further analyze the off-equatorial case and test the results of Sec.\il\ref{Sef:aomega}
 in other kinds of geometries.
In a future analysis,  we shall analyze other axisymmetric  spacetimes admitting  Killing horizons and consider
the possible thermodynamic implications of the results discussed here, particularly,
 in relation to the possibility of formulating the  \textbf{BH} thermodynamic laws   in terms of metric bundles.
We also point out that metric bundles and horizons remnants seem to be related to the concept of
pre-horizon regimes.
There is a pre-horizon regime
in the spacetime  when  there are  mechanical effects allowing circular orbit  observers, which can
recognize the close presence of an event horizon.
This concept was introduced in \cite{de-Felice1-frirdtforstati}
and detailed for the Kerr geometry in \cite{de-FeliceKerr,de-Felice-anceKerr}.
The pre-horizon was identified and analyzed in \cite{de-Felice-first-Kerr}, leading
to the conclusion that a gyroscope would observe a memory of the static case
in Kerr metric. Clearly, this aspect could be of  relevance during the gravitational
collapse \cite{de-Felice3,de-Felice-mass,de-Felice4-overspinning,Chakraborty:2016mhx}.

{We now summarize the main aspects of our analysis focusing on the possible interpretation of the main results.}

{
(\emph{i}) We identified   bottlenecks essentially as \emph{horizon remnants}. Similar  interpretations have been presented in
\cite{de-Felice1-frirdtforstati,de-FeliceKerr,de-Felice-anceKerr,de-Felice-first-Kerr}, by using the concept of pre-horizons,
 and in \cite{Tanatarov:2016mcs}, by analyzing the so-called whale  diagrams.  These structures could play an important role
for describing the formation of black holes and for testing the possible existence of naked  singularities.
Notably, the concept of remnants,  as expressed here, refers  to and evokes  a sort of spacetime ``plasticity'',
 which naturally led  to the introduction in Sec.\il(\ref{Sef:aomega})   of the concept of the  extended plane.
In this new frame, we found several properties  emerging from and affecting  the spacetime geometries when we  consider an entire
family of  solutions    as a  \emph{unique geometric object}.
}

{
(\emph{ii})  Considering the Kerr family  as a single object,  the  geometric {quantities} (for example,
the horizons) defined for a single solution acquire a completely different significance when considered for the entire family.
}

{
(\emph{iii}) Considering  metric bundles with \textbf{WNS}, we found that the inner horizon is confined in the metric bundle framework.
 }

{
(\emph{{iv}}) We proved the existence of a connection between black holes and naked singularities. To each \textbf{BH} corresponds
the pair (\textbf{WNS}, \textbf{SNS}) or  the pair (\textbf{BH}, \textbf{SNS}). This correspondence is important for the definition of the Killing horizons.  }

{
\emph{ (v)}  We proved that  \textbf{WNSs} (\textbf{SNSs}) are necessary for the construction of the inner (outer) Killing horizon.
This result   could shed light on the physical meaning  of \textbf{NSs} solutions.}
}

{To conclude, we present a schematic summary of the main results presented in this work.
\begin{itemize}
\item Analysis of \emph{Killing throats} and definition \emph{Killing bottlenecks} for particular naked singularities--Section\il\ref{Sec:bottlem}. To define Killing throats, we study the  limiting photons surfaces  $r_{s}^{\pm}$ and frequencies $\omega_{\pm}$ which are
defined in Eq.\il(\ref{Eq:rspm}) and Eq.\il(\ref{Eq:b-y-proc}), respectively. We also interpret bottlenecks as \emph{horizons remnants} in
weak naked singularities.
\item \emph{Inner horizon confinement}--Sec.\il\ref{Sec:bottlem}. In Eq.\il(\ref{Eq:mart-re}), we identify the  photon orbits   characterized by the  horizons  orbital frequencies. The
    radii $r^{\mp}_{\mp}$  represent   the set  of photon orbits  with   frequencies  $\omega_H^{\pm}$ at
the $\textbf{BH}$ horizons---Eq.\il(\ref{Eq:mart-re})--Fig.\il\ref{Fig:MaSuorVIRMa}.
    The inner horizon confinement,  according to  the constraints on  $r^{-}_{-}$, is in agreement with the confinement
		of the metric bundles containing  \textbf{BHs} and \textbf{WNSs} (see Sec.\il\ref{Sef:aomega}).
\item Definition   of the  \emph{extended plane} $\pi_a$ and \emph{metric bundles} $g_{\omega}$ in Sec.\il\ref{Sef:aomega}.
Metric bundles are curves in $\pi_a$  tangent to the horizons characterized by the limiting photon frequency
$\omega_+$ or $\omega_-$.
 \item
\emph{Tangency condition  and horizon construction in the extended plane}.
We demonstrate that the horizon curve corresponds to  the envelope surface of the metric bundles.
All the metric bundles are tangent to the horizon curve and all the points of the horizons are associated to one and only one metric bundle tangent to that point-- Figs.\il\ref{Fig:pesoMa}, \ref{Fig:PostMerge},
 \ref{Fig:manet} and  Tables \ref{Table:ABCDmodels} and \ref{Table:number-cicle}. Consequently,
 all the limiting photon orbital frequencies (on the equatorial plane)  are \textbf{all and only} the frequencies of the horizon curve in
 $\pi_a$, related to two metric bundles, with frequencies $\omega_0$ and  $\varpi_\pm$, respectively.
Particularly, through the relationship between metric bundles and horizons in
$\pi_a^+$, a \textbf{NS} or \textbf{BH}   metric  can be parameterized in terms of the horizon frequency identified by the corresponding metric bundle. The inner  and outer  horizons in $\pi_a^+$ correspond to envelope surfaces.
We analyze the  lines $a=$constant (i.e. a single spacetime of the metric  bundle) and $r=$constant in the extended
plane--Fig.\il\ref{Fig:SoOv} and classify the singularities according to the horizon construction
as shown  in  Fig.\il(\ref{Fig:RoyalS}), i.e.,
 strong naked singularities, $\textbf{SNS}=\textbf{SNS}^+\cup\textbf{SNS}^-$,  having  $a_0>2M$
with
$\textbf{SNS}^+$ for $a_0>4M$ and  $\textbf{SNS}^-$ for $a_0\in[2M,4M[$;
\textbf{WNS}--weak naked singularities  with  $a_0\in]M,2M[$ and
$\textbf{BH}=\textbf{BH}^+\cup\textbf{BH}^-$ with   $\textbf{BH}^+$ for $a\in[a_{g}^1,M]$, $a_{g}^1=3/4 M$ and
$\textbf{BH}^-$ for $a\in[0,a_{g_1}]$.
\item
Demonstration of the \emph{confinement} of metric bundles with origin in \textbf{BHs} and \textbf{WNSs} and
 identification and study of the \emph{ corresponding bundles}.
 In Table\il\ref{Table:number-cicle}, we  proved the confinement of the bundles in the extended  plane delimited by the
inner horizon curve. The horizon and bundle  frequencies are related by the  relation $\omega_0^+\omega_0^-=1/4$.
\item
\emph{Properties of horizons and bundles}.
For the entire family of Kerr geometries, we established the relations between  metric bundles  and horizons. We found two relations which can be specified in detail as follows.
Relations \textbf{I}: {\small \texttt{ $\omega_0^{-1}\equiv a_0^{\pm}/M=({2 r_{\pm}(a_g)}/{a_g})\equiv \omega_H^{-1}(a_g)$,  $\omega_H^+(r_g,a_g)=\omega_0=Ma_0^{-1}$, $\omega_H^-(r^{\prime}_g,a_g)=\omega^{\prime}_0=M/a_0^{\prime}$
where
 $r^{\prime}_g\in r_-$ ($r_+=r_g$,  $r_-=r_g^{\prime}$)}}--Fig.\il(\ref{Fig:RoyalS}).
Relation \textbf{II}: {\small \texttt{$\omega^{\prime}_0=({4 \omega_0})^{-1}$, $ \omega_H^+\omega_H^-={1}/{4}$, (or equivalently $a_0^+(a_g)a_0^-(a_g)=4M^2$), $a_0^{\pm}/M=(2 r_{\pm}(a_g))/{a_g}$  where $a=a_0$ and  $a=a_p$}}--Fig.\il\ref{Fig:pesoMa}.
\item
Properties of specific spacetimes.
We studied the metric bundles corresponding to a   single \textbf{BH} spacetime (with equal tangent spin  $a_g$).
We analyzed the Kerr spacetime in terms of metric bundles--Sec.\il\ref{Sef:aomega}.
\item Proof of the \textbf{BH-NS} relation through the properties of the corresponding metric bundles.
\item
Analysis of the frequencies  $\varpi_\pm$  of the spacetime by using the  maximum crossing of two metric bundles--Eq.\il(\ref{Eq:gen2-dat}).
We proved he existence of the corresponding metric bundles  and analyzed its  significance for the horizon construction and properties of a \textbf{BH} spacetime--Tables\il\ref{Table:number-cicle} and \il \ref{Table:ABCDmodels};
Figs.\il\ref{Fig:RoyalS},\ref{Fig:pesoMa},\ref{Fig:PostMerge},\ref{Fig:PostMerge},\ref{Fig:manet},\ref{Fig:Bollet} and
Eqs.\il(\ref{Eq:aptriplet}) and (\ref{Eq:fast-can}).
\item
Identification of the  Killing throats and bottlenecks and  metric bundles in the static  and charged \emph{Reissner Nordstr\"om} solution and in the axisymmetric, electrically charged, \emph{Kerr-Newmann spacetime}.
We proved that the  bending  (curvature) of the Kerr metric bundles in $\pi_a$ is  related to the  frame-dragging
of the  spinning spacetimes. We noticed the different roles played by  the electric and rotational charges --Eq.\il(\ref{Eq:KN.RN.shown}).  We studied the metric bundles $
(Q_{\omega}^{\pm})^2$ in terms of the electric charge  $Q$--Eq.\il(\ref{Eq:Q-para-metric}). In Sec.\il\ref{Eq:off-e}, we analyzed
the off-equatorial case of the  Kerr  and Kerr-Newman geometries.
In Sec.\il\ref{Sec:appendix-area}, we  considered the relations between the areas of the horizon  and of the metric bundles regions in the extended plane.
\end{itemize}}

\begin{acknowledgements}
D.P. acknowledges partial support from the Junior GACR grant of the Czech Science Foundation No:16-03564Y.
{D.P. is grateful to Donato Bini, Fernando de Felice and Andrea Geralico for  discussing  many aspects of this work.}
This work was partially supported  by UNAM-DGAPA-PAPIIT, Grant No. 111617, and by the Ministry of Education and Science of RK,
Grant No. BR05236322 and AP05133630.
\end{acknowledgements}

\appendix
\section{Analysis of the  Kerr  and Kerr-Newman geometries: the off-equatorial case}\label{Eq:off-e}
In this appendix, we summarize the analysis presented in Secs. \ref{Sec:bottlem}, \ref{Sef:aomega} and \ref{Sec:RN-KN-cases} for
off-equatorial stationary observers in Kerr and Kerr-Newman spacetimes.  We focus on the presentation of the corresponding plots which contain all the relevant information for this case.

\begin{figure}[h!]
\centering
\begin{tabular}{lccr}
\includegraphics[scale=.35]{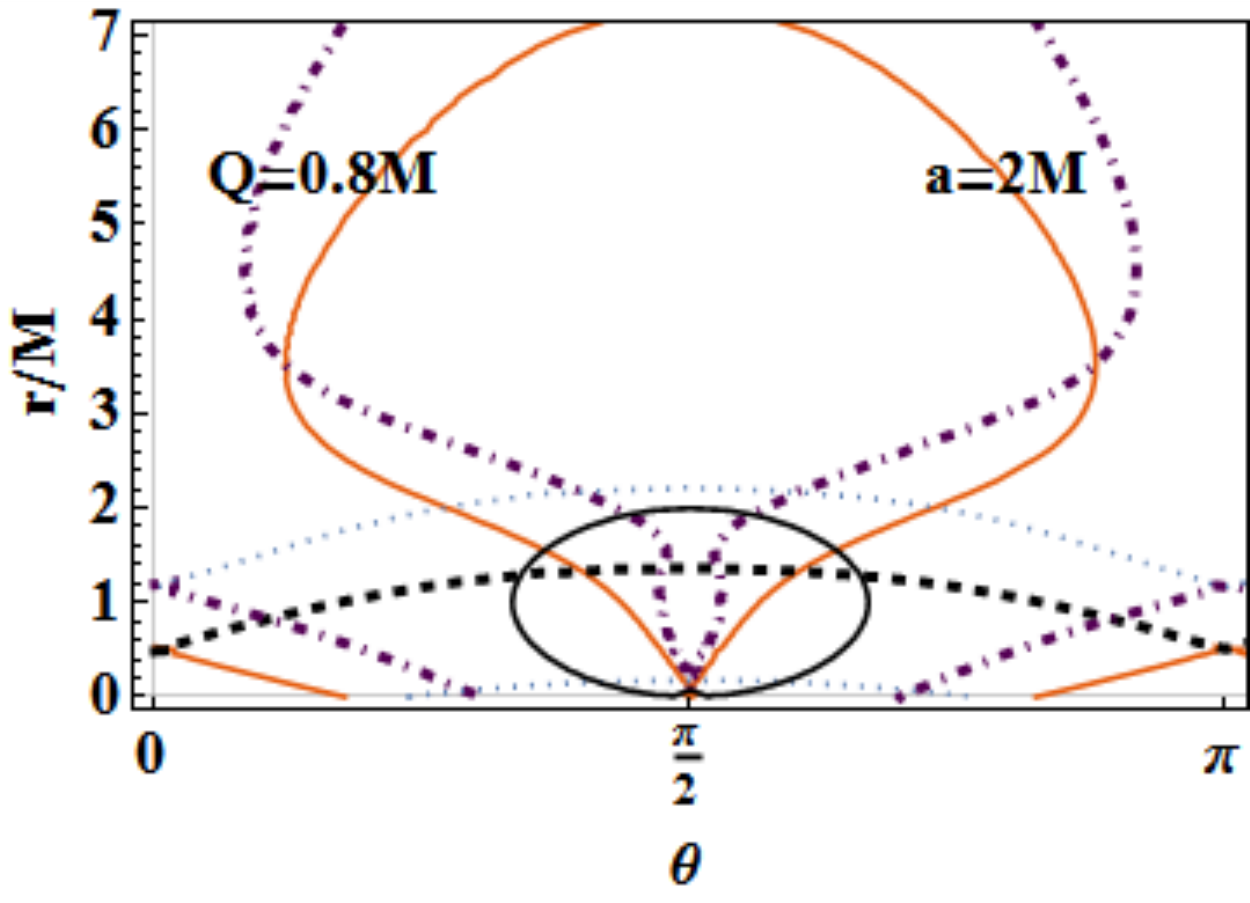}
\includegraphics[scale=.35]{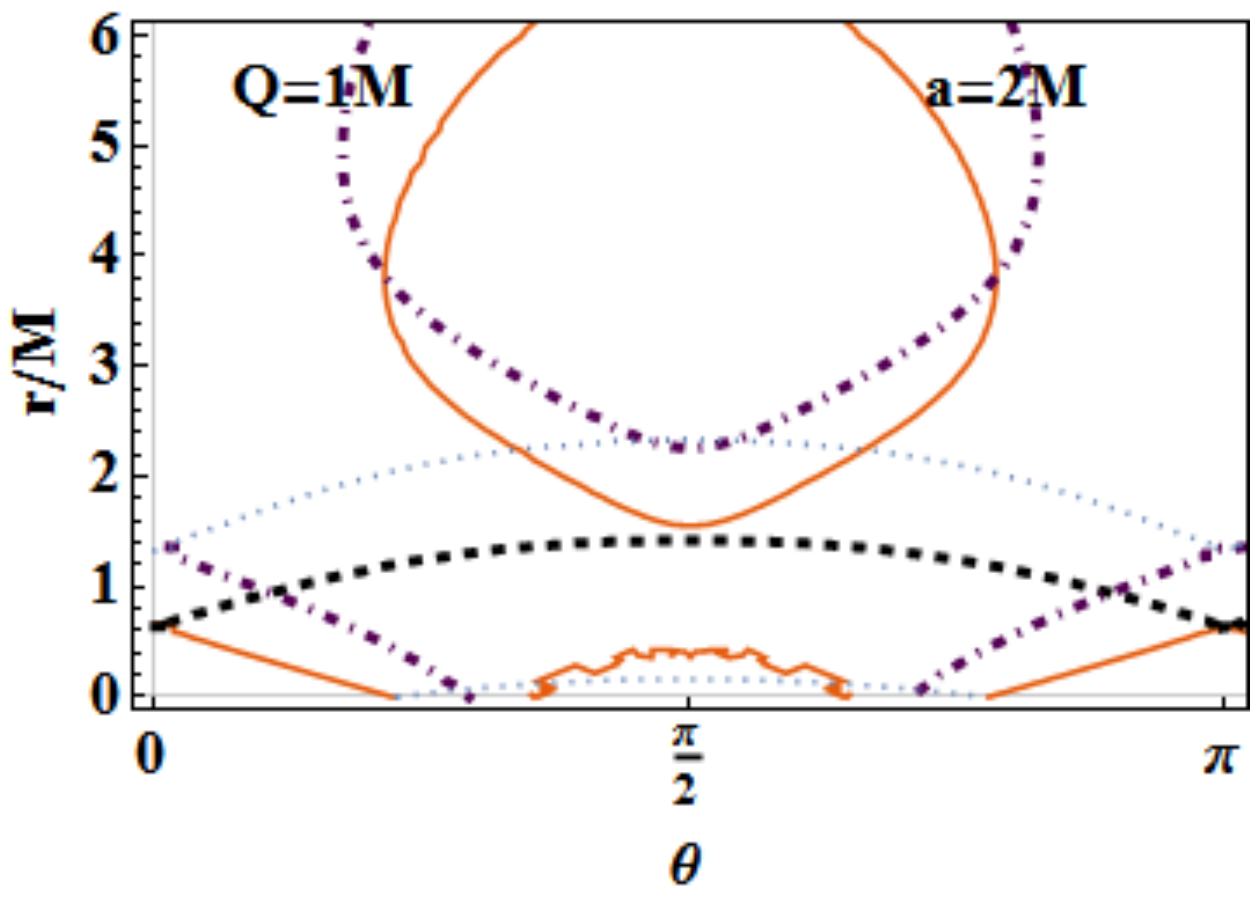}
\includegraphics[scale=.35]{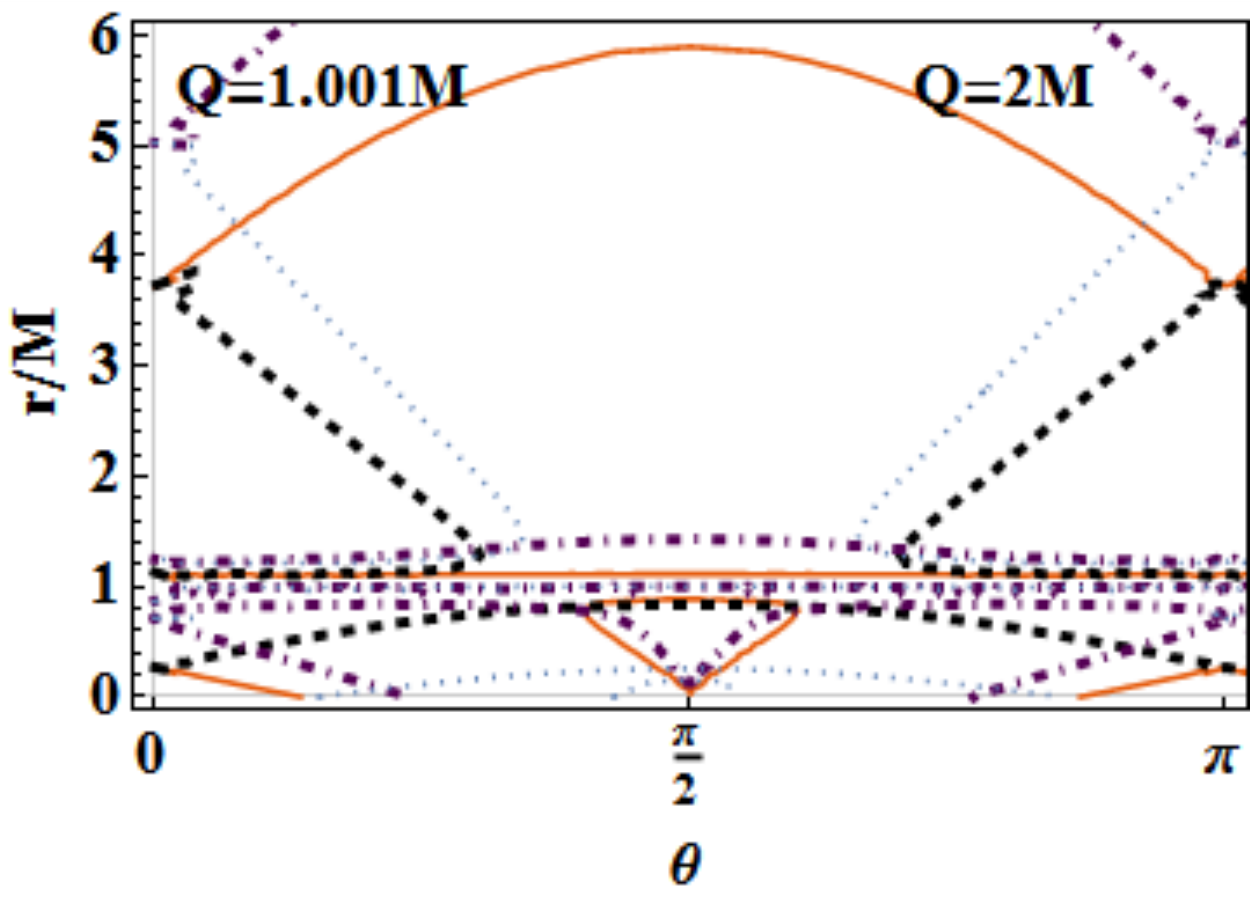}
\includegraphics[scale=.35]{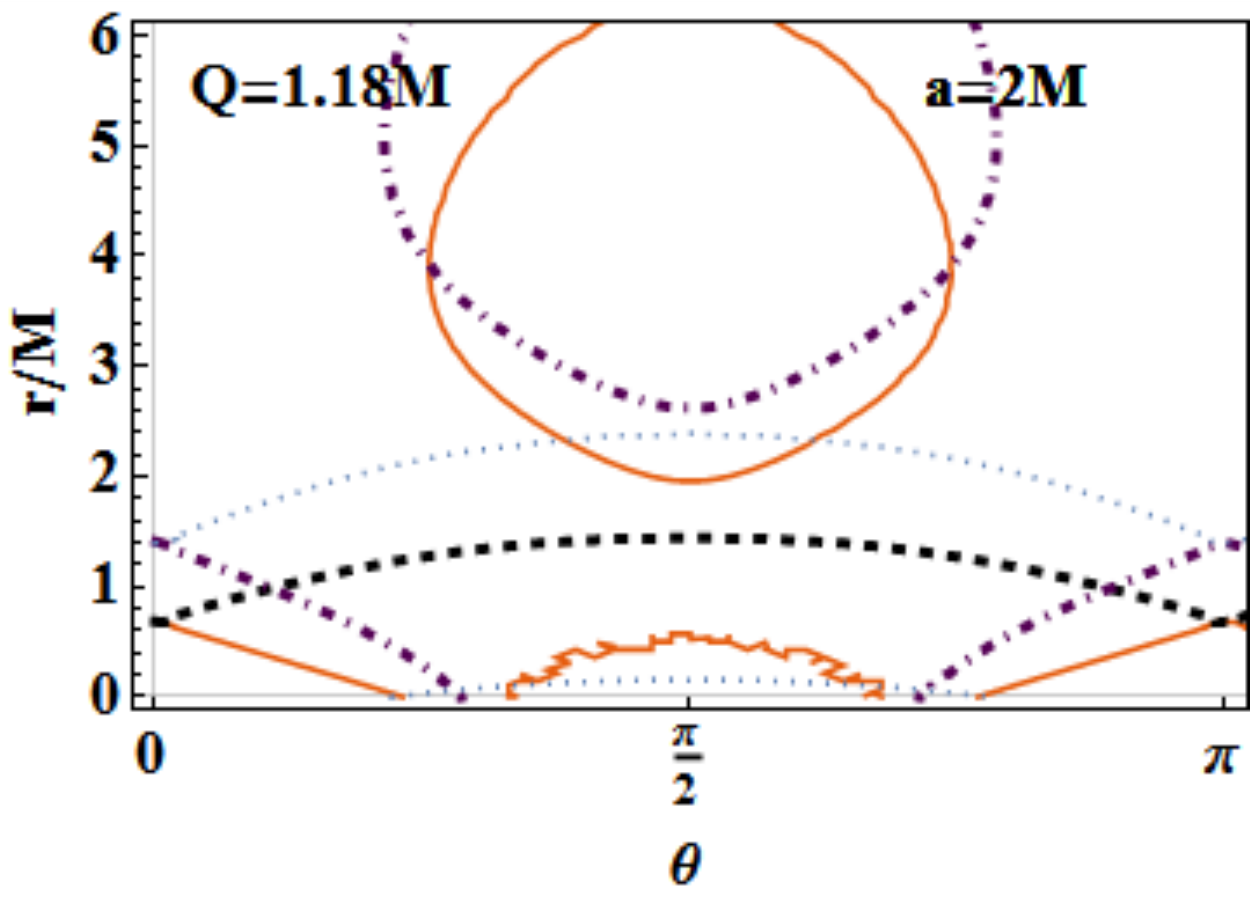}
\\
\includegraphics[scale=.35]{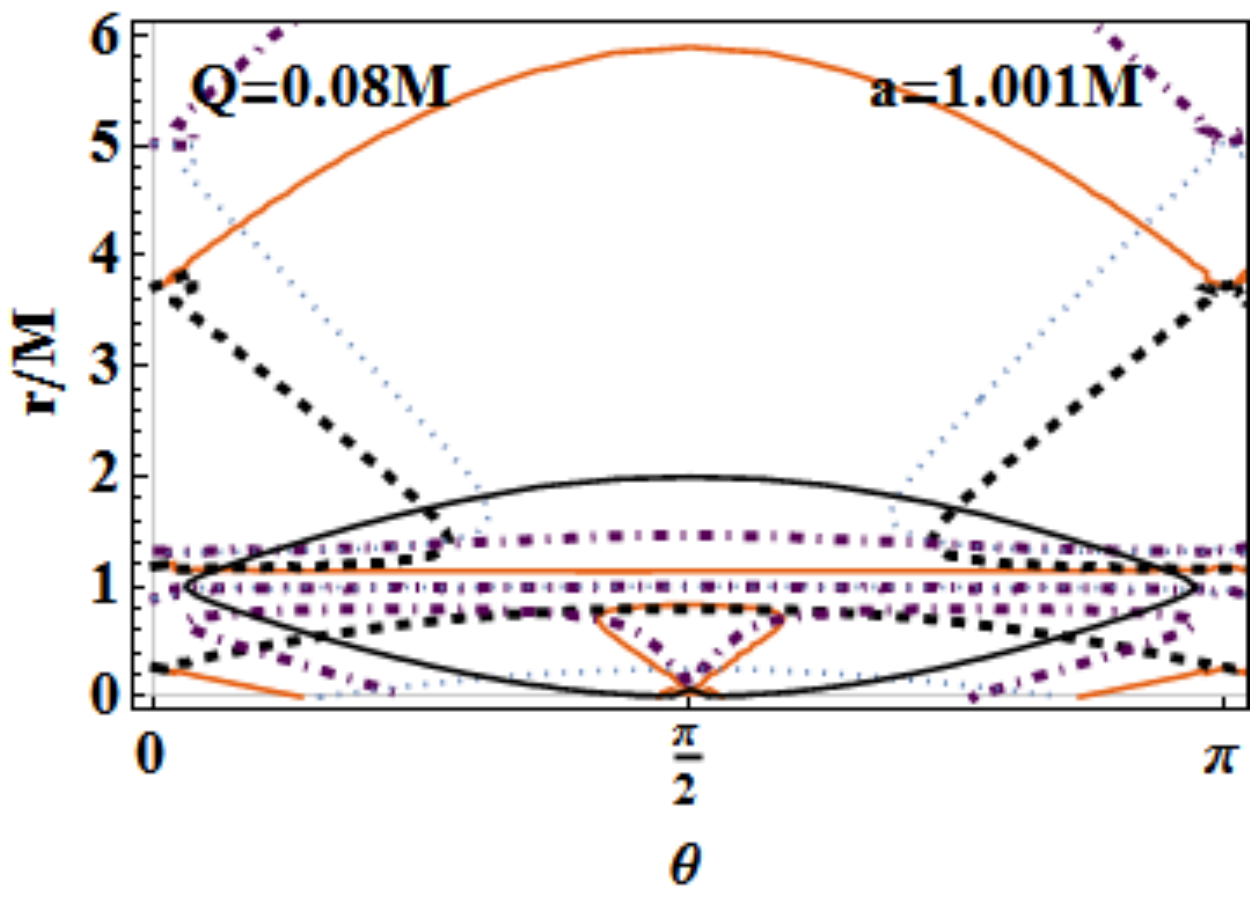}
\includegraphics[scale=.35]{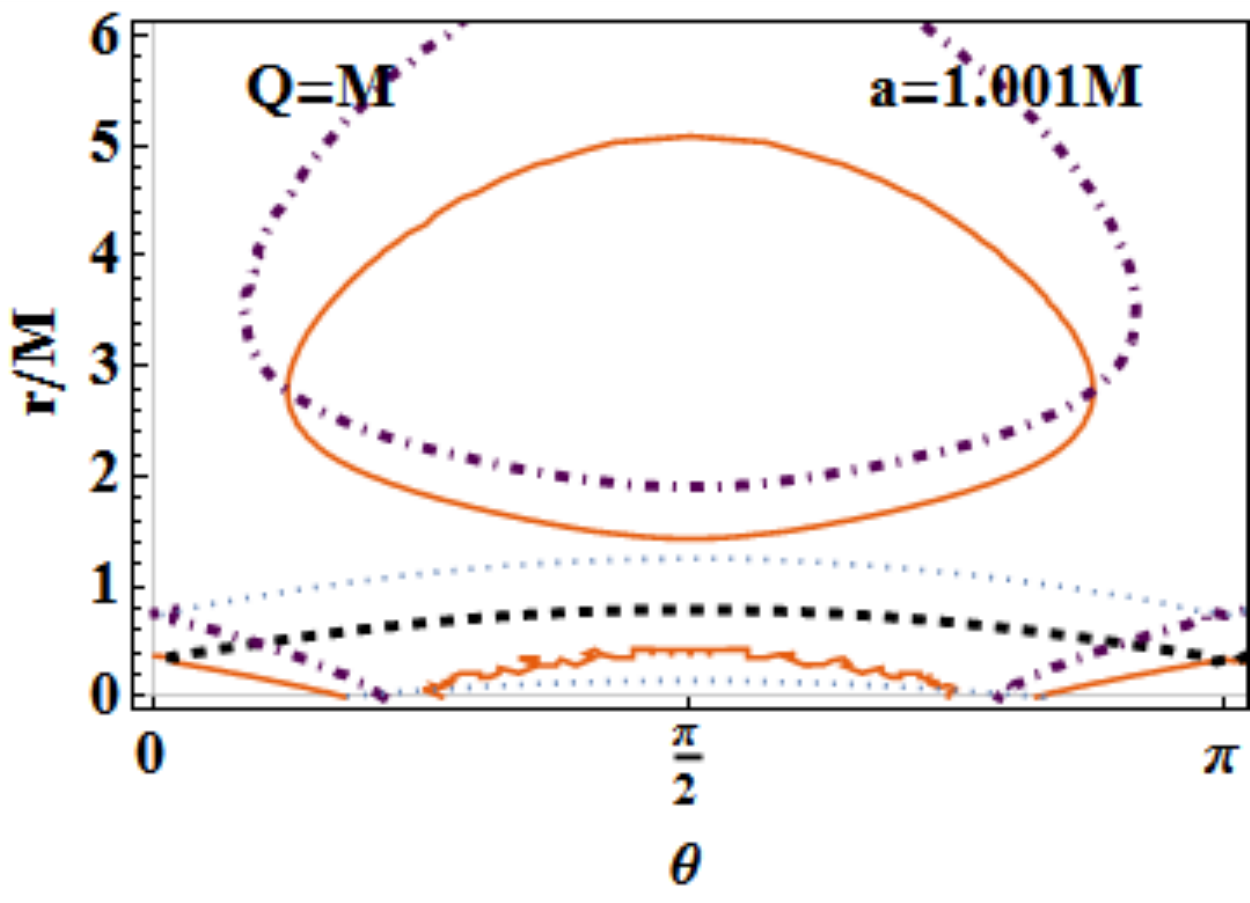}
\includegraphics[scale=.35]{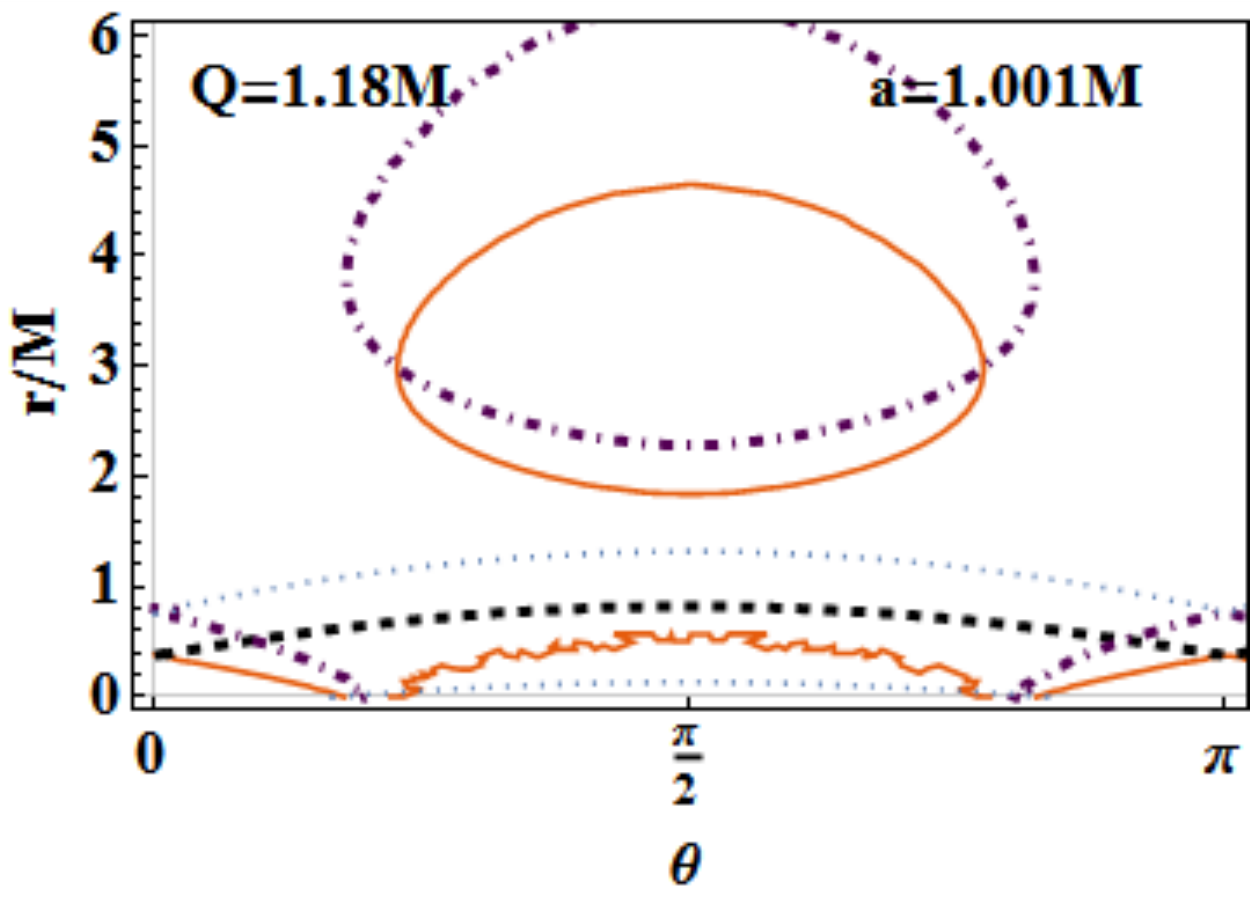}
\includegraphics[scale=.35]{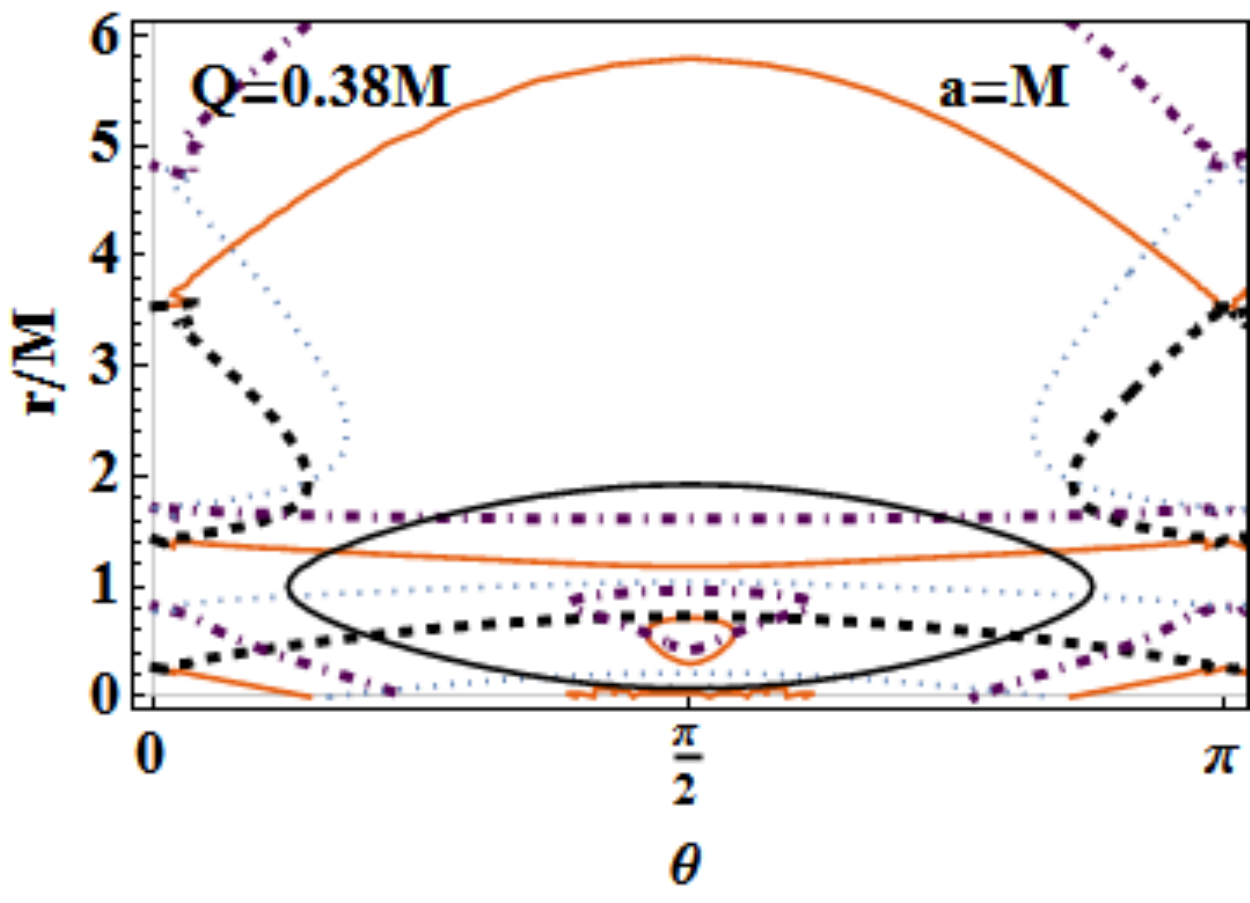}
\\
\includegraphics[scale=.35]{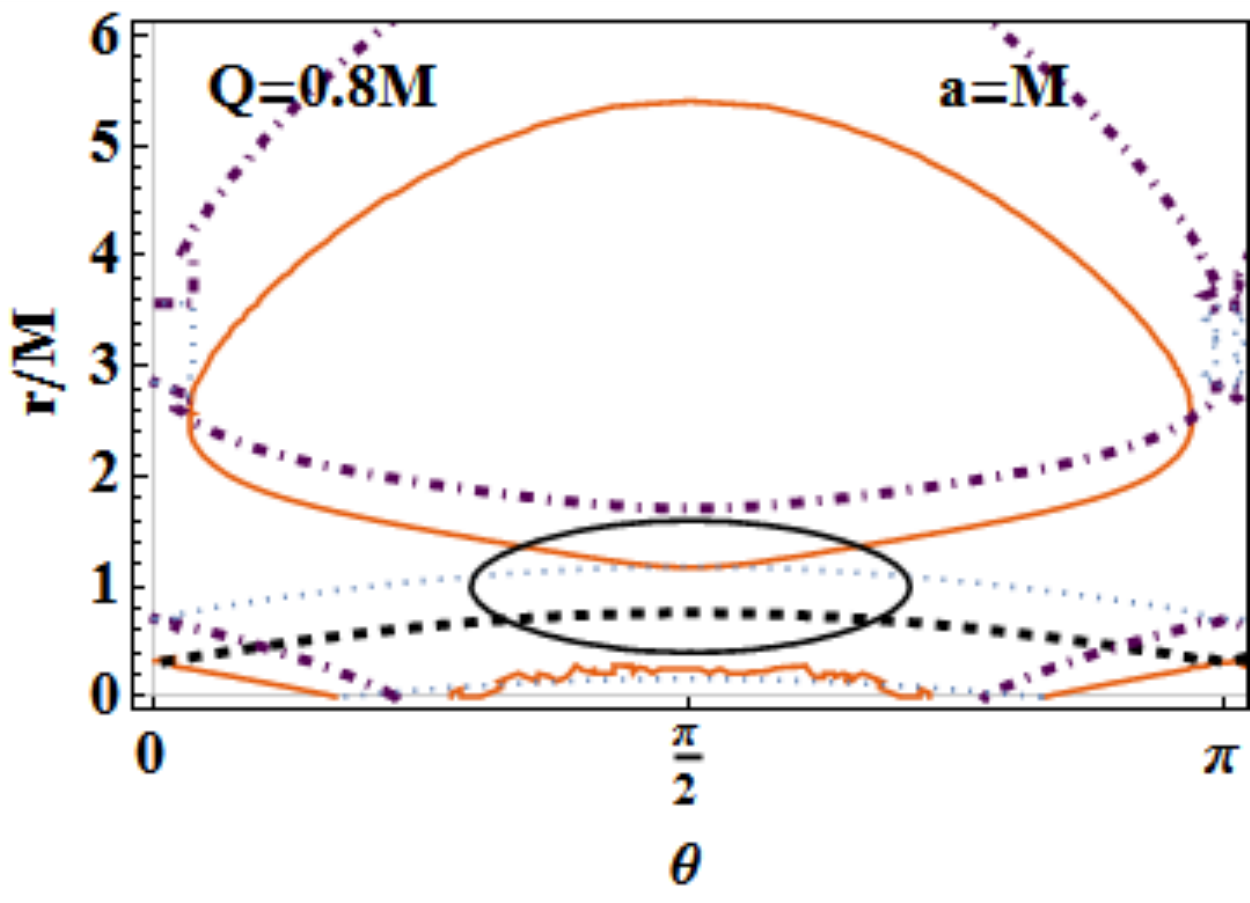}
\includegraphics[scale=.35]{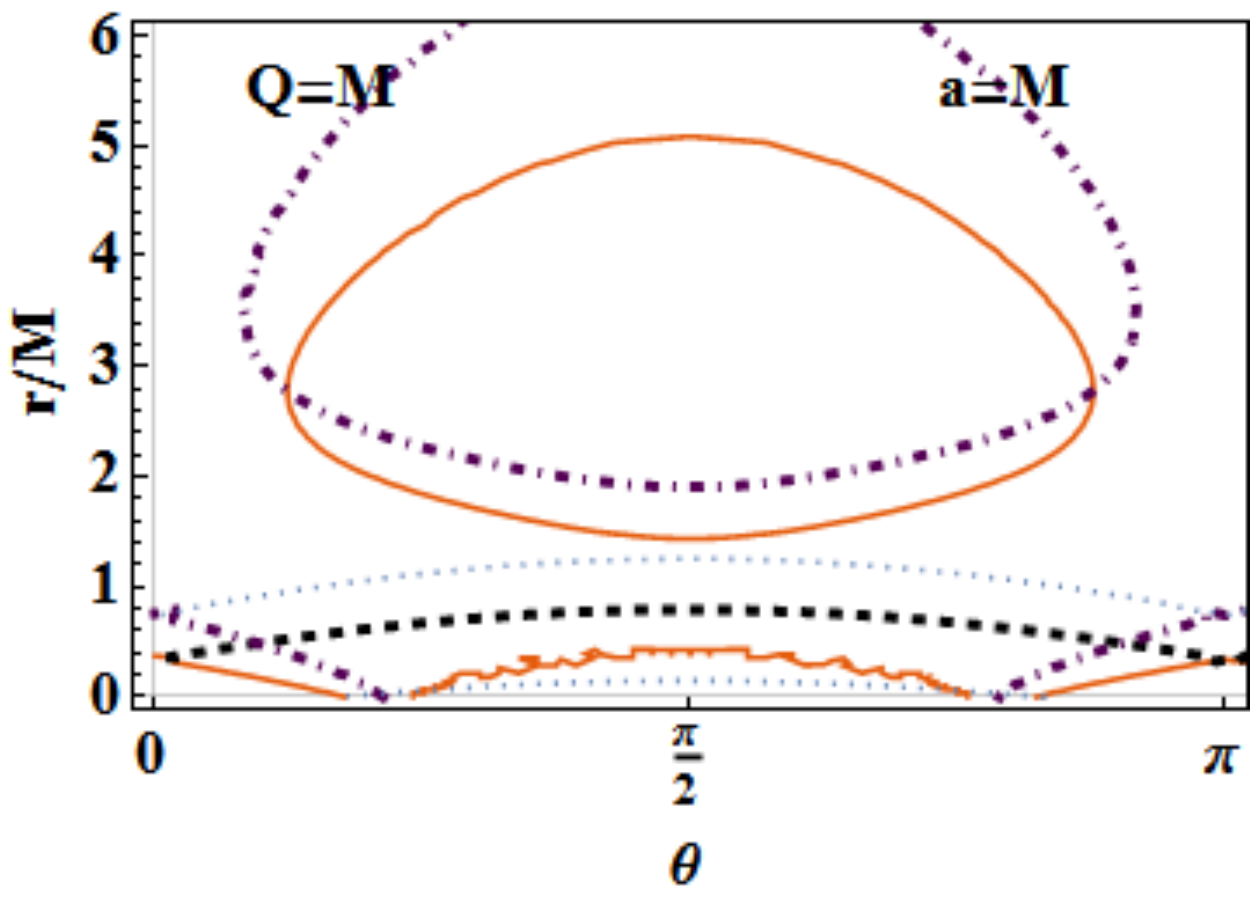}
\includegraphics[scale=.35]{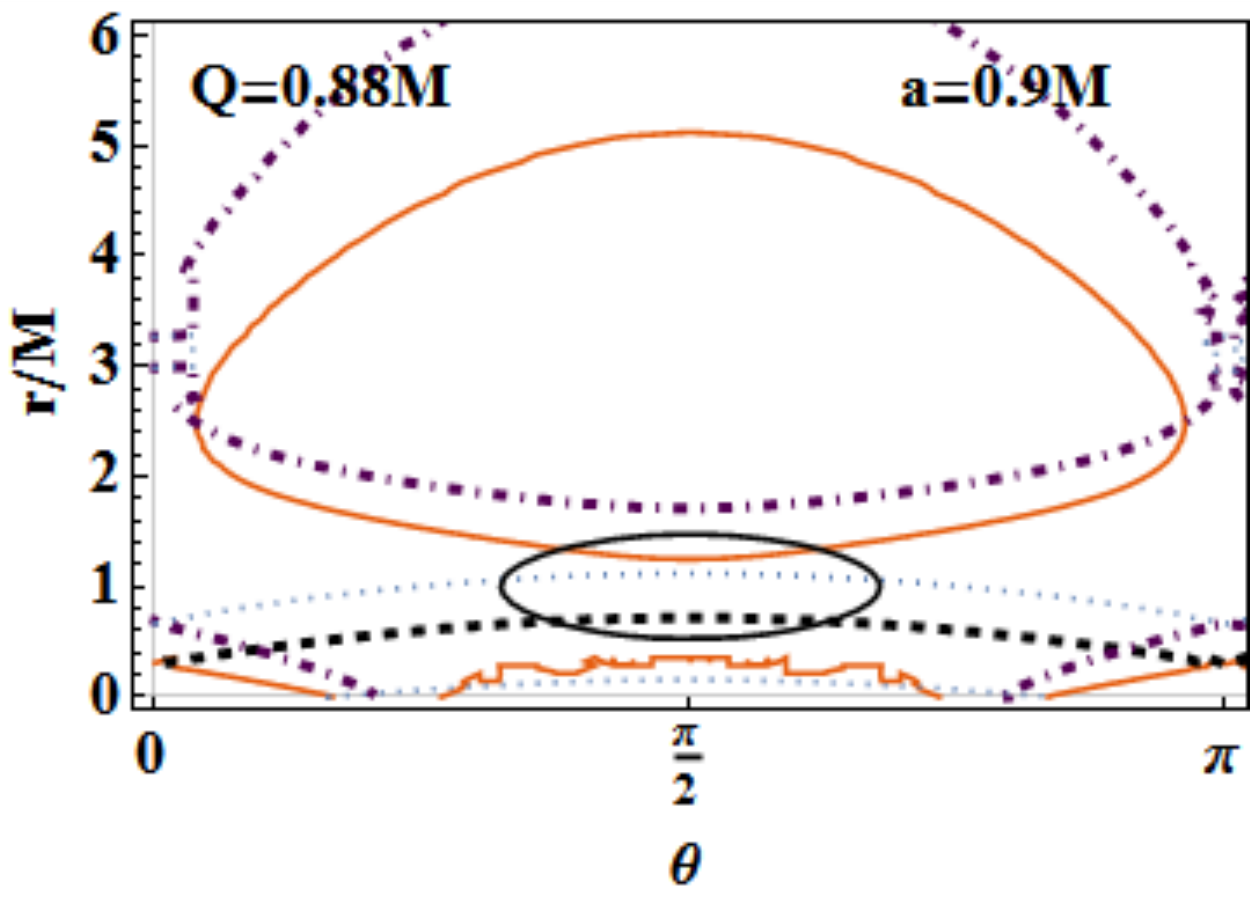}
\includegraphics[scale=.35]{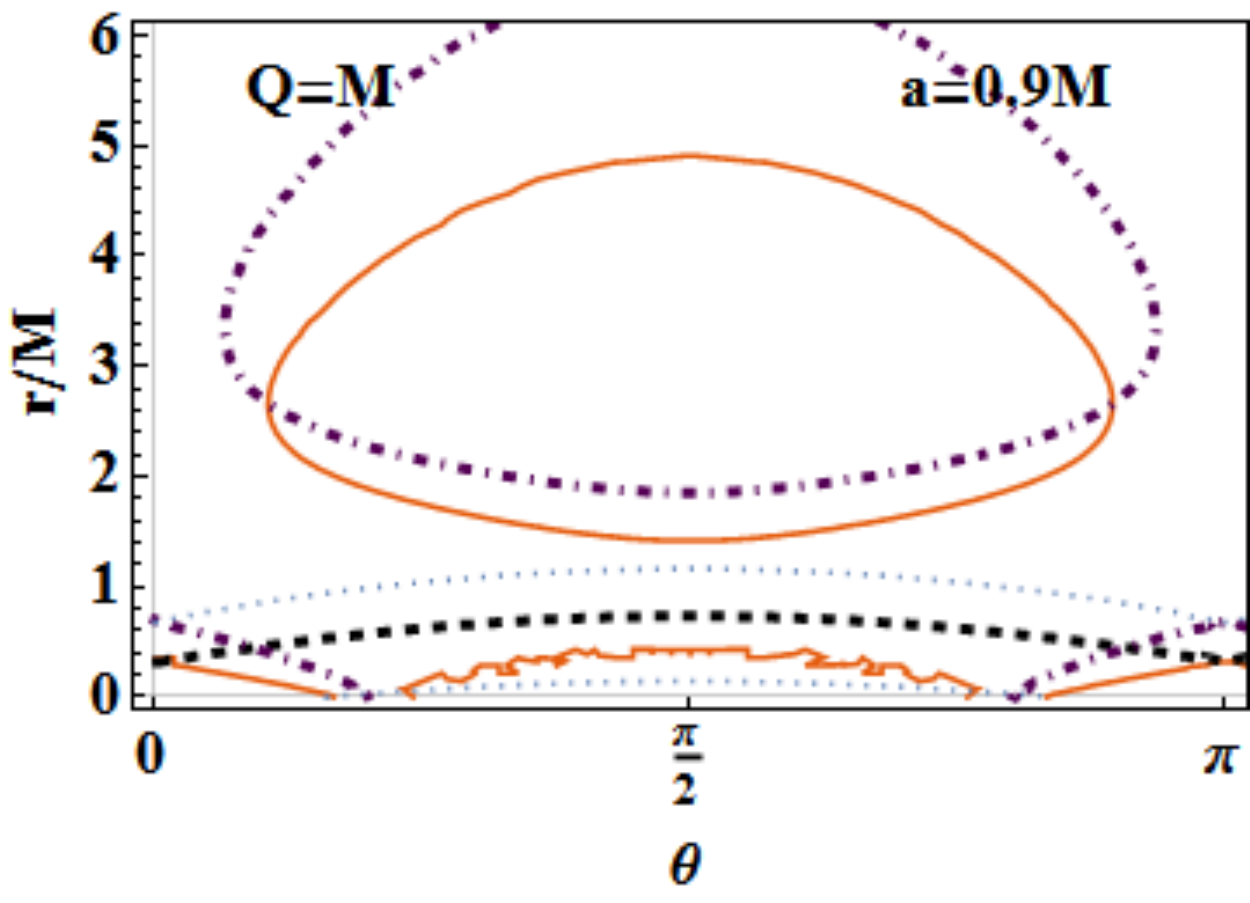}
\\
\includegraphics[scale=.35]{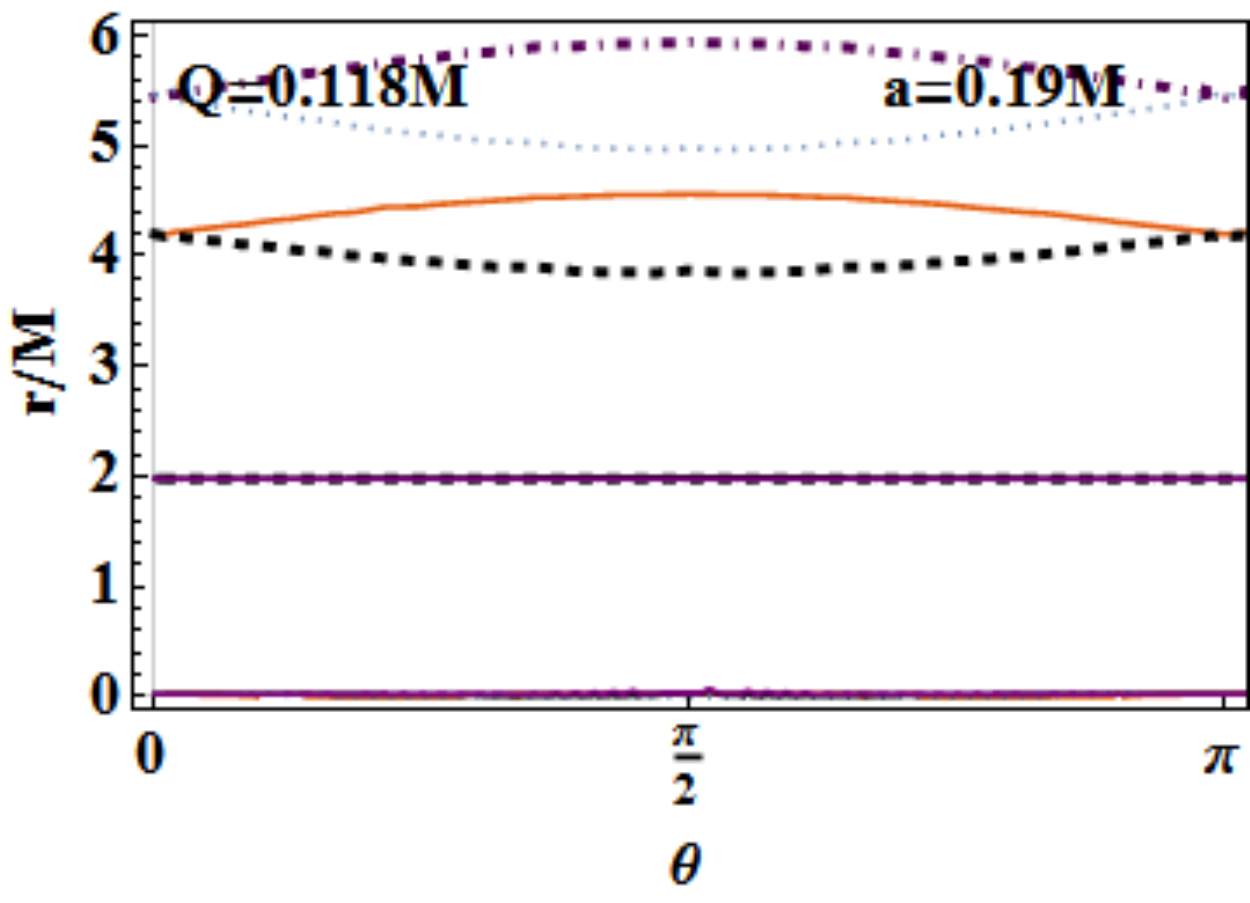}
\includegraphics[scale=.35]{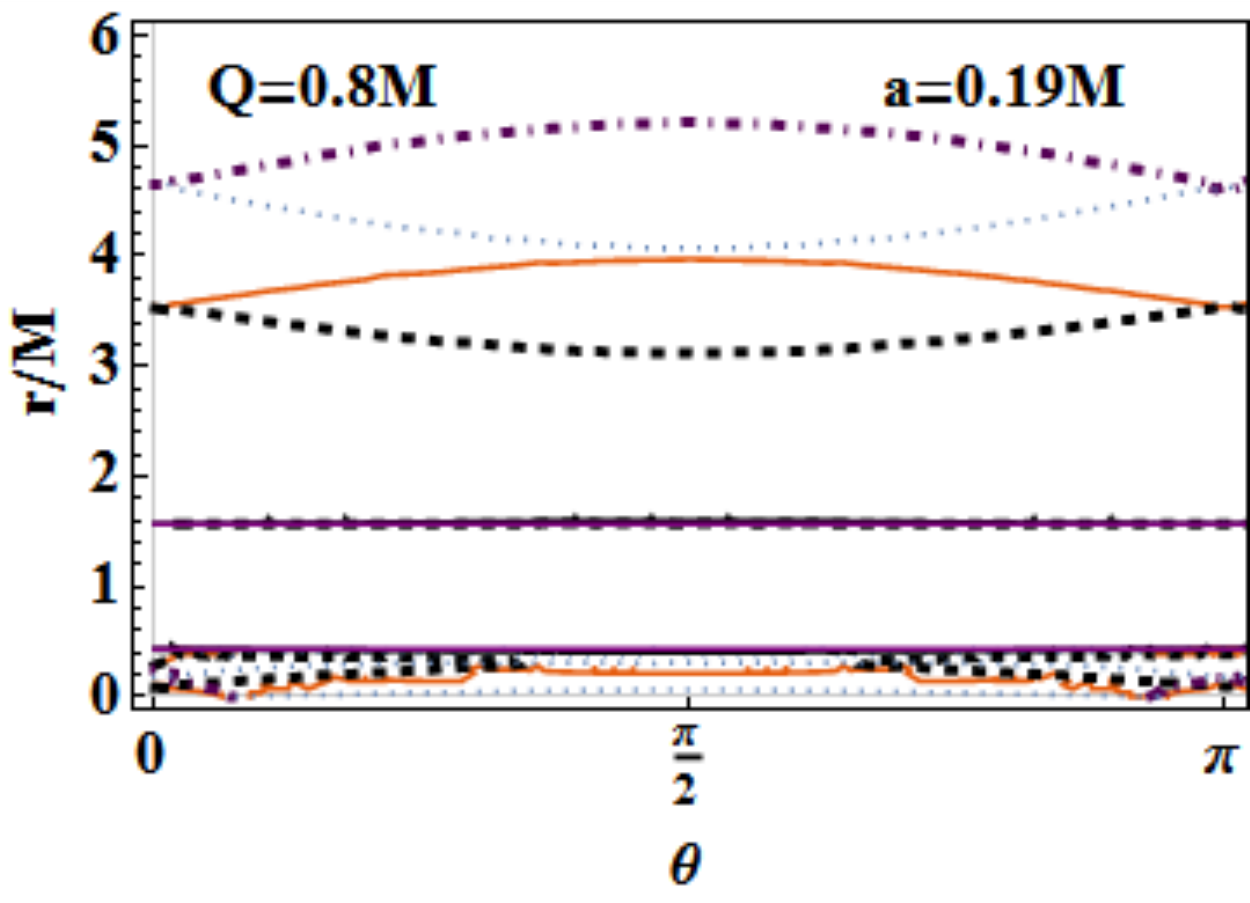}
\includegraphics[scale=.35]{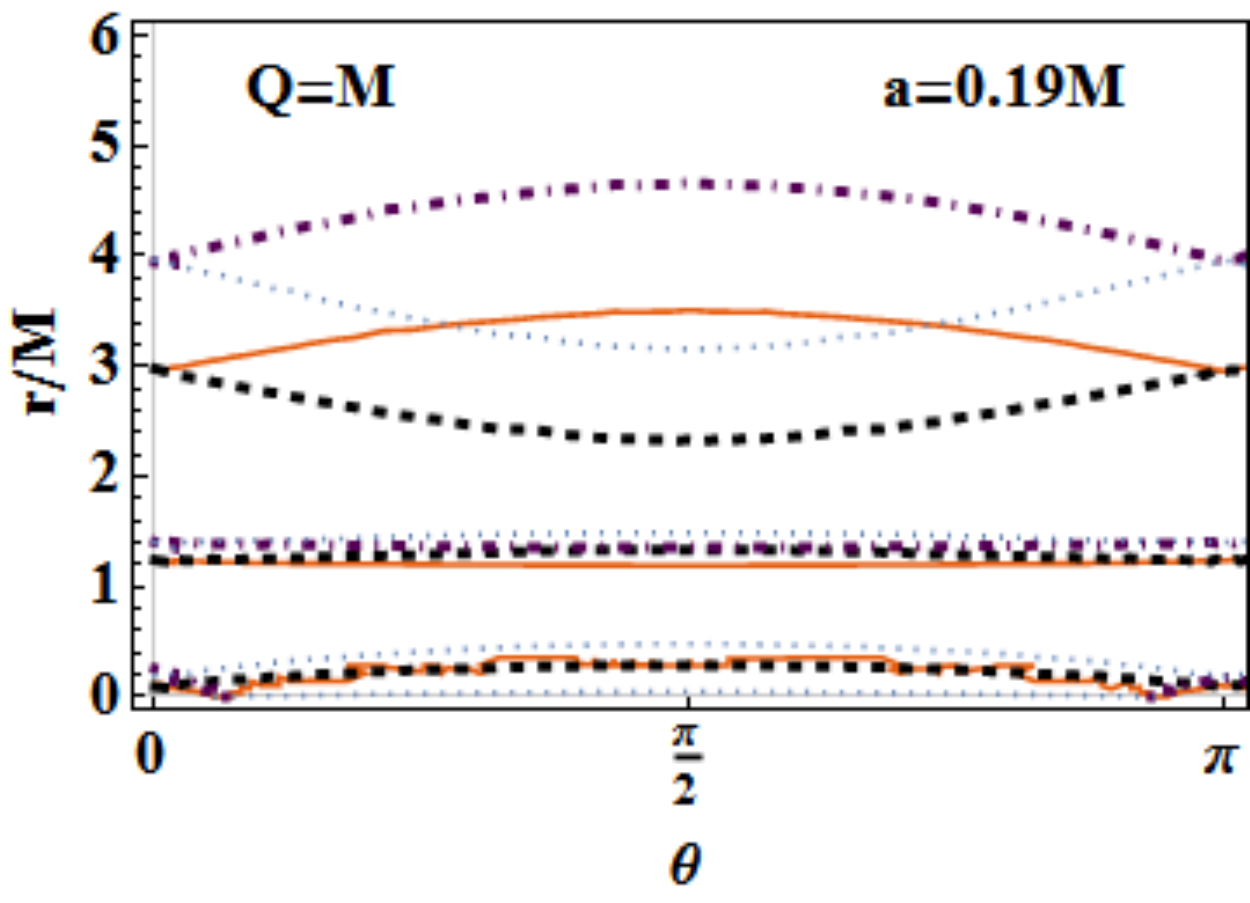}
\end{tabular}
\caption[font={footnotesize,it}]{Kerr-Newman  spacetimes: Off-equatorial analysis.
 Solutions  of $\partial^2_r\omega_{\pm}=0$ (orange and dashed curves)  and  $\partial^3_r\omega_{\pm}=0$ (dotted and dotted-dashed curves), defining the Killing bottleneck of  naked singularities.
Solutions of $g_{tt}=0$ (black), defining the static limits, and $g_{rr}^{-1}=0$
(red) defining the horizons.
 See also Fig.\il(\ref{Fig:spossov10}) for the case  $Q=0$ (Kerr spacetimes).}
\label{Fig:AGpossov15}
\end{figure}
\begin{figure}
  \includegraphics[width=3.95cm]{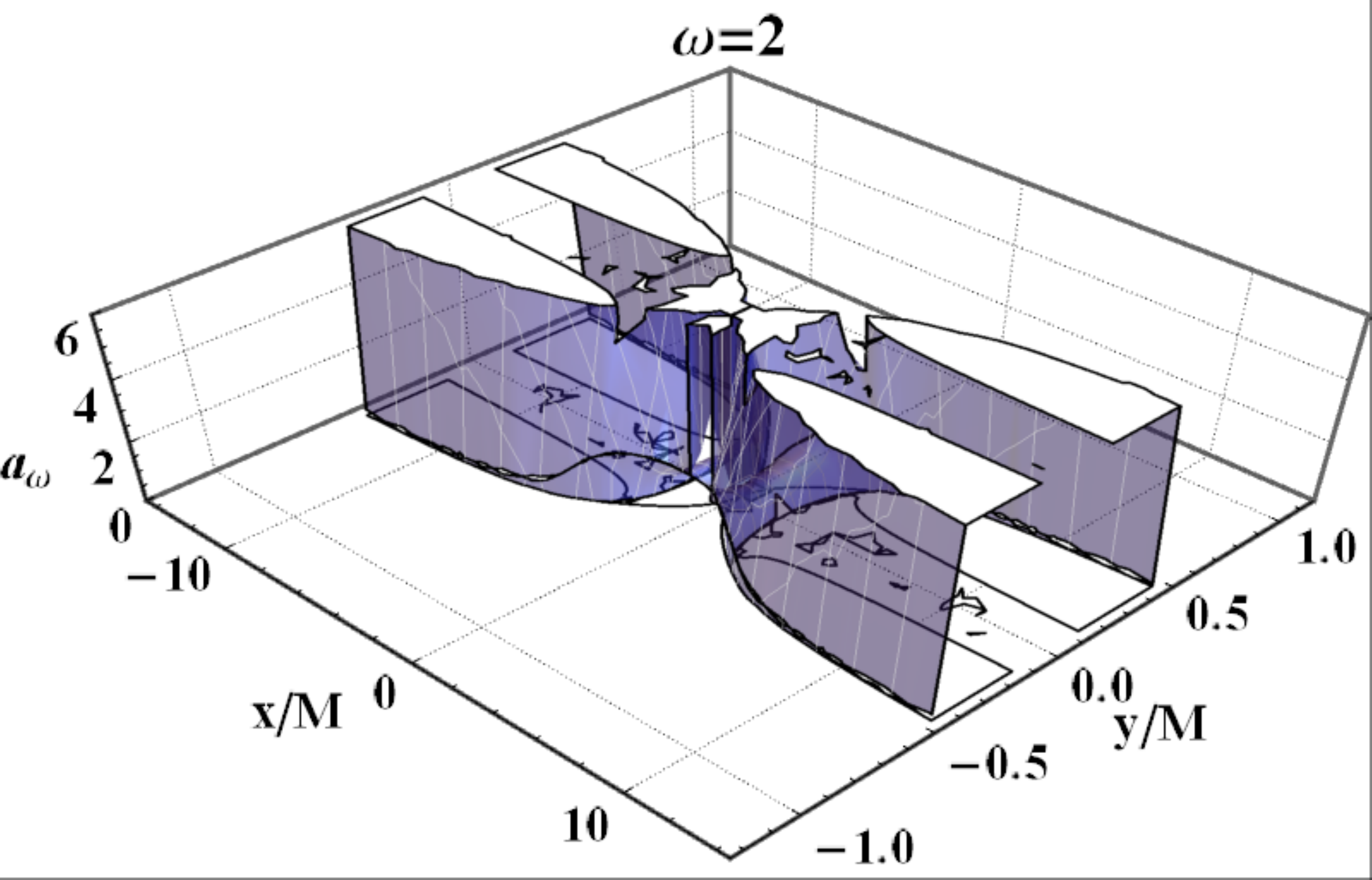}
   \includegraphics[width=3.95cm]{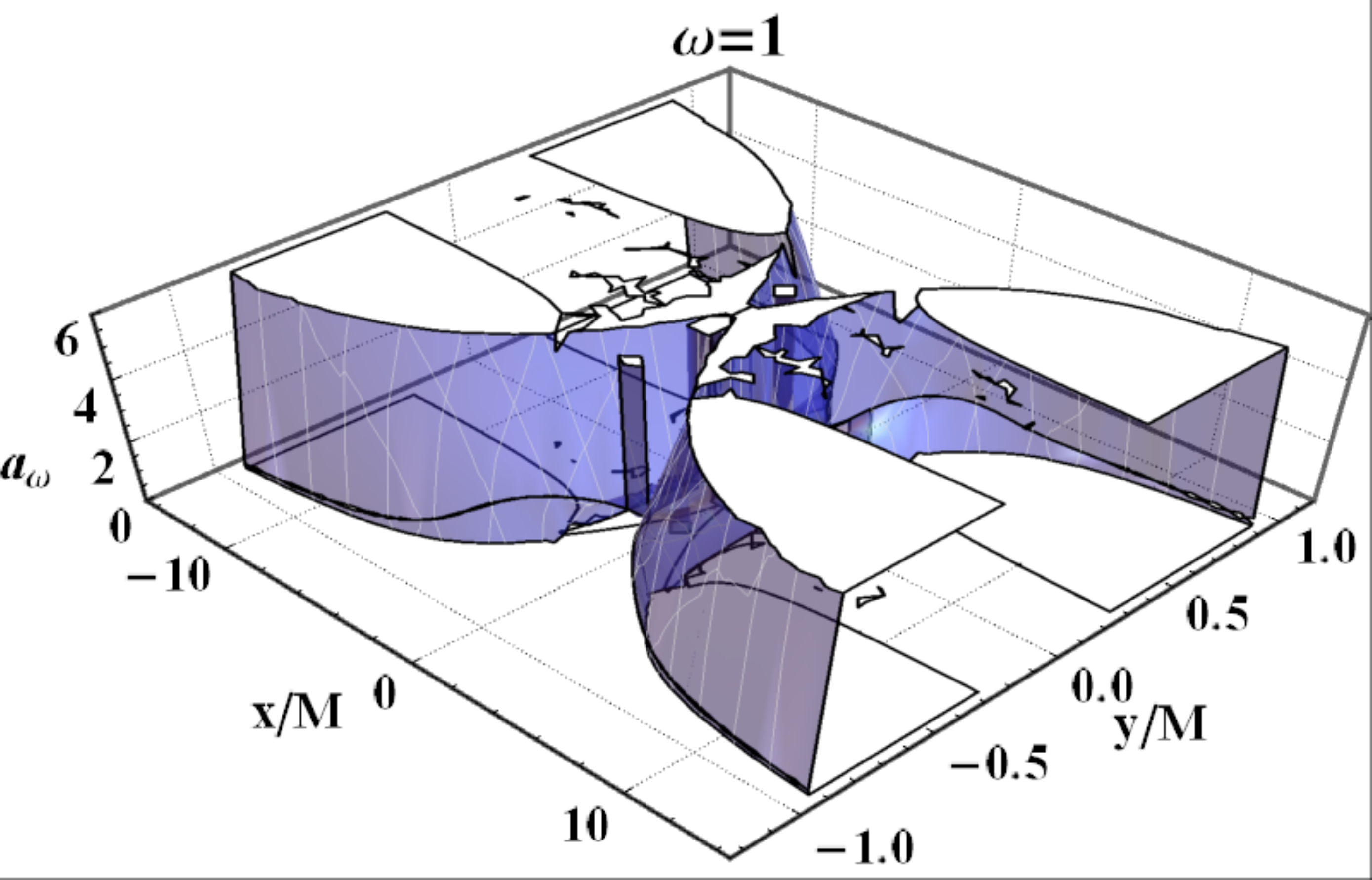}
   \includegraphics[width=3.95cm]{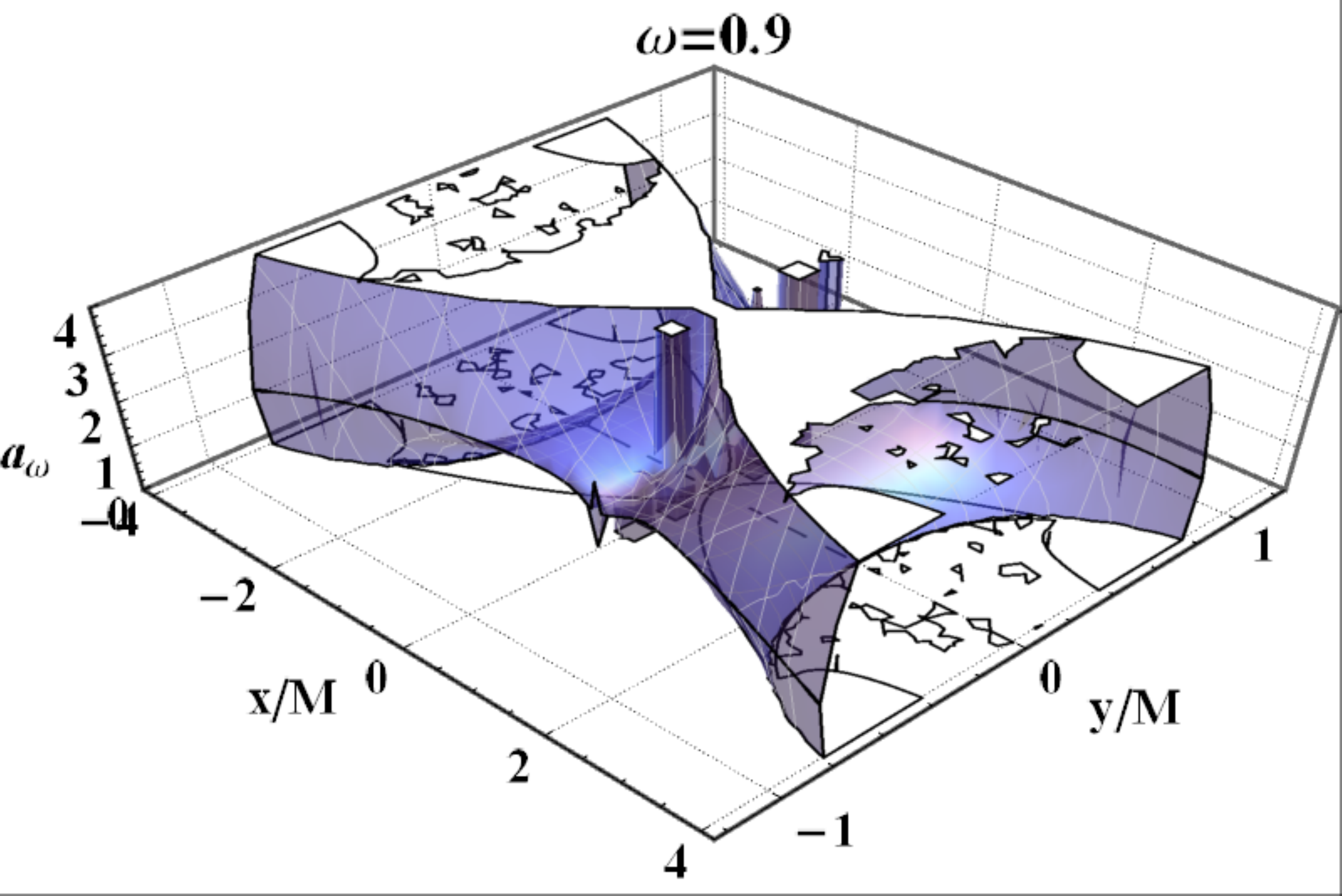}
   \includegraphics[width=3.95cm]{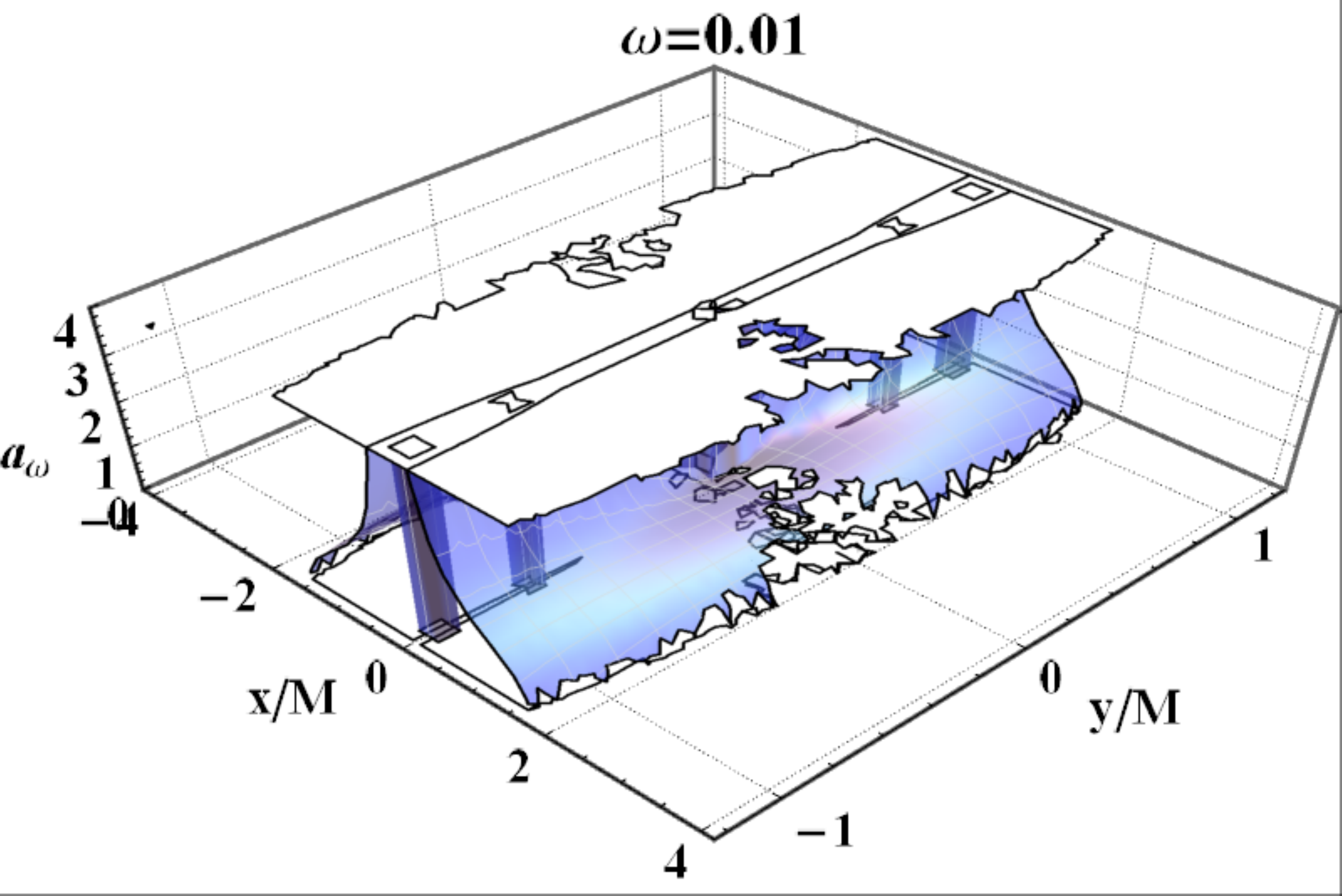}
  \caption{Kerr spacetimes: Off-equatorial analysis. The surfaces $a_{\omega}^{\pm}(r,\theta)$ as functions of the radial distance $r$
	in Cartesian coordinates $(x,y)$ for different frequencies  values $\omega$, including \textbf{BH}s and \textbf{NS}s --see also
	Figs.\il\ref{Fig:spossov10}, \ref{Fig:possov131} and \ref{Fig:possov132}. The spin functions are in Eq.\il(\ref{Eq:esa.tt}) for $Q=0$.}\label{FIG:fortCarl8}
\end{figure}
\begin{figure}[h!]
\begin{tabular}{lccr}
\includegraphics[scale=.35]{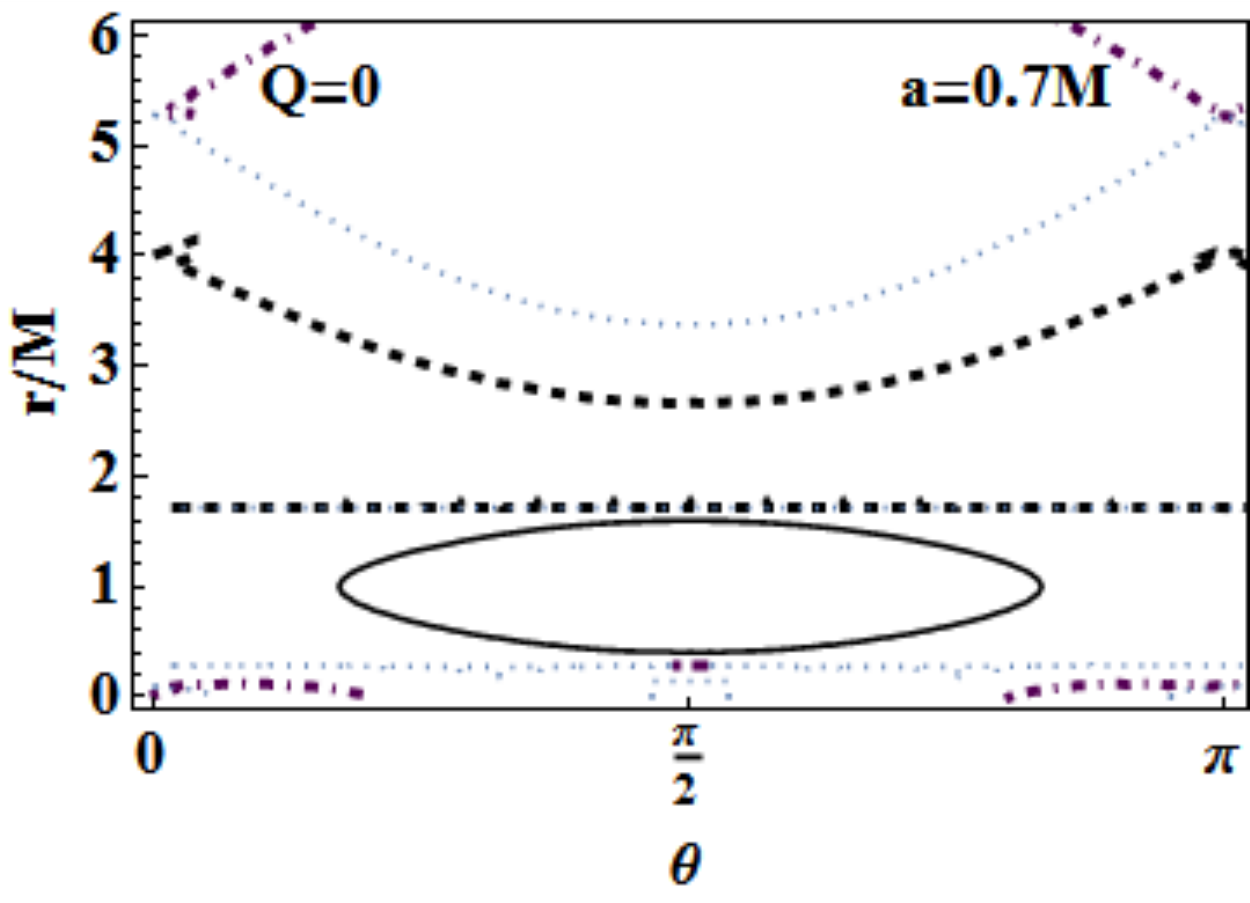}
\includegraphics[scale=.35]{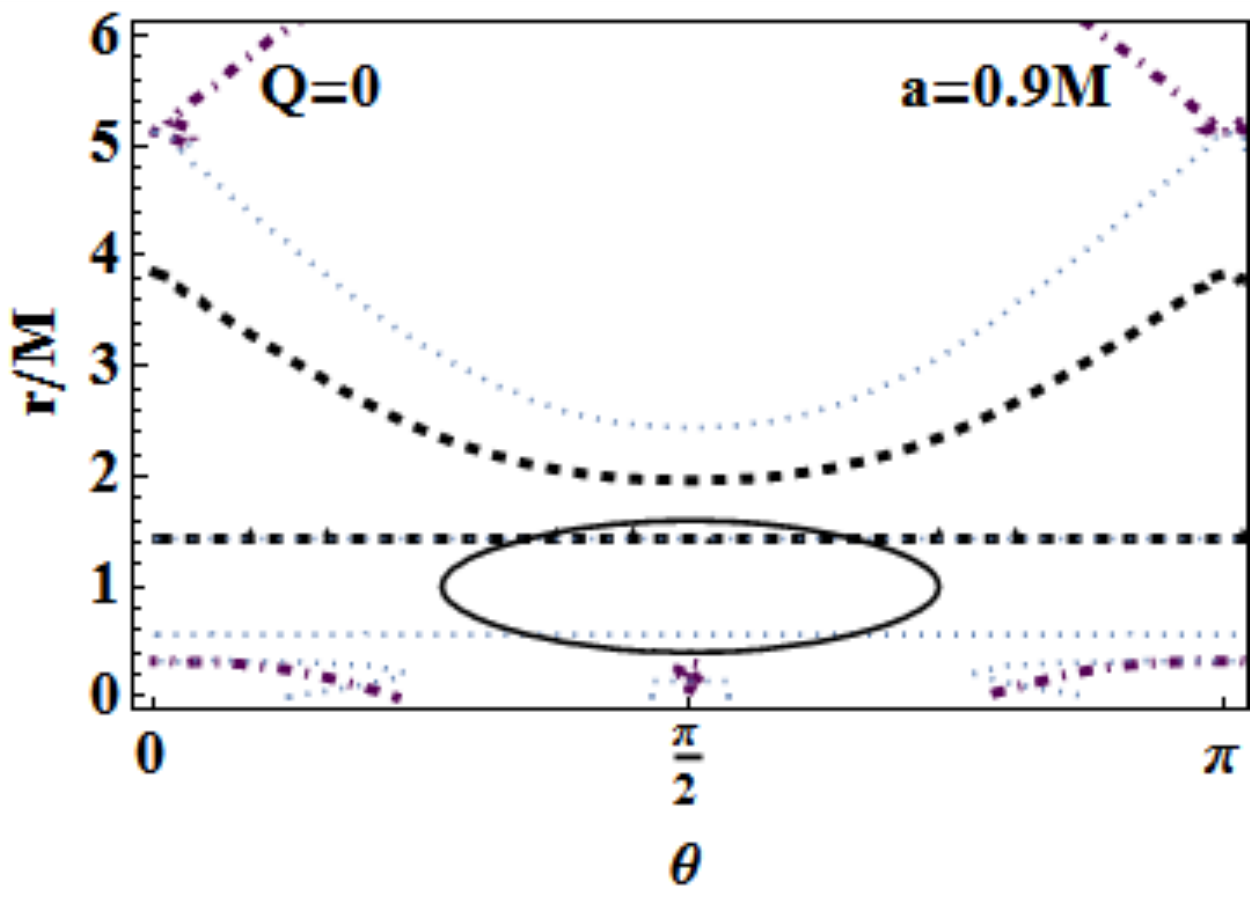}
\includegraphics[scale=.35]{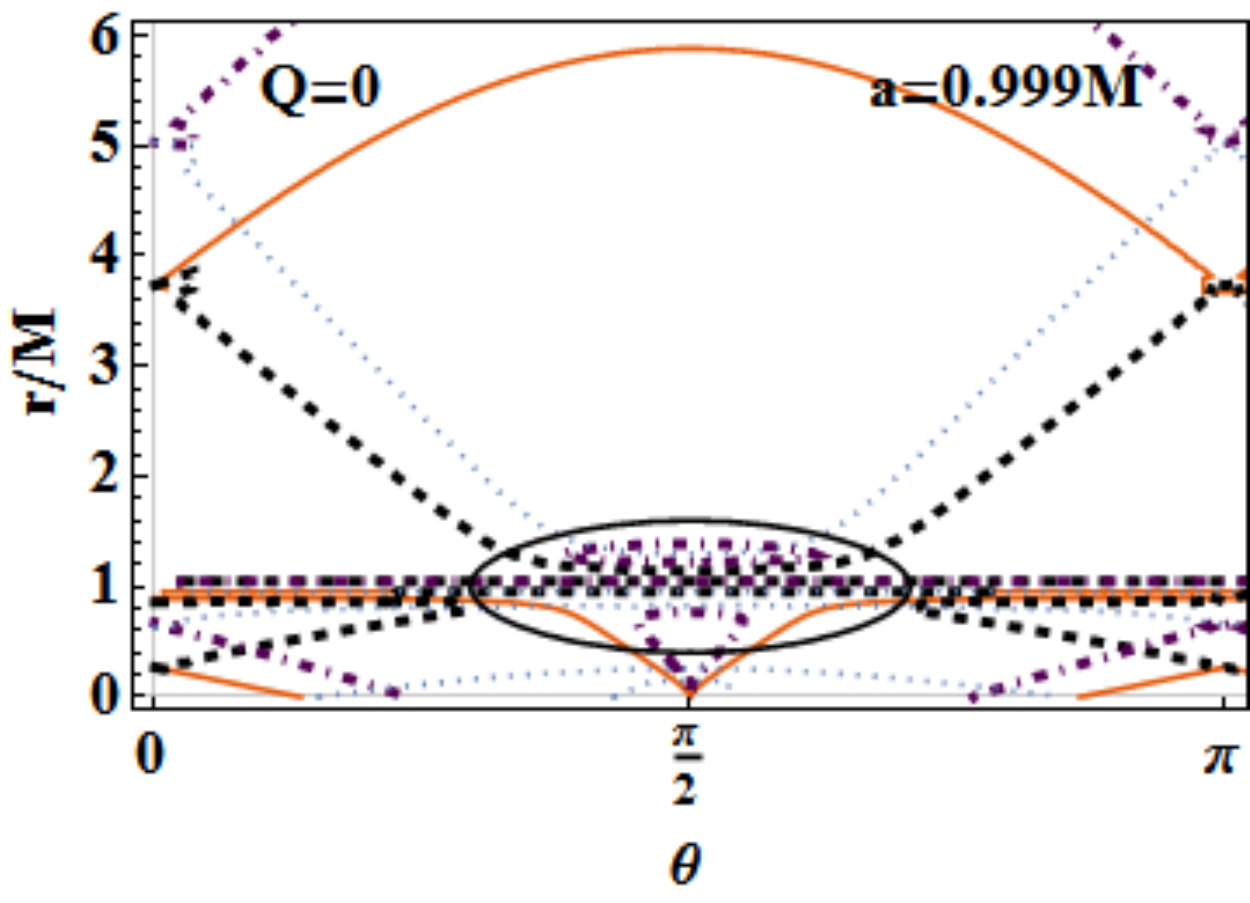}
\includegraphics[scale=.35]{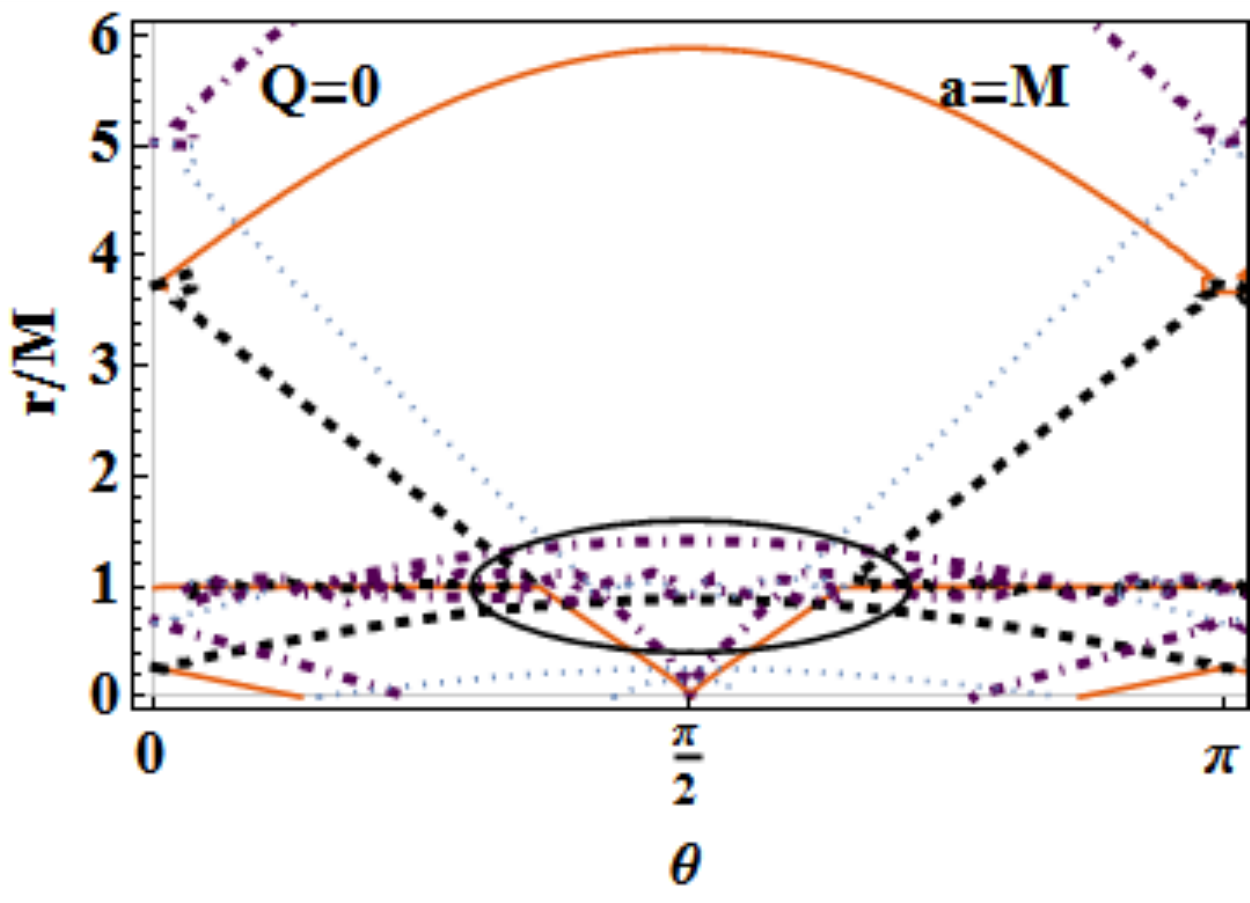}\\
\includegraphics[scale=.35]{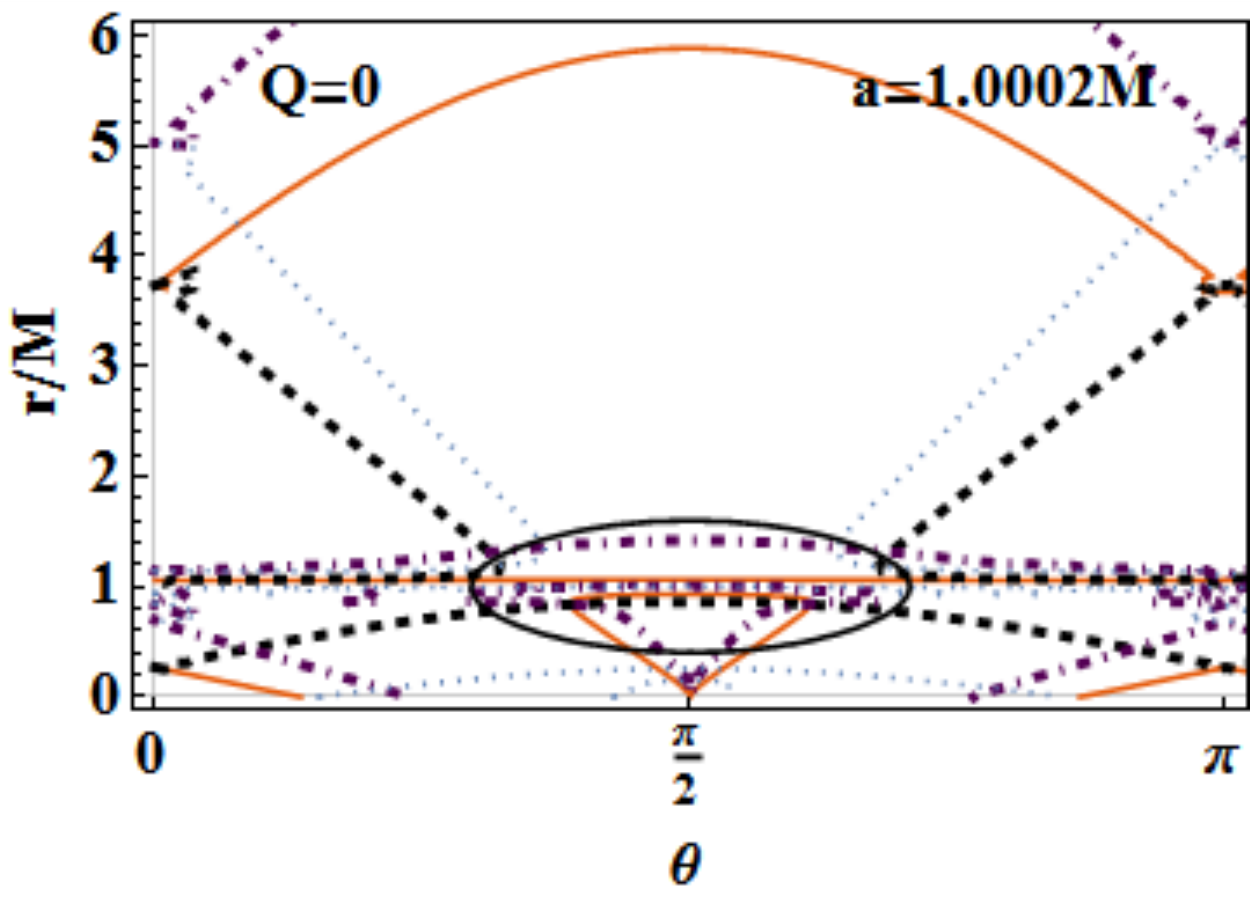}
\includegraphics[scale=.35]{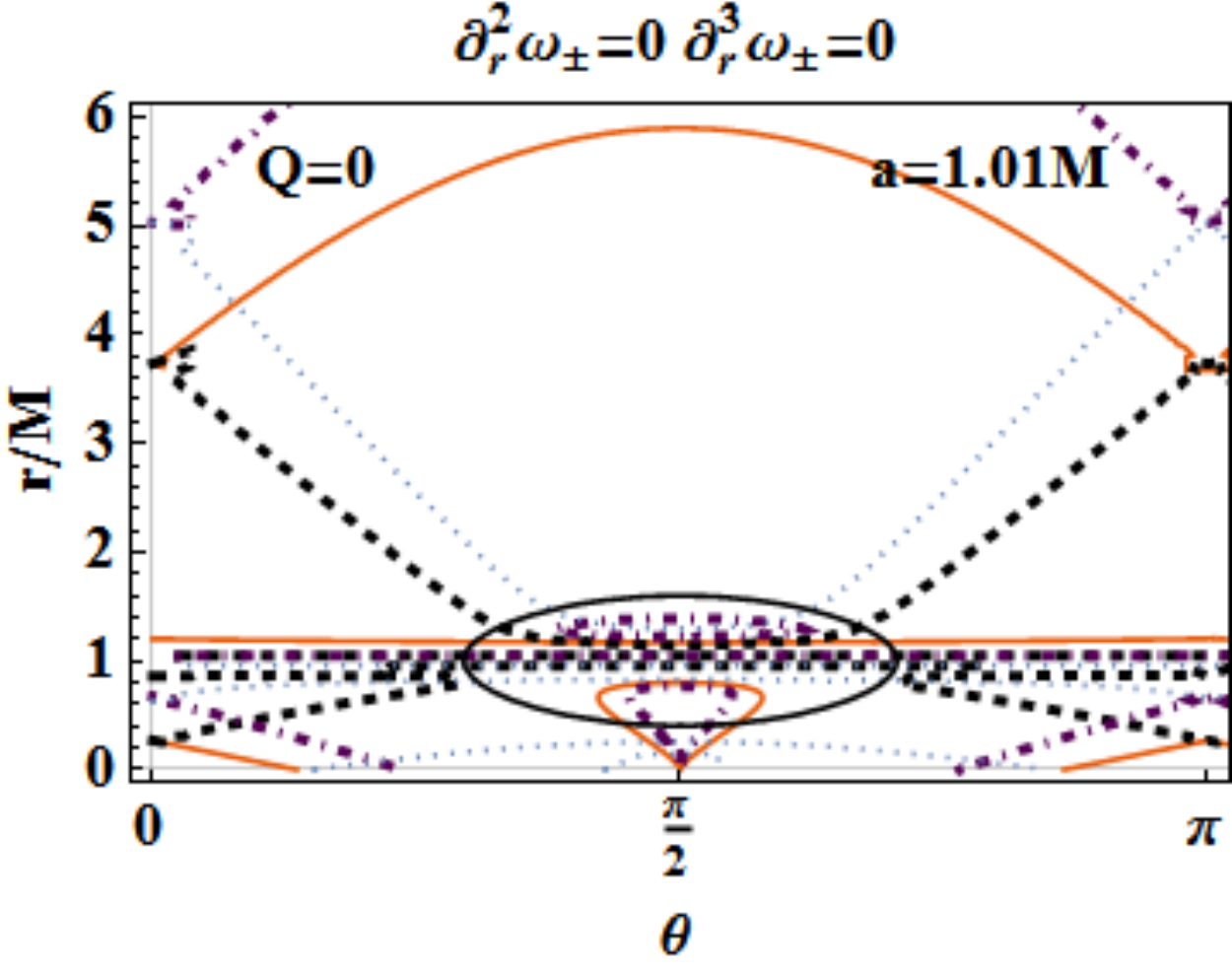}
\includegraphics[scale=.35]{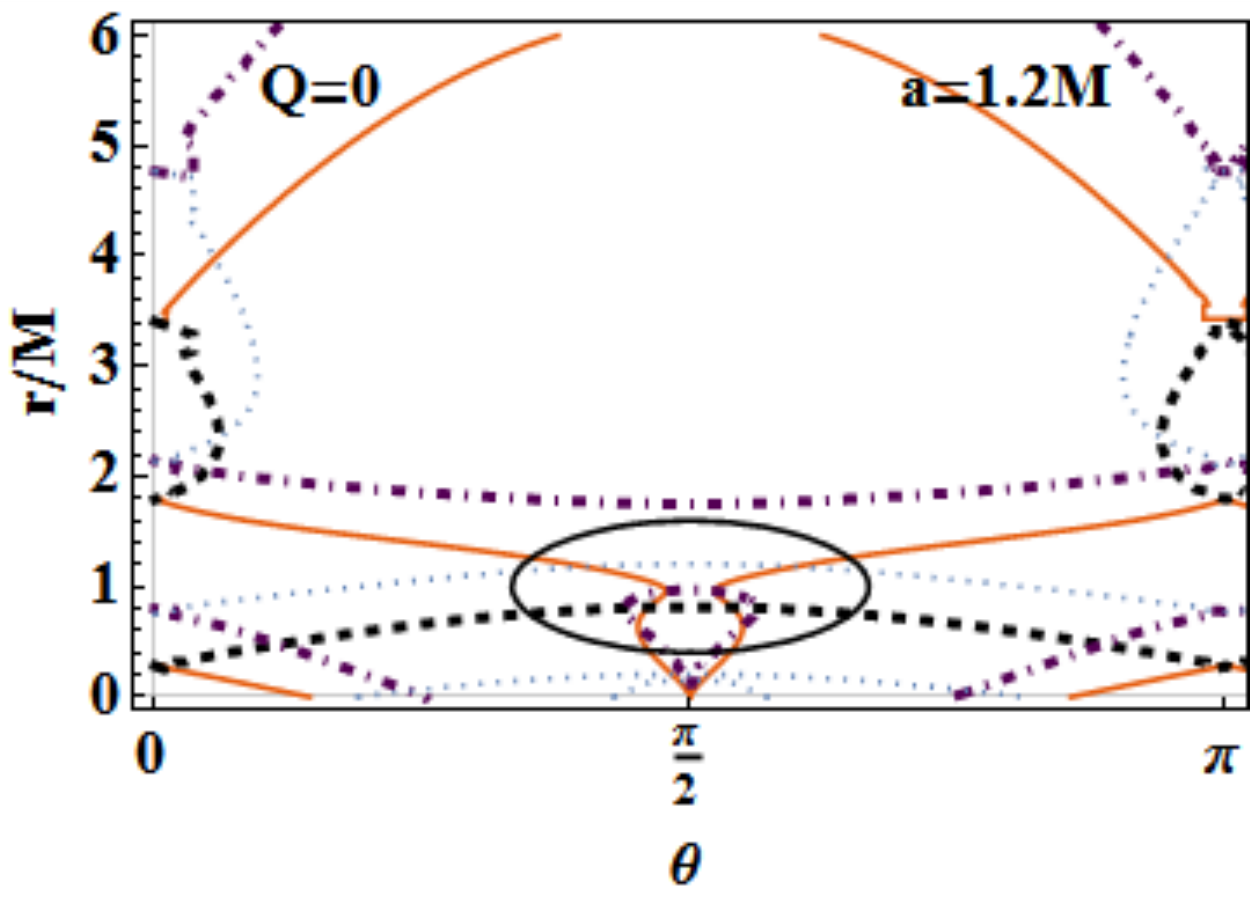}
\includegraphics[scale=.35]{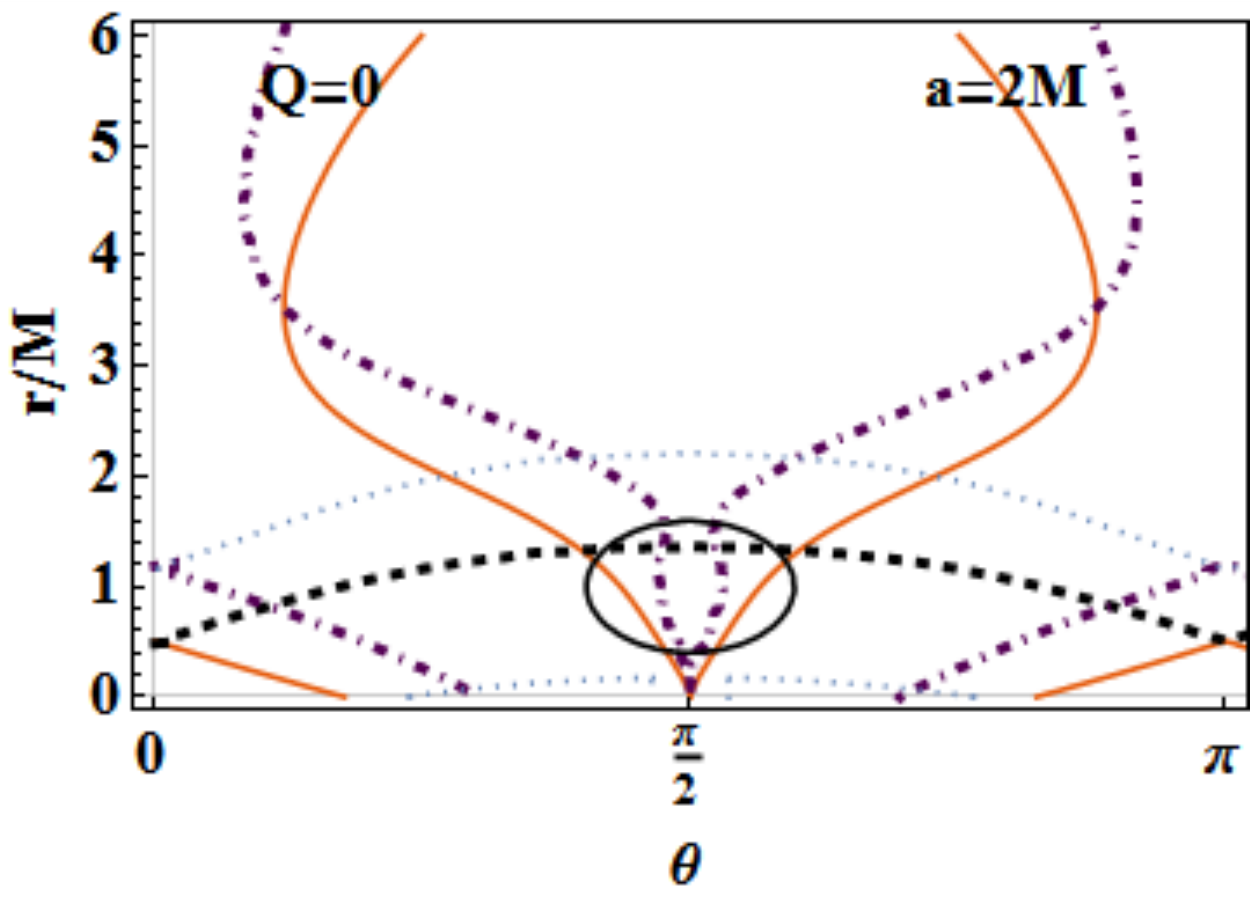}\\
\includegraphics[scale=.35]{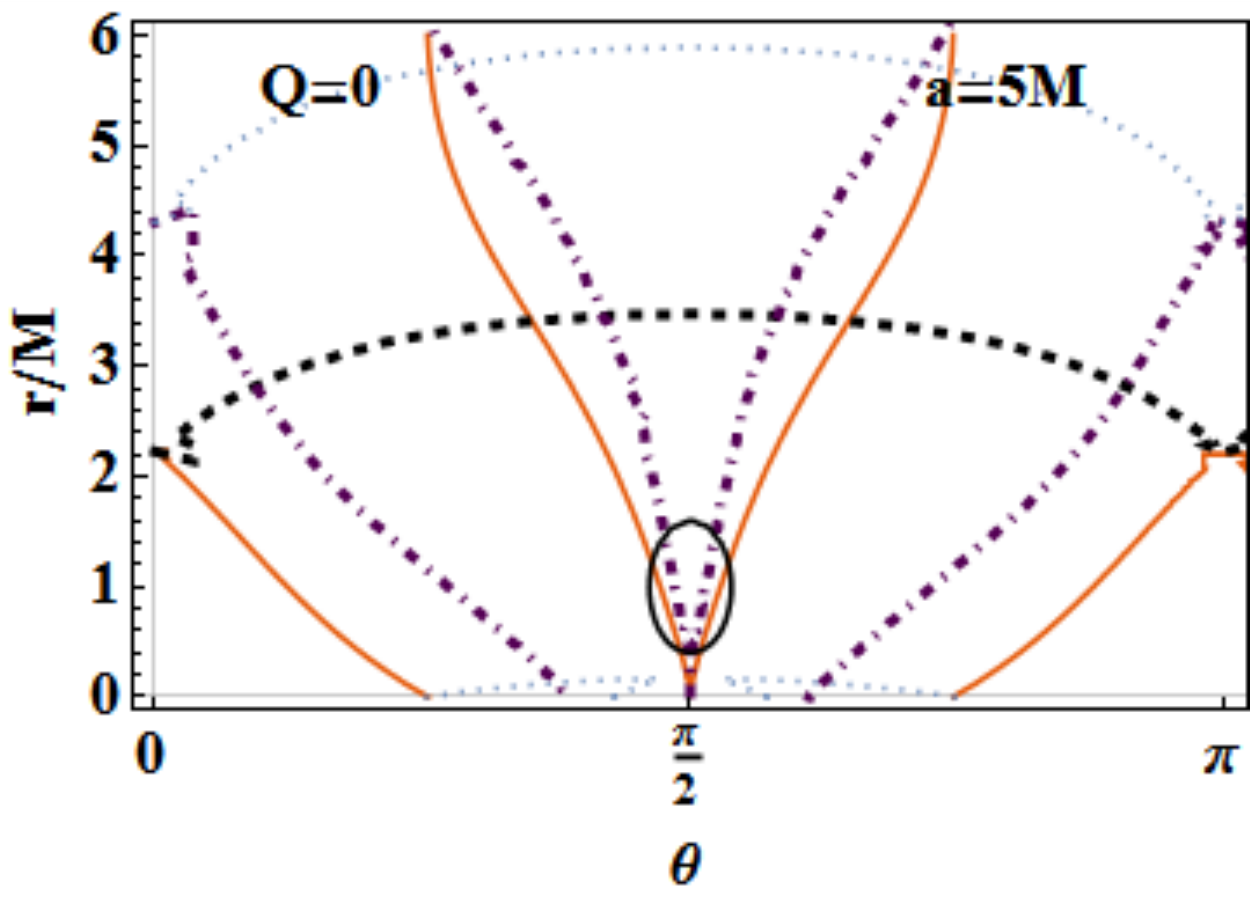}
\includegraphics[scale=.35]{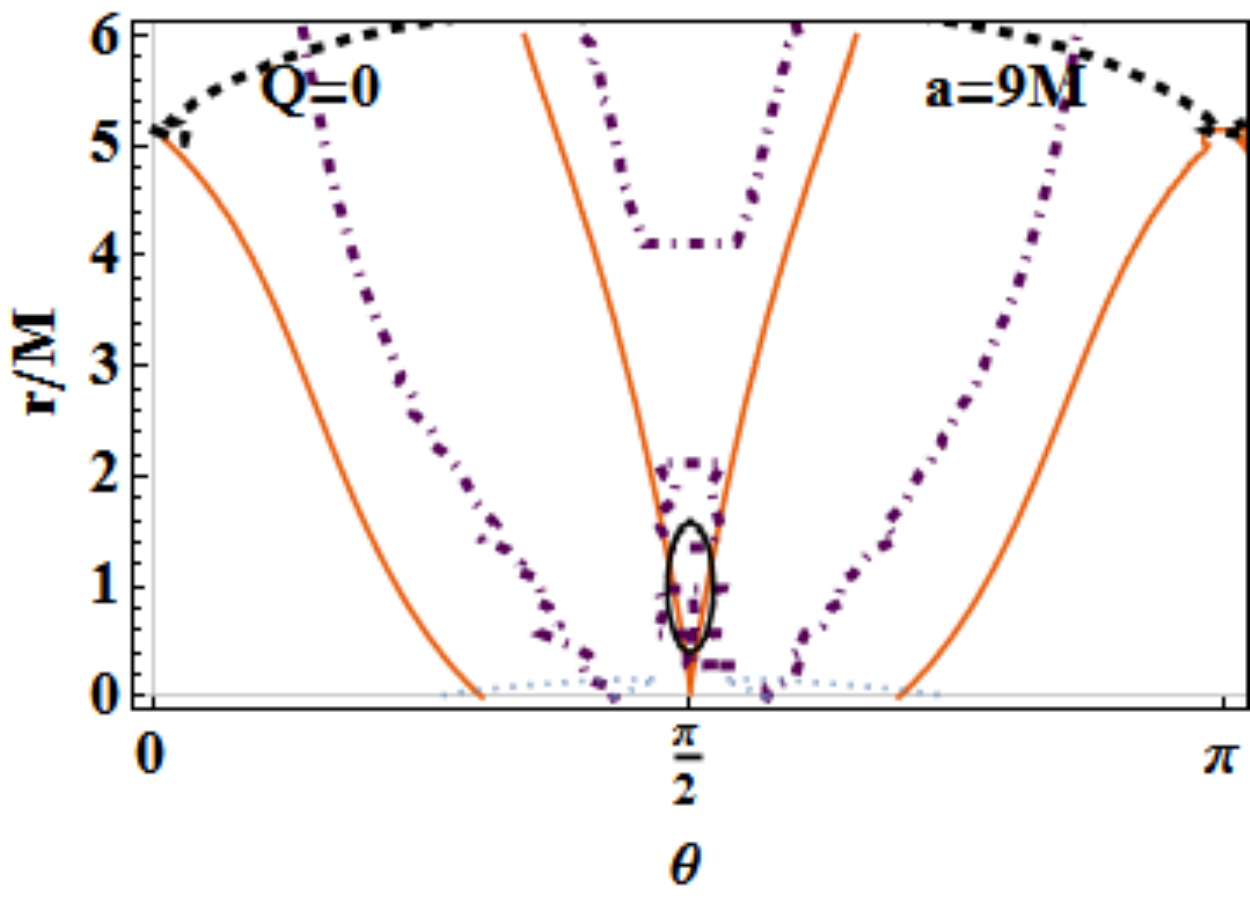}
\end{tabular}
\caption[font={footnotesize,it}]{Kerr geometries: Off-equatorial case. Solutions  of $\partial^2_r\omega_{\pm}=0$ (orange and dashed curves)
 and  $\partial^3_r\omega_{\pm}=0$ (dotted and dotted-dashed curves), defining the Killing bottlenecks of  naked singularities,
$g_{tt}=0$ (black), defining the static limits, and $g_{rr}^{-1}=0$ (red), defining the horizons.
 See also Fig.\il(\ref{Fig:AGpossov15}) for the case  $Q=0$ (Kerr spacetimes). }
\label{Fig:spossov10}
\end{figure}
\section{Areas of the horizon  and of the metric bundles regions in $\pi_a^+$}\label{Sec:appendix-area}
In this section, we analyze  the area  of the regions of $\pi_a^+$ in Fig.\il\ref{Fig:SoOv}
bounded by the curves $a_{\omega}^{\pm}$  for $\omega=$constant, between $\mathbf{\mathcal{{P}}_\mathcal{{S}}}$
and $\mathcal{P}_\mathcal{L}$,  including  the  entire collection of spacetimes $\textbf{g}\in]g_{\omega}^-,g_{\omega}^+[$
bounded   by $(g_{\omega}^-,g_{\omega}^+)$. We compare this region with the (inaccessible) section in
$\pi_a^+$  bounded by the horizons $r_{\pm}$.

First, note that each region bounded by $g_{\omega}^{\pm}$ can be decomposed into other  non-disjoint metric bundles.
In fact, as can be seen in Fig.\il\ref{Fig:SoOv}, the metric bundles $g_{\omega}^{\pm}$ cross each other  in $\pi_a^+$. This
corresponds to the fact that for a fixed point $\mathrm{p}=(a/M, r/M)\in \pi_a^+$, different frequencies are possible, i.e., a
light-like particle  can have different  orbital frequencies corresponding to the  two solutions $\omega_{\pm}$.
To explore this aspect and also the \textbf{BH}-\textbf{NS}  connection, we  introduce the radii
\bea\label{Eq:rom-grad-mira}
r_{\beta}/M\equiv \frac{2}{4 \omega ^2+1},\quad r_{\nu}/M\equiv \frac{1-\omega }{\omega },\quad r_{\pi}/M\equiv \frac{\sqrt{\frac{1}{\omega ^2}+\frac{6}{\omega }-7} \omega +\omega -1}{2 \omega } \ ,
\eea
which are plotted  in Fig.\il\ref{Fig:CORESUNO} as functions of the frequency $\omega$.
It is clear that the
functions
 $\{r_{\beta},r_{\nu}, r_{\pi}\}$ are limiting  radii.
A generalization of this study is  also discussed in  Sec.\il\ref{Sec:RN-KN-cases},
where we consider the Reissner-Nordstr\"om and Kerr-Newman geometries.

\begin{figure}
  \includegraphics[width=5.7cm]{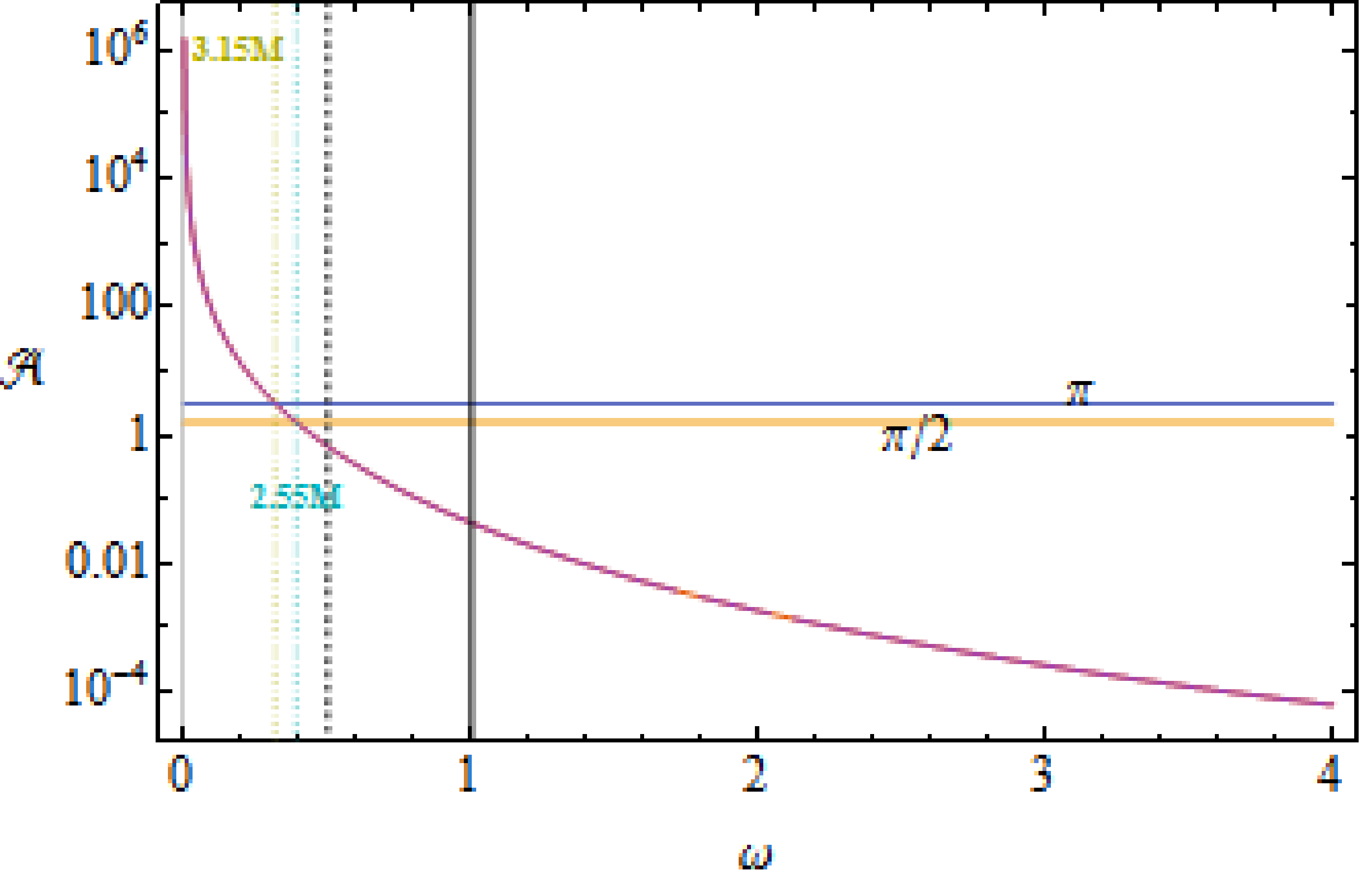}
    \includegraphics[width=5.7cm]{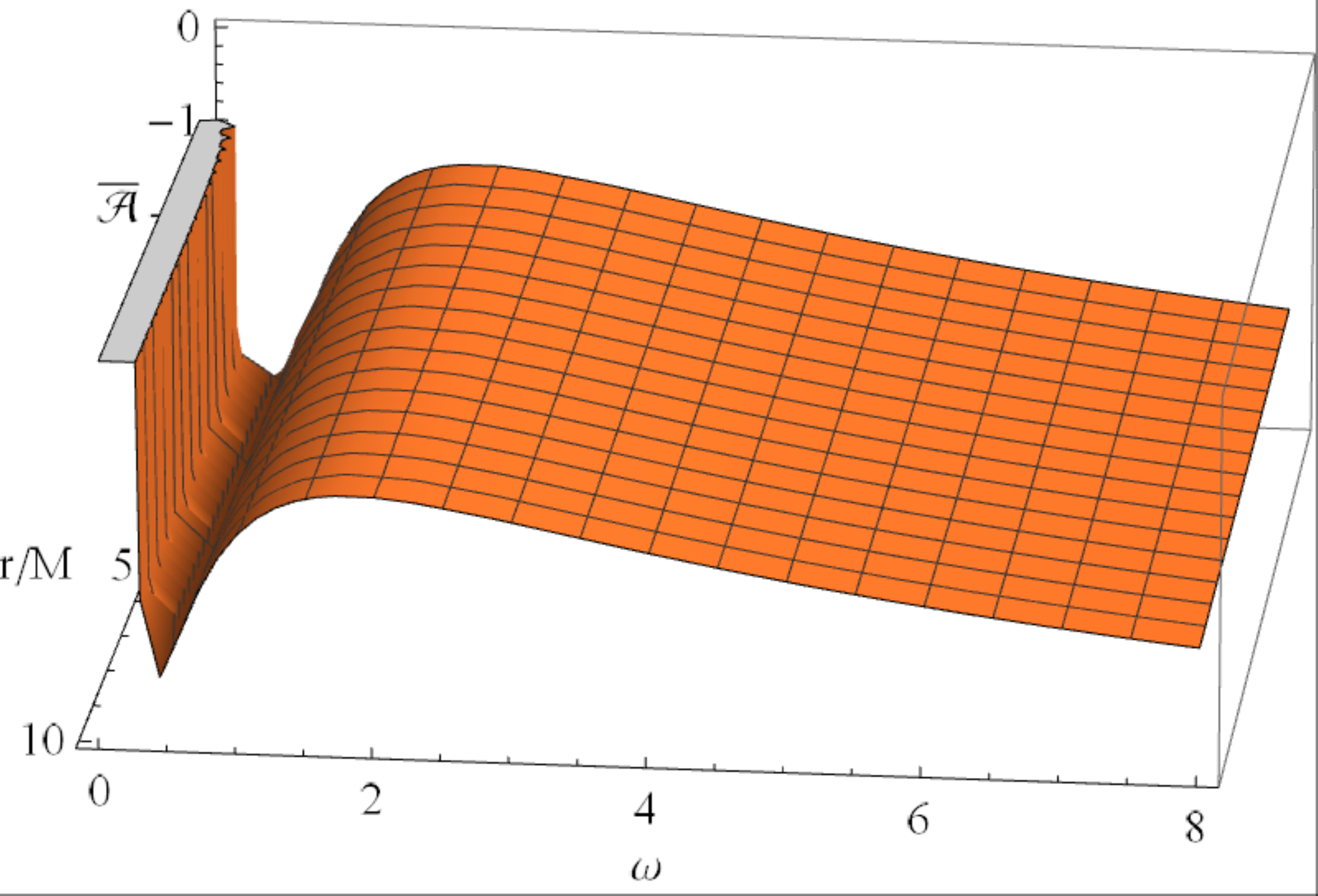}
      \includegraphics[width=5.7cm]{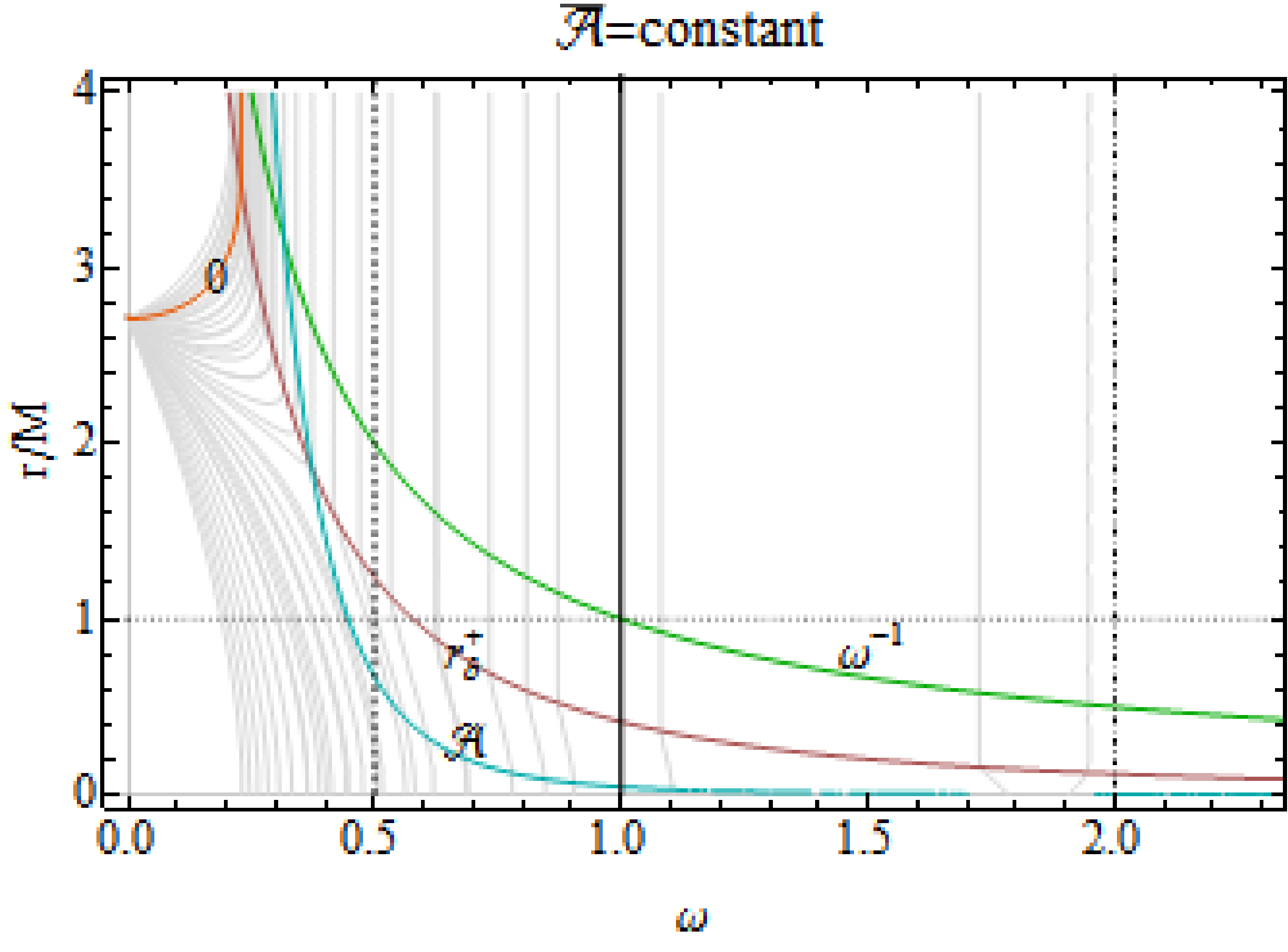}
  \caption{The areas $\mathcal{A}$ of the regions of the extended plane $\pi_a$ bounded by $a_{\omega}^{\pm}$ of Fig.\il\ref{Fig:SoOv}.
	The areas $\mathcal{A}$, $\bar{\mathcal{A}}$ and $\mathcal{A}_{r_{\pm}}=\pi/2$ of the region bounded by the horizons $r_{\pm}$ in
	$\pi_a^+$-region $a\geq>0$-$\mathcal{A}_{r_{\pm}}^{+}$ are defined in Eq.\il(\ref{Eq:ren4Marx-Bon}).
	Left panel: Area $\mathcal{A}$ as a function of the frequency $\omega$. The areas  $\mathcal{A}_{r_{\pm}}$  and
   $2\mathcal{A}_{r_{\pm}}$  are also shown. Frequencies corresponding to  solutions
   $\mathcal{A}_{r_{\pm}}=\mathcal{A}$  and $2\mathcal{A}_{r_{\pm}}=\mathcal{A}$ are pointed with vertical lines.
The	corresponding  geometries $(a=M/\omega)$ are shown. Center panel: Area $\mathcal{A}$ as a function of the frequency $\omega$ and radius $r/M$. The role of the frequencies $\omega=2$ and $\omega=0.5$ is indicated. Right panel: Curves $\mathcal{A}=$constant as functions of the frequency $\omega$ and radius $r/M$. The curves $\mathcal{A}=0$  are red colored. The
area $\mathcal{A}$, radius $r_{\partial}^+$ and $\omega^{-1}$
   are also plotted. Negative areas correspond to solutions $a<0$. See also Fig.\il\ref{FIG:SPlotooc} for a $3D$ representation of these regions.}\label{Fig:BA3D}
\end{figure}

The areas $\mathcal{A}$ correspond to  the  regions of the extended plane $\pi_a^+$ bounded by $a_{\omega}^{\pm}$ and
are confronted here with the areas of the region bounded by the horizons $\mathcal{A}_{r_{\pm}}^{+}=\pi/2$.
An analysis of these areas is shown in   Figs.\il\ref{Fig:BA3D}. We can write the areas $\mathcal{A}$  as functions of
the frequency $\omega$ of the metric bundles $g_{\omega}^{\pm}$ or, equivalently, the spin origins $a_0=M/\omega$,
as follows
 \bea\label{Eq:ren4Marx-Bon}
 &&
  \mathcal{A}(\omega)=|\left.f_{\mathcal{A}}(r,\omega)\right|_{r_{\partial^+}}-
 \left.f_{\mathcal{A}}(r,\omega)\right|_{0}| \ ,
\\&&\nonumber
 f_{\mathcal{A}}(r)\equiv\frac{\omega  \left[c_1 \omega +4 \log \left[(r+2) \omega^2+\Omega +1\right]+(r-3) \Omega \right]+i \left(1-3 \omega^2\right) \log (2 [\Omega -i (r+1) \omega ])-4 \omega  \log (r+2)}{\omega^2},\\&&
 \mathcal{A}=\frac{1}{\omega^2}\left[3 \omega -4 \omega  \log \left(r_{\partial}^++2\right)+i \left(3 \omega ^2-1\right) \log (2-2 i \omega )+\right.
 \\\nonumber
 &&\left.i \log \left(-2 i r_{\partial}^+ \omega \right)+\omega  \left\{-3 i \omega  \log \left(-2 i r_{\partial}^+ \omega \right)-4 \log \left(\omega ^2+1\right)+4 \log \left[\left(r_{\partial}^++2\right) \omega ^2+1\right]\right\}\right] \ ,
 \eea
where dimensionless quantities $r\rightarrow r/M$ have been used.
We also define the quantities
$\bar{\mathcal{A}}(\omega)\equiv\left.f_{\mathcal{A}}(r)\right|_{c_1=0}$ and $\Omega\equiv\sqrt{1-r (r+2) \omega ^2}$, where
 $f_{\mathcal{A}}(r,\omega)$ is a function of the frequency $\omega=M/a_0$ of the metric bundle and of the radius $r$. Moreover,  $c_1$ is an integration constant. 
   From Fig.\il\ref{Fig:BA3D}, it is clear that the area $\mathcal{A}$ is a decreasing function of the frequency $\omega$, which is in agreement with the results of Fig.\il\ref{Fig:SoOv}. Indeed, the metric bundles shrink at the  origins $a_0<M$,
	that is, for frequencies $\omega_0=M/a>M$, where $g_{\omega}^{\pm}$ are all contained in $r\in[0,r_-]$. Viceversa, the area  grows as the spin-mass ratio of the \textbf{NS} increases.
The right panel of Fig.\il\ref{Fig:BA3D}  shows an area $\mathcal{A}=$constant with respect to the frequency $\omega$ and the radius $r/M$.
 Special cases correspond to the  limiting geometries $\mathcal{P}_{\odot}$  and
 $\mathcal{P}_{\otimes}$.
Finally, the evaluation of the areas  in  $\pi_a^+$ takes into account the curvature of the curves in the plane. Therefore, it is necessary to consider some relative quantities  reported below and represented  in  Fig.\il(\ref{Fig:BA3D}).
 Considering the quantities
\bea&&
\Delta^\pm a_{\omega}\equiv a_{\omega}^+\pm a_{\omega}^-;
\quad\mbox{where}\quad
\Delta^+ a_{\omega}=\frac{4M}{(r+2M) \omega },\quad \Delta^- a_{\omega}=\frac{2 \sqrt{r^2  \omega ^2 \left[M^2-r (r+2M)  \omega ^2\right]}}{(r+2M)  \omega ^2M},
\eea
and using $r_{\partial}^+$ in Eq.\il(\ref{Eq:rela-a-partialv}),
we can obtain the area $\mathcal{A}$  of the regions bounded by the curves  $a_{\omega}^{\pm}$,
between the points  $\mathcal{P_S}$ and $\mathcal{P_L}$.
The curves bending  the area $\mathcal{A}$ are related  to solutions of    the equation $\partial_{\omega}^2\Delta^- a_{\omega}=0$,
which is solved   for $r/M={\sqrt{{(3 \omega^2+2)}/{\omega^2}}}/({\sqrt{3}}-1)$, while the only  solutions $\partial_{r/M}^2\Delta^- a_{\omega}=0$  are for the frequencies  $\omega ={\sqrt{2}M^2}/({\sqrt{r (r+2M) (r+3M)M}})$ in the frequency range $\omega\in]0, 1.78015[$, where the  distance between the two curves  is extreme.
 These quantities, considering the variation of $\Delta^- a_{\omega}$ with respect to the frequencies and the radius $r$ respectively, are related to the curvature of the   $\Delta^- a_{\omega}$, where the extreme radius as function of the frequency $\omega$ is
\bea\label{Eq:test-cris}
&&
\frac{r_{\Delta^+_\omega}^{1}}{M}=\frac{1}{3} \left(\sqrt[3]{\frac{27}{ \omega ^2}+3 \sqrt{\frac{81}{ \omega ^4}+\frac{60}{ \omega ^2}-27}+10}+\sqrt[3]{\frac{27}{ \omega ^2}-3 \sqrt{\frac{81}{ \omega ^4}+\frac{60}{ \omega ^2}-27}+10}-5\right)\quad \omega\in[0,1.178]
 \\&&
\frac{r_{\Delta^+_\omega}^{2}}{M}=\frac{1}{3} \left(2 \sqrt{7} \cos \left[\frac{1}{3} \cos ^{-1}\left(\frac{10  \omega ^2+27}{7 \sqrt{7}  \omega ^2}\right)\right]-5\right),\quad \omega>1.178.
\eea
%



\end{document}